\documentstyle[12pt,epsfig]{report}
\def\TL{\hfil$\displaystyle{##}$}
\def\TR{$\displaystyle{{}##}$\hfil}
\def\TC{\hfil$\displaystyle{##}$\hfil}
\def\TT{\hbox{##}}
\def\seqalign#1#2{\vcenter{\openup1\jot
  \halign{\strut #1\cr #2 \cr}}}
\def\lbldef#1#2{\expandafter\gdef\csname #1\endcsname {#2}}
\def\eqn#1#2{\lbldef{#1}{(\ref{#1})}%
\begin{equation} #2 \label{#1} \end{equation}}
\def\eqalign#1{\vcenter{\openup1\jot
    \halign{\strut\span\TL & \span\TR\cr #1 \cr
   }}}
\def\eno#1{(\ref{#1})}
\def\href#1#2{#2}  
\def\ads{{\it AdS}}
\def\adsp{{\it AdS}$_{p+2}$}

\newcommand{\beq}{\begin{equation}}
\newcommand{\eeq}{\end{equation}}
\newcommand{\ber}{\begin{eqnarray}}
\newcommand{\eer}{\end{eqnarray}}

\def\del{{\partial}}
\def\vev#1{\left\langle #1 \right\rangle}
\def\cn{{\cal N}}
\def\co{{\cal O}}
\newfont{\Bbb}{msbm10 scaled 1200}     
\newcommand{\mathbb}[1]{\mbox{\Bbb #1}}
\def\IC{{\mathbb C}}
\def\IR{{\mathbb R}}
\def\IZ{{\mathbb Z}}
\def\RP{{\bf RP}}
\def\CP{{\bf CP}}
\def\Poincare{{Poincar\'e }}
\def\tr{{\rm tr}}
\def\tp{{\tilde \Phi}}

\def\comment#1{}
\def\fixit#1{}
\def\tf#1#2{{\textstyle{#1 \over #2}}}
\def\mop#1{\mathop{\rm #1}\nolimits}
\def\tr{\mop{tr}}
\def\Vol{\mop{Vol}}

\def\sqr#1#2{{\vcenter{\vbox{\hrule height.#2pt
         \hbox{\vrule width.#2pt height#1pt \kern#1pt
            \vrule width.#2pt}
         \hrule height.#2pt}}}}
\def\square{\mathop{\mathchoice\sqr56\sqr56\sqr{3.75}4\sqr34\,}\nolimits}
\def\idget{$\sqr55$\hskip-0.5pt}
\def\endrow{\hskip0.5pt\cr\noalign{\vskip-1.5pt}}
\def\endyoung{\hskip0.5pt\cr}

\def \ov {\over}
\newcommand{\beqar}{\begin{eqnarray}}
\newcommand{\cN}{{\cal N}}
\newcommand{\cO}{{\cal O}}
\newcommand{\cA}{{\cal A}}
\newcommand{\cT}{{\cal T}}
\newcommand{\cF}{{\cal F}}

\newcommand{\cR}{{\cal R}}
\newcommand{\cW}{{\cal W}}
\newcommand{\eeqar}{\end{eqnarray}}

\def\tr{{\rm Tr}}

\begin{document}
\baselineskip=15.5pt
\pagestyle{plain}
\setcounter{page}{1}
\begin{titlepage}

\begin{center}
\vspace*{-3cm} \today \hfill CERN-TH/99-122\\
hep-th/9905111 \hfill HUTP-99/A027\\ 
{}~{} \hfill  LBNL-43113\\
{}~{} \hfill RU-99-18\\
{}~{} \hfill UCB-PTH-99/16\\

\vskip 1cm

{\LARGE {\bf Large $N$ Field Theories, \\
 ~\\ 
String Theory and Gravity}}

\vskip 1 cm

{\large Ofer Aharony,$^1$ Steven S. Gubser,$^2$
Juan Maldacena,$^{2,3}$ \\
 ~\\
Hirosi Ooguri,$^{4,5}$ and Yaron Oz$^6$}

\vskip .8cm
   ${}^1$  Department of Physics and Astronomy,
    Rutgers University, \\
Piscataway, NJ 08855-0849, USA

\medskip

  ${}^2$ Lyman Laboratory of Physics,
    Harvard University, Cambridge, MA  02138, USA

\medskip

  ${}^3$ School of Natural Sciences, Institute for Advanced Study, 
Princeton, NJ 08540

\medskip

  ${}^4$ Department of Physics,
     University of California, Berkeley, CA 94720-7300, USA

\medskip

${}^5$ 
      Lawrence Berkeley National Laboratory, 
MS 50A-5101, 
 Berkeley, CA 94720, USA

\medskip

${}^6$ Theory Division, CERN,
CH-1211, Geneva 23, Switzerland 

\vskip 0.5cm

{\tt oferah@physics.rutgers.edu, ssgubser@bohr.harvard.edu, 
 malda@pauli.harvard.edu, hooguri@lbl.gov, yaron.oz@cern.ch}

\vspace{5mm}

{\bf Abstract}
\end{center}

\noindent
We review the holographic correspondence between field theories
and string/M theory, focusing on the relation between
compactifications of string/M theory on Anti-de Sitter spaces and
conformal field theories.  We review the background for this 
correspondence
and discuss its motivations and the evidence for its correctness. We
describe the main results that have been derived from the
correspondence in the regime that the field theory is approximated 
by classical or semiclassical gravity.
 We focus on the case of the $\cn=4$ supersymmetric
gauge theory in four
dimensions, but we discuss also field theories in other dimensions,
conformal and non-conformal, with or without supersymmetry, and in
particular the relation to QCD. We also discuss some implications 
for black hole physics.

\vskip .3cm

\centerline{({\sl To be published in Physics Reports})}

\end{titlepage}

\newpage
\tableofcontents
\newpage


\chapter{Introduction}
\label{ChapIntro}

\section{General Introduction and Overview}
\label{introduction}

The microscopic description of nature as presently
understood and verified by experiment involves 
quantum field theories. All particles are excitations of 
some field. These particles are pointlike and they 
interact locally with other particles. 
Even though quantum field theories 
 describe nature at the distance scales we observe,
there are strong indications that new
elements will be involved 
 at very short distances (or very high energies), distances
of the order of the
Planck scale. 
The reason is that at those distances (or energies) quantum gravity 
effects become important. It has not been possible to quantize
gravity following the usual perturbative methods. 
Nevertheless, one
 can incorporate quantum gravity in a consistent quantum 
theory by giving up 
the notion that particles are pointlike and assuming that 
the fundamental objects in the theory are strings, namely one-dimensional
extended objects \cite{Green:1987sp,joebook}.
 These strings can oscillate, and there is 
a spectrum of energies, or masses, for these oscillating strings.
The oscillating strings  look like localized, particle-like
excitations to a low energy observer.  
So, a single oscillating string can effectively give rise to many
types of
particles, depending on its state of oscillation. All string theories
include a particle with zero mass and spin two. 
Strings can interact by splitting and joining interactions.
The only consistent interaction for massless spin two 
particles is that  of gravity. Therefore, 
any string theory will contain gravity. 
The structure of string theory is highly constrained. String theories
do not make sense in an arbitrary number of dimensions or on any 
arbitrary geometry. Flat space string theory exists (at least in
perturbation theory) only in 
ten dimensions. Actually, 10-dimensional string theory is described
by a string which also has fermionic excitations and gives rise
to a supersymmetric theory.\footnote{One could consider a  string with no
fermionic excitations, the so called ``bosonic'' string. It  lives in
26 dimensions and contains tachyons, signaling an instability of the
theory.}
String theory is then a candidate for a quantum theory of gravity.
One can get down to four dimensions by considering string theory on 
$\IR^4\times M_6$ where $M_6$ is some six dimensional compact manifold. 
Then, low energy interactions are determined by the geometry
of $M_6$. 

Even though this is the motivation usually given for string theory
nowadays, it is not how string theory was originally discovered. 
String theory was discovered in an attempt to describe 
the large number of mesons and hadrons that were experimentally 
discovered in the 1960's. The idea was to view all these particles 
as different oscillation modes of a string. 
The string idea described well some features of the 
hadron spectrum. For example,
 the mass of the lightest hadron with a given spin 
obeys a  relation like $m^2 \sim T J^2 + const $. This is explained 
simply by assuming that the mass and angular momentum come
from a rotating, relativistic string of tension $T$. 
It was later discovered that hadrons and mesons  are actually
made of quarks
and that they are described by QCD. 

QCD is a gauge theory based
on the group $SU(3)$. This is sometimes stated by saying that quarks
have three  colors. QCD is asymptotically free, meaning that the 
effective
coupling constant decreases as the energy increases. At low energies
QCD becomes strongly coupled and it is not easy to perform calculations. 
One possible approach is to use numerical simulations on the lattice.
This is at present the best available tool to do calculations in 
QCD at low energies. It was suggested by 't Hooft that 
the theory might simplify when the number of colors $N$ is large
 \cite{'tHooft:1974jz}.
The hope was that one could solve exactly the theory with 
$N = \infty$, and then one could do an expansion in $1/N = 1/3$. 
Furthermore, as explained in the next section, the diagrammatic expansion
of the field theory suggests that the large $N$  theory 
is a free string theory and that the string coupling constant is $1/N$.  
If the case with $N=3$ is similar to the case with $N=\infty$
then this explains why the string model gave the correct relation 
between the mass and the angular momentum. In this way 
the large $N$ limit connects gauge  theories with string theories. 
The 't Hooft argument, reviewed below, is very general, so it suggests
that different kinds of gauge theories will correspond to different 
string theories. In this review we will study this correspondence between
string theories and the large $N$ limit of field theories. We 
will see that the strings arising in the large $N$ limit of field 
theories are the same as the strings describing quantum gravity.
Namely, string theory in some backgrounds, including quantum gravity,
is equivalent (dual) to a field theory.

We said above that  strings are not consistent in four flat
dimensions. Indeed, if one wants to quantize a four dimensional
string theory an anomaly appears that forces the introduction 
of an extra field, sometimes called the 
``Liouville'' field \cite{Polyakov:1981rd}. 
This field on the string worldsheet may be interpreted as an extra 
dimension, so that the
strings effectively move in five dimensions. One might
qualitatively think of this new field as the ``thickness'' of the string.
If this is the case, why do we say that the string  moves in 
five dimensions? The reason is that, like any string theory, this 
theory will
contain gravity, and the gravitational theory will live in as
many dimensions as the number of fields we have on the string. 
It is crucial then that the five dimensional geometry is curved, so that
it can correspond to a four dimensional field theory, as described in
detail below. 

The argument that gauge theories are related to string theories in the 
large $N$ limit is very general and is
valid for basically any gauge theory.
In particular we could consider a gauge theory where the coupling
does not run (as a function of the energy scale). 
Then, the theory is  conformally invariant.
It is quite hard to find quantum field theories that are conformally
invariant. In supersymmetric theories it is sometimes possible to 
prove exact conformal invariance. A simple example, which will 
be the main example
 in this review, is the supersymmetric $SU(N)$ (or $U(N)$)
gauge theory in four
dimensions
with four spinor supercharges ($\cn=4$).  
Four is the maximal possible number of supercharges for a field theory
in four dimensions. Besides the gauge fields (gluons) 
this theory contains also four
fermions and six scalar fields in the adjoint representation of the
gauge group. 
The Lagrangian of such theories is completely 
determined by supersymmetry. There is a global $SU(4)$ $R$-symmetry that
rotates the six scalar fields and the four fermions.   
 The conformal group in four dimensions is $SO(4,2)$, including 
the usual \Poincare transformations as well as scale
transformations and special conformal transformations (which include
the inversion symmetry
$x^\mu \to x^\mu/x^2$). 
These symmetries of the field theory should be reflected in the dual
string theory. The simplest way for this to happen is if 
the five dimensional geometry has
these symmetries. Locally there is only one space 
with $SO(4,2)$ isometries: five dimensional
Anti-de-Sitter space, or $AdS_5$. Anti-de Sitter space is the maximally 
symmetric solution of Einstein's equations with a negative
cosmological constant.  
In this supersymmetric case we expect the strings to also be
 supersymmetric. We said that 
 superstrings move in ten dimensions. Now that
we have added one more dimension it is not surprising any more
to add five more to get to a ten dimensional space. 
Since the gauge theory has an $SU(4)\simeq SO(6)$ global symmetry 
 it is rather natural that the extra five dimensional space
should be a five sphere, $S^5$. So, we conclude that 
${\cal N}=4$ $U(N)$ Yang-Mills theory
could be the same as ten dimensional superstring
theory on $AdS_5 \times S^5 $ \cite{Maldacena:1997re}.
Here we have presented a very 
heuristic argument for this equivalence; later we will be more precise
and give more evidence for this correspondence. 

The relationship we described between gauge theories and string theory 
on Anti-de-Sitter
spaces was motivated by studies of D-branes and black holes in
strings theory. D-branes are solitons in string theory
 \cite{Polchinski:1995mt}. 
They come in various dimensionalities. If they have zero spatial
dimensions they are like ordinary localized, particle-type soliton
solutions, analogous to the 't Hooft-Polyakov 
\cite{'tHooft:1974qc,Polyakov:1974ek} 
monopole in gauge theories. These are called D-zero-branes. 
If they have one extended dimension they are called D-one-branes
or D-strings. They are much heavier than ordinary fundamental strings
when the string coupling is small. In fact, the tension of all D-branes
is proportional to $1/g_s$, where $g_s$ is the string coupling constant.
D-branes are defined in string perturbation theory in a very simple
way: they are surfaces where open strings can end. These open
strings have some massless modes, which describe the oscillations 
of the branes, a gauge field living on the brane, and their
fermionic partners. If we have $N$ coincident branes the open strings
can start and end on different branes, so they carry two indices
that run from one to $N$. 
This in turn implies that the low energy dynamics is 
described by a $U(N)$ gauge theory. 
 D-$p$-branes are charged under $p+1$-form gauge potentials, in the
same way that a 0-brane (particle) can be charged under 
a one-form gauge potential (as in electromagnetism). 
These $p+1$-form gauge potentials have $p+2$-form field strengths, and
they are part of the massless closed
string modes, which belong to the supergravity (SUGRA) multiplet 
containing the massless fields in flat
space string theory (before
we put in any D-branes). If we now add D-branes they generate a 
flux of the corresponding field strength, and this flux in turn 
contributes to the stress energy tensor so the geometry 
becomes curved. Indeed it is possible to find solutions of
the supergravity equations carrying these fluxes. 
Supergravity is the low-energy limit of string theory, and it is
believed that these solutions may be extended to solutions of the full
string theory. These solutions
are very similar to extremal charged black hole solutions in general
relativity, except that in this case they are black branes
with $p$ extended spatial dimensions. Like black holes they contain
event horizons. 

If we consider a set of $N$ coincident 
D-3-branes the near horizon geometry turns out to be
$AdS_5\times S^5$. On the other hand, the low energy 
dynamics on their worldvolume is governed by a $U(N)$ gauge theory
with ${\cal N} =4$ supersymmetry \cite{Witten:1996im}. 
These two pictures of D-branes are perturbatively valid for
different regimes in the space of possible coupling constants. Perturbative
field theory is valid when $g_s N$ is small, 
while the low-energy gravitational description is perturbatively 
valid when
the radius of curvature is much larger than the string scale,
which turns out to imply that $g_s N$ should be very large. 
As an object is brought closer and closer to the 
 black brane  horizon
its energy measured by an outside observer is redshifted, due to the
large gravitational potential, and the energy seems to be very small. 
On the other hand low energy excitations on the branes are governed
by the Yang-Mills theory. So, it becomes natural to conjecture that 
Yang-Mills theory at strong coupling is describing
the near horizon region of the black brane, whose geometry
is $AdS_5\times S^5$.
The first indications that this is the case came from calculations
of low energy graviton absorption cross sections 
\cite{Klebanov:1997kc,Gubser:1997yh,Gubser:1997se}. It was
noticed there that the calculation done using gravity and the 
calculation done using super Yang-Mills theory agreed. 
These calculations, in turn, were inspired by similar calculations
 for coincident D1-D5 branes. In this case the
near horizon geometry involves $AdS_3\times S^3$ and the low energy
field theory living on the D-branes is a 1+1 dimensional conformal
field theory.
In this D1-D5 case 
there were numerous calculations that agreed between the field theory
and gravity. First black hole entropy for extremal black holes was
calculated in terms of the field theory in \cite{Strominger:1996sh},
and then 
agreement was shown for near extremal black holes 
\cite{Callan:1996dv,Horowitz:1996fn} and
for absorption cross sections 
\cite{Das:1996wn,Dhar:1996vu,Maldacena:1997ix}. 
More generally, we will see that
correlation functions in the gauge theory can be calculated 
using the string theory (or gravity for large $g_s N$)
description, by considering the propagation
of particles between different points in the boundary of $AdS$, 
the points where operators are inserted 
\cite{Gubser:1998bc,Witten:1998qj}. 

Supergravities on $AdS$ spaces  were 
 studied very  extensively,
see 
\cite{Salam:1989fm,Duff:1986hr} for reviews. See also
\cite{Boonstra:1997dy,Sfetsos:1998xs} for earlier hints of the
 correspondence.

One of the main points of this review 
will be that the strings coming from gauge theories
are very much like the ordinary superstrings that have been 
studied during the last 20 years. The only particular feature is that
they are moving on a curved geometry 
(anti-de Sitter space) which has a boundary at spatial infinity. 
The boundary is at an infinite spatial distance, but a light ray 
can go to the boundary and come back in finite time. Massive particles
can never get to the boundary. The radius of curvature of 
Anti-de Sitter space 
 depends on $N$  so  that large $N$ corresponds to a large radius
of curvature. Thus, by taking $N$ to be large 
we can make the curvature as small as we want. 
The theory in $AdS$ includes gravity, since any string theory 
includes gravity. So in the end we claim that there is an equivalence
between a gravitational 
theory and a field theory. However, the mapping between the gravitational
and field theory degrees of freedom is quite non-trivial since
the field theory lives in a lower dimension. In some sense the 
field theory (or at least the set of 
local observables in the field theory)
lives on the boundary of spacetime. 
One could argue that in general any quantum gravity theory in $AdS$
defines a conformal field theory (CFT) ``on the boundary''. 
In some sense the situation is similar to the correspondence
between three dimensional Chern-Simons theory and a WZW model on
the boundary \cite{Witten:1989hf}. This is a topological theory in three 
dimensions that induces a normal (non-topological) 
 field theory on the boundary.
A theory which includes gravity
is in some sense topological since one is integrating
over all metrics and therefore the theory does not depend on the
metric. Similarly, in a quantum gravity theory we do not
have any local observables. 
Notice that when we say that the theory includes
``gravity on $AdS$'' we are considering
any finite energy excitation, even black holes in $AdS$. 
So this is really a sum over all spacetimes that are asymptotic to 
$AdS$ at the boundary. This is analogous to the usual 
flat space discussion of quantum gravity,
where asymptotic flatness is required, but
the spacetime could have any topology as long as it is asymptotically 
flat. The asymptotically $AdS$ case as well as the asymptotically 
flat cases are special in the sense that one can choose a natural
time and an associated Hamiltonian to define the quantum theory. 
Since black holes might be present this time coordinate is not 
necessarily
globally well-defined, but it is certainly well-defined at infinity. 
If we assume that the conjecture we made above is valid, then the $U(N)$ 
Yang-Mills theory 
gives a non-perturbative definition of string theory on $AdS$.
And, by taking the limit $N\to \infty$, we can extract the (ten
dimensional string theory) flat space
physics, a procedure  which is in principle (but not in detail) 
similar to the one used in matrix theory \cite{Banks:1997vh}. 

The fact that the field theory lives in a lower dimensional space 
blends in perfectly with some previous speculations about quantum 
gravity. It was suggested \cite{'tHooft:1993gx,Susskind:1995vu}
 that quantum gravity theories
should be holographic, in the sense that physics in some region 
can be described by a theory at the boundary with no more than 
one degree of freedom per Planck area. This ``holographic'' 
principle comes from thinking about the Bekenstein bound which 
states that the maximum amount of entropy in some region 
is given by the area of the region in Planck units 
\cite{Bekenstein:1994dz}. 
The reason for this bound is that otherwise black hole 
formation could violate the second law of thermodynamics.
We will see that the correspondence between field theories
and string theory on $AdS$ space (including gravity) 
is a concrete realization of this holographic principle. 

The review is organized as follows. 

In the rest of the introductory chapter, we present background material.
In section 1.2, we present the   
't Hooft large $N$ limit and its indication that gauge theories may be 
dual to string theories.
In section 1.3, we review the $p$-brane supergravity solutions.
We discuss D-branes, their worldvolume theory and their relation to the p-branes.
We discuss greybody factors and their calculation for
black holes built out of D-branes.  

In chapter 2, we review conformal field theories and $AdS$ spaces.
In section 2.1, we give a brief description of conformal field theories.
In section 2.2, we summarize
the geometry of $AdS$ spaces and gauged supergravities.

In chapter 3,
we ``derive'' the correspondence between supersymmetric
Yang Mills theory and string theory on 
$AdS_5 \times S^5$ from the physics of D3-branes in
string theory. We define, in section 3.1, the correspondence 
between fields in 
the string theory and operators of the conformal field theory and
the prescription for the computation of correlation functions.
We also point out that the correspondence gives an explicit
holographic description of gravity. 
In section 3.2,
we review the direct tests of the duality, including
matching the spectrum of chiral primary operators and some correlation functions
and anomalies. Computation
of correlation functions is reviewed in section 3.3.
The isomorphism of the Hilbert spaces of string theory on $AdS$ spaces and 
of CFTs is decribed in section 3.4.
We describe
how to introduce Wilson loop operators in section 3.5. 
In section 3.6, we analyze  finite temperature theories 
and the thermal phase transition.

In chapter 4, we review other topics involving $AdS_5$. 
In section 4.1, we consider some
other gauge theories that arise from D-branes at orbifolds, orientifolds,
or conifold points. In section 4.2,
we review  how baryons and instantons arise in the string theory
description. In section 4.3,
we study  some deformations of the CFT and how they arise in the
string theory description. 

In chapter 5, we describe a similar correspondence involving
1+1 dimensional CFTs and $AdS_3$ spaces. We also describe 
the relation of these results to black holes in five dimensions. 

In chapter 6, we consider  other examples of the 
AdS/CFT correspondence as well as non conformal and non supersymmetric
cases. In section 6.1, we analyse the M2 and M5 branes theories, 
and go on to describe situations that are not conformal, 
realized on the worldvolume
of various Dp-branes, and the ``little string theories'' on the
worldvolume of NS 5-branes.
In section 6.2, we describe an approach to studying theories
that are  confining and have a behavior similar
to QCD in three and four dimensions. We discuss confinement, $\theta$-vacua, 
the mass spectrum and other dynamical aspects of these theories. 

Finally, the last chapter is devoted to a summary and discussion.

Other reviews  of this subject are 
\cite{DiVecchia:1999du,Douglas:1999ww,Petersen:1999zh,Klebanov:1999ku}.


\section{Large $N$ Gauge Theories as String Theories}
\label{largen}

The relation between gauge theories and string theories has been an
interesting topic of research for over three decades. String theory
was originally developed as a theory for the strong interactions, due
to various string-like aspects of the strong interactions, such as
confinement and Regge behavior. It was later realized that there is
another description of the strong interactions, in terms of an $SU(3)$
gauge theory (QCD), which is consistent with all experimental data to
date. However, while the gauge theory description is very useful for
studying the high-energy behavior of the strong interactions, it is
very difficult to use it to study low-energy issues such as
confinement and chiral symmetry breaking (the only current method for
addressing these issues in the full non-Abelian gauge theory is by
numerical simulations). In the last few years many examples of the
phenomenon generally known as ``duality'' have been discovered, in
which a single theory has (at least) two different descriptions, such
that when one description is weakly coupled the other is strongly
coupled and vice versa (examples of this phenomenon in two dimensional
field theories have been known for many years). One could hope that a
similar phenomenon would apply in the theory of the strong
interactions, and that a ``dual'' description of QCD exists which
would be more appropriate for studying the low-energy regime where the
gauge theory description is strongly coupled.

There are several indications that this ``dual'' description could be
a string theory. QCD has in it string-like objects which are the flux
tubes or Wilson lines. If we try to separate a quark from an
anti-quark, a flux tube forms between them (if $\psi$ is a quark
field, the operator ${\bar \psi}(0) \psi(x)$ is not gauge-invariant
but the operator ${\bar \psi}(0) P\exp(i\int_0^x A_\mu dx^\mu)
\psi(x)$ is gauge-invariant). In many ways these flux tubes behave
like strings, and there have been many attempts to write down a string
theory describing the strong interactions in which the flux tubes are
the basic objects. It is clear that such a stringy description would
have many desirable phenomenological attributes since, after all, this
is how string theory was originally discovered. The most direct
indication from the gauge theory that it could be described in terms
of a string theory comes from the 't Hooft large $N$ limit
\cite{'tHooft:1974jz}, which we will now describe in detail.

Yang-Mills (YM) theories in four dimensions have no dimensionless
parameters, since the gauge coupling is dimensionally transmuted into
the QCD scale $\Lambda_{QCD}$ (which is the only mass scale in these
theories). Thus, there is no obvious perturbation expansion that can
be performed to learn about the physics near the scale
$\Lambda_{QCD}$. However, an additional parameter of $SU(N)$ gauge
theories is the integer number $N$, and one may hope that the gauge
theories may simplify at large $N$ (despite the larger number of
degrees of freedom), and have a perturbation expansion in terms of the
parameter $1/N$. This turns out to be true, as shown by 't Hooft based
on the following analysis (reviews of large $N$ QCD may be found in 
\cite{Coleman:1980nk,Manohar:1998xv}).

First, we need to understand how to scale the coupling $g_{YM}$ as we
take $N \to \infty$. In an asymptotically free theory, like pure YM
theory, it is natural to scale $g_{YM}$ so that $\Lambda_{QCD}$
remains constant in the large $N$ limit. The beta function equation
for pure $SU(N)$ YM theory is
\eqn{betaym}{\mu {dg_{YM}\over d\mu} = -{11\over 3} N {{g_{YM}^3}
\over {16 \pi^2}} + {\cal O}(g_{YM}^5),}
so the leading terms are of the same order for large $N$ if we take
$N \to \infty$ while keeping $\lambda \equiv g_{YM}^2 N$ fixed (one
can show that the higher order terms are also of the same order in
this limit). This is known as the {\it 't Hooft limit}. 
The same behavior is valid if we include also matter
fields (fermions or scalars) in the adjoint representation, as long as
the theory is still asymptotically free. If the theory is conformal,
such as the $\cn=4$ SYM theory which we will discuss in detail below,
it is not obvious that the limit of constant $\lambda$ is the only one
that makes sense, and indeed we will see that other limits, in which
$\lambda \to \infty$, are also possible. However, the limit of constant
$\lambda$ is still a particularly interesting limit
and we will focus on it in the remainder of this chapter.

Instead of focusing just on the YM theory, let us describe a general
theory which has some fields $\Phi_i^a$, where $a$ is an index in the
adjoint representation of $SU(N)$, and $i$ is some label of the field
(a spin index, a flavor index, etc.). Some of these fields can be
ghost fields (as will be the case in gauge theory). We will assume
that as in the YM theory (and in the $\cn=4$ SYM theory), the 3-point
vertices of all these fields are proportional to $g_{YM}$, and the
4-point functions to $g_{YM}^2$, so the Lagrangian is of the schematic
form 
\eqn{schemlag}{{\cal L} \sim \tr(d\Phi_i d\Phi_i) + g_{YM}
c^{ijk} \tr(\Phi_i \Phi_j \Phi_k) + g_{YM}^2 d^{ijkl} \tr(\Phi_i
\Phi_j \Phi_k \Phi_l),} 
for some constants $c^{ijk}$ and $d^{ijkl}$
(where we have assumed that the interactions are
$SU(N)$-invariant; mass terms can also be added and do not change the 
analysis). Rescaling the fields by $\tp_i \equiv g_{YM} \Phi_i$,
the Lagrangian becomes 
\eqn{newschemlag}{{\cal L} \sim {1\over g_{YM}^2} \left[ \tr(d\tp_i
d\tp_i) + c^{ijk} \tr(\tp_i \tp_j \tp_k) + d^{ijkl} \tr(\tp_i \tp_j
\tp_k \tp_l) \right],}
with a coefficient of $1/ g_{YM}^2 = N/ \lambda$ in front
of the whole Lagrangian.

\begin{figure}[htb]
\begin{center}
\epsfxsize=4in\leavevmode\epsfbox{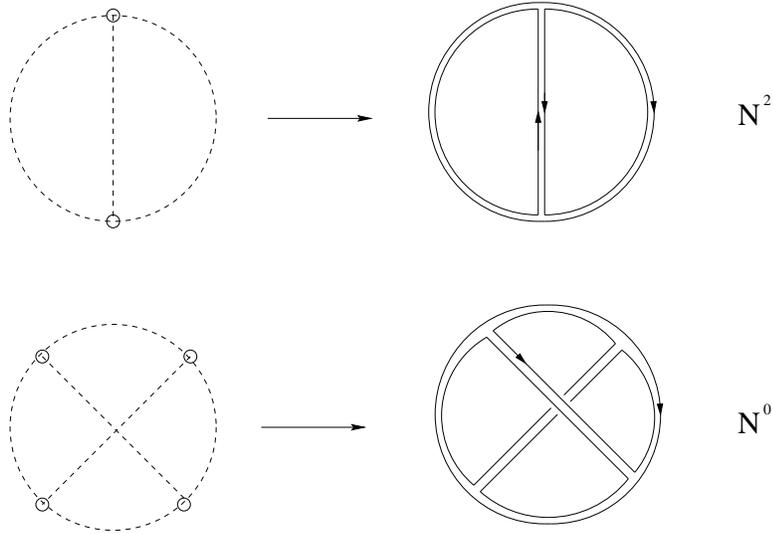}
\end{center}
\caption{Some diagrams in a field theory with adjoint fields in the
standard representation (on the left) and in the double line
representation (on the right). The dashed lines are propagators for
the adjoint fields, the small circles represent interaction vertices,
and  solid lines carry indices in the fundamental
representation.}
\label{thooft}
\end{figure}

Now, we can ask what happens to correlation functions in the limit of
large $N$ with constant $\lambda$. Naively, this is a classical limit
since the coefficient in front of the Lagrangian diverges, but in fact
this is not true since the number of components in the fields also
goes to infinity in this limit. We can write the Feynman diagrams of
the theory \eno{newschemlag} in a double line notation, in which an
adjoint field $\Phi^a$ is represented as a direct product of a fundamental
and an anti-fundamental field, $\Phi^i_j$, as in figure \ref{thooft}.
The interaction vertices we wrote are all consistent with this sort of
notation. The propagators are also consistent with it in a $U(N)$ 
theory; in an $SU(N)$ theory there is a small mixing term
\eqn{propsun}{\vev{\Phi^i_j \Phi^k_l} \propto (\delta^i_l \delta^j_k -
{1\over N} \delta^i_j \delta^k_l),} 
which makes the expansion slightly more complicated, but this involves
only subleading terms in the large $N$ limit so we will neglect this
difference here. Ignoring the second term the propagator for the
adjoint field is (in terms of the index structure) like that of a
fundamental-anti-fundamental pair.  Thus, any Feynman diagram of
adjoint fields may be viewed as a network of double lines. Let us
begin by analyzing vacuum diagrams (the generalization to adding
external fields is simple and will be discussed below). In such a
diagram we can view these double lines as forming the edges in a
simplicial decomposition (for example, it could be a triangulation)
of a surface, if we view each single-line loop
as the perimeter of a face of the simplicial decomposition. 
The resulting
surface will be oriented since the lines have an orientation (in one
direction for a fundamental index and in the opposite direction for an
anti-fundamental index). When we compactify space by adding a point at
infinity, each diagram thus corresponds to a compact, closed, oriented
surface.

What is the power of $N$ and $\lambda$ associated with such a
diagram? From the form of \eno{newschemlag} it is clear that each
vertex carries a coefficient proportional to $N/ \lambda$, while
propagators are proportional to $\lambda / N$. Additional powers of
$N$ come from the sum over the indices in the loops, which gives a
factor of $N$ for each loop in the diagram (since each index has $N$
possible values). Thus, we find that a diagram with $V$ vertices, $E$
propagators (= edges in the simplicial decomposition) 
and $F$ loops (= faces in
the simplicial decomposition) comes with a coefficient proportional to
\eqn{euler}{N^{V-E+F} \lambda^{E-V} = N^{\chi} \lambda^{E-V},}
where $\chi \equiv V-E+F$ is the Euler character of the surface
corresponding to the diagram. For closed oriented surfaces, $\chi =
2-2g$ where $g$ is the genus (the number of handles) of the
surface.\footnote{We are discussing here only connected diagrams, for
disconnected diagrams we have similar contributions from each
connected component.}  Thus, the perturbative expansion of any diagram 
in the field theory may be
written as a double expansion of the form 
\eqn{thooftexp}{\sum_{g=0}^{\infty}
N^{2-2g} \sum_{i=0}^{\infty} c_{g,i} \lambda^i = \sum_{g=0}^{\infty}
N^{2-2g} f_g(\lambda),} 
where $f_g$ is some polynomial in $\lambda$
(in an asymptotically free theory the $\lambda$-dependence will turn into
some $\Lambda_{QCD}$-dependence but the general form is similar; infrared
divergences could also lead to the appearance of terms which are not 
integer powers of
$\lambda$). 
In the large $N$ limit we see that any computation will be
dominated by the surfaces of maximal $\chi$ or minimal genus, which
are surfaces with the topology of a sphere (or equivalently a plane). 
All these {\it planar
diagrams} will give a contribution of order $N^2$, while all other
diagrams will be suppressed by powers of $1/N^2$. For example, the
first diagram in figure \ref{thooft} is planar and proportional to
$N^{2-3+3}=N^2$, while the second one is not and is proportional to 
$N^{4-6+2}=N^0$. We presented our
analysis for a general theory, but in particular it is true for any gauge
theory coupled to adjoint matter fields, like the $\cn=4$ SYM
theory. The rest of our discussion will be limited mostly to gauge
theories, where only gauge-invariant ($SU(N)$-invariant) objects are
usually of interest.

The form of the expansion \eno{thooftexp} is the same as one finds in
a perturbative theory with closed oriented strings, if we identify
$1/N$ as the string coupling constant\footnote{In the conformal case,
where $\lambda$ is a free parameter, there is actually a freedom of
choosing the string coupling constant to be $1/N$ times any function
of $\lambda$ without changing the form of the expansion, and this will
be used below.}. Of course, we do not really see
any strings in the expansion, but just diagrams with holes in them;
however, one can hope that in a full non-perturbative description of
the field theory the holes will ``close'' and the surfaces of the Feynman
diagrams will become actual closed surfaces. The analogy of
\eno{thooftexp} with perturbative string theory is one of
the strongest motivations for believing that field theories and string
theories are related, and it suggests that this relation would be more
visible in the large $N$ limit where the dual string theory may be
weakly coupled. However, since the analysis was based on perturbation
theory which generally does not converge, it is far from a rigorous
derivation of such a relation, but rather an indication that it might
apply, at least for some field theories (there are certainly also
effects like instantons which are non-perturbative in the $1/N$
expansion, and an exact matching with string theory would require a
matching of such effects with non-perturbative effects in string
theory).

The fact that $1/N$ behaves as a coupling constant in the large $N$
limit can also be seen directly in the field theory analysis of the 't
Hooft limit. While we have derived the behavior \eno{thooftexp} only
for vacuum diagrams, it actually holds for any correlation function of
a product of gauge-invariant fields $\vev{\prod_{j=1}^n G_j}$ such
that each $G_j$ cannot be written as a product of two gauge-invariant
fields (for instance, $G_j$ can be of the form ${1\over N}\tr(\prod_i
\Phi_i)$). We can study such a correlation function by adding to the
action $S \to S + N\sum g_j G_j$, and then, if $W$ is the sum of
connected vacuum diagrams we discussed above (but now computed with
the new action), 
\eqn{genvev}{\vev{\prod_{j=1}^n G_j} = (iN)^{-n} \left[ {{\del^n W}\over
{\prod_{j=1}^n \del g_j}} \right]_{g_j=0}.}
Our analysis of the vacuum diagrams above holds also for these
diagrams, since we put in additional vertices with a factor of $N$,
and, in the double line representation, each of the operators we
inserted becomes a vertex of the simplicial decomposition of the
surface (this would not
be true for operators which are themselves products, and which would
correspond to more than one vertex). Thus, the leading contribution to
$\vev{\prod_{j=1}^n G_j}$ will come from planar diagrams with $n$
additional operator insertions, leading to
\eqn{largencorr}{\vev{\prod_{j=1}^n G_j} \propto N^{2-n}}
in the 't Hooft limit. We see that (in terms of powers of $N$) the
2-point functions of the $G_j$'s come out to be canonically
normalized, while 3-point functions are proportional to $1/N$, so
indeed $1/N$ is the coupling constant in this limit (higher genus
diagrams do not affect this conclusion since they just add higher
order terms in $1/N$). In the string theory analogy the operators
$G_j$ would become vertex operators inserted on the string
world-sheet. For asymptotically free confining theories (like QCD) one
can show that in the large $N$ limit they have an infinite spectrum of
stable particles with rising masses (as expected in a free string
theory). Many additional properties of the large $N$ limit are
discussed in
\cite{Migdal:1977nu,Coleman:1980nk} and other references.

The analysis we did of the 't Hooft limit for $SU(N)$ theories with
adjoint fields can easily be generalized to other cases. Matter in the
fundamental representation appears as single-line propagators in the
diagrams, which correspond to boundaries of the corresponding
surfaces. Thus, if we have such matter we need to sum also over
surfaces with boundaries, as in open string theories. For $SO(N)$ or
$USp(N)$ gauge theories we can represent the adjoint representation as
a product of two fundamental representations (instead of a fundamental
and an anti-fundamental representation), and the fundamental
representation is real, so no arrows appear on the propagators in the
diagram, and the resulting surfaces may be non-orientable. Thus, these
theories seem to be related to non-orientable string theories
\cite{Cicuta:1982fu}. We will not discuss these cases in detail here,
some of the relevant aspects will be discussed in section
\ref{orientifolds} below.

Our analysis thus far indicates that gauge theories may be dual to
string theories with a coupling proportional to $1/N$ in the 't Hooft
limit, but it gives no indication as to precisely which string theory
is dual to a particular gauge theory. For two dimensional gauge
theories much progress has been made in formulating the appropriate
string theories
\cite{Gross:1993tu,Gross:1993hu,Minahan:1993sk,Gross:1993yt,Naculich:1993ve,
Ramgoolam:1993hh,Cordes:1994sd,Horava:1999wf}, 
but for four dimensional gauge
theories there was no concrete construction of a corresponding string
theory before the results reported below, since the planar diagram
expansion (which corresponds to the free string theory) is very
complicated. Various direct approaches towards constructing the
relevant string theory were attempted, many of which were based on the
loop equations \cite{Makeenko:1979pb} for the Wilson loop observables
in the field theory, which are directly connected to a string-type
description.

Attempts to directly construct a string theory equivalent to a four
dimensional gauge theory are plagued with the well-known problems of
string theory in four dimensions (or generally below the critical
dimension). In particular, additional fields must be added on the
worldsheet beyond the four embedding coordinates of the string to
ensure consistency of the theory. In the standard quantization of four
dimensional string theory an additional field called the Liouville
field arises \cite{Polyakov:1981rd}, 
which may be interpreted as a fifth space-time
dimension. Polyakov has suggested \cite{Polyakov:1997tj,
Polyakov:1998ju} that such a five dimensional string theory
could be related to four dimensional gauge theories if the couplings
of the Liouville field to the other fields take some specific
forms. As we will see, the AdS/CFT correspondence realizes this idea,
but with five additional dimensions (in addition to the radial
coordinate on AdS which can be thought of as a generalization of the
Liouville field), leading to a standard (critical) ten dimensional
string theory.

\section{Black $p$-Branes} 
\label{black_pbranes}

The recent insight into the connection between
large $N$ field theories and string theory has emerged
from the study of $p$-branes in string theory. The $p$-branes 
were originally found as classical solutions to supergravity,
which is the low energy limit of string theory. 
Later it was pointed out by Polchinski that D-branes give
their full string theoretical description. Various comparisons 
of the two descriptions led to the discovery of the AdS/CFT 
correspondence.

\subsection{Classical Solutions}

String theory has a variety of classical solutions corresponding
to extended black holes
\cite{Gibbons:1982ih,Gibbons:1988ps,Callan:1989hs,Dabholkar:1989jt,
Dabholkar:1990yf,Duff:1991wv,Garfinkle:1991qj,Callan:1991dj,Callan:1991ky,
Horowitz:1991cd,Duff:1992pe}. Complete descriptions of all
possible black hole solutions would be beyond the scope of this 
review, and we will discuss here only illustrative examples
corresponding to parallel D$p$ branes. 
For a more extensive review of extended objects in
string theory, see \cite{Duff:1996zn,Peet:1997es}. 

Let us consider type II string theory in ten dimensions,
and look for a black hole solution carrying 
 electric charge with respect 
to the Ramond-Ramond (R-R) $(p+1)$-form $A_{p+1}$ 
\cite{Gibbons:1988ps,Garfinkle:1991qj,Horowitz:1991cd}.
In type IIA (IIB) theory, $p$ is even (odd). 
The theory contains also magnetically charged 
$(6-p)$-branes, which are  electrically charged under the dual 
$  dA_{7-p}= * dA_{p+1}$ potential.
Therefore, 
R-R charges have to be quantized according to the Dirac quantization
condition. 
To find the solution, we start with the low energy effective
action in the string frame,
\beq
  S =\frac{1}{(2\pi)^7 l_s^8} \int d^{10} x \sqrt{-g} \left( e^{-2\phi}
\left({\cal R} + 4 (\nabla \phi)^2 \right) 
- \frac{2}{(8-p)!} F_{p+2}^2 \right),
\label{action}
\eeq
where $l_s$ is the string length, related to the string tension
$(2\pi \alpha')^{-1}$ as $\alpha' = l_s^2$, and 
$F_{p+2}$ is the field strength of the $(p+1)$-form
potential,  $F_{p+2} = dA_{p+1}$. In the self-dual case of $p=3$
we work directly with the equations of motion.  
We then look for a solution corresponding to
a $p$-dimensional electric source of charge $N$ for $A_{p+1}$,
by requiring the Euclidean symmetry $ISO(p)$ in $p$-dimensions:
\beq
  ds^2 = ds_{10-p}^2 + e^\alpha \sum_{i=1}^p dx^i dx^i.
\eeq
Here $ds_{10-p}^2$ is a Lorentzian-signature metric 
in $(10-p)$-dimensions. We also assume that the metric is
spherically symmetric in $(10-p)$ dimensions with the  R-R source
at the origin,
\beq
   \int_{S^{8-p}} ~^* F_{p+2} = N ,
\eeq
where $S^{8-p}$ is the $(8-p)$-sphere surrounding the source. 
By using the Euclidean symmetry $ISO(p)$, we can reduce
the problem to the one of finding a spherically symmetric
charged black hole solution in $(10-p)$ dimensions \cite{Gibbons:1988ps,
Garfinkle:1991qj,Horowitz:1991cd}. The resulting metric, in the string
frame, is given by
\beq
 ds^2 = - \frac{f_+(\rho)}{\sqrt{f_-(\rho)}} dt^2
+  \sqrt{f_-(\rho)}
   \sum_{i=1}^p dx^i dx^i 
+  \frac{f_-(\rho)^{-\frac{1}{2} -\frac{5-p}{7-p}}}{f_+(\rho)} d\rho^2
 + r^2  f_-(\rho)^{\frac{1}{2}-\frac{5-p}{7-p}} d\Omega_{8-p}^2 ,
\label{solution}
\eeq
with the dilaton field,
\beq
e^{-2\phi} = g_s^{-2} f_-(\rho)^{-\frac{p-3}{2}},
\eeq
where
\beq
  f_{\pm}(\rho) = 1 - \left(\frac{r_\pm}{\rho} \right)^{7-p},
\eeq
and $g_s$ is the asymptotic string coupling constant. 
The parameters $r_+$ and $r_-$ are related to the mass $M$ 
(per unit volume) and 
the RR charge $N$ of the solution by
\beq
M =  \frac{1}{(7-p) (2 \pi)^7 d_p l_P^8}
  \left((8-p) r_+^{7-p} - r_-^{7-p} \right),~~~
N = \frac{1}{d_pg_s l_s^{7-p}}(r_+ r_-)^{\frac{7-p}{2}} ,
\label{masscharge}
\eeq
where $l_P = g_s^{\frac{1}{4}} l_s$ is the 10-dimensional Planck
length and $d_p$ is a numerical factor,
\beq
  d_p = 2^{5-p} \pi^{\frac{5-p}{2}} \Gamma\left( \frac{7-p}{2}
 \right).
\eeq

The metric in the Einstein frame, $(g_{\cal E})_{\mu\nu}$, 
is defined by multiplying
the string frame metric $g_{\mu\nu}$ by
$\sqrt{ g_se^{-\phi}}$
in (\ref{action}), so that the action takes the standard
Einstein-Hilbert form, 
\beq
S= \frac{1}{(2\pi)^7 l_P^8}\int d^{10} x 
 \sqrt{-g_{\cal E}} ( {\cal R}_{\cal E} - {1 \over 2} (\nabla \phi)^2 +
\cdots ).
\eeq
The Einstein frame metric has a horizon at $\rho=r_+$. For
$p\leq 6$, there is also a curvature singularity at $\rho=r_-$.
When $r_+ > r_-$, the singularity is covered by the horizon
and the solution can be regarded as a black hole. 
When $r_+ < r_-$, there is a timelike naked singularity
and the Cauchy problem is not well-posed. 

The situation is subtle in the critical case $r_+ = r_-$. 
If $p \neq 3$, the horizon
and the singularity coincide and there is a ``null'' singularity\footnote{
This is the case for $p < 6$. For $p=6$, the singularity is timelike
as one can see from the fact that it can be lifted to the Kaluza-Klein 
monopole in 11 dimensions.}. 
Moreover, the dilaton either diverges or vanishes at $\rho=r_+$. 
This singularity, however, is milder than in the case of $r_+ < r_-$,
and the supergravity description is still valid up to 
a certain distance from the singularity. The situation is
much better for $p=3$. In this case, the dilaton is constant.
Moreover, the $\rho=r_+$ surface is regular even when $r_+=r_-$,
allowing a smooth analytic extension beyond $\rho=r_+$  \cite{Gibbons:1995vm}.

According to (\ref{masscharge}), for a fixed value of $N$, 
the mass $M$ is an increasing
function of $r_+$. The condition $r_+ \geq r_-$ for 
the absence of the timelike naked singularity therefore translates into
an inequality between the mass $M$ and 
the R-R charge $N$, of the form
\beq
   M \geq  \frac{N}{(2 \pi)^p g_s l_s^{p+1}} .
\label{bps}
\eeq
The solution whose mass $M$ is at the lower bound of this inequality
is called an {\it extremal $p$-brane}. On the other hand, 
when $M$ is strictly greater than that, we have
a {\it non-extremal black $p$-brane}. It is called {\it black}
since there is an event horizon for $r_+ > r_-$. 
The area of the black hole horizon goes 
to zero in the extremal limit $r_+ = r_-$. Since
the extremal solution with $p \neq 3$ 
has a  singularity,
the supergravity description breaks down
near $\rho = r_+$ and we need to use the full string
theory.  The D-brane construction discussed below
will give exactly such a description. 
The inequality (\ref{bps}) is 
also the BPS bound with respect to the 10-dimensional 
supersymmetry, and the extremal solution $r_+=r_-$ 
preserves one half of the supersymmetry
in the regime where we can trust the supergravity
description. This suggests that the extremal $p$-brane
is a ground state of the black $p$-brane for a given charge
$N$. 

The extremal limit $r_+ = r_-$ of the solution (\ref{solution}) is
given by
\beq
  ds^2 = \sqrt{f_+(\rho)} \left( -dt^2 + \sum_{i=1}^p dx^i dx^i \right)
 + f_+(\rho)^{-\frac{3}{2} - \frac{5-p}{7-p}} d\rho^2
 + \rho^2 f_+(\rho)^{\frac{1}{2} - \frac{5-p}{7-p}} d\Omega_{8-p}^2.
\label{extremal}
\eeq
In this limit, the symmetry of the metric is enhanced from
the Euclidean group $ISO(p)$ to the Poincar\'e group
$ISO(p,1)$. This fits well with the interpretation that
the extremal solution corresponds to the ground state 
of the black $p$-brane.
To describe the geometry of the extremal solution outside of
the horizon,  it is often useful to define a new coordinate $r$ by
\beq
  r^{7-p} \equiv \rho^{7-p} - r_+^{7-p},
\eeq
and introduce the isotropic coordinates, $r^a = r \theta^a$
($a=1,...,9-p; ~\sum_a (\theta^a)^2 = 1$). 
The metric and the dilaton for the extremal $p$-brane are
then written as
\beq
ds^2 = \frac{1}{\sqrt{H(r)}}
 \left(-dt^2 + \sum_{i=1}^p dx^idx^i \right)
 + \sqrt{H(r)}
   \sum_{a=1}^{9-p} dr^a dr^a ,
\label{another}
\eeq
\beq
  e^{\phi} = g_s H(r)^{\frac{3-p}{4}},
\label{dilaton}
\eeq
where
\beq
  H(r) = \frac{1}{f_+(\rho)} = 1 + \frac{r_+^{7-p}}{r^{7-p}},~~~~~~~
r_+^{7-p} = d_p g_s N l_s^{7-p}.
\eeq
The horizon is now located at $r=0$. 

In general, (\ref{another}) and (\ref{dilaton}) give
a solution to the supergravity equations of motion for any
function $H(\vec{r})$ which is a harmonic function in the
$(9-p)$ dimensions which are transverse to the $p$-brane.  
For example, we may consider a more general solution, of the form
\beq
   H(\vec{r}) = 1 + \sum_{i=1}^k 
\frac{r_{(i)+}^{7-p}}{|\vec{r}-\vec{r}_i|^{7-p}},~~~~~~~r_{(i)+}^{7-p}
= d_p g_s N_i l_s^{7-p}.
\label{multicentered}
\eeq
This is called a multi-centered solution and represents
parallel extremal $p$-branes located at $k$ different locations,
$\vec{r}=\vec{r}_i$ ($i=1,\cdots,k$), each of which carries
$N_i$ units of the R-R charge. 

So far we have discussed the black $p$-brane using the classical
supergravity. This description is appropriate when the curvature
of the $p$-brane geometry is small compared to the string scale,
so that stringy corrections are negligible. Since the strength 
of the curvature is characterized by $r_+$, this requires
 $r_+ \gg l_s$. 
To suppress string loop corrections, 
the effective string coupling $e^{\phi}$ also needs
to be kept small. When $p=3$,
the dilaton is constant and we can make it small
everywhere in the $3$-brane geometry by setting
$g_s < 1$, namely $l_P <l_s$. If $g_s>1$ we might need to do an 
$S$-duality, $g_s \rightarrow 1/g_s$, first. 
 Moreover, in this case it is known that 
the metric (\ref{another}) can be analytically
extended beyond the horizon $r=0$, and that
the maximally extended metric is
geodesically complete and without a singularity
\cite{Gibbons:1995vm}. 
The strength of the curvature is then
uniformly bounded by $r_+^{-2}$. 
To summarize, for $p=3$, the supergravity approximation is valid
when
\beq
   l_P < l_s \ll r_+.
\eeq
Since $r_+$ is related to the R-R charge $N$ as
\beq
     r_+^{7-p} = d_p g_s N l_s^{7-p},
\eeq
this can also be expressed
as
\beq
    1 \ll g_s N < N.
\label{hier1}
\eeq
For  $p \neq 3$, the metric is singular at $r=0$,
and the supergravity description is valid only 
in a limited region of the spacetime. 

\subsection{D-Branes}

\begin{figure}[htb]
\begin{center}
\epsfxsize=4.5in\leavevmode\epsfbox{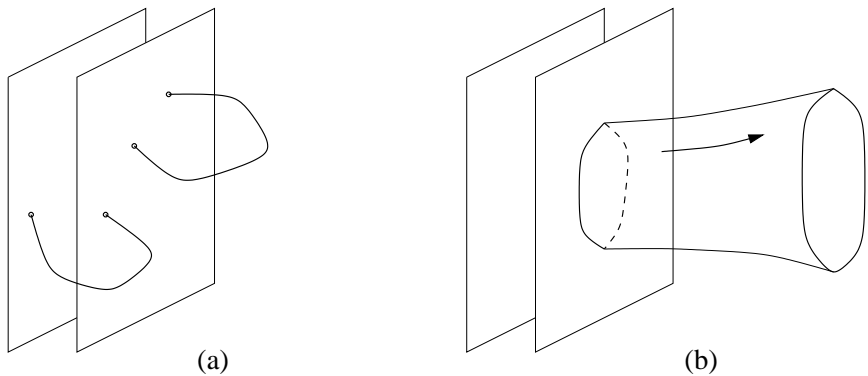}
\end{center}
\caption{(a) The D-brane is where open strings can end.
(b) The D-brane is a source
of closed strings.}
\label{F9}
\end{figure}

Alternatively, the extremal $p$-brane can be
described as a D-brane. For a review of D-branes,
see \cite{Polchinski:1996na}.
The D$p$-brane is a $(p+1)$-dimensional hyperplane
in spacetime where an open string can end. 
By the worldsheet duality, this means that 
the D-brane is also a source of closed strings (see Fig. \ref{F9}).
In particular, it can carry the R-R charges.  
It was shown in  \cite{Polchinski:1995mt} that,
if we put $N$ D$p$-branes on top of each other,
the resulting $(p+1)$-dimensional hyperplane carries 
exactly $N$ units of the $(p+1)$-form charge. 
On the worldsheet of a type II string, the left-moving degrees
of freedom 
and the right-moving degrees of freedom 
carry separate spacetime supercharges.
Since the open string boundary condition identifies
the left and right movers,  
the D-brane breaks at least one half of the spacetime 
supercharges. In type IIA (IIB) string theory, 
precisely one half
of the supersymmetry is preserved if $p$ is even
(odd). This is consistent with the types of R-R charges
that appear in the theory. Thus, the D$p$-brane is a BPS 
object in string theory which carries 
exactly the same charge as the black $p$-brane
solution in supergravity. 
  
It is believed that the extremal $p$-brane in supergravity
and the D$p$-brane are two different descriptions of the same
object. The D-brane uses the string worldsheet and, therefore,
is a good description in string perturbation theory. 
When there are $N$ D-branes on top of each other,
the effective loop expansion parameter for the open strings is
$g_s N$ rather than $g_s$, since each open string boundary loop 
ending on the D-branes
comes with the Chan-Paton factor $N$ as well as the
string coupling $g_s$. Thus, the D-brane description
is good when $g_s N \ll 1$. 
This is complementary to the regime (\ref{hier1}) where
the supergravity description is appropriate. 

The low energy effective theory of open strings on 
the D$p$-brane is the $U(N)$ gauge theory in $(p+1)$ dimensions
with $16$ supercharges \cite{Witten:1996im}. 
The theory has $(9-p)$ scalar fields
$\vec{\Phi}$ in the adjoint representation of $U(N)$. 
If the vacuum expectation value $\langle \vec{\Phi} \rangle$ 
has $k$ distinct eigenvalues\footnote{There is a potential
$\sum_{I,J} \tr [\Phi^I, \Phi^J]^2$ for the scalar fields,
so expectation values of the matrices
$\Phi^I$ ($I=1,\cdots,9-p$) minimizing the potential
are simultaneously diagonalizable.}, with $N_1$ identical 
eigenvalues $\vec{\phi}_1$, $N_2$ identical eigenvalues $\vec{\phi}_2$
and so on, the gauge group $U(N)$ is broken 
to $U(N_1) \times \cdots \times U(N_k)$. 
This corresponds to the situation when $N_1$ D-branes are at 
$\vec{r}_1 = \vec{\phi}_1 l_s^2$, $N_2$ D$p$-branes are
at $\vec{r}_2= \vec{\phi}_2 l_s^2$, and so on. 
In this case, there are massive
$W$-bosons for the broken gauge groups. The 
$W$-boson in the bi-fundamental representation of
$U(N_i) \times U(N_j)$ comes from the open string stretching 
between the D-branes at $\vec{r}_i$ and $\vec{r}_j$, 
and the mass of the W-boson
is proportional to the Euclidean distance $|\vec{r}_i - \vec{r}_j|$
between the $D$-branes. It is important to note that the same result 
is obtained if we use the supergravity solution for the multi-centered 
$p$-brane (\ref{multicentered}) and compute the mass of
the string going from $\vec{r}_i$ to $\vec{r}_j$, since
the factor $H(\vec{r})^{\frac{1}{4}}$ from
 the metric in the $\vec{r}$-space (\ref{another})
is cancelled by the redshift factor 
$H(\vec{r})^{-\frac{1}{4}}$ when converting
the string tension into energy. Both
the D-brane description and the supergravity solution give
the same value of the W-boson mass, since it is determined
by the BPS condition.

\subsection{Greybody Factors and Black Holes} 
\label{gbFactorsBH}

An important precursor to the AdS/CFT correspondence was the calculation of
greybody factors for black holes built out of D-branes.  It was noted in
\cite{Callan:1996dv} that Hawking radiation could be mimicked by processes
where two open strings collide on a D-brane and form a closed string which
propagates into the bulk.  The classic computation of Hawking (see, for
example, \cite{Hawking:1974df} for details) shows in a semi-classical
approximation that the differential rate of spontaneous emission of
particles of energy $\omega$ from a black hole is
  \eqn{HawkingRadiation}{
   d\Gamma_{\rm emit} = {v 
\sigma_{\rm absorb} \over e^{\omega/T_H} \pm 1} 
     {d^n k \over (2\pi)^n} \ ,
  }
where $v$ is the velocity of the emitted particle in the 
transverse directions, and
 the sign in the denominator is minus for bosons and plus for
fermions.  We use $n$ to denote the number of spatial dimensions around the
black hole (or if we are dealing with a black brane, it is the number of
spatial dimensions perpendicular to the world-volume of the brane).  $T_H$
is the Hawking temperature, and $\sigma_{\rm absorb}$ is the cross-section
for a particle coming in from infinity to be absorbed by the black hole.
In the differential emission rate, the emitted particle is required to have
a momentum in a small region $d^n k$, and $\omega$ is a function of
$k$. To obtain a total emission rate we
would integrate \HawkingRadiation\ over all $k$.

If $\sigma_{\rm absorb}$ were a constant, then \HawkingRadiation\ tells us
that the emission spectrum is the same as that of a blackbody.  Typically,
$\sigma_{\rm absorb}$ is not constant, but varies appreciably over the
range of finite $\omega/T_H$.  The consequent deviations from the pure
blackbody spectrum have earned $\sigma_{\rm absorb}$ the name ``greybody
factor.''  A successful microscopic account of black hole thermodynamics
should be able to predict these greybody factors.  In \cite{Das:1996wn} and
its many successors, it was shown that the D-branes provided an account of
black hole microstates which was successful in this respect.

Our first goal will be to see how greybody factors are computed in the
context of quantum fields in curved spacetime.  The literature on this
subject is immense.  We refer the reader to \cite{Matzner} for an overview
of the General Relativity literature, and to
\cite{Maldacena:1997ix,Gubser:1997yh,Peet:1997es} and references therein
for a first look at the string theory additions.

In studying scattering of particles off of a black hole (or any fixed
target), it is convenient to make a partial wave expansion.  For
simplicity, let us restrict the discussion to scalar fields.  Assuming that
the black hole is spherically symmetric, one can write the asymptotic
behavior at infinity of the time-independent scattering solution as
  \eqn{PartialWave}{\eqalign{
   \phi(\vec{r}) &\sim e^{ikx} + 
f(\theta) {e^{ikr} \over r^{n/2}}  \cr
    &\sim \sum_{\ell = 0}^\infty \tf{1}{2} \tilde{P}_\ell(\cos \theta)
     {S_\ell e^{i k r} + (-1)^\ell i^n e^{-i k r} \over
      (i k r)^{n/2}} \ ,
  }}
 where $x = r\cos \theta$.  The term $e^{ikx}$ represents the incident
wave, and the second term in the first line represents the scattered wave.
The $\tilde{P}_\ell(\cos\theta)$ are generalizations of Legendre
polynomials.  The absorption probability for a given
partial wave is given by
$P_\ell = 1-|S_\ell|^2$.  An application of the Optical
Theorem leads to the absorption cross section \cite{Gubser:1997qr}
  \eqn{OptTheorem}{
   \sigma_{\rm abs}^\ell = 
    {2^{n-1} \pi^{{n-1 \over 2}} \over k^n}
    \Gamma\left( {n-1 \over 2} \right) \left( \ell + {n-1 \over 2} \right) 
     {\ell+n-2 \choose \ell} P_\ell \ .
  }
 Sometimes the absorption probability $P_\ell$ is called the greybody
factor.

The strategy of absorption calculations in supergravity is to solve a
linearized wave equation, most often the Klein-Gordon
 equation $\square \phi =
0$, using separation of variables, $\phi = e^{-i\omega t}
P_\ell(\cos\theta) R(r)$.  Typically the radial function cannot be
expressed in terms of known functions, so some approximation scheme is
used, as we will explain in more detail below.  Boundary conditions are
imposed at the black hole horizon corresponding to infalling matter.  Once
the solution is obtained, one can either use the asymptotics \PartialWave\
to obtain $S_\ell$ and from it $P_\ell$ and $\sigma_{\rm abs}^\ell$, or
compute the particle flux at infinity and at the horizon and note that
particle number conservation implies that $P_\ell$ is their ratio.

One of the few known universal results is that for $\omega/T_H \ll 1$,
$\sigma_{\rm abs}$ for an $s$-wave massless scalar approaches the horizon
area of the black hole \cite{Das:1997we}.  This result holds for any
spherically symmetric black hole in any dimension.  For $\omega$ much
larger than any characteristic curvature scale of the geometry, one can use
the geometric optics approximation to find $\sigma_{\rm abs}$.

We will be interested in the particular black hole geometries for which
string theory provides a candidate description of the microstates.  Let us
start with $N$ coincident D3-branes, where the low-energy world-volume
theory is $d=4$ ${\cal N}=4$ $U(N)$ gauge theory.  The equation of motion
for the dilaton is $\square\phi = 0$ where $\square$ is the laplacian for
the metric 
  \eqn{D3Metric}{
   ds^2 = \left( 1 + {R^4 \over r^4} \right)^{-1/2} 
    \left( -dt^2 + dx_1^2 + dx_2^2 + dx_3^2 \right) + 
    \left( 1 + {R^4 \over r^4} \right)^{1/2} 
    \left( dr^2 + r^2 d\Omega_5^2 \right) \ .
  }
 It is convenient to change radial variables: $r = R e^{-z}$, $\phi =
e^{2z} \psi$.  The radial equation for the $\ell^{\rm th}$ partial wave is
  \eqn{LthPartial}{
   \left[ \partial_z^2 + 2 \omega^2 R^2 \cosh 2z - (\ell+2)^2 \right]
    \psi_\ell(z) = 0 \ ,
  }
 which is precisely Schrodinger's equation with a potential $V(z) = - 2
\omega^2 R^2 \cosh 2z$.  The absorption probability is precisely the
tunneling probability for the barrier $V(z)$: the transmitted wave at large
positive $z$ represents particles falling into the D3-branes.  At leading
order in small $\omega R$, the absorption probability for the $\ell^{\rm
th}$ partial wave is
  \eqn{LowCross}{
   P_\ell = {4\pi^2 \over (\ell+1)!^4 (\ell+2)^2} 
    \left( {\omega R \over 2} \right)^{8 + 4\ell} \ .
  }
 This result, together with a recursive algorithm for computing all
corrections as a series in $\omega R$, was obtained in \cite{Gubser:1998iu}
from properties of associated Mathieu functions, which are the solutions of
\LthPartial.  An exact solution of a radial equation in terms of known
special functions is rare.  We will therefore present a standard
approximation technique (developed in \cite{Unruh:1976fm} and applied to
the problem at hand in \cite{Klebanov:1997kc}) which is sufficient to
obtain the leading term of \LowCross.  Besides, for comparison with string
theory predictions we are generally interested only in this leading term.

The idea is to find limiting forms of the radial equation which can be
solved exactly, and then to match the limiting solutions together to
approximate the full solution.  Usually a uniformly good approximation can
be found in the limit of small energy.  The reason, intuitively speaking,
is that on a compact range of radii excluding asymptotic infinity and the
horizon, the zero energy solution is nearly equal to solutions with very
small energy; and outside this region the wave equation usually has a
simple limiting form.  So one solves the equation in various regions and
then matches together a global solution.

It is elementary to show that this can be done for \LthPartial\ using two
regions:
  \eqn{MatchSoln}{\seqalign{\span\TT \qquad & \span\TR}{
   far region: $z \gg \log \omega R$ \qquad & 
    \eqalign{& \left[ \partial_z^2 + 
      \omega^2 R^2 e^{2z} - (\ell+2)^2 \right] \psi = 0  
       \cr\noalign{\vskip-0.5\jot}
      & \quad \psi(z) = H^{(1)}_{\ell+2}(\omega R e^z)}
     \cr\noalign{\vskip2\jot}
   near region: $z \ll -\log \omega R$ \qquad & 
    \eqalign{& \left[ \partial_z^2 + 
      \omega^2 R^2 e^{-2z} - (\ell+2)^2 \right] \psi = 0  
       \cr\noalign{\vskip-0.5\jot}
      & \quad \psi(z) = a J_{\ell+2}(\omega R e^{-z})}
  }}
 It is amusing to note the $\IZ_2$ symmetry, $z \to -z$, which exchanges
the far region, where the first equation in \MatchSoln\ is just free
particle propagation in flat space, and the near region, where the second
equation in \MatchSoln\ describes a free particle in $AdS_5$.  This
peculiar symmetry was first pointed out in \cite{Klebanov:1997kc}.  It
follows from the fact that the full D3-brane metric comes back to itself,
up to a conformal rescaling, if one sends $r \to R^2/r$.  A similar duality
exists between six-dimensional flat space and $AdS_3 \times S^3$ in the
D1-D5-brane solution, where the Laplace equation again can be solved in
terms of Mathieu functions \cite{ghUnp,Cvetic:1999fv}.  To our knowledge
there is no deep understanding of this ``inversion duality.''

 For low energies $\omega R \ll 1$, the near and far regions overlap in a
large domain, $\log \omega R \ll z \ll -\log \omega R$, and by comparing
the solutions in this overlap region one can fix $a$ and reproduce the
leading term in \LowCross.  It is possible but tedious to obtain the
leading correction by treating the small terms which were dropped from the
potential to obtain the limiting forms in \MatchSoln\ as perturbations.
This strategy was pursued in \cite{Gubser:1998kv,Taylor-Robinson:1998tk}
before the exact solution was known, and in cases where there is no exact
solution.  The validity of the matching technique is discussed in
\cite{Matzner}, but we know of no rigorous proof that it holds in all the
circumstances in which it has been applied.

The successful comparison of the $s$-wave dilaton cross-section in
\cite{Klebanov:1997kc} with a perturbative calculation on the D3-brane
world-volume was the first hint that Green's functions of ${\cal N}=4$
super-Yang-Mills theory could be computed from supergravity.  In
summarizing the calculation, we will follow more closely the conventions of
\cite{Gubser:1997yh}, and give an indication of the first application of
non-renormalization arguments \cite{Gubser:1997se} to understand why the
agreement between supergravity and perturbative gauge theory existed
despite their applicability in opposite limits of the 't~Hooft coupling.

Setting normalization conventions so that the pole in the propagator of the
gauge bosons has residue one at tree level, we have the following action
for the dilaton plus the fields on the brane:
  \eqn{BBAction}{
   S = {1 \over 2\kappa^2} \int d^{10} x \sqrt{g} \, \left[ {\cal R} - 
    \tf{1}{2} (\partial\phi)^2 + \ldots \right] +
    \int d^4 x \, \left[ -\tf{1}{4} e^{-\phi} \tr F_{\mu\nu}^2 + 
      \ldots \right] \ ,
  }
 where we have omitted other supergravity fields, their interactions with
one another, and also terms with the lower spin fields
in the gauge theory action.  A
plane wave of dilatons with energy $\omega$ and momentum perpendicular to
the brane is kinematically equivalent on the world-volume to a massive
particle which can decay into two 
gauge bosons through the coupling ${1 \over
4} \phi \tr F_{\mu\nu}^2$.  In fact, the absorption cross-section is given
precisely by the usual expression for the decay rate into $k$ particles:
  \eqn{CrossFeynman}{
   \sigma_{\rm abs} = {1 \over S_f} {1 \over 2\omega} 
    \int {d^3 p_1 \over (2\pi)^3 2\omega_1} \ldots
     {d^3 p_k \over (2\pi)^3 2\omega_k} (2\pi)^4 \delta^4(P_f - P_i)
     \overline{|{\cal M}|}^2 \ .
  }
 In the Feynman rules for ${\cal M}$, a factor of $\sqrt{2\kappa^2}$
attaches to an external dilaton line on account of the non-standard
normalization of its kinetic term in \BBAction.  This factor gives
$\sigma_{\rm abs}$ the correct dimensions: it is a length to the fifth
power, as appropriate for six transverse spatial dimensions.  In
\CrossFeynman, $\overline{|{\cal M}|}^2$ indicates summation over
distinguishable processes: in the case of the $s$-wave dilaton there are
$N^2$ such processes because of the number of gauge bosons.  One easily
verifies that $\overline{|{\cal M}|}^2 = N^2 \kappa^2 \omega^4$.  $S_f$ is
a symmetry factor for identical particles in the final state: in the case
of the $s$-wave dilaton, $S_f = 2$ because the outgoing gauge bosons are
identical.

Carrying out the $\ell=0$ calculation explicitly, one finds
  \eqn{CrossFinal}{
   \sigma_{\rm abs} = {N^2 \kappa^2 \omega^3 \over 32\pi} \ ,
  }
 which, using \OptTheorem\ and the relation between $R$ and $N$, 
can be shown to be in precise agreement with the
leading term of $P_0$ in \LowCross.  This is now understood to be due to a
non-renormalization theorem for the two-point function of the operator
${\cal O}_4 = {1 \over 4} \tr F^2$.

\begin{figure}
   \vskip0cm
   $$
    \sum_X \left| 
       \ooalign{\lower0.25in\hbox{\smash{\psfig{figure=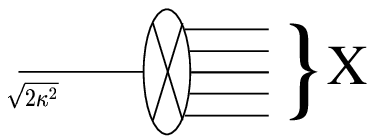}}}}\ 
      \right|^2 = 
     {2\kappa^2 \over i} {\rm Disc} \, 
       \ooalign{\lower0.34in\hbox{\smash{\psfig{figure=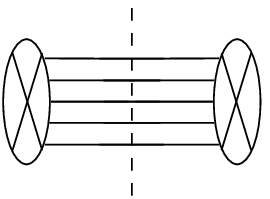}}}}
   $$
   \vskip0cm
 \caption{An application of the optical theorem.}\label{OptTh}
  \end{figure}

To understand the connection with two-point functions, note that an
absorption calculation is insensitive to the final state on the D-brane
world-volume.  The absorption cross-section is therefore related to the
discontinuity in the cut of the two-point function of the operator to which
the external field couples.  To state a result of some generality, let us
suppose that a scalar field $\phi$ in ten dimensions couples to a gauge
theory operator through the action
  \eqn{SintDef}{
   S_{\rm int} = \int d^4 x \, \partial_{y_{i_1}} \cdots 
    \partial_{y_{i_\ell}} \phi(x,y_i)\Big|_{y_i = 0} 
    {\cal O}^{i_1 \ldots i_\ell}(x) \ ,
  }
 where we use $x$ to denote the four coordinates parallel to the
world-volume and $y_i$ to denote the other six.  An example where this
would be the right sort of coupling is the $\ell^{\rm th}$ partial wave of
the dilaton \cite{Gubser:1997yh}.  The $\ell^{\rm th}$ partial wave
absorption cross-section for a particle with initial momentum $p = \omega
(\hat{t} + \hat{y}_1)$ is obtained by summing over all final states
that could be created by the operator ${\cal O}^{1\ldots
1}$:\footnote{There is one restriction on the final states: for a process
to be regarded as an $\ell^{\rm th}$ partial wave absorption cross-section,
$\ell$ units of angular momentum must be picked up by the brane.  Thus
${\cal O}^{i_1 \ldots i_\ell}$ must transform in the irreducible
representation which is the $\ell^{\rm th}$ traceless symmetric power of
the ${\bf 6}$ of $SO(6)$.}
  \eqn{SigmaPartial}{\eqalign{
   \sigma_{\rm abs} &= {1 \over 2\omega} \sum_n \prod_{i=1}^n {1 \over S_f}
    {d^3 p_i \over (2\pi)^3 2\omega_i} (2\pi)^4 \delta^4(P_f - P_i)
    \overline{|{\cal M}|}^2  \cr
    &= {2\kappa^2 \omega^\ell \over 2i\omega} {\rm Disc} \, 
     \int d^4 x \, e^{i p \cdot x} \langle {\cal O}^{1\ldots 1}(x)
      {\cal O}^{1\ldots 1}(0) \rangle \Big|_{p=(\omega,0,0,0)} \ .
  }}
 In the second equality we have applied the Optical Theorem (see
figure~\ref{OptTh}).  
  \comment{had to hack at this one with ooalign.  Anyone with TeX tricks is
welcome to show me a better way.  SSG}
 The factor of $2\kappa^2$ is the square of the external leg factor for the
incoming closed string state, which was included in the invariant amplitude
${\cal M}$.  The factor of $\omega^\ell$ arises from acting with the $\ell$
derivatives in \SintDef\ on the incoming plane wave state.  The symbol
${\rm Disc}$ indicates that one is looking at the unitarity cut in the
two-point function in the $s$ plane, where $s = p^2$.  The two-point
function can be reconstructed uniquely from this cut, together with some
weak conditions on regularity for large momenta.  Results analogous to
\SigmaPartial\ can be stated for incoming particles with spin, only it
becomes more complicated because a polarization must be specified, and the
two-point function in momentum space includes a polynomial in $p$ which
depends on the polarization.

Expressing absorption cross-sections in terms of two-point functions helps
illustrate why there is ever agreement between the tree-level calculation
indicated in \CrossFeynman\ and the supergravity result, which one would
{\it a priori} expect to pick up all sorts of radiative corrections.
Indeed, it was observed in \cite{Gubser:1997se} that the $s$-wave graviton
cross-section agreed between supergravity and tree-level gauge theory
because the correlator $\langle T_{\alpha\beta} T_{\gamma\delta} \rangle$
enjoys a non-renormalization theorem.  One way to see that there must be
such a non-renormalization theorem is to note that conformal Ward
identities relate this two-point function to $\langle T^\mu_\mu
T_{\alpha\beta} T_{\gamma\delta} \rangle$ (see for example
\cite{Erdmenger:1997yc} for the details), and supersymmetry in turn relates
this anomalous three-point function to the anomalous VEV's of the
divergence of R-currents in the presence of external sources for them.  The
Adler-Bardeen theorem \cite{Adler:1969er} protects these anomalies, hence
the conclusion.

Another case which has been studied extensively is a system consisting of
several D1 and D5 branes.  The D1-branes are delocalized on the four extra
dimensions of the D5-brane, which are taken to be small, so that the total
system is effectively 1+1-dimensional.  We will discuss the physics of this
system more extensively in chapter~\ref{ChapAdS3}, and the reader can also
find background material in \cite{Peet:1997es}.  For now our purpose is to
show how supergravity absorption calculations relate to finite-temperature
Green's functions in the 1+1-dimensional theory.

Introducing momentum along the spatial world-volume (carried by open
strings attached to the branes), one obtains the following ten-dimensional
metric and dilaton:
  \eqn{TenSol}{\eqalign{
   ds_{10,{\rm str}}^2 &= H_1^{-1/2} H_5^{-1/2} \left[ -dt^2 + dx_5^2 + 
    {r_0^2 \over r^2} (\cosh\sigma dt + \sinh\sigma dx_5)^2 \right]  \cr 
      & \qquad + 
    H_1^{1/2} H_5^{-1/2} \sum_{i=1}^4 dy_i^2 +
    H_1^{1/2} H_5^{1/2} 
    \left[ \left( 1 - {r_0^2 \over r^2} \right)^{-1} dr^2 + 
    r^2 d\Omega_3^2 \right]  \cr
   e^{\phi-\phi_\infty} &= H_1^{1/2} H_5^{-1/2}  \cr
   H_1 &= 1 + {r_1^2 \over r^2} \qquad H_5 = 1 + {r_5^2 \over r^2} \ .
  }}
 The quantities $r_1^2$, $r_5^2$, and $r_K^2 = r_0^2 \sinh^2\sigma$ are
related to the number of D1-branes, the number of D5-branes, and the net
number of units of momentum in the $x_5$ direction, respectively.  The
horizon radius, $r_0$, is related to the non-extremality.  For details, see
for example \cite{Maldacena:1997ix}.  If $r_0 = 0$ then there are only
left-moving open strings on the world-volume; if $r_0 \neq 0$ then there
are both left-movers and right-movers.  The Hawking temperature can be
expressed as ${2 \over T_H} = {1 \over T_L} + {1 \over T_R}$, where
  \eqn{TLR}{
   T_L = {1 \over \pi} {r_0 e^\sigma \over 2 r_1 r_5} \qquad
   T_R = {1 \over \pi} {r_0 e^{-\sigma} \over 2 r_1 r_5} \ .
  }
 $T_L$ and $T_R$ have the interpretation of temperatures of the left-moving
and right-moving sectors of the 1+1-dimensional world-volume theory.  There
is a detailed and remarkably successful account of the Bekenstein-Hawking
entropy using statistical mechanics in the world-volume theory.  It was
initiated in \cite{Strominger:1996sh}, developed in a number of subsequent
papers, and has been reviewed in \cite{Peet:1997es}.

The region of parameter space which we will be interested in is
  \eqn{DiluteGas}{
   r_0,r_K \ll r_1,r_5
  }
 This is known as the dilute gas regime.  The total energy of the open
strings on the branes is much less than the constituent mass of either the
D1-branes or the D5-branes.
 We are also interested in low energies
$\omega r_1, \omega r_5 \ll 1$, but $\omega/T_{L,R}$ can be arbitrary
thanks to \DiluteGas , \TLR . 
  The D1-D5-brane system is not stable because
left-moving open strings
   can run into right-moving open string and form a
closed string: indeed, this is exactly the process we aim to quantify.
Since we have collisions of left and right moving excitations we
expect that the answer will contain the left and right moving occupation
factors, so that the emission rate is 
\eqn{emission}{
d\Gamma = g^2_{eff} { 1 \over (e^{ \omega \over 2 T_L} -1) }
 { 1 \over (e^{ \omega \over 2 T_R} -1) }{d^4k \over (2\pi)^4 }
}
where $g_{eff}$ is independent of the temperature and measures the 
coupling of the open strings to the closed strings. 
The functional form of \emission\ seems, at first sight,
 to be different from \HawkingRadiation . But 
in order to compare them we should calculate 
the absorption cross section
appearing in \HawkingRadiation .


Off-diagonal gravitons $h_{y_1 y_2}$
(with $y_{1,2}$ in the compact directions) reduce
to scalars in six dimensions which obey the massless Klein
Gordon  equation.  These
so-called 
minimal scalars have been the subject of the most detailed study.
We will consider only the $s$-wave and we take the momentum along the 
string to be zero.   Separation of variables leads to the
radial equation
  \eqn{RadialLaplace1}{\eqalign{
   & \left[ {h \over r^3} \partial_r hr^2 \partial_r + 
     \omega^2 f \right] R = 0 \ , \cr
  & ~~h = 1 - {r_0^2 \over r^2},~~~~f =
 \left( 1 + {r_1^2\over r^2} \right) 
 \left( 1 + {r_5^2\over r^2} \right)  \left( 1 + {r_0^2\sinh^2 \sigma
\over r^2} \right)  ~ .
  }}
 Close to the horizon, a convenient radial variable is $z = h = 1 -
r_0^2/r^2$.  The matching procedure can be summarized as follows:
  \eqn{MatchSolnAgain1}{\seqalign{\span\TT \qquad & \span\TR}{
   far region:  & 
    \eqalign{& \left[ {1 \over r^3} \partial_r r^3 \partial_r + 
      \omega^2 \right] R = 0  
       \cr\noalign{\vskip-0.5\jot}
      & \quad R = A {J_1(\omega r) \over r^{3/2}},}
     \cr\noalign{\vskip2\jot}
   near region:  & 
    \eqalign{& \left[ z(1-z) \partial_z^2 + 
      \left( 1 - i {\omega \over 2\pi T_H} \right) (1-z) \partial_z + 
      {\omega^2 \over 16 \pi^2 T_L T_R} \right] 
       z^{i\omega \over 4\pi T_H} R = 0
       \cr\noalign{\vskip-0.5\jot}
      & \quad R = z^{-{i \omega \over 4\pi T_H}}
       F\left( -i {\omega \over 4\pi T_L}, -i {\omega \over 4\pi T_R};
        1 - i {\omega \over 2\pi T_H}; z \right).  }
  }}
 After matching the near and far regions together and comparing the
infalling flux at infinity and at the horizon, one arrives at
  \eqn{SigmaSwave}{
   \sigma_{\rm abs} = \pi^3 r_1^2 r_5^2 \omega
    {e^{\omega \over T_H} - 1 \over 
     \left( e^{\omega \over 2 T_L} - 1 \right) 
     \left( e^{\omega \over 2 T_R} - 1 \right)} \ .
  }
This has precisely the right form to ensure the matching of 
\emission\ with \HawkingRadiation\  (note that for a massless particle
with no momentum along the black string $v =1$ in \HawkingRadiation ).
It is possible to be more precise and  calculate the coefficient in 
\emission\  and actually check that the matching is precise 
\cite{Das:1996wn}. We leave 
this to chapter \ref{ChapAdS3}.

\begin{figure}[htb]
\begin{center}
\epsfxsize=3.5in\leavevmode\epsfbox{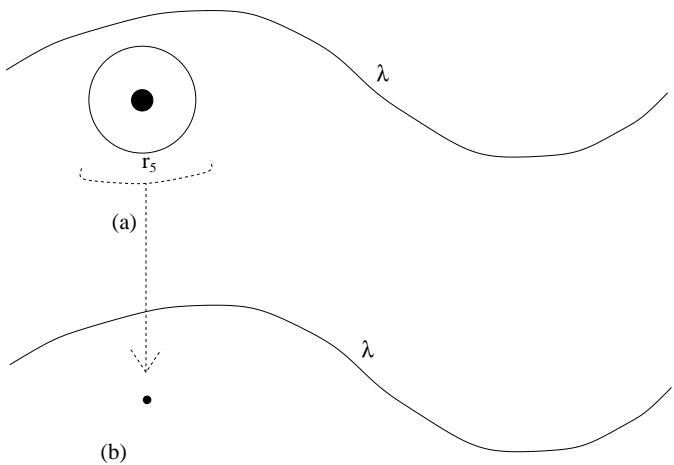}
\end{center}
\caption{ Low energy dynamics of extremal or near-extremal black branes. 
$r_5$ denotes the typical gravitational size of the system, namely
the position where the metric significantly deviates from 
Minkowski space. The Compton wavelength of 
the particles we scatter is much larger than the gravitational size,
$ \lambda \gg r_5$. In this situation we replace the whole black 
hole geometry (a)  by a point-like system in the transverse coordinates 
 with localized excitations (b).
These excitations are the ones described by the field theory living on
the brane. 
}
\label{nature}
\end{figure} 

Both
in the D3-brane analysis and in the D1-D5-brane analysis, one can see that
all the interesting physics is resulting from the near-horizon region: the
far region wave-function describes free particle propagation.  For quanta
whose Compton wavelength is much larger than the size of the black hole,
the effect of the far region is merely to set the boundary conditions in
the near region. See figure \ref{nature}. 
This provides a motivation for the prescription for
computing Green's functions, to be described in
section \ref{correlators}: as the calculations of this section
demonstrate, cutting out the near-horizon region of the supergravity
geometry and replacing it with the D-branes leads to an identical response
to low-energy external probes.

\chapter{Conformal Field Theories and AdS Spaces}
\label{ChapCFT}

\section{Conformal Field Theories} 
\label{cft}

Symmetry principles, and in particular Lorentz and \Poincare
invariance, play a major role in our understanding of quantum field
theory. It is natural to look for possible generalizations of
\Poincare invariance in the hope that they may play some role in
physics; in \cite{Coleman:1967ad} 
it was argued that for theories with a non-trivial
S-matrix there are no such bosonic generalizations. An interesting
generalization of \Poincare invariance is the addition of a scale
invariance symmetry linking physics at different scales (this is
inconsistent with the existence of an S-matrix since it does not allow
the standard definition of asymptotic states). Many interesting field
theories, like Yang-Mills theory in four dimensions, are
scale-invariant; generally this scale invariance does not extend to
the quantum theory (whose definition requires a cutoff which
explicitly breaks scale invariance) but in some special cases (such as
the $d=4,\cn=4$ supersymmetric Yang-Mills theory) it does, and even
when it does not (like in QCD) it can still be a useful tool, leading
to relations like the Callan-Symanzik equation. It was realized in the
past 30 years that field theories generally exhibit a renormalization
group flow from some scale-invariant (often free) UV fixed point
to some scale-invariant (sometimes trivial) IR fixed point, and
statistical mechanics systems also often have non-trivial IR
scale-invariant fixed points. Thus, studying scale-invariant theories
is interesting for various physical applications.

It is widely believed that unitary interacting scale-invariant
theories are always invariant under the full conformal group, which is
a simple group including scale invariance and \Poincare
invariance. This has only been proven in complete generality for two
dimensional field theories
\cite{Zamolodchikov:1986gt,Polchinski:1988dy}, but there are no known
counter-examples. In this section we will review the conformal group
and its implications for field theories, focusing on implications
which will be useful in the context of the AdS/CFT
correspondence. General reviews on conformal field theories may be
found in \cite{Mack:1988nf,Fradkin:1996is,Fradkin:1997df} 
and references therein.

\subsection{The Conformal Group and Algebra}
\label{conformal_group}

The conformal group is the group of transformations which preserve the
form of the metric up to an arbitrary scale factor, $g_{\mu
\nu}(x) \to \Omega^2(x) g_{\mu \nu}(x)$ (in this section greek letters
will correspond to the space-time coordinates,
$\mu,\nu=0,\cdots,d-1$). It is the minimal group that includes the
\Poincare group as well as the inversion symmetry $x^\mu \to x^\mu /
x^2$.

The conformal group of Minkowski space\footnote{More precisely, some of
these transformations can take finite points in Minkowski space to
infinity, so they should be defined on a compactification of Minkowski
space which includes points at infinity.} is generated by the
\Poincare transformations, the scale transformation
\eqn{scale}{x^\mu \to \lambda x^\mu,}
and the special conformal transformations
\eqn{specconf}{x^\mu \to {{x^\mu + a^\mu x^2} \over {1 + 2 x^\nu a_\nu
+ a^2 x^2}}.}
We will denote the generators of these transformations by $M_{\mu
\nu}$ for the Lorentz transformations, $P_\mu$ for translations, $D$
for the scaling transformation \eno{scale} and $K_\mu$ for the
special conformal transformations \eno{specconf}. The vacuum of a
conformal theory is annihilated by all of these generators.
They obey the conformal algebra
\eqn{algebra}{\eqalign{[M_{\mu\nu},P_\rho]=-i(\eta_{\mu\rho}P_\nu-
\eta_{\nu\rho}P_\mu); &\qquad
[M_{\mu\nu},K_\rho]=-i(\eta_{\mu\rho}K_\nu-\eta_{\nu\rho}K_\mu); \cr
[M_{\mu\nu},M_{\rho\sigma}]=-i\eta_{\mu\rho}M_{\nu\sigma} 
\pm permutations; &\qquad
[M_{\mu\nu},D]=0; \qquad
[D,K_\mu]=iK_\mu; \cr
[D,P_\mu]=-iP_\mu; &\qquad
[P_\mu,K_\nu]=2iM_{\mu\nu}-2i\eta_{\mu\nu}D, }}
with all other commutators vanishing.
This algebra is isomorphic to the algebra of $SO(d,2)$, and can be put
in the standard form of the $SO(d,2)$ algebra (with signature
$-,+,+,\cdots,+,-$) with generators $J_{ab}$ ($a,b=0,\cdots,d+1$)
by defining
\eqn{bigso}{J_{\mu\nu} = M_{\mu\nu};\qquad J_{\mu d} = {1\over 2}
(K_\mu - P_\mu);\qquad J_{\mu (d+1)} = {1\over 2}(K_\mu +
P_\mu);\qquad J_{(d+1)d} = D.}
For some applications it is useful to study the conformal theory in
Euclidean space; in this case the conformal group is 
$SO(d+1,1)$,\footnote{Strictly speaking, $SO(d+1,1)$ is the connected 
component of the conformal group which includes the identity, and it does 
not include $x^\mu \to x^\mu / x^2$.  We will hereafter ignore such 
subtleties.}
 and
since $\IR^d$ is conformally equivalent to $S^d$ the field theory on
$\IR^d$ (with appropriate boundary conditions at infinity) is isomorphic
to the theory on $S^d$. Much of what we say below will apply also to
the Euclidean theory.

In the special case of $d=2$ the conformal group is larger, and in
fact it is infinite dimensional. The special aspects of this case will
be discussed in chapter \ref{ChapAdS3} where they will be needed.

\subsection{Primary Fields, Correlation Functions, and Operator Product
Expansions}
\label{primaries}

The interesting representations (for physical applications) of the
conformal group involve operators (or fields) which are eigenfunctions
of the scaling operator $D$ with eigenvalue $-i\Delta$ ($\Delta$ is
called the {\it scaling dimension} of the field). This means that
under the scaling transformation \eno{scale} they transform as
$\phi(x) \to \phi^\prime(x) = \lambda^{\Delta} \phi(\lambda x)$. 
The commutation
relations \eno{algebra} imply that the operator $P_\mu$
raises the dimension of the field, while the operator $K_\mu$ lowers
it. In unitary field theories there is a lower bound on the dimension
of fields (for scalar fields it is $\Delta \geq (d-2)/2$ which is the
dimension of a free scalar field), and, therefore, each representation
of the conformal group which appears must have some operator of lowest
dimension, which must then be annihilated by $K_\mu$ (at
$x=0$). Such operators are called {\it primary operators}. The action
of the conformal group on such operators is given by \cite{Mack:1969rr}
\eqn{confaction}{\eqalign{
[P_\mu, \Phi(x)] &= i\del_\mu \Phi(x),\cr
[M_{\mu\nu}, \Phi(x)] &= [i(x_\mu \del_\nu - x_\nu \del_\mu) +
\Sigma_{\mu\nu}] \Phi(x), \cr
[D,\Phi(x)] &= i(-\Delta + x^\mu \del_\mu) \Phi(x),\cr
[K_\mu, \Phi(x)] &= [i(x^2\del_\mu - 2x_\mu x^\nu \del_\nu + 2x_\mu
\Delta) - 2x^\nu \Sigma_{\mu\nu}] \Phi(x), }}
where $\Sigma_{\mu\nu}$ are the matrices of a finite dimensional
representation of the Lorentz group, acting on the indices of the
$\Phi$ field. The representations of the conformal group corresponding
to primary operators are classified by the Lorentz representation and
the scaling dimension $\Delta$ (these determine the Casimirs of the
conformal group).  These representations include the primary field and
all the fields which are obtained by acting on it with the generators
of the conformal group (specifically with $P_\mu$).  Since the
operators in these representations are eigenfunctions of $D$, they
cannot in general 
be eigenfunctions of the Hamiltonian $P_0$ or of the mass
operator $M^2 = -P^\mu P_\mu$ (which is a Casimir operator of the
\Poincare group but not of the conformal group); in fact, they have a
continuous spectrum of $M^2$ ranging from $0$ to $\infty$ (there are
also representations corresponding to free massless fields which have
$M^2=0$).

Another possible classification of the representations of the
conformal group is in terms of its maximal compact subgroup, which is
$SO(d)\times SO(2)$. The generator of $SO(2)$ is $J_{0(d+1)} = {1\over
2}(K_0+P_0)$, and the representations of the conformal group described
above may be decomposed into representations of this subgroup. This is
useful in particular for the oscillator constructions of the
representations of superconformal algebras 
\cite{Gunaydin:1982yq,Bars:1983ep,Gunaydin:1985fk,Gunaydin:1985vz,
Gunaydin:1985wc,Gunaydin:1986tc,Gunaydin:1987hb}, which we will not
describe in detail here (see \cite{Minic:1999eq} for a recent review).
This subgroup is also useful in the radial quantization of the
conformal field theory on $S^{d-1}\times \IR$, which will be related to
AdS space in global coordinates.

Since the conformal group is much larger than the \Poincare group, it
severely restricts the correlation functions of primary fields, which
must be invariant under conformal transformations. 
It has been shown by
Luscher and Mack \cite{Luscher:1974ez} that the Euclidean Green's
functions of a CFT may be analytically continued to Minkowski space,
and that the resulting Hilbert space carries a unitary representation
of the Lorentzian conformal group. The formulas we will write here for
correlation functions apply both in Minkowski and in Euclidean space.
It is easy to show using the
conformal algebra that the 2-point functions of fields of different
dimension vanish, while for a single scalar field of scaling dimension
$\Delta$ we have
\eqn{twopoint}{\vev{\phi(0)\phi(x)} \propto 
{1\over |x|^{2\Delta}} \equiv {1\over (x^2)^\Delta}.}
3-point functions are also determined (up to a constant) by the
conformal group to be of the form
\eqn{threepoint}{\vev{\phi_i(x_1) \phi_j(x_2) \phi_k(x_3)} = {c_{ijk}
\over 
|x_1-x_2|^{\Delta_1+\Delta_2-\Delta_3}
|x_1-x_3|^{\Delta_1+\Delta_3-\Delta_2}
|x_2-x_3|^{\Delta_2+\Delta_3-\Delta_1}}. }
Similar expressions (possibly depending on additional constants) arise
for non-scalar fields.
With 4 independent $x_i$ one can construct two combinations of the $x_i$
(known as harmonic ratios)
which are conformally invariant, so the correlation function can be any
function of these combinations; for higher $n$-point functions there are
more and more independent functions which can appear in the
correlation functions. Many other properties of conformal field
theories are also easily determined using the conformal invariance;
for instance, their equation of state is necessarily of the form $S =
c V (E / V)^{(d-1)/d}$ for some constant $c$.

The field algebra of any conformal field theory includes the
energy-momentum tensor $T_{\mu\nu}$ which is an operator of dimension
$\Delta=d$; the Ward identities of the conformal algebra relate
correlation functions with $T$ to correlation functions without
$T$. Similarly, whenever there are global symmetries, their
(conserved) currents $J_\mu$ are necessarily operators of dimension
$\Delta=d-1$. The scaling dimensions of other operators are not
determined by the conformal group, and generally they receive quantum
corrections. For any type of field there is, however, a lower bound on
its dimension which follows from unitarity; as mentioned above, for
scalar fields the bound is $\Delta \geq (d-2)/2$, where equality can
occur only for free scalar fields.

A general property of local field theories is the existence of an {\it
operator product expansion} (OPE). As we bring two operators
$\co_1(x)$ and $\co_2(y)$ to the same point, their product creates a
general local disturbance at that point, which may be expressed as a
sum of local operators acting at that point; in general all operators
with the same global quantum numbers as $\co_1 \co_2$ may appear. The
general expression for the OPE is $\co_1(x) \co_2(y) \to \sum_n
C_{12}^n(x-y) \co_n(y)$, where this expression should be understood as
appearing inside correlation functions, and the coefficient functions
$C_{12}^n$ do not depend on the other operators in the correlation
function (the expression is useful when the distance to all other
operators is much larger than $|x-y|$). In a conformal theory, the
functional form of the OPE coefficients is determined by conformal
invariance to be $C_{12}^n(x-y) = c_{12}^n /
|x-y|^{\Delta_1+\Delta_2-\Delta_n}$, where the constants $c_{12}^n$
are related to the 3-point functions described above. The leading
terms in the OPE of the energy-momentum tensor with primary fields are
determined by the conformal algebra. For instance, for a scalar
primary field $\phi$ of dimension $\Delta$ in four dimensions,
\eqn{emope}{T_{\mu \nu}(x) \phi(0) \propto \Delta \phi(0) \del_\mu
\del_\nu({1\over {x^2}}) + \cdots.}

One of the basic properties of conformal field theories is the
one-to-one correspondence between local operators $\co$ and states
$|\co \rangle$ in the radial quantization of the theory. In radial
quantization the time coordinate is chosen to be the radial direction
in $\IR^d$, with the origin corresponding to past infinity, so that
the field theory lives on $\IR \times S^{d-1}$. The Hamiltonian in
this quantization is the operator $J_{0(d+1)}$ mentioned above. An
operator $\co$ can then be mapped to the state $|\co \rangle =
\lim_{x\to 0} \co(x) |0\rangle$. Equivalently, the state may be viewed
as a functional of field values on some ball around the origin, and
then the state corresponding to $\co$ is defined by a functional
integral on a ball around the origin with the insertion of the
operator $\co$ at the origin. The inverse mapping of states to
operators proceeds by taking a state which is a functional of field
values on some ball around the origin and using conformal invariance
to shrink the ball to zero size, in which case the insertion of the
state is necessarily equivalent to the insertion of some local
operator.

\subsection{Superconformal Algebras and Field Theories}
\label{superconfalg}

Another interesting generalization of the \Poincare algebra is the
supersymmetry algebra, which includes additional fermionic operators
$Q$ which anti-commute to the translation operators $P_\mu$. It is
interesting to ask whether supersymmetry and the conformal group can
be joined together to form the largest possible simple algebra
including the \Poincare group; it turns out that in some dimensions and
for some numbers of supersymmetry charges this is indeed
possible. The full classification of superconformal algebras was
given by Nahm \cite{Nahm:1978tg}; 
it turns out that superconformal algebras exist
only for $d\leq 6$. In addition to the generators of the conformal
group and the supersymmetry, superconformal algebras include two other
types of generators. There are fermionic generators $S$ (one for each
supersymmetry generator) which arise in the commutator of $K_\mu$ with
$Q$, and there are (sometimes) R-symmetry generators forming some Lie
algebra, which appear in the anti-commutator of $Q$ and
$S$ (the generators $Q$ and $S$ are in the fundamental representation
of this Lie algebra). 
Schematically (suppressing all indices), 
the commutation relations of the superconformal
algebra include, in addition to \eno{algebra}, the relations
\eqn{superalgebra}{\eqalign{
[D,Q] &= -{i\over 2}Q;\qquad [D,S] = {i\over 2}S;\qquad [K,Q] \simeq
S;\qquad [P,S] \simeq Q;\cr
\{Q,Q\} &\simeq P;\qquad \{S,S\} \simeq K;\qquad \{Q,S\} \simeq
M+D+R. }}
The exact form of the commutation relations is different for different
dimensions (since the spinorial representations of the conformal group
behave differently) and for different R-symmetry groups,
and we will not write them explicitly here.

For free field theories without gravity, which do not include fields
whose spin is bigger than one, the maximal possible number of
supercharges is 16 (a review of field theories with this number of
supercharges appears in \cite{Seiberg:1997ax}); it is believed that
this is the maximal possible number of supercharges also in
interacting field theories. Therefore, the maximal possible number of
fermionic generators in a field theory superconformal algebra is
32. Superconformal field theories with this number of supercharges
exist only for $d=3,4,6$ ($d=1$ may also be possible but there are no
known examples). For $d=3$ the R-symmetry group is $Spin(8)$ and the
fermionic generators are in the $\bf(4,8)$ of $SO(3,2)\times Spin(8)$;
for $d=4$ the R-symmetry group is $SU(4)$ and\footnote{Note that this
is different from the other $\cn$-extended superconformal algebras in
four dimensions which have a $U(\cn)$ R-symmetry.} the fermionic
generators are in the $\bf(4,4)+\bf({\overline 4},{\overline 4})$ of
$SO(4,2)\times SU(4)$; and for $d=6$ the R-symmetry group is
$Sp(2)\simeq SO(5)$ and the fermionic generators are in the $\bf(8,4)$
representation of $SO(6,2)\times Sp(2)$.

Since the conformal algebra is a subalgebra of the superconformal
algebra, representations of the superconformal algebra split into
several representations of the conformal algebra. Generally a primary
field of the superconformal algebra, which is (by definition)
annihilated (at $x=0$) by the generators $K_\mu$ and $S$, will include
several primaries of the conformal algebra, which arise by acting with
the supercharges $Q$ on the superconformal primary field. The
superconformal algebras have special representations corresponding to
{\it chiral primary operators}, which are primary operators which are
annihilated by some combination of the supercharges. These
representations are smaller than the generic representations,
containing less conformal-primary fields. A special property of chiral
primary operators is that their dimension is uniquely determined by
their R-symmetry representations and cannot receive any quantum
corrections. This follows by using the fact that all the $S$
generators and some of the $Q$ generators annihilate the field, and
using the $\{Q,S\}$ commutation relation to compute the eigenvalue of
$D$ in terms of the Lorentz and R-symmetry representations
\cite{Kac:1977hp,Dobrev:1987qz,Dobrev:1985qv,Seiberg:1997ax,
Minwalla:1998ka}. The
dimensions of non-chiral primary fields of the same representation are
always strictly larger than those of the chiral primary fields. A
simple example is the $d=4,\cn=1$ superconformal algebra (which has a
$U(1)$ R-symmetry group); in this case a chiral multiplet (annihilated
by $\overline Q$) which is a primary is also a chiral primary, and the
algebra can be used to prove that the dimension of the scalar
component of such multiplets is $\Delta={3\over 2}R$ where $R$ is the
$U(1)$ R-charge. A detailed description of the structure of chiral
primaries in the $d=4,\cn=4$ algebra will appear in section \ref{tests}.

When the R-symmetry group is Abelian, we find a bound of the form
$\Delta \geq a|R|$ for some constant $a$, ensuring that there is no
singularity in the OPE of two chiral ($\Delta=aR$) or anti-chiral
($\Delta=a|R|=-aR$) operators. On the other hand, when the R-symmetry
group is non-Abelian, singularities can occur in the OPEs of chiral
operators, and are avoided only when the product lies in particular
representations.

\section{Anti-de Sitter Space} 
\label{adsgeom}

\subsection{Geometry of Anti-de Sitter Space}

In this section, we will review some geometric facts about 
anti-de Sitter space. One of the important facts is
the relation between the conformal compactifications of $AdS$
and of flat space. In the case of the Euclidean signature
metric, it is well-known that the flat
space $\IR^n$ can be compactified to the $n$-sphere $S^n$ by 
{\it adding a point at infinity}, and a conformal
field theory is naturally defined on $S^n$. On the other hand, 
the $(n+1)$-dimensional hyperbolic space, which is the Euclidean
version of $AdS$ space, can be conformally
mapped into the $(n+1)$-dimensional disk $D_{n+1}$.
Therefore the boundary
of the compactified hyperbolic space is the compactified
Euclid space. A similar correspondence holds in the 
case with the Lorentzian signature metric, as we will see below. 


\subsubsection{Conformal Structure of Flat Space}
\label{confflatspace}

One of the basic features  of the {\it AdS}/CFT correspondence
is the identification of the isometry group of \adsp\ with the conformal
symmetry of flat Minkowski space $\IR^{1,p}$. Therefore, it would be
appropriate to start our discussion by reviewing the conformal structure of 
flat space. 

\bigskip
\noindent $\circ$ $\IR^{1,1}$

\medskip

\begin{figure}[htb]
\begin{center}
\epsfxsize=2.5in\leavevmode\epsfbox{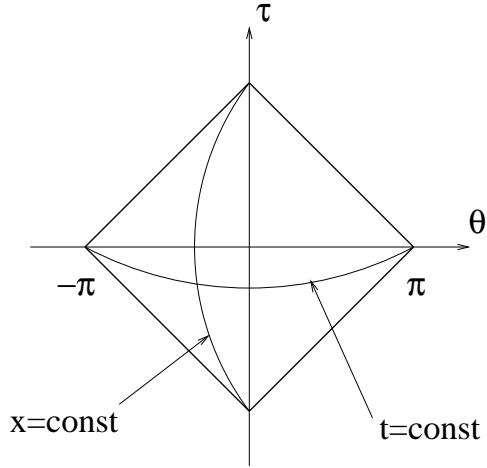}
\end{center}
\caption{Two-dimensional Minkowski space is conformally mapped
into the interior of the rectangle.}
\label{F1}
\end{figure}

We begin with two-dimensional Minkowski
space $\IR^{1,1}$:
\beq
  ds^2 = - dt^2 + dx^2,~~~~~~~~(-\infty < t, x < + \infty).
\label{flat1}
\eeq
This metric can be rewritten by the following coordinate
transformations
\ber
  ds^2 & = & - du_+ du_- ~,~~~~~~~~~(u_\pm = t \pm x) \nonumber \\
          & = & \frac{1}{4\cos^2\tilde{u}_+\cos^2 \tilde{u}_-}
            (-d\tau^2 + d\theta^2) ~,~~~~~
 (u_\pm = \tan \tilde{u}_\pm; ~
\tilde{u}_\pm  = (\tau \pm \theta)/2).
\label{flat2}
\eer
In this way, the Minkowski space is conformally mapped into 
the interior of the compact region, $| \tilde u_\pm| < \pi/2$, 
as shown in figure \ref{F1}. Since light ray  trajectories are 
invariant under a conformal rescaling of the metric, this
provides a convenient way to express the causal structure of
$\IR^{1,1}$. The new coordinates $(\tau, \theta)$ are well
defined at the asymptotical regions of the flat space. Therefore,
the conformal compactification is used to give a
rigorous definition of {\it asymptotic flatness} of spacetime
--- a spacetime is called asymptotically flat if it has the same
boundary structure as that of the flat space after conformal
compactification. 
\begin{figure}[htb]
\begin{center}
\epsfxsize=4in\leavevmode\epsfbox{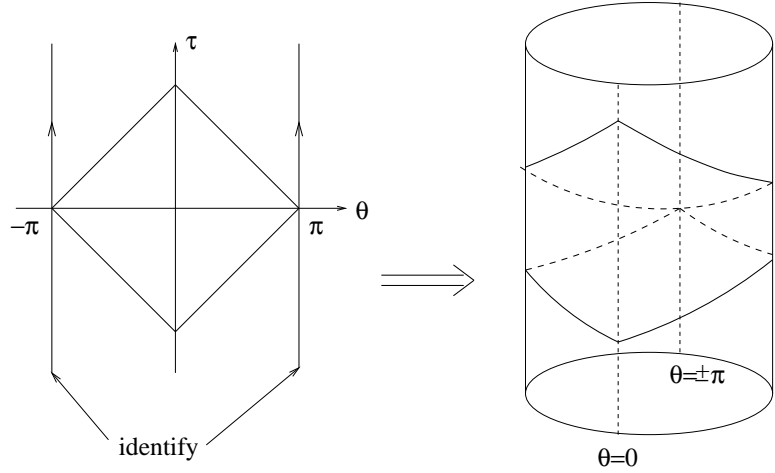}
\end{center}
\caption{The rectangular region can be embedded in
a cylinder, with $\theta=\pi$ and $\theta=-\pi$ being identified.}
\label{F2}
\end{figure} 

The two corners of the rectangle at
$(\tau, \theta) = (0, \pm \pi)$ correspond
to the spatial infinities $x=\pm \infty$
in the original coordinates. By identifying these
two points, we can embed the rectangular image of $\IR^{1,1}$ 
in a cylinder $\IR \times S^1$ as shown in figure \ref{F2}.
It was proven by L\"uscher and Mack  \cite{Luscher:1974ez} that
correlation functions of
a conformal field theory (CFT) on $\IR^{1,1}$ can be analytically
continued to the entire cylinder. 

As we saw in section \ref{cft}, 
the global conformal symmetry of $\IR^{1,1}$ is
$SO(2,2)$, which is
generated by the 6 conformal Killing vectors
$\partial_\pm, u_\pm \partial_\pm, u_\pm^2 \partial_\pm$.
The translations 
along the cylinder $\IR \times S^1$ are expressed as their linear
combinations
\beq
 \frac{\partial}{\partial \tau} \pm
  \frac{\partial}{\partial \theta}=   
\frac{\partial}{\partial \tilde{u}_\pm} = 
 (1 + u_\pm^2) \frac{\partial}{\partial u_\pm}.
\label{cylindertranslation}
\eeq
In the standard form of $SO(2,2)$ generators, $J_{ab}$,
given in section \ref{cft}, they 
correspond to $J_{03}$ and $J_{12}$, 
and generate the maximally compact subgroup 
$SO(2) \times SO(2)$ of $SO(2,2)$.

\bigskip
\noindent $\circ$ $\IR^{1,p}$ with $p \geq 2$

\medskip
It is straightforward to extend
the above analysis to higher dimensional Minkowski
space:
\beq
  ds^2 = -dt^2 + dr^2 + r^2 d\Omega_{p-1}^2,
\label{flat3}
\eeq
where $d\Omega_{p-1}$ is the line element on the unit
sphere $S^{p-1}$. A series of coordinate changes transforms
this as
\ber
  ds^2 & = & - du_+ du_- + \frac{1}{4}(u_+-u_-)^2 d\Omega_{p-1}^2
~,~~~~~~~~ (u_\pm = t \pm r) \nonumber \\
  & = & \frac{1}{\cos^2\tilde{u}_+\cos^2\tilde{u}_-}
                  \left(- d\tilde{u}_+d\tilde{u}_-
   + \frac{1}{4} \sin^2(\tilde{u}_+ - \tilde{u}_-) d\Omega_{p-1}^2 \right)
           ~,~~~~~~~(u_\pm = \tan \tilde{u}_\pm) \nonumber \\
     & = & \frac{1}{4\cos^2\tilde{u}_+\cos^2\tilde{u}_-}
            (-d\tau^2 + d\theta^2 + 
            \sin^2 \theta d \Omega_{p-1}^2), ~~~~~~~~
(\tilde{u}_\pm = (\tau \pm \theta)/2).
\label{flat4}
\eer

\begin{figure}[htb]
\begin{center}
\epsfxsize=2.0in\leavevmode\epsfbox{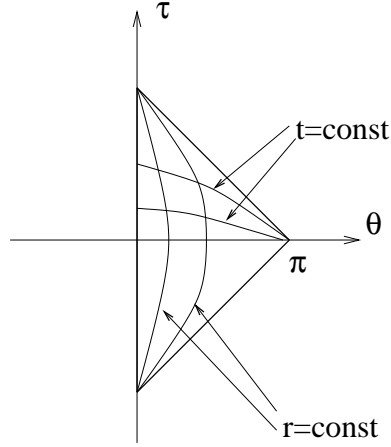}
\end{center}
\caption{The conformal transformation maps the 
$(t,r)$ half plane into a triangular region in
the $(\tau,\theta)$ plane. }
\label{F3}
\end{figure} 

As shown in figure \ref{F3}, 
the $(t,r)$ half-plane (for a fixed point on $S^{p-1}$)
is mapped into a triangular region in the $(\tau,\theta)$ plane.
The conformally scaled metric 
\beq
ds'^2 = -d\tau^2 + d\theta^2 + \sin^2 \theta d \Omega_{p-1}^2
\label{flat5}
\eeq 
can be analytically continued outside
of the triangle, and the maximally extended space with
\beq
   0 \leq \theta \leq \pi, ~~-\infty < \tau < + \infty,
\eeq
has the
geometry of $\IR \times S^p$ (Einstein static universe),
where $\theta=0$ and $\pi$ corresponds
to the north and south poles of $S^p$. This is a natural
generalization of the conformal embedding of $\IR^{1,1}$ into
$\IR \times S^1$ that we saw in the $p=1$ case.  
  
Since
\beq
  \frac{\partial}{\partial \tau} = \frac{1}{2} (1 + u_+^2) 
\frac{\partial}{\partial u_+}
      + \frac{1}{2}(1+u_-^2) \frac{\partial}{\partial u_-},
\eeq
the generator $H$ of the global time translation on $\IR \times S^p$
is identified with the linear combination
\beq
  H = \frac{1}{2}(P_0 + K_0) = J_{0,p+2},
\label{globaltime}
\eeq
where $P_0$ and $K_0$ are translation and special conformal
generators,
\beq
  P_0: {1 \over 2}\left( \frac{\partial}{\partial u_+}
 + \frac{\partial}{\partial u_-} \right), ~~
  K_0:  {1 \over 2}\left( u_+^2 \frac{\partial}{\partial u_+}
 + u_-^2 \frac{\partial}{\partial u_-} \right)
\eeq
on $\IR^{1,p}$ defined in section \ref{cft}. 
The generator $H = J_{0,p+2}$ corresponds to the $SO(2)$ part 
of the maximally compact subgroup $SO(2) \times SO(p+1)$ 
of $SO(2,p+1)$. Thus the subgroup $SO(2) \times
SO(p+1)$ (or to be precise its universal cover)
of the conformal group $SO(2,p+1)$ can be identified
with  the isometry of the Einstein static universe $\IR \times S^p$.
The existence of the generator $H$ also guarantees
that a correlation function of a CFT on $\IR^{1,p}$ can
be analytically extended to the entire Einstein static universe
$\IR \times S^p$.

\begin{figure}[htb]
\begin{center}
\epsfxsize=2.5in\leavevmode\epsfbox{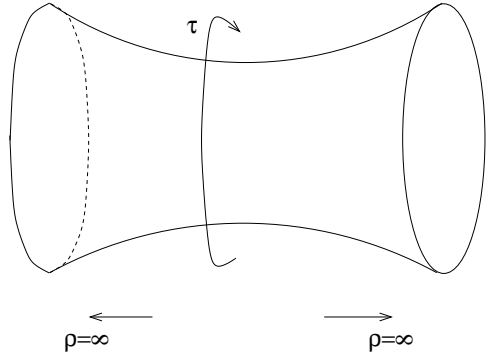}
\end{center}
\caption{\adsp\ is realized as a hyperboloid in $\IR^{2,p+1}$.
The hyperboloid has closed timelike curves along the $\tau$
direction. To obtain a causal space, we need to unwrap the
circle to obtain a simply connected space.}
\label{F3.5}
\end{figure}

\subsubsection{Anti-de Sitter Space}


The $(p+2)$-dimensional anti-de Sitter space (\adsp)
can be represented as the hyperboloid 
\beq
   X_{0}^2 + X_{p+2}^2 - \sum_{i=1}^{p+1} X_i^2  = R^2, 
\label{hyperboloid}
\eeq 
in the flat $(p+3)$-dimensional space with metric
\beq
  ds^2 = - dX_{0}^2 - dX_{p+2}^2 + \sum_{i=1}^{p+1} dX_i^2.
\label{metric1}
\eeq 
By construction, the space has the isometry $SO(2,p+1)$,
and it is homogeneous and isotropic. 

Equation (\ref{hyperboloid}) can be solved by setting
\ber
   X_{0} &=& R \cosh \rho\ \cos \tau, 
 ~~~~~ X_{p+2} = R \cosh \rho\ \sin \tau, 
  \nonumber \\
   X_i &=& R \sinh \rho\ \Omega_{i} ~~~(i=1,\cdots,p+1; 
 \sum_i \Omega_i^2 = 1).
\label{globalcoord}
\eer
Substituting this into
(\ref{metric1}), we obtain the metric on \adsp\ as
\beq
  ds^2 = R^2 ( -\cosh^2\rho\ d\tau^2 + d\rho^2 + \sinh^2\rho\
                    d\Omega^2 ).
\label{metric2}
\eeq
By taking $0 \leq \rho$ and $0 \leq \tau < 2 \pi$
the solution (\ref{globalcoord}) covers the entire hyperboloid
once. Therefore, $(\tau,\rho, \Omega_i)$ are called the global
coordinates of \ads.
Since the metric behaves near $\rho = 0$ as
$ds^2 \simeq R^2 ( -d\tau^2 + d \rho^2 + \rho^2\ d\Omega^2)$,
the hyperboloid has the topology of $S^1 \times \IR^{p+1}$, with
$S^1$ representing closed timelike curves in the $\tau$ direction.
To obtain a causal spacetime, we can simply unwrap the circle $S^1$
(i.e. take $-\infty < \tau < \infty$ with no identifications)
and obtain the universal covering of the hyperboloid without closed
timelike curves. In this paper, when we refer to \adsp ,
we only consider this universal covering space. 

The isometry group $SO(2,p+1)$ of \adsp\ has the maximal compact subgroup
$SO(2) \times SO(p+1)$.  
From the above construction, it is clear that the $SO(2)$ part 
represents
the constant translation in the $\tau$ direction, and
the $SO(p+1)$ gives rotations of $S^p$.

\begin{figure}[htb]
\begin{center}
\epsfxsize=3in\leavevmode\epsfbox{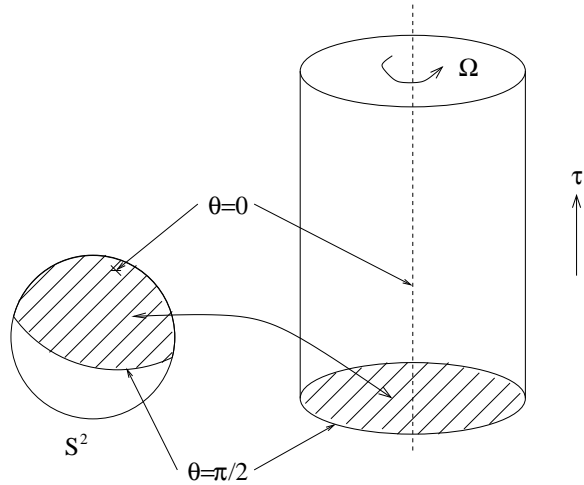}
\end{center}
\caption{\ads$_3$ can be conformally mapped into one half
of the Einstein static universe $\IR \times S^2$.}
\label{F4}
\end{figure} 

To study the causal structure of \adsp, it is convenient to introduce
a coordinate $\theta$ related to $\rho$ by $\tan \theta = \sinh \rho$
($0 \leq \theta < \pi/2$). The metric (\ref{metric2}) then takes
the form
\beq
   ds^2 = \frac{R^2}{\cos^2 \theta}
        ( - d\tau^2 + d\theta^2 + \sin^2 \theta\ d\Omega^2).
\label{metric3}
\eeq
The causal structure of the spacetime does not change by a conformal
rescaling on the metric. Multiplying the metric by 
$R^{-2} \cos^2 \theta$, it
becomes 
\beq
 ds'^2 =  - d\tau^2 + d\theta^2 + \sin^2 \theta\ d\Omega^2.
\label{metric4}
\eeq
This is the metric of the Einstein static universe, which
also appeared, with the dimension lower by one, in
the conformal compactification of $\IR^{1,p}$ (\ref{flat5}). 
This time, however,  
the coordinate $\theta$ takes values in $0 \leq \theta < \pi/2$, rather
than $0 \leq \theta < \pi$ in (\ref{flat5}). Namely, \adsp\ can be 
conformally mapped
into {\it one half} of the Einstein static universe; the spacelike
hypersurface of constant $\tau$ is a $(p+1)$-dimensional
hemisphere.  The equator at $\theta = \pi/2$ is a boundary of the space
with the topology of $S^p$, as shown in figure \ref{F4} in the case of
$p=1$. (In the case of $AdS_2$, the coordinate $\theta$ ranges
$-\pi/2 \leq \theta \leq \pi/2$ since $S^0$ consists of two points.)
As in the case of the flat space discussed earlier, the
conformal compactification is a convenient way to describe the
asymptotic regions of \ads . In general, if a spacetime can
be conformally compactified into a region which has the same
boundary structure as one half of the Einstein static universe,
the spacetime is called {\it asymptotically AdS}. 

Since the boundary extends in the timelike
direction labeled by $\tau$, 
we need to specify a boundary condition on
the $\IR \times S^p$ at $\theta = \pi/2$ in order
to make the Cauchy problem well-posed on \ads\ \cite{Avis:1978yn}. 
It turns out that the boundary of \adsp , or to be precise the
boundary of the conformally compactified \adsp , is identical
to the conformal compactification of the $(p+1)$-dimensional
Minkowski space. This fact plays an essential role
in the \adsp/CFT$_{p+1}$ correspondence.

\begin{figure}[htb]
\begin{center}
\epsfxsize=3in\leavevmode\epsfbox{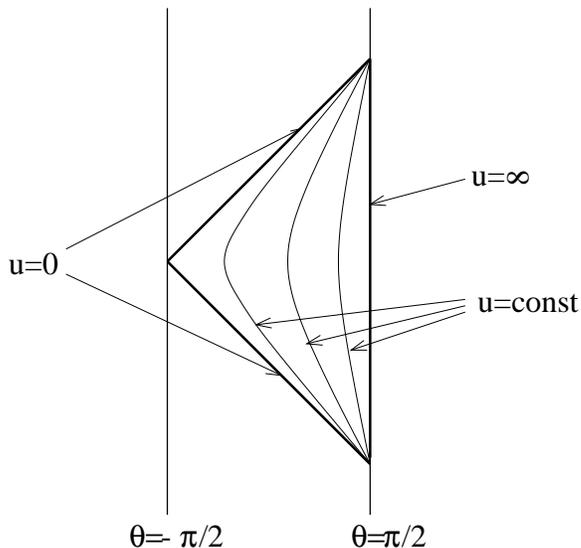}
\end{center}
\caption{\ads$_2$ can be conformally mapped into $\IR \times [-\pi/2,\pi/2]$.
The $(u,t)$ coordinates cover the triangular region.}
\label{F5}
\end{figure} 

In addition to the global parametrization (\ref{globalcoord})
of \ads, there is another set of coordinates $(u,t,\vec{x})$ ($0 < u,
\vec{x} \in \IR^p$) which will be useful
later. It is defined by
\ber
   X_{0} &=& \frac{1}{2u}\left( 1 + u^2 (R^2 + \vec{x}^{~2} - t^2)
   \right), ~~~~~ 
   X_{p+2} = Rut, \nonumber \\
  X^{i} &=& Ru x^i~~~(i=1,\cdots,p), \nonumber \\
   X^{p+1} &=& \frac{1}{2u}\left( 1 - u^2(R^2 - \vec{x}^{~2} +
     t^2)    \right)  .
\label{poincarecoord}
\eer
These coordinates cover one half of the hyperboloid
(\ref{hyperboloid}), as shown in figure \ref{F5} in the case
of $p=0$. Substituting this into (\ref{metric1}), we obtain
another form of the \adsp\ metric
\beq
   ds^2 = R^2 \left( \frac{du^2}{u^2} + u^2 ( -dt^2 + d \vec{x}^{~2})
   \right).
\label{metric5}
\eeq
The coordinates $(u, t, \vec{x})$ are called the Poincar\'e
coordinates. In this form of the metric, the subgroups 
$ISO(1,p)$ and $SO(1,1)$ of the $SO(2,p+1)$ isometry are manifest,
where $ISO(1,p)$ is the Poincar\'e transformation on $(t, \vec{x})$
and $SO(1,1)$ is 
\beq
    (t, \vec{x}, u) \rightarrow (c t, c \vec{x}, c^{-1} u), ~~~
     c>0.
\eeq
In the \ads/CFT correspondence,
this is identified with the dilatation $D$ in the conformal symmetry
group of $\IR^{1,p}$.

\begin{figure}[htb]
\begin{center}
\epsfxsize=1.8in\leavevmode\epsfbox{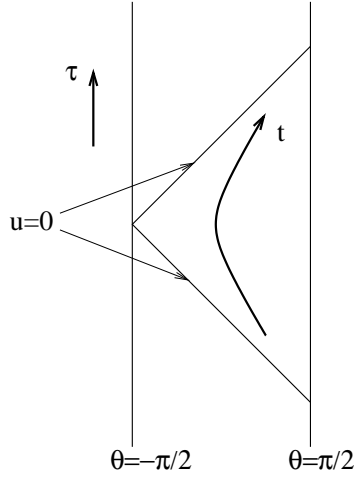}
\end{center}
\caption{The timelike Killing vector $\partial_t$ is depicted
in the $AdS_2$ case. The vector $\partial_t$
becomes a null vector at $u=0$.}
\label{F6}
\end{figure} 

It is useful to compare the two expressions, (\ref{metric2}) and
(\ref{metric5}), for the metric
of \adsp. In (\ref{metric2}), the norm of the timelike 
Killing vector $\partial_\tau$ is everywhere non-zero. In particular,
it has a constant norm in the conformally rescaled metric
(\ref{metric3}). For this reason, $\tau$ is called the global time
coordinate of \ads. On the other hand, the timelike Killing vector
$\partial_t$ in (\ref{metric5}) becomes null at $u=0$ (Killing
horizon), as depicted in figure \ref{F6} in the $AdS_2$ case. 

\subsubsection{Euclidean Rotation}


Since \adsp\ has the global time coordinate $\tau$ and
the metric (\ref{metric2}) is static with respect to $\tau$,
quantum field theory on \adsp\ (with an appropriate boundary condition
at spatial infinity) allows the Wick rotation
in $\tau$, $e^{i\tau H} \rightarrow e^{- \tau_E H}$. From
(\ref{globalcoord}), one finds that
the Wick rotation $\tau \rightarrow \tau_E = -i\tau$ is expressed
in the original coordinates $(X_{0},  \vec{X}, X_{p+2})$ on the hyperboloid
as $X_{p+2} \rightarrow X_E = -i X_{p+2}$, and the space becomes
\ber
    & & X_{0}^2 - X_E^2 - \vec{X}^2 = R^2, \nonumber \\ 
    & & ds_E^2 = -dX_{0}^2 + dX_E^2 + d \vec{X}^2 .
\label{eads}
\eer    

We should point out that the same space is obtained by
rotating the time coordinate $t$ of the Poincar\'e coordinates
(\ref{poincarecoord}) as $t \rightarrow t_E = -i t$, even
though the Poincar\'e coordinates cover only a part of the entire
\ads\ (half of the hyperboloid).  This is analogous to 
the well-known fact in
flat Minkowski space that the Euclidean rotation of the time 
coordinate $t$ in the Rindler space $ds^2 = -r^2 dt^2 + dr^2$ gives 
the flat Euclidean plane $\IR^2$, even though the Rindler coordinates
$(t,r)$ cover only a $1/4$ of the entire Minkowski space $\IR^{1,1}$.

 In the coordinates
$(\rho, \tau_E, \vec{\Omega}_p)$ and $(u, t_E, \vec{x})$,
the Euclidean metric is expressed as
\ber
   ds_E^2 & = & R^2 \left( \cosh^2 \rho\ d\tau_E^2 + d\rho^2
              + \sinh^2 \rho \ d \Omega_p^2 \right) \nonumber \\
     & = & R^2 \left( \frac{du^2}{u^2} + u^2 (dt_E^2 + d\vec{x}^2 )
     \right).
\label{eads2}
\eer
In the following, we also use another, trivially equivalent,
form of the metric, obtained from the above by
setting $u =1/y$ in (\ref{eads2}), giving
\eqn{poincusual}{
ds^2 = R^2 \left( { dy^2 + dx_1^2 + \cdots + dx^2_{p+1} \over y^2 } \right).
}

\begin{figure}[htb]
\begin{center}
\epsfxsize=4.5in\leavevmode\epsfbox{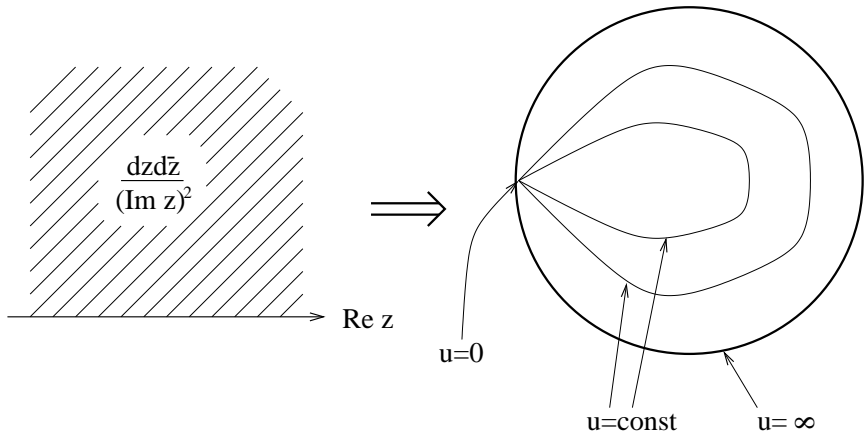}
\end{center}
\caption{The Euclidean $AdS_2$ is the upper half plane with
the Poincar\'e metric. It can be mapped into a disk, where
the infinity of the upper half plane is mapped to a point
on the boundary of the disk.}
\label{F7}
\end{figure} 

The Euclidean \adsp\ is useful for various practical
computations in field theory. For theories on flat space,
it is well-known that 
correlation functions $\langle \phi_1 \cdots \phi_n \rangle$
of fields on the Euclidean space are related,
by the Wick rotation, to the $T$-ordered 
correlation functions $\langle 0 | T (\phi_1 \cdots \phi_n) | 
0 \rangle$ in the Minkowski space. 
The same is true in the anti-de Sitter space if the theory has
a positive definite Hamiltonian with respect to the global
time coordinate $\tau$. Green functions of free fields on 
\adsp\ have been computed in  
\cite{Burgess:1985ti,Inami:1985wu} using this method.

The Euclidean \adsp\ can be mapped into a $(p+2)$-dimensional
disk. In the coordinates $(u, t_E, \vec{x})$,
$u=\infty$ is the sphere $S^{p+1}$
at the boundary with one point removed. The full boundary
sphere is recovered by adding a point corresponding to 
$u=0$ (or equivalently $\vec{x} = \infty$). This
is shown in figure \ref{F7} in the case of \ads$_2$,
for which $z=t_E + i/u$ gives a complex coordinate on
the upper-half plane. By adding a point at infinity,
the upper-half plane is compactified into a disk. 
 In the Lorentzian case,
$u=0$ represented the Killing horizon giving the boundary
of the $(u,t,\vec{x})$ coordinates. Since the $u=0$ plane
is null in the Lorentzian case, it shrinks to a point
in the Euclidean case.

\subsection{Particles and Fields in Anti-de Sitter Space}
\label{pfinads}

Massive particles, moving along geodesics,
 can never get to the boundary of $AdS$. 
On the other hand, since the Penrose diagram of $AdS$ is a cylinder,
 light rays
can go to the boundary and back in finite time, as observed by
an observer moving along a geodesic in AdS.
More precisely, the light ray will reflect if suitable boundary 
conditions are set for the fields propagating in $AdS$. 

Let us first consider the case of a scalar field propagating in 
\adsp. The 
field equation
\beq
  \left(\Delta - m^2 \right) \phi = 0
\label{kg}
\eeq
has stationary wave solutions
\beq
    \phi = e^{i\omega \tau} G(\theta) Y_l(\Omega_p),
\eeq
where $Y_l(\Omega_p)$ is a spherical harmonic, which 
is an eigenstate of the Laplacian on $S^p$ with
an eigenvalue $l(l+p-1)$, and  
$G(\theta)$ is given by the hypergeometric function
\beq
   G(\theta) = (\sin \theta)^l (\cos \theta)^{\lambda_\pm}
 ~_2F_1 \left(a,b,c; \sin \theta\right),
\label{hyper}
\eeq
with
\ber
   a & = & \frac{1}{2}(l+\lambda_\pm - \omega R), \nonumber \\
   b & = & \frac{1}{2}(l+\lambda_\pm + \omega R), \nonumber \\
   c & = & l + \frac{1}{2}(p+1),
\eer
and
\beq
   \lambda_\pm = \frac{1}{2} (p+1) \pm \frac{1}{2}\sqrt{(p+1)^2 + 4
     (m R)^2 }.
\label{branch}
\eeq
The energy-momentum tensor
\beq
   T_{\mu\nu} = 2 \partial_\mu \phi \partial_\nu \phi
     -g_{\mu\nu}\left( (\partial \phi)^2 + m^2  \phi^2 \right)
   + \beta(g_{\mu\nu} \Delta - D_\mu D_\nu + R_{\mu\nu}) \phi^2
\eeq
is conserved for any constant value of $\beta$. The value of
$\beta$ is determined by the coupling of the scalar curvature
to $\phi^2$, which on \ads\ has the same effect as  the mass term in the
wave equation (\ref{kg}). The choice of $\beta$ for
each scalar field depends on the theory we are considering.
The total energy $E$ of the scalar field fluctuation,
\beq
   E = \int d^{p+1} x \sqrt{-g} T^0_0,
\eeq
is conserved only if the energy-momentum flux through the 
boundary at $\theta=\pi/2$ vanishes,
\beq
\int_{S^p} d\Omega_p \sqrt{g} n_i T^i_{0|\theta = \pi/2} 
= 0.
\eeq
This requirement reduces to the boundary condition
\beq
   (\tan \theta)^p \left[ (1-2\beta) \partial_{\theta}
   + 2\beta \tan \theta \right] \phi^2 \rightarrow 0
 ~~~(\theta \rightarrow \pi/2).
\eeq
Going back to the stationary wave solution (\ref{hyper}), 
this is satisfied if and only if either $a$ or $b$ in (\ref{hyper})
is an integer. If we require the energy $\omega$
to be real, we find
\beq
  |\omega| R = \lambda_\pm + l + 2n,~~~~(n=0,1,2,\cdots).
\label{energyquanta}
\eeq  
This is possible only when $\lambda$ defined by (\ref{branch}) is
real. Consequently, the mass is bounded from below as
\beq
  - \frac{1}{4}(p+1)^2 \leq m^2 R^2  .
\label{posbound} \eeq
This is known as the Breitenlohner-Freedman bound 
\cite{Breitenlohner:1982bm,Breitenlohner:1982jf}. Note
that a negative (mass)$^2$ is allowed to a certain extent.
The Compton wavelength for these possible tachyons is comparable
to the curvature radius of $AdS$. 
If $m^2>-(p-1)(p+3)/4R^2$, we should choose 
$\lambda_+ $ in (\ref{energyquanta})
since this solution is normalizable while the solution with $\lambda_-$
is not.
If $m^2\leq -(p-1)(p+3)/4R^2$, both solutions are
normalizable and there are two different
quantizations of the scalar field on $AdS$ space.
Which quantization to choose is often determined by
requiring symmetry. See 
\cite{Breitenlohner:1982jf,Hawking:1983m,Mezincescu:1984iu}
for discussions of boundary conditions in supersymmetric
theories. 
In general, all solutions to the wave equation form a single 
$SO(2,p+1)$ highest weight representation. The highest weight state
is the lowest energy solution \cite{Heidenreich:1982rz}. 
Since $SO(2,p+1)$ acts on $AdS$ as
isometries, the action of its generators 
on the solutions is given by 
first order differential operators.


\subsection{Supersymmetry in Anti-de Sitter Space}
\label{susyinads}

The $SO(2,p+1)$ isometry group of \adsp\ has a supersymmetric 
generalization called an \ads\ supergroup. To understand the supersymmetry on 
\ads , it would be useful to start with the simple supergravity 
with a cosmological constant $\Lambda$. 
In four dimensions, for example, 
the action of the ${\cal N}=1$ theory is \cite{Townsend:1977qa}
\beq
  S = \int d^4 x \left(
      - \sqrt{g} ({\cal R} - 2\Lambda) + \frac{1}{2}
      \epsilon^{\mu\nu\rho\sigma} \bar{\psi}_\mu \gamma^5
   \gamma_\nu \tilde{D}_\rho \psi_\sigma \right),
\label{simpleaction}
\eeq
where 
\beq
 \tilde{D}_\mu = D_\mu + \frac{i}{2} \sqrt{\frac{\Lambda}{3}} \gamma_\mu
\eeq
and $D_\mu$ is the standard covariant derivative. The local
supersymmetry transformation rules for the vierbein $V_{a\mu}$
and the gravitino $\psi_\mu$ are
\ber
  \delta V_{a\mu} & = & -i \bar{\epsilon}(x) \gamma_a \psi_\mu,
  \nonumber \\
  \delta \psi_\mu & = & \tilde{D}_\mu \epsilon(x).
\label{localsusy}
\eer 

A global supersymmetry of a given supergravity background is
determined by requiring that the gravitino variation is annihilated,
$\delta \psi_\mu = 0$. The resulting condition on $\epsilon(x)$,
\beq
  \tilde{D}_\mu \epsilon = \left(D_\mu + \frac{i}{2} 
\sqrt{\frac{\Lambda}{3}} \gamma_\mu
  \right) \epsilon = 0,
\label{killing}
\eeq
is known as the Killing spinor equation. The integrability of
this equation requires
\beq
 [\tilde{D}_\mu, \tilde{D}_\nu ] \epsilon
 = \frac{1}{2} ({\cal R}_{\mu\nu\rho\sigma} \sigma^{\rho\sigma}
        - \frac{2}{3} \Lambda \sigma_{\mu\nu}) \epsilon = 0,
\label{integrable}
\eeq
where
\beq
\sigma_{\mu\nu} = \frac{1}{2} \gamma_{[\mu,} \gamma_{\nu]}.
\eeq
Since \ads\ is maximally symmetric, the curvature obeys
\beq
   {\cal R}_{\mu\nu\rho\sigma} = {1 \over R^2} (g_{\mu\rho}g_{\nu\sigma}
        - g_{\mu\sigma}g_{\nu\rho}) ~,
\eeq
where $R$ is the size of the hyperboloid defined by (\ref{hyperboloid}).
Thus, if we choose the curvature of \ads\ to be $\Lambda = 3/R^2$ 
(this is necessary for \ads\ to be a classical solution of
(\ref{simpleaction})), the integrability condition (\ref{integrable})
is obeyed for any spinor $\epsilon$. Since the Killing spinor equation
(\ref{killing}) is a first order equation, this means that there
are as many solutions to the equation as the number of independent
components of the spinor. Namely, \ads\ preserves as many supersymmetries
as flat space. 

The existence of supersymmetry implies that, with an appropriate set
of boundary conditions, the supergravity theory on \ads\ is stable
with its energy bounded from below. The supergravity theories
on \ads\ typically contains scalar fields with negative (mass)$^2$.
However they all satisfy the bound (\ref{posbound}) 
\cite{Mezincescu:1984iu,Mezincescu:1985ev}. 
The issue of the boundary condition and
supersymmetry in \ads\ was further studied in \cite{Hawking:1983m}.  
A non-perturbative 
proof of the stability of $AdS$ is given in \cite{Gibbons:1983aq},
 based on a generalization of 
Witten's proof \cite{Witten:1981mf} of the positive energy theorem
in flat space \cite{Schoen:1982re}.

\subsection{Gauged Supergravities and Kaluza-Klein Compactifications}
\label{kkcatalogue}

Extended supersymmetries in \adsp\ with $p = 2,3,4,5$ are 
classified by Nahm \cite{Nahm:1978tg} (see also \cite{Kac:1977em}) as
\ber
   AdS_4 &:& ~~ OSp({\cal N} | 4), ~~ {\cal N} = 1,2,\cdots 
        \nonumber \\
   AdS_5 &:& ~~ SU(2,2|{\cal N}/2), ~~ {\cal N}=2,4,6,8  \nonumber \\
   AdS_6 &:& ~~ F(4) \nonumber \\
   AdS_7 &:& ~~OSp(6,2 |{\cal N}), ~~ {\cal N}=2,4 .
\eer
For \adsp\ with $p>5$, there is no simple \ads\ supergroup.
These extended supersymmetries are realized as global symmetries of 
gauged supergravity on \adsp . The AdS/CFT correspondence identifies
them with the superconformal algebras discussed in section \ref{superconfalg}.
Gauged supergravities are supergravity
theories with non-abelian gauge fields in the 
supermultiplet of the graviton. Typically the cosmological
constant is negative and \adsp\ is a natural background geometry.
Many of them are related to Kaluza-Klein compactification of
the supergravities in 10 and 11 dimensions. A complete catalogue of
gauged supergravities in dimensions $\leq 11$ is found in 
\cite{Salam:1989fm}. Here we list some of them.

\medskip
\noindent
$\circ$ $\underline{AdS_7}$

\smallskip
The gauged supergravity in 7 dimensions
has global supersymmetry $OSp(6,2|{\cal N})$. 
The maximally supersymmetric case of ${\cal N}=4$
constructed in \cite{Pernici:1984xx} 
contains a Yang-Mills field with a gauge group $Sp(2) \simeq SO(5)$. 
The field content of this theory can be derived from a
truncation of the spectrum of
the Kaluza-Klein compactification of the 11-dimensional
supergravity to 7 dimensions,
\eqn{seventofour}{
\IR^{11} \rightarrow AdS_7 \times S^4.}
The 11-dimensional supergravity has the Lagrangian
\eqn{elevensugra}{{\cal L} = \sqrt{g} \left( \frac{1}{4} R 
  - \frac{1}{48} F_{\mu\nu\rho\sigma} F^{\mu\nu\rho
\sigma} \right) + \frac{1}{72} A \wedge F \wedge F
+ {\rm fermions},}
where $A$ is a 3-form gauge field and $F = dA$.
It was pointed out by Freund and Rubin \cite{Freund:1980xh}
that there is a natural way  to ``compactify''
the theory to 4 or 7 dimensions. 
We have put the word ``compactify'' in quotes since we will
see that typically the size of the compact dimensions is comparable
to the radius of curvature of the non-compact dimensions. 
To compactify the theory to 7 dimensions, 
the ansatz of Freund and Rubin sets
the 4-form field strength $F$ to be proportional 
to the volume element on a 4-dimensional subspace $M_4$. 
The Einstein equation, which includes the contribution of $F$ to the
energy-momentum tensor,
implies a positive curvature on $M_4$ and a
constant negative curvature on the non-compact dimensions,
$i.e.$ they are $AdS_7$. 

The maximally symmetric case is obtained by considering
$M_4=S^4$. Since there is no cosmological constant
in 11 dimensions, the radius $R$ of $S^4$ is
proportional to the curvature radius of $AdS_7$. 
By the Kaluza-Klein mechanism, 
the $SO(5)$ isometry of $S^4$ becomes the
gauge symmetry in 7 dimensions. The spherical harmonics
on $S^4$ give an infinite tower of Kaluza-Klein particles 
on \ads$_7$. 
A truncation of this spectrum to include only the graviton
supermultiplet gives the spectrum of the $\cn=4$ $SO(5)$ gauged
supergravity on \ads$_7$. It has been believed 
that this is a consistent truncation of the full theory, and 
very recently it was shown in \cite{vN:99ct} that this
is indeed the case. In general, there
are subtleties in the consistent truncation procedure, 
which will be discussed
in more detail in the next subsection. There are also 
other ${\cal N}=4$ theories with non-compact gauge groups $SO(p,q)$ with
$p+q=5$ \cite{Pernici:1985zw}.

The seven dimensional ${\cal N}=2$ gauged supergravity with gauge
group $Sp(1) \simeq SU(2)$ was constructed in \cite{Townsend:1983kk}.
In this case, one can have also a matter theory with possibly another
gauge group $G$. It is not known whether a matter theory of arbitrary
$G$ with arbitrary coupling constant can be coupled to gauged
supergravity. The Kaluza-Klein compactification of 10-dimensional
${\cal N}=1$ supergravity, coupled to ${\cal N}=1$ super Yang-Mills,
on $S^3$ gives a particular example. In this case, ten dimensional
anomaly cancellation requires particular choices of $G$.

\medskip
\noindent
$\circ$ $\underline{AdS_6}$

The 6-dimensional anti-de Sitter supergroup $F(4)$ is
realized by the ${\cal N}=4$ gauged supergravity with
gauge group $SU(2)$.
It was predicted to exist in \cite{DeWitt:1982wm} and
constructed in \cite{Romans:1986tw}.
It was
conjectured in \cite{Ferrara:1998gv} to be related
to a compactification of the ten dimensional massive type IIA supergravity theory.
The relevant compactification
of the massive type IIA supergravity is constructed as
a fibration
of  \ads$_6$ over $S^4$ \cite{us:1999ma}. The form of the ten dimensional space is
called a 
{\it warped product}
\cite{vanNieuwenhuizen:1985ri} and it is the most general one that has the 
$AdS$
isometry group \cite{van:1983xx}.
The $SU(2)$ gauge group of  
the 6-dimensional ${\cal N}=4$ gauged supergravity is associated with an $SU(2)$ subgroup
of the  $SO(4)$ isometry group of the compact part of the ten dimensional
space.

\medskip
\noindent
$\circ$ $\underline{AdS_5}$

In 5 dimensions, there are ${\cal N} = 2,4,6$ and $8$ gauged
supergravities with supersymmetry $SU(2,2|{\cal N}/2)$.
The gauged ${\cal N}=8$ supergravity was constructed 
in \cite{Pernici:1985ju,Gunaydin:1986cu}. 
It has the gauge group $SU(4) \simeq SO(6)$ and
the global symmetry $E_6$. 
This theory can be
derived by a truncation of the
compactification of $10$-dimensional type IIB
supergravity on $S^5$ using the Freund-Rubin ansatz, i.e.
setting the self-dual 5-form field strength $F^{(5)}$ 
to be proportional to the volume form of $S^5$
\cite{Schwarz:1983qr,Gunaydin:1985fk,Kim:1985ez}. By the Einstein equation,
the strength of $F^{(5)}$ determines the radius of $S^5$
and the cosmological constant $R^{-2}$ of \ads$_5$.

This case is of particular interest; as we will see below,
the $AdS$/CFT correspondence claims that 
it is dual to the large $N$ (and large $g_{YM}^2 N$) limit of 
${\cal N}=4$ supersymmetric $SU(N)$ gauge theory in four dimensions. 
The complete Kaluza-Klein mass spectrum of
the IIB supergravity theory on $AdS_5 \times S^5$ was obtained in 
\cite{Gunaydin:1985fk,Kim:1985ez}.
One of the interesting features
of the Kaluza-Klein spectrum (in this case as well as in the other
cases discussed in this section) is that the frequency $\omega$ of
stationary wave solutions is quantized. For example, the masses of the
scalar fields in the Kaluza-Klein tower are all of the form
$(m R)^2  = \tilde{l}(\tilde{l} + 4)$, where $\tilde{l}$ is
an integer bounded from below. Substituting this into (\ref{branch})
with $p=3$, we obtain
\beq
 \lambda_\pm = 2 \pm |\tilde{l} + 2|.
\eeq
Therefore, the frequency $\omega$ given by (\ref{energyquanta})
takes values in integer multiples of $1/R$: 
\beq
  |\omega| R = 2 \pm | \tilde{l} + 2| + l + 2n, ~~~(n=0,1,2,\cdots).
\eeq
This means that all the scalar fields in the supergravity
multiplet are periodic in $\tau$ with the period $2\pi$, i.e.
the scalar fields are single-valued on the original hyperboloid
(\ref{hyperboloid}) before taking the universal covering. 
This applies to all other fields in the supermultiplet as well,
with the fermions obeying the Ramond boundary condition around
the timelike circle. 

The fact that the frequency $\omega$ is quantized has its origin
in supersymmetry. The supergravity particles in 10 dimensions
are BPS objects and preserve one half of the supersymmetry.
This property is preserved under the Kaluza-Klein
compactification on $S^5$. The notion of the BPS particles in the case
of \ads\ supergravity is clarified in \cite{Freedman:1984na} 
and it is shown, in the context of theories in
4 dimensions, that it leads to the
quantization of $\omega$. In the \ads/CFT correspondence, 
this is dual to the fact that chiral primary operators
do not have  anomalous dimensions. 

On the other hand, energy levels of other states, such as stringy
states or black holes, are not expected to be quantized as 
the supergravity modes are. Thus, the full string theory does not make
sense on the hyperboloid but only on its universal cover
without the closed timelike curve.

The ${\cal N}=4$ gauged supergravity with gauge group $SU(2) \times
U(1)$ was constructed in
\cite{Romans:1986ps}. Various ${\cal N}=2$ theories were
constructed in \cite{D'Auria:1982yi,Gunaydin:1984bi,Gunaydin:1984nt,
Gunaydin:1985ak}.

\medskip
\noindent
$\circ$ $\underline{AdS_4}$

In four dimensions, some of the possible $AdS$ supergroups are
$OSp({\cal N}|4)$ with ${\cal N}=1,2,4$ and $8$. $\cn=8$ is the
maximal supergroup that corresponds to a supergravity theory.
The ${\cal N}=8$ gauged supergravity with $SO(8)$
gauge group was constructed in \cite{deWit:1982eq,
deWit:1982ig}. This theory (like the other theories discussed in
this section) has a highly
non-trivial potential for scalar fields, whose extrema
were analyzed in \cite{Warner:1983vz,Warner:1984du}. It was
shown in \cite{deWit:1985iy}  that the extremum with
${\cal N}=8$ supersymmetry corresponds to a truncation of the
compactification of  11-dimensional supergravity
on $AdS_4\times S^7$. Some of the other extrema can also be
identified with truncations of
compactifications of the 11-dimensional theory. 
For a review of the 4-dimensional compactifications
of 11-dimensional supergravity, see
\cite{Duff:1986hr}.

\medskip
\noindent
$\circ$ $\underline{AdS_3}$

Nahm's classification does not include this case since
the isometry group $SO(2,2)$ of $AdS_3$ is not a simple group
but rather the direct product of two $SL(2,\IR)$ factors.
The supergravity theories associated with  the $AdS_3$
supergroups $OSp(p|2) \times OSp(q|2)$
were constructed in \cite{Achucarro:1989gm}
and studied more recently in \cite{Nishimura:1998ud}. They
can be regarded as the Chern-Simons gauge theories of gauge group
$OSp(p|2) \times OSp(q|2)$. Therefore, they are topological
field theories without local degrees of freedom. The case
of $p=q=3$ is obtained, for example, by a truncation of
the Kaluza-Klein
compactification of the 6-dimensional $\cn=(2,0)$ supergravity
on $S^3$. In addition to $OSp(p|2)$, several other
supersymmetric extensions of $SL(2,\IR)$ are known, such as:
\eqn{extensions}{
SU({\cal N}|1,1),~G(3),~F(4),~D(2,1,\alpha).}
Their representations are studied extensively in the context
of two-dimensional superconformal field theories.

\subsection{Consistent Truncation of Kaluza-Klein Compactifications}
\label{ConsistentTruncation}

Despite the fact that the equations of motion for type~IIB
supergravity in
ten dimensions are known, it turns out to be difficult to extract any
simple form for the equations of motion of fluctuations around
its five-dimensional Kaluza-Klein compactification on $S^5$.
The
spectrum of this compactification
is known from the work of \cite{Kim:1985ez,Gunaydin:1985fk}.
It
is a general feature of compactifications involving anti-de Sitter
space
that the positively curved compact part has a radius of curvature on
the
same order as the negatively curved anti-de Sitter part.  As a result,
the
positive $\hbox{(mass)}^2$ of Kaluza-Klein modes is of the same order
as
the negative $\hbox{(mass)}^2$ of tachyonic modes.  Thus there is no
low-energy limit in which one can argue that all but finitely many
Kaluza-Klein harmonics decouple.  This was a traditional worry for all
compactifications of eleven-dimensional supergravity on squashed
seven-spheres.

\begin{figure}
      \vskip0cm
   \centerline{\psfig{figure=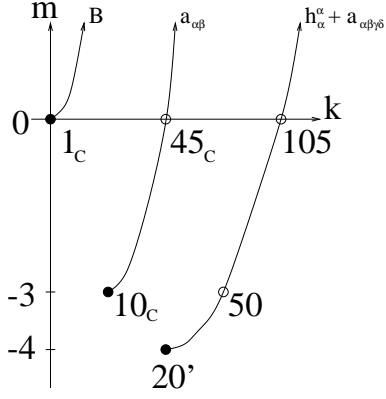,width=2in}}
   \vskip0cm
 \caption{The low-lying scalar fields in the Kaluza-Klein reduction of
type~IIB supergravity on $S^5$.  The filled dots indicate fields which
are
kept in the truncation to gauged supergravity. We also indicate
schematically the ten-dimensional origin of the
scalars.}\label{figCssg}
  \end{figure}

However, fairly compelling evidence exists (\cite{deWit:1987iy} and
references therein) 
that the reduction of eleven-dimensional supergravity on $S^7$ can be
{\it
consistently truncated} to four-dimensional ${\cal N}=8$ gauged
supergravity.  This is an exact statement about the equations of
motion,
and does not rely in any way on taking a low-energy limit.  Put
simply, it
means that any solution of the truncated theory can be lifted to a
solution
of the untruncated theory.  Charged black hole metrics in anti-de
Sitter
space provide a non-trivial example of solutions that can be lifted to
the
higher-dimensional theory
\cite{Chamblin:1999tk,Cvetic:1999ne,Cvetic:1999xp}.  There is a belief
but
no proof that a similar truncation may be made from ten-dimensional
type~IIB supergravity on $S^5$ to five-dimensional ${\cal N}=8$
supergravity.  To illustrate how radical a truncation this is, we
indicate
in figure~\ref{figCssg} the five-dimensional scalars that are kept
(this is
a part of one of the figures in \cite{Kim:1985ez}).  Note that not all
of
them are $SO(6)$ singlets.  Indeed, the fields which are kept are
precisely the
superpartners of the massless graviton under the supergroup
$SU(2,2|4)$,
which includes $SO(6)$ as its R-symmetry group.

The historical route to gauged supergravities was as an elaboration of
the
ungauged theories, and only after the fact were they argued to be
related
to the Kaluza-Klein reduction of higher dimensional theories on
positively
curved manifolds.  In ungauged $d=5$ $\cn=8$
supergravity, the scalars parametrize
the
coset $E_{6(6)}/USp(8)$ (following \cite{Cremmer:1980gs} we use here
$USp(8)$ to denote the unitary version of the symplectic group with a
four-dimensional Cartan subalgebra).  The spectrum of gauged
supergravity
is almost the same: the only difference is that twelve of the vector
fields
are dualized into anti-symmetric two-forms.  Schematically, we write
this
as
  \eqn{SpectrumSplit}{
\begin{picture}(150,70)(-15,-40)
\thicklines
\put(0,0){$g_{\mu\nu} \qquad \psi^a_\mu \qquad A^{ab}_\mu \qquad
  \chi^{abc} \qquad
 \phi^{abcd}$}
\put(-3,15){$1$}
\put(37,15){$8$}
\put(73,15){$27$}
\put(114,15){$48$}
\put(156,15){$42$}
\put(81,-5){\line(1,-1){10}}
\put(81,-5){\line(-1,-1){10}}
\put(58,-30){$A_{\mu\,IJ}$}
\put(91,-30){$B^{\,I\alpha}_{\mu\nu}$}
\put(53,-45){$15$}
\put(106,-45){$12$}
\end{picture}
  }
 Lower-case Roman indices are the eight-valued indices of the
 fundamental
of $USp(8)$.  Multiple $USp(8)$ indices in \SpectrumSplit\ are
antisymmetrized and the symplectic trace parts removed.  The upper-case
Roman indices $I$, $J$ are the six-valued indices of the vector
representation of $SO(6)$, while the index $\alpha$ indicates a
 doublet of
the $SL(2,\IR)$ which descends directly from the $SL(2,\IR)$
 global
symmetry of type~IIB supergravity.  These groups are embedded into
$E_{6(6)}$ via the chain
  \eqn{EmbedSO}{
   E_{6(6)} \supset SL(6,\IR) \times SL(2,\IR) \supset
    SO(6) \times SL(2,\IR) \ .
  }

The key step in formulating gauged supergravities is to introduce
minimal
gauge couplings into the Lagrangian for all fields which are charged
under
the subgroup of the global symmetry group that is to be gauged.  For
instance, if $X_I$ is a scalar field in the vector representation of
$SO(6)$, one makes the replacement
  \eqn{GaugeIt}{
   \partial_\mu X_I \to D_\mu X_I = \partial_\mu X_I - g A_{\mu\,IJ}
   X^J
  }
 everywhere in the ungauged action.  The gauge coupling $g$ has
 dimensions
of energy in five dimensions, and one can eventually show that $g=2/R$
where $R$ is the radius of the $S^5$ in the $AdS_5 \times S^5$
geometry.
The replacement \GaugeIt\ spoils supersymmetry, but it was shown in
\cite{Gunaydin:1986cu,Pernici:1985ju} that a supersymmetric Lagrangian
can
be recovered by adding terms at $O(g)$ and $O(g^2)$.  The full
Lagrangian
and the supersymmetry transformations can be found in these
references.  It
is a highly non-trivial claim that this action, with its beautiful
non-polynomial structure in the scalar fields, represents a consistent
truncation of the reduction of type~IIB supergravity on $S^5$.  This
is not
entirely implausible, in view of the fact that the $SO(6)$ isometry of
the
$S^5$ becomes the local gauge symmetry of the truncated theory.
Trivial
examples of consistent truncation include situations where one
restricts to
fields which are invariant under some subgroup of the gauge group.
For
instance, the part of ${\cal N}=8$ five-dimensional supergravity
invariant
under a particular $SU(2) \subset SO(6)$ is ${\cal N}=4$ gauged
supergravity coupled to two tensor multiplets \cite{Freedman:1999gp}.
A similar trunction to ${\cal N}=6$ supergravity was considered in
\cite{Ferrara:1998zt}.

The $O(g^2)$ term in the Lagrangian is particularly interesting: it is
a
potential $V$ for the scalars.  $V$ is an $SO(6) \times SL(2,\IR)$
invariant function on the coset manifold $E_{6(6)}/USp(8)$.  It
involves
all the $42$ scalars except the dilaton and the axion.  Roughly
speaking,
one can think of the $40$ remaining scalars as parametrizing a
restricted
class of deformations of the metric and 3-form fields on the
$S^5$, and of $V$ as measuring the
response of
type~IIB supergravity to these deformations.  If the scalars are
frozen to
an extremum of $V$, then the value of the potential sets the
cosmological constant in five dimensions.  The associated conformal
field theories were discussed in 
\cite{Distler:1998gb,Girardello:1998pd,Khavaev:1998fb}.  The known
extrema can be classified by the subset of the $SO(6)$ global
R-symmetry group that is preserved.

\chapter{AdS/CFT Correspondence}
\label{ChapCorrBasic}

\section{The Correspondence}
\label{correspondence}

In this section we will present an argument connecting type IIB string
theory compactified on $AdS_5\times S^5$ to ${\cal N} =4 $
super-Yang-Mills theory
\cite{Maldacena:1997re}.  Let us start with type IIB string theory in
flat, ten dimensional Minkowski space. Consider $N$ parallel D3 branes
that are sitting together or very close to each other (the precise
meaning of ``very close'' will be defined below). The D3 branes are
extended along a $(3+1)$ dimensional plane in $(9+1)$ dimensional
spacetime.  String theory on this background contains two kinds of
perturbative excitations, closed strings and open strings. The closed
strings are the excitations of empty space and the open strings
end on the D-branes and describe 
excitations of the D-branes.
 If we consider the system at low energies, energies lower
than the string scale $1/l_s$, then only the massless string states
can be excited, and we can write an effective Lagrangian describing
their interactions. The closed string massless states give a gravity
supermultiplet in ten dimensions, and their low-energy effective
Lagrangian is that of type IIB supergravity. The open string massless
states give an $\cn=4$ vector supermultiplet in $(3+1)$ dimensions,
and their low-energy effective Lagrangian is that of $\cn=4$ $U(N)$
super-Yang-Mills theory
\cite{Witten:1996im,joebook}. 

The complete effective action of the massless modes will have
the form
\eqn{lowenergy}{
S = S_{\rm bulk} + S_{\rm brane} + S_{\rm int}. }  $S_{\rm bulk}$ is
the action of ten dimensional supergravity, plus some higher
derivative corrections.  Note that the Lagrangian \lowenergy\ involves
only the massless fields but it takes into account the effects of
integrating out the massive fields. It is not renormalizable (even for
the fields on the brane), and it should only be understood as an
effective description in the Wilsonian sense, i.e. we integrate out
all massive degrees of freedom but we do not integrate out the
massless ones. The brane action $S_{\rm brane}$ is defined on the
$(3+1)$ dimensional brane worldvolume, and it contains the ${\cal N} =
4 $ super-Yang-Mills Lagrangian plus some higher derivative
corrections, for example terms of the form $\alpha'^2 \tr(F^4) $.
Finally, $S_{\rm int}$ describes the interactions between the brane
modes and the bulk modes. The leading terms in this interaction
Lagrangian can be obtained by covariantizing the brane action,
introducing the background metric for the brane
\cite{Leigh:1989jq}.  

We can expand the bulk action as a free quadratic part describing the
propagation of free massless modes (including the graviton), plus some
interactions which are proportional to positive powers of the square
root of the Newton constant.  Schematically we have
\eqn{expans}{
 S_{bulk} \sim {1 \over 2 \kappa^2} \int \sqrt{g} {\cal R} \sim  
\int (\partial h)^2 + \kappa (\partial h)^2 h + \cdots,} 
where we have written the metric as
$g = \eta + \kappa h $. We indicate explicitly
the dependence on the graviton, but the other terms in the Lagrangian,
involving other fields, can be expanded in a similar way.  Similarly,
the interaction Lagrangian $S_{int}$ is proportional to positive powers of
$\kappa $.  If we take the low energy limit, all interaction terms
proportional to $\kappa $ drop out. This is the well known fact that
gravity becomes free at long distances (low energies).  

In order to
see more clearly what happens in this low energy limit it is
convenient to keep the energy fixed and send $l_s \to 0$ ($\alpha' \to
0$) keeping all the dimensionless parameters fixed, including the
string coupling constant and $N$.  In this limit the coupling 
$\kappa \sim g_s
\alpha'^2 \to 0$, so that the interaction Lagrangian relating the bulk
and the brane vanishes. In addition all the higher derivative terms
in the brane action vanish, leaving just the pure ${\cal N} = 4$ $U(N)$
gauge theory in $3+1 $ dimensions, which is known to be a conformal field
theory. And, the supergravity theory in the
bulk becomes free.  So, in this low energy limit we have two decoupled
systems. On the one hand we have free gravity in the bulk and on the
other hand we have the four dimensional gauge theory.  

Next, we consider
the same system from a different point of view.  D-branes are massive
charged objects which act as a source for the various supergravity
fields. As shown in section \ref{black_pbranes} we can find a D3 brane solution
\cite{Horowitz:1991cd} of supergravity, of the form
\eqn{dthree}{ \eqalign{
ds^2 & = f^{-1/2} ( -dt^2 + dx_1^2 + dx_2^2 + dx_3^2 ) + f^{1/2}
(dr^2 + r^2 d\Omega_5^2 )~,
\cr
F_5 &= (1 + * ) dt dx_1 dx_2 dx_3 df^{-1} ~,
\cr
f &= 1 + { R^4 \over r^4 }~,~~~~~~~~ R^4 \equiv  
4 \pi  g_s \alpha'^2 N ~. }}
Note that since $g_{tt}$ is non-constant,
the energy $E_p$ of an object as measured by an observer 
at a constant  position $r$ and the energy $E$ measured by an observer at
infinity are related  by the redshift factor
\eqn{redshift}{
 E = f^{-1/4} E_p ~.
}
 This means that the same object brought closer
and closer to $r =0$ would appear to have lower and lower energy 
for the observer at infinity. 
Now we take the low energy limit in the background described  by equation
\dthree. There are two kinds of low energy excitations (from the point
of view of an observer at infinity). 
We can have  massless particles propagating in the bulk region
with wavelengths that becomes very large, or we can have any kind
of excitation that we bring closer and closer to 
$r=0$. In the low energy limit these two types of excitations 
decouple from each other. The bulk massless particles decouple from
the near horizon region (around $r=0$)
 because the low energy absorption cross section goes like
$ \sigma \sim  \omega^3 R^8 $ \cite{Klebanov:1997kc,Gubser:1997yh},
 where $\omega $ is the 
energy. 
This can be understood from the 
fact that in this limit the wavelength of the particle becomes 
much bigger than  the typical gravitational size of the brane 
(which is of order $R$). 
Similarly, the excitations that live very close to $r = 0$ find it 
harder and harder to climb the gravitational potential and
escape to the asymptotic region. In conclusion, the low energy 
theory consists of two decoupled pieces, one is free bulk supergravity
and the second is the near horizon region of the geometry. 
In the near horizon region, $r \ll R $, we can approximate 
$f \sim R^4/r^4$, and
the geometry becomes 
\eqn{nearhor}{
ds^2 = { r^2 \over R^2 } ( -dt^2 + dx_1^2 + dx_2^2 + dx_3^2 ) + R^2
{ dr^2 \over r^2 } + R^2  d\Omega_5^2,
}
which is the geometry of $AdS_5 \times S^5$. 

We see that both from the point of view of a field theory of open
strings living on the brane, and from the point of view of the
supergravity description, we have two decoupled theories in the
low-energy limit. In both cases one of the decoupled systems is
supergravity in flat space.  So, it is natural to identify the second
system which appears in both descriptions. Thus, we are led to the
conjecture that {\it ${\cal N} =4 $ $U(N)$ super-Yang-Mills theory in
$3+1$ dimensions is the same as (or dual to) type IIB superstring
theory on $AdS_5\times S^5$} \cite{Maldacena:1997re}.

We could be a bit more precise about the near horizon limit and 
how it is being taken. Suppose that we take $\alpha' \to  0$, as 
we did when we discussed the field theory living on the brane. 
We want to keep fixed the energies of the objects in the throat 
(the near-horizon region) in string units,
so that we can consider arbitrary excited
string states there. This implies that
 $\sqrt{\alpha'} E_p \sim {\rm fixed} $. 
For small $\alpha'$  \redshift\  reduces to 
$E \sim E_p r/\sqrt{\alpha'} $. Since we want to keep 
fixed the energy measured from infinity, which is  the
way energies are measured in the field theory, we need to take 
$r \to 0$ keeping $r / \alpha' $ fixed.  
It is then convenient to define a new variable $U \equiv r / \alpha'$, so
that the metric becomes
\eqn{metricu}{
ds^2 = \alpha' \left[ {U^2 \over \sqrt{4 \pi g_s N} }
( - dt^2 + dx_1^2 + dx_2^2 + dx_3^2 ) +
\sqrt{4 \pi g_s N} { d U^2 \over U^2}  + \sqrt{4 \pi g_s N} d \Omega_5^2
\right].
}

This can also be seen by considering a D3 brane sitting at $\vec r$. 
As discussed in section \ref{black_pbranes} this corresponds to 
giving a vacuum expectation value to one of the scalars in the 
Yang-Mills theory. When we take the $\alpha' \to 0$ limit we want
to keep the mass of the ``$W$-boson'' fixed. This mass, which is the
mass of the string stretching between the branes sitting at 
$\vec r =0$ and the one at $\vec r$, is proportional to $  U = 
r/\alpha'$, 
so this quantity should remain fixed in the decoupling limit. 

A $U(N)$ gauge theory is essentially equivalent to a free $U(1)$
vector multiplet times an $SU(N)$ gauge theory, up to some $\IZ_N$
identifications (which affect only global issues).  In the dual string
theory all modes interact with gravity, so there are no decoupled
modes. Therefore, the bulk $AdS$ theory is describing the $SU(N)$ part
of the gauge theory.  In fact we were not precise when we said that
there were two sets of excitations at low energies, the excitations in
the asymptotic flat space and the excitations in the near horizon
region. There are also some zero modes which live in the region
connecting the ``throat'' (the near horizon region) 
with the bulk, which correspond to the $U(1)$
degrees of freedom mentioned above. The $U(1)$ vector supermultiplet
includes six scalars which are related to the center of mass motion of
all the branes
\cite{Gibbons:1993sv}.  From the $AdS$ point of view these zero 
modes live
at the boundary, and it looks like we might or might not decide to
include them in the $AdS$ theory. Depending on this choice we could have
a correspondence to an $SU(N)$ or a $U(N)$ theory. 
The $U(1)$ center of mass degree of freedom
is related to the topological theory of $B$-fields on $AdS$ 
\cite{Witten:1998wy}; if one imposes local boundary conditions for
these $B$-fields at the boundary of $AdS$ one finds a $U(1)$ gauge 
field living at the boundary \cite{seibergprivate},
 as is familiar in Chern-Simons theories
\cite{Witten:1989hf,Elitzur:1989nr}.
These modes living
at the boundary are sometimes called singletons (or doubletons) 
\cite{Fronsdal:1982gq,Freedman:1984na,Pilch:1984xy,Gunaydin:1985wc,%
Gunaydin:1986tc,Gunaydin:1986cs,Bergshoeff:1988jm,Bergshoeff:1988jx,%
Bergshoeff:1989uc}.

As we saw in section \ref{adsgeom}, Anti-de-Sitter space has a large group
of isometries, which is $SO(4,2)$ for the case at hand.  This is the
same group as the conformal group in $3+1$ dimensions.  Thus, the fact
that the low-energy field theory on the brane is conformal is
reflected in the fact that the near horizon geometry is Anti-de-Sitter
space.  We also have some supersymmetries.
The number of supersymmetries is twice that of the full solution 
(\ref{dthree})
containing the asymptotic region \cite{Gibbons:1993sv}.  This doubling
of supersymmetries is viewed in the field theory as a consequence of
superconformal invariance (section \ref{susyinads}),
 since the superconformal
algebra has twice as many fermionic generators as the corresponding
Poincare superalgebra. 
We also have an $SO(6)$
symmetry which rotates the $S^5$. This can be identified with the
$SU(4)_R$ R-symmetry group of the field theory. In fact, the whole
supergroup is the same for the $\cn=4$ field theory and the $AdS_5
\times S^5 $ geometry, so both sides of the conjecture
have the same spacetime symmetries. We will discuss in more detail the
matching between the two sides of the correspondence in section
\ref{tests}.

In the above derivation the field theory is naturally defined on
$\IR^{3,1}$, but we saw in section \ref{confflatspace} that we could also
think of the conformal field theory as defined on $S^3 \times \IR$ by
redefining the Hamiltonian. Since the isometries of $AdS$ are in one
to one correspondence with the generators of the conformal group of the
field theory, we can conclude that this new Hamiltonian 
${1\over 2}(P_0+K_0)$ can be
associated on $AdS$ to the generator of translations in global time.
This formulation of the conjecture is more useful since in the global
coordinates there is no horizon.  When we put the field theory on
$S^3$ the Coulomb branch is lifted and there is a unique ground
state. This is due to the fact that the scalars $\phi^I$ in the field
theory are conformally coupled, so there is a term of the form $\int
d^4 x \tr(\phi^2) {\cal R}$ in the Lagrangian, where $\cal R$ is the
curvature of the four-dimensional space on which the theory is
defined. Due to the positive curvature of $S^3$ this leads to a mass
term for the scalars \cite{Witten:1998qj}, lifting the moduli space.

The parameter $N$ appears on the string theory side as the flux
of the five-form Ramond-Ramond field strength on the $S^5$,
\eqn{flux}{
\int_{S^5}  F_5  = N. 
}
  From the physics of D-branes we know that 
 the Yang-Mills coupling is 
related to the string coupling
through \cite{Polchinski:1995mt,Douglas:1995bn}
\eqn{couplingcon}{
\tau \equiv 
{ 4\pi i  \over g_{YM}^2 } + { \theta \over 2 \pi}   = 
  { i\over g_s} 
+  { \chi \over 2 \pi } ~, } 
where we have also included the relationship of the $\theta$ angle to
the expectation value of the RR scalar $\chi$. We have written the
couplings in this fashion because both the gauge theory and the string
theory have an $SL(2,\IZ)$ self-duality symmetry under which $\tau \to
(a \tau +b) / (c \tau + d)$ (where $a,b,c,d$ are integers with
$ad-bc=1$). In fact, $SL(2,\IZ)$ is a conjectured strong-weak coupling
duality symmetry of type IIB string theory in flat space
\cite{Hull:1995ys}, and it should also be a symmetry in the present
context since all the fields that are being turned on in the $AdS_5 \times
S^5$ background (the metric and the five form field strength) are
invariant under this symmetry.  The connection between the $SL(2,\IZ)$
duality symmetries of type IIB string theory and ${\cal N} =4$ SYM was
noted in
\cite{Tseytlin:1996it,Green:1996qg,Douglas:1996du}. 
The string theory seems to have a parameter that does not appear in
the gauge theory, namely $\alpha'$, which sets the string tension and
all other scales in the string theory. However, this is not really a
parameter in the theory if we do not compare it to other scales in the
theory, since only relative scales are meaningful.  In fact, only the
ratio of the radius of curvature to $\alpha'$ is a parameter, but not
$\alpha'$ and the radius of curvature independently. Thus, $\alpha'$
will disappear from any final physical quantity we compute in this
theory.  It is sometimes
convenient, especially when one is doing gravity calculations, to set
the radius of curvature to one.  This can be achieved by writing the
metric as $ds^2 = R^2 d{\tilde s}^2$, and rewriting everything in terms
of $\tilde g$. With these conventions $ G_N \sim 1/N^2$ and $\alpha'
\sim 1/\sqrt{g_s N} $.  This implies that any quantity calculated purely
in terms of the gravity solution, without including stringy effects,
will be independent of $g_s N$ and will depend only on $N$.  $\alpha'$
corrections to the gravity results give corrections which are
proportional to powers of $1/\sqrt{g_s N}$.

Now, let us address the question of
the validity of various approximations. The analysis of loop diagrams
in the field theory shows that
we can trust the perturbative analysis in the Yang-Mills theory when
\eqn{pert}{
g^2_{YM} N \sim g_s N \sim { R^4 \over l_s^4 } \ll 1 .
}
Note that we need $g_{YM}^2 N$ to be small and not just $g_{YM}^2$.
On the other hand, the classical gravity description becomes reliable when
the radius of curvature $R$ of $AdS$ and of $S^5$ becomes large compared
to the string length,
\eqn{gravity}{
{ R^4 \over l_s^4 } \sim g_s N \sim  g^2_{YM} N \gg 1.
}
We see that the gravity regime \gravity\ and the perturbative 
field theory regime \pert\ are perfectly incompatible. In this fashion
we avoid any obvious contradiction due to  the fact that
the two theories look very different. This is the reason that 
this correspondence is called a ``duality''. The two theories are
conjectured to be exactly the same, but when one side is weakly coupled
the other is strongly coupled and vice versa. This makes the correspondence
both hard to prove and useful,
as we can solve a strongly coupled gauge theory via classical supergravity.
Notice that in \pert \gravity\ we implicitly assumed that 
$g_s <1$. If $g_s > 1$ we can perform an $SL(2,\IZ)$ duality transformation
and get conditions similar to \pert \gravity\ 
but with $g_s \to 1/g_s$. So, we cannot get into the gravity regime
\gravity\ by taking $N$ small  ($N=1,2,..$)
 and $g_s$ very large, since in that case
the D-string becomes light and renders the gravity approximation 
invalid. Another way to see this is to note that the radius of
curvature in Planck units is $R^4/l_p^4 \sim N$.
So, it is always necessary, but not sufficient, to have
large $N$ in order to have a weakly coupled supergravity description. 

One might wonder why the above argument was not a proof rather than
a conjecture. It is not a proof because we did not treat the string
theory non-perturbatively (not even  non-perturbatively in $\alpha'$).
We could also consider different forms of the conjecture. 
In its weakest form the gravity description would be valid
for large $g_s N$, but the full string theory on $AdS$ might not agree with
the field theory. A not so weak form would say that the conjecture 
is valid even for finite $g_s N$, but only in the $N \to \infty $ limit
(so that the $\alpha'$ corrections would agree with the field theory,
but the $g_s$ corrections may not).
The strong form of the conjecture, which is the most interesting one
and which we will assume here, 
is that the two theories are exactly the same for all values 
of $g_s$ and $N$. 
In this conjecture the spacetime is only required to be asymptotic
to $AdS_5\times S^5$ as we approach the boundary. In the interior we
can have all kinds of processes; gravitons, highly excited fundamental
string states, D-branes, black holes, etc. Even the topology of
spacetime can change in the interior. The Yang-Mills theory is
supposed to effectively sum over all spacetimes which are asymptotic 
to $AdS_5\times S^5$. This is  completely analogous to the usual 
conditions of asymptotic flatness. We can have black holes and all kinds
of topology changing processes, as long as spacetime is asymptotically 
flat. In this case asymptotic flatness is replaced by the asymptotic
$AdS$ behavior.

\subsection{Brane Probes and Multicenter Solutions}
\label{multicenter_sols}

The moduli space of vacua of the $\cn=4$ $U(N)$ 
gauge theory is $ (\IR^6)^N/S_N$, parametrizing the positions
of the $N$ branes in the six dimensional transverse space. 
In the supergravity solution one can replace 
\eqn{multicenter}{
f \propto { N \over r^4 } \rightarrow 
\sum_{i=1}^N { 1 \over | \vec r - \vec r_i |^4 },
}
and still have a solution to the supergravity equations.  We see that
if $|\vec r| \gg |\vec r_i |$ then the two solutions are basically the
same, while when we go to $r \sim r_i $ the solution starts looking
like the solution of a single brane. Of course, we cannot trust the
supergravity solution for a single brane (since the curvature in
Planck units is proportional to a negative power of $N$). What we can
do is separate the $N$ branes into groups of $N_i$ branes with
$g_s N_i \gg 1$ for all $i$. 
Then we can trust the gravity solution everywhere.

Another possibility is to separate just one brane (or a small number
of branes) from a group of $N$ branes. Then we can view this brane as
a D3-brane in the $AdS_5$ background which is generated by the other
branes (as described above).  A string stretching between the brane
probe and the $N$ branes appears in the gravity description as a
string stretching between the D3-brane and the horizon of $AdS$.  This
seems a bit surprising at first since the proper distance to the
horizon is infinite. However, we get a finite result for the energy of
this state once we remember to include the redshift factor.  The
D3-branes in $AdS$ (like any D3-branes in string theory) are
described at low energies by the Born-Infeld action, which is the
Yang-Mills action plus 
some higher derivative corrections.
This seems to contradict, at first sight, the fact that
the dual field theory (coming from the original branes) is just the pure
Yang-Mills theory.
In order to understand this point more precisely let us write
explicitly the bosonic part of the Born-Infeld action for a D-3 brane
in $AdS$ \cite{Leigh:1989jq},
\eqn{borninf}{
\eqalign{
S = - { 1 \over (2 \pi)^3 g_s \alpha'^2 } & \int d^4 x  f^{-1}
 \left[ \right. 
\cr
& ~~~~
\left. 
\sqrt{- \det( \eta_{\alpha \beta} + f \partial_\alpha
r  \partial_\beta r +  r^2 f  g_{ij}\partial_\alpha
\theta^i\partial_\beta\theta^j + 
  2 \pi \alpha' \sqrt{f} F_{\alpha \beta } )} -1 \right]~,  
\cr          
 ~~~~f = { 4 \pi g_s \alpha'^2 N \over r^4 } & ~,
}}
where $\theta^i$ are angular coordinates on the 5-sphere.           
We can easily check that if we define a new coordinate
 $U= r / \alpha'$, then
all the $\alpha'$ dependence drops out of this action. 
Since $U$ (which has dimensions of energy) 
corresponds to the mass of the W bosons in this configuration,
it is the natural way to express the Higgs 
expectation value that breaks $U(N+1)$ to $U(N)\times U(1)$.
In fact,  the action \borninf\ is precisely the  low-energy 
effective action in the field theory for
the massless $U(1)$ degrees of freedom, that we
obtain after integrating out the massive degrees of freedom (W bosons). 
We can expand \borninf\ in powers of $\partial U$ and
we see that the quadratic term does not have any correction, which
is consistent with the non-renormalization theorem for ${\cal N} =4 $ 
super-Yang-Mills \cite{Seiberg:1994aj}.
 The $(\partial U)^4 $ term has only a one-loop 
correction, 
 and this is also consistent with another non-renormalization
theorem \cite{Dine:1997nq}.
This one-loop correction can be evaluated explicitly
in the gauge theory and the result agrees with the supergravity result
\cite{Douglas:1997yp}. 
It is possible to argue, using broken conformal invariance, 
that all terms in \borninf\  are determined by the $(\partial U)^4 $ 
term \cite{Maldacena:1997re}. 
Since the massive degrees of freedom that we are integrating  out
have a mass proportional to $U$, the action \borninf\ makes sense
as long as the energies involved are much smaller than $U$. 
In particular, we need $\partial U /U \ll U $. Since \borninf\
has the form ${\cal L}( g_s N (\partial U)^2/U^4 )$, the higher order terms
in \borninf\ could become important in the supergravity regime,
when $g_s N \gg 1 $. The Born Infeld action \borninf , as always,
makes sense only when the curvature of the brane is small, but
the deviations from a straight flat brane could be large. In this
regime we can keep the non-linear terms in \borninf\ while
we still neglect the massive string modes and similar effects.
Further gauge theory calculations for effective actions of 
D-brane probes include \cite{Douglas:1998tk,Das:1999ij,Das:1999fx}.

\subsection{The Field $\leftrightarrow$ Operator Correspondence}
\label{field_operator}

A conformal field theory does not have asymptotic states or  an
S-matrix, so the natural objects to consider are operators. 
For example, in ${\cal N} =4 $ super-Yang-Mills we 
have a deformation by a
marginal operator which changes the value of the coupling 
constant. Changing the coupling constant in the field theory is
related by (\ref{couplingcon}) to
changing the coupling constant in the string theory, which is then 
related to the expectation value of the dilaton. 
The expectation value of the dilaton is set by the boundary condition 
for the dilaton at infinity. So, changing the gauge theory coupling
constant  corresponds to changing the boundary 
value of the dilaton. More precisely, let us denote by ${\cal O}$ the
corresponding operator. We can consider adding 
 the term $\int d^4 x \phi_0(\vec x) {\cal O}(\vec x) $ to the 
Lagrangian (for simplicity we assume that such a term was not
present in the original Lagrangian, otherwise we consider 
$\phi_0(\vec x)$ to be the total coefficient of ${\cal O}(\vec x)$ in the
Lagrangian). 
According to the discussion above,
it is natural to assume that this will change the boundary 
condition of the dilaton at the boundary of $AdS$ to (in
the coordinate system (\ref{poincusual}))
$ \phi(\vec x, z )|_{z = 0} = \phi_0(\vec x)$. 
More precisely, as argued in \cite{Gubser:1998bc,Witten:1998qj},
 it is natural to propose that
\eqn{genera}{
\langle e^{\int d^4 x \phi_0(\vec x) {\cal O}(\vec x) } \rangle_{CFT}
= {\cal Z}_{string} 
\Bigg[ \phi(\vec x, z)\Big|_{z = 0 } = \phi_0(\vec x)
\Bigg],
}
where the left hand side is the generating function of correlation 
functions in the field theory, i.e. $\phi_0$ is an arbitrary function
and we can calculate correlation functions of ${\cal O}$ by taking 
functional derivatives with respect to $\phi_0$ and then setting
$\phi_0 =0$. The right hand side is the full partition function
of string theory with the boundary condition that 
the field $\phi$ has the value $\phi_0$ on the boundary of $AdS$. 
Notice that $\phi_0$ is a function of the four variables parametrizing
the boundary of $AdS_5$.

A  formula like \genera\ is valid in general, for any field $\phi$. 
Therefore, each field propagating on AdS space 
is in a one to one correspondence with 
an operator in the field theory.
There is a relation between the mass  of the field $\phi$ and 
the scaling dimension of the operator in the conformal field theory. 
Let us describe this more generally in $AdS_{d+1}$.
The wave 
equation in Euclidean space for  a field of mass $m$ has 
two independent solutions, 
which behave like $z^{d - \Delta } $ and $z^{\Delta}$
for small $z$ (close to the boundary of $AdS$),
where 
\eqn{dimenmass}{
\Delta = {d\over 2} + \sqrt{ {d^2 \over 4} + R^2 m^2 } .
}
Therefore, in order to get consistent behavior for a massive field, 
the boundary condition on the
field in the right hand side of \genera\ should in general be changed to
\eqn{bcond}{
\phi(\vec x , \epsilon) = \epsilon^{d - \Delta } \phi_0(\vec x),
}
and eventually we would take the limit where $\epsilon \to 0$. 
Since $\phi$ is dimensionless, we see that $\phi_0$ has dimensions
of $[{\rm length}]^{\Delta - d}$ which implies, through the 
left hand side of \genera, that the associated operator ${\cal  O}$
has dimension $\Delta$ (\ref{dimenmass}). A more detailed derivation of 
this relation will be given in section \ref{correlators},
where we will verify that the two-point correlation function
of the operator ${\cal O}$ behaves as that of an operator of dimension
$\Delta$ \cite{Gubser:1998bc,Witten:1998qj}. 
A similar relation between fields on AdS and operators in the field
theory exists also for non-scalar fields, including fermions and tensors
on AdS space.



\begin{figure}[htb]
\begin{center}
\epsfxsize=3.5in\leavevmode\epsfbox{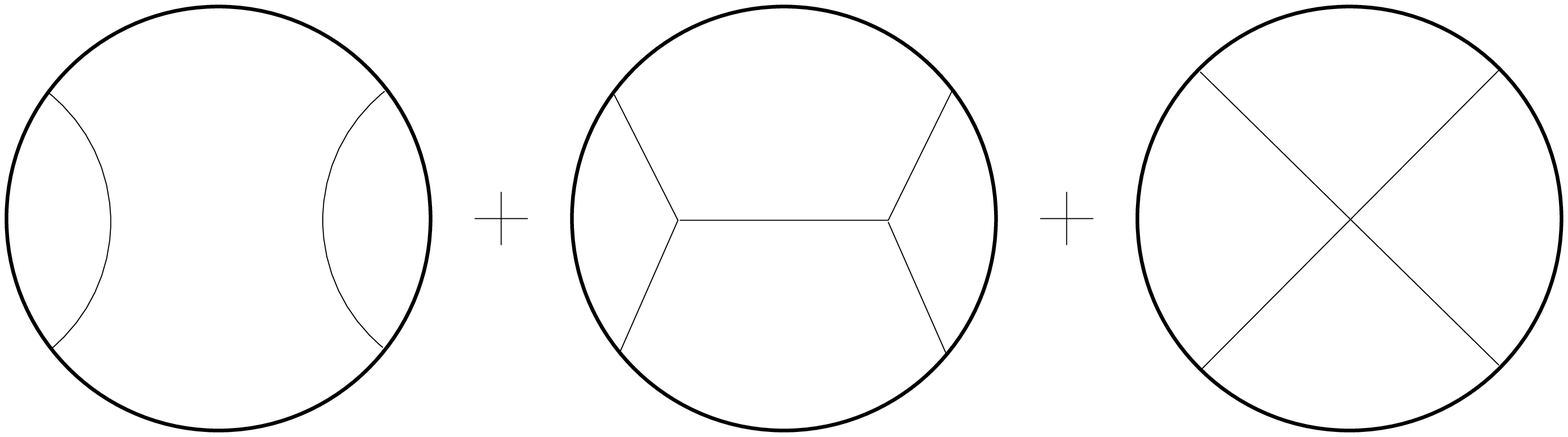}
\end{center}
\caption{
 Correlation functions can be calculated (in the large $g_s N$ limit) in
terms of supergravity Feynman diagrams. Here we see the leading
contribution coming from a disconnected diagram plus connected pieces
involving interactions of the supergravity fields in the bulk of $AdS$.  At
tree level, these diagrams and those related to them by crossing are the
only ones that contribute to the four-point function.
}
\label{dia}
\end{figure}

Correlation functions in the gauge theory can be computed from \genera\ by
differentiating with respect to $\phi_0$.  Each differentiation brings down
an insertion ${\cal O}$, which sends a $\phi$ particle (a closed string
state) into the bulk.  Feynman diagrams can be used to compute the
interactions of particles in the bulk.  In the limit where classical
supergravity is applicable, the only diagrams that contribute are the
tree-level diagrams of the gravity theory (see for instance
figure~\ref{dia}).

This method of defining the correlation functions of a field theory
which is dual to a gravity theory in the bulk of AdS space is quite
general, and it applies in principle to any theory of gravity
\cite{Witten:1998qj}.  Any local field theory contains the stress
tensor as an operator. Since the correspondence described above
matches the stress-energy tensor with the graviton, this implies that
the $AdS$ theory includes gravity.  It should be a well defined
quantum theory of gravity since we should be able to compute loop
diagrams.  String theory provides such a theory. But if a new way of
defining quantum gravity theories comes along we could consider those
gravity theories in $AdS$, and they should correspond to some
conformal field theory ``on the boundary''.  In particular, we could
consider backgrounds of string theory of the form $AdS_5 \times M^5$
where $M^5$ is any Einstein manifold
\cite{Kehagias:1998gn,Gubser:1999vd,Romans:1985an}.  Depending on the
choice of $M^5$ we get different dual conformal field theories,
as discussed in section \ref{other_backgrounds}.  Similarly,
this discussion can be extended to any $AdS_{d+1}$ space,
corresponding to a conformal field theory in $d$ spacetime dimensions
(for $d>1$). We will discuss examples of this in section \ref{adsmore}.

\subsection{Holography}
\label{holography}

In this section we will describe how the AdS/CFT correspondence
gives a holographic description of physics in $AdS$ spaces. 

Let us start by explaining the Bekenstein bound, which states that the
maximum entropy in a region of space is $S_{max} = { \rm Area}/4 G_N$
\cite{Bekenstein:1994dz}, where the area is that of the boundary of
the region.  Suppose that we had a state with more entropy than
$S_{max}$, then we show that we could violate the second law of
thermodynamics.  We can throw in some extra matter such that we form a
black hole. The entropy should not decrease. But if a black hole forms
inside the region its entropy is just the area of its horizon, which
is smaller than the area of the boundary of the region (which by our
assumption is smaller than the initial entropy).  So, the second law
has been violated.

Note that this bound implies that the number of degrees of freedom
inside some region grows  as the area of the boundary of a 
region and not like the volume
of the region. In standard quantum field theories this is certainly 
not possible. Attempting to understand this behavior leads to 
the ``holographic
principle'', which states that in a quantum gravity theory 
all physics within some volume
can be described in terms of some theory on the boundary which 
has less than one degree of freedom per Planck area 
\cite{'tHooft:1993gx,Susskind:1995vu} (so that its entropy satisfies the
Bekenstein bound). 

In the AdS/CFT correspondence we are describing physics in the bulk of
$AdS$ space by a field theory of one less dimension (which can be thought
of as living
on the boundary), so it looks like holography. However, it is hard
to check what the number of degrees of freedom per Planck area is, since
the   theory, being 
conformal,  has an infinite number of degrees of freedom, and the 
area of the boundary of AdS space is also infinite. 
Thus, in order to compare things properly we should introduce a cutoff on
the number of degrees of freedom in the field theory and see what
it corresponds to  in the gravity theory. 
For this  purpose let us write the metric of $AdS$ as 
\eqn{adscav}{
ds^2 = R^2 \left[ - \left( { 1 + r^2 \over 1 - r^2 } \right)^2 dt^2 +
{ 4 \over ( 1 - r^2 )^2 }( dr^2 + r^2 d\Omega^2 ) \right].
}
In these coordinates the boundary of $AdS$ is at $r= 1$.
We saw above that when we calculate correlation functions 
we have to specify boundary conditions at $r = 1-\delta $ and then
take the limit of $\delta \to 0$. 
It is clear by studying the action of the conformal group on 
\Poincare coordinates that the radial position plays the role of
some energy scale, since we approach the boundary when we do a 
conformal transformation that localizes objects in the CFT. 
So, the limit $\delta \to 0$ corresponds to going to the UV of the
field theory.
When we are close to the boundary we could also use the \Poincare 
coordinates
\eqn{poinc}{
ds^2 = R^2 { -dt^2 + d{\vec x}^2 + dz^2 \over z^2},
}
in which the boundary is at $z=0$.
 If we consider a particle or wave propagating in 
\poinc\ or \adscav\
 we see that its motion is independent of $R$ in the supergravity
approximation. Furthermore,  if we are in Euclidean space  and
we have a wave that has some spatial extent 
$\lambda$ in the $\vec x$ directions, it will also have an 
extent $\lambda$ in the $z$ direction. This can be seen from \poinc\
by eliminating $\lambda $ through the change of variables 
$ x \to \lambda x $, $ z \to \lambda z$. 
This implies that a cutoff at 
\eqn{iruv}{
z \sim \delta 
} 
corresponds to 
a UV cutoff in the field theory at distances $\delta$, with 
no factors of $R$ ($\delta$ here is dimensionless, in the field theory
it is measured in terms of the radius of the $S^4$ or $S^3$ that the
theory lives on). Equation \iruv\ is called the UV-IR 
relation \cite{Susskind:1998dq}.

 Consider the case of ${\cal N} = 4 $ 
SYM on a three-sphere of radius one.
 We can estimate the number of degrees of freedom in the field theory
with a UV cutoff $\delta$. We get
\eqn{entroft}{
S \sim N^2 \delta^{-3},
}
since the number of cells into which we divide the three-sphere is of
order $1/\delta^3$.
In the gravity solution  \adscav\ 
the area in Planck units of 
the surface at $r = 1 - \delta $, for $\delta \ll 1$, is 
\eqn{areagrav}{
{ { \rm Area} \over 4 G_N} = { V_{S^5} R^3 \delta^{-3} \over 4 G_N} 
\sim N^2 \delta^{-3}.
}
Thus, we see that the AdS/CFT correspondence saturates the holographic bound
\cite{Susskind:1998dq}. 

One could be a little suspicious of the statement that gravity 
in $AdS$ is holographic, since it does not seem to be saying much
because in $AdS$ space
the volume and the boundary area of a given region scale in the
same fashion as we increase the size of the region. 
In fact, {\it any } field theory in $AdS$ would
be holographic in the sense that the number of degrees of 
freedom within some (large enough) 
volume is proportional to the area (and also to the volume). What makes 
this case different is that we have the additional 
parameter $R$, and then we can take $AdS$ spaces of 
different radii (corresponding to different values of
$N$ in the SYM theory),
and then we can ask whether the number of
degrees of freedom goes like the volume or the area, since 
these have a different dependence on $R$. 

One might  get confused by the fact that the surface $r =1-\delta$
is really nine dimensional as opposed to four dimensional. From the
form of the full metric on $AdS_5\times S^5$ 
we see that as we take $\delta \to 0$ the physical size of
four of the dimensions
of this nine dimensional space grow, while the other five, the $S^5$, 
remain constant. So, we see that the theory on this nine dimensional 
surface  becomes effectively four dimensional, since we need to multiply 
the metric by a factor that goes to zero as we approach the boundary
in order to define a finite  metric
for the four dimensional gauge theory. 

Note that even though it is often said that the field theory 
is defined on the boundary of $AdS$, it actually describes 
all the physics that is going on inside $AdS$.
When we are thinking in the $AdS$ picture it is incorrect to 
consider {\it at the same time} an additional field theory living
at the boundary\footnote{Except possibly for a small number of
singleton fields.}. Different regions of $AdS$ space, which are at
different radial positions, correspond to 
physics at different energy scales in the field theory. 
It is interesting that depending on what boundary we take, $\IR^{3+1}$ 
(in the \Poincare coordinates) or $S^3\times \IR$ (in the global coordinates),
we can either have a horizon or not have one. The 
presence of a horizon in the  $\IR^{3+1}$ case is related to the 
fact that the theory has no mass gap and we can have
excitations at arbitrarily low energies. This will always happen
when we have a horizon, since by bringing a particle close to a horizon
its energy becomes arbitrarily small. We are talking about the
energy  measured with respect to the time associated to the
Killing vector that vanishes at the horizon. In the $S^3 $ case there
is no horizon, and correspondingly the theory has a gap. In this
case the field theory
has a discrete spectrum since it is in finite volume. 

\begin{figure}[htb]
\begin{center}
\epsfxsize=1.5in\leavevmode\epsfbox{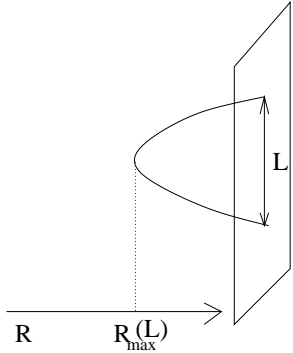}
\end{center}
\caption{
Derivation of the IR/UV relation by considering a spatial 
geodesic ending at two points on the boundary. 
}
\label{geodesic}
\end{figure}

Now let us consider the UV/IR correspondence in spaces that are not
$AdS$, like the ones which correspond to the field theories 
living on D-$p$-branes
with $p \not = 3$ (see section \ref{dpbranes}). 
A simple derivation involves  considering
 a classical spatial geodesic
that ends on the boundary at two points separated by a distance $L$ in 
field theory units (see figure \ref{geodesic}). 
 This geodesic goes into the bulk, and it 
has  a point at which the distance to the boundary is maximal.
Let us call this point $r_{max}(L)$. Then, one formulation of 
the UV/IR relation is
\eqn{uvirgen}{
 r = r_{max}(L) \leftrightarrow L.
}
A similar criterion arises if we consider the wave equation 
instead of classical geodesics \cite{Peet:1998wn};
of course both are the 
same since a classical geodesic arises as a limit of the wave equation
for very massive particles. 

Since the radial  direction arises holographically, it is not obvious
at first sight that the theory will be causal in the bulk.
Issues of causality in the holographic description of the spacetime
physics were discussed in 
\cite{Kabat:1999yq,Balasubramanian:1999ri,Horowitz:1999gf,%
Giddings:1999qu}. 

This holographic description has implications for the physics 
of black holes. This description
should therefore explain how the singularity inside black holes
should be treated (see \cite{Horowitz:1998pq}).
Holography also  implies that black hole evolution is 
unitary since the boundary theory is unitary.  
It is not totally clear, from the gravity point of view, how
the information comes back out or where it is stored
 (see \cite{Lowe:1999pk} for a discussion).
Some speculations about  holography and a 
 new uncertainty principle were 
discussed in \cite{Minic:1998nu}.

\section{Tests of the AdS/CFT Correspondence} 
\label{tests}

In this section we review the direct tests of the AdS/CFT
correspondence. In section \ref{correspondence} 
we saw how string theory on $AdS$
defines a partition function which can be used to define a field
theory. Here we will review the evidence showing that this field
theory is indeed the same as the conjectured dual field theory. We
will focus here only on tests of the correspondence between the
$\cn=4$ $SU(N)$ SYM theory and the type IIB string theory compactified
on $AdS_5\times S^5$; most of the tests described here can
be generalized also to cases in other dimensions and/or with less
supersymmetry, which will be described below.

As described in section \ref{correspondence}, 
the AdS/CFT correspondence is a
strong/weak coupling duality. In the 't Hooft large $N$ limit, it
relates the region of weak field theory coupling $\lambda=g_{YM}^2 N$
in the SYM theory to the region of high curvature (in string units) in
the string theory, and vice versa. Thus, a direct comparison of
correlation functions is generally not possible, since (with our
current knowledge) we can only compute most of them perturbatively in
$\lambda$ on the field theory side and perturbatively in
$1/\sqrt{\lambda}$ on the string theory side. For example, as
described below, we can compute the equation of state of the SYM
theory and also the quark-anti-quark potential both for small
$\lambda$ and for large $\lambda$, and we obtain different answers,
which we do not know how to compare since we can only compute them
perturbatively on both sides. A similar situation arises also in many
field theory dualities that were analyzed in the last few years (such
as the electric/magnetic $SL(2,\IZ)$ duality of the $\cn=4$ SYM theory
itself), and it was realized that there are several properties of
these theories which do not depend on the coupling, so they can be
compared to test the duality. These are:

\begin{itemize}
\item{} The global symmetries of the theory, which cannot change as we
change the coupling (except for extreme values of the coupling). As
discussed in section \ref{correspondence}, 
in the case of the AdS/CFT correspondence we
have the same supergroup $SU(2,2|4)$ (whose bosonic subgroup is
$SO(4,2)\times SU(4)$) as the global symmetry of both theories. Also,
both theories are believed to have a non-perturbative $SL(2,\IZ)$
duality symmetry acting on their coupling constant $\tau$. These are
the only symmetries of the theory on $\IR^4$. Additional $\IZ_N$
symmetries arise when the theories are compactified on
non-simply-connected manifolds, and these were also successfully
matched in \cite{Aharony:1998qu,Witten:1998wy}\footnote{Unlike 
most of the other tests described
here, this test actually tests the finite $N$ duality and not just the
large $N$ limit.}.

\item{} Some correlation functions, which are usually related to
anomalies, are protected from any quantum corrections and do not
depend on $\lambda$. The matching of these correlation functions will
be described in section \ref{anomalies} below.

\item{} The spectrum of chiral operators does not change as the
coupling varies, and it will be compared in section \ref{chiralops}
below.

\item{} The moduli space of the theory also does not depend on the
coupling. In the $SU(N)$ field theory the moduli space is
$\IR^{6(N-1)}/S_N$, parametrized by the eigenvalues of six commuting
traceless $N\times N$ matrices. On the AdS side it is not clear
exactly how to define the moduli space. As described in section 
\ref{multicenter_sols},
there is a background of string theory corresponding to any point in
the field theory moduli space, but it is not clear how to see that
this is the exact moduli space on the string theory side (especially
since high curvatures arise for generic points in the moduli space).

\item{} The qualitative behavior of the theory upon deformations by
relevant or marginal operators also does not depend on the coupling
(at least for chiral operators whose dimension does not depend on the
coupling, and in the absence of phase transitions). This will be
discussed in section \ref{deformations}. 

There are many more
qualitative tests of the correspondence, such as the existence of
confinement for the finite temperature theory \cite{Witten:1998zw},
which we will not discuss in this section. We will also not discuss
here tests involving the behavior of the theory on its moduli space
\cite{Douglas:1998tk,Gonzalez-Rey:1998uh,Das:1999ij}.

\end{itemize}

\subsection{The Spectrum of Chiral Primary Operators}
\label{chiralops}

\subsubsection{The Field Theory Spectrum}
\label{fieldspect}

The $\cn=4$ supersymmetry algebra in $d=4$ has four generators
$Q_\alpha^A$ (and their complex conjugates ${\bar
Q}_{\dot{\alpha}A}$), where $\alpha$ is a Weyl-spinor index
(in the $\bf 2$ of the $SO(3,1)$ Lorentz group) and $A$ is an index in
the $\bf 4$ of the $SU(4)_R$ R-symmetry group (lower indices $A$ will be
taken to transform in the $\bf {\bar 4}$ representation). 
They obey the algebra
\eqn{nfouralgebra}{\eqalign{
\{Q_{\alpha}^A, {\bar Q}_{\dot{\alpha}B} \} &= 2(\sigma^\mu)_{\alpha
\dot{\alpha}} P_\mu \delta^A_B, \cr
\{Q_\alpha^A, Q_\beta^B \} &= \{{\bar Q}_{\dot{\alpha}A}, {\bar
Q}_{\dot{\beta}B} \} = 0, }
}
where $\sigma^i$ ($i=1,2,3$) are the Pauli matrices and
$(\sigma^0)_{\alpha \dot{\alpha}}=-\delta_{\alpha \dot{\alpha}}$
(we use the conventions of Wess and Bagger \cite{Bagger}).

$\cn=4$ supersymmetry in four dimensions has a unique multiplet which
does not include spins greater than one, which is the vector
multiplet. It includes a vector field $A_\mu$ ($\mu$ is a vector
index of the $SO(3,1)$ Lorentz group), four complex
Weyl fermions $\lambda_{\alpha A}$ (in the $\bf {\bar 4}$ of
$SU(4)_R$), and six real scalars
$\phi^{I}$ (where $I$ is an index in the $\bf 6$ of $SU(4)_R$). The
classical action of the supersymmetry generators on these fields is
schematically given (for on-shell fields) by
\eqn{nfouraction}{\eqalign{
[Q_\alpha^A, \phi^I] &\sim \lambda_{\alpha B}, \cr
\{Q_\alpha^A, \lambda_{\beta B}\} &\sim
(\sigma^{\mu \nu})_{\alpha \beta} F_{\mu \nu} + \epsilon_{\alpha
\beta} [\phi^I, \phi^J], \cr
\{Q_\alpha^A, {\bar \lambda}_{\dot{\beta}}^B \} &\sim
({\sigma}^\mu)_{\alpha \dot{\beta}} {\cal D}_\mu \phi^I, \cr
[Q_\alpha^A, A_\mu] &\sim ({\sigma}_\mu)_{\alpha \dot{\alpha}} {\bar
\lambda}_{\dot{\beta}}^A \epsilon^{\dot{\alpha} \dot{\beta}}, 
}}
with similar expressions for the action of the $\bar Q$'s, where
$\sigma^{\mu \nu}$ are the generators of the Lorentz group in the
spinor representation, ${\cal D}_\mu$ is the covariant derivative,
the field strength
$F_{\mu \nu} \equiv [{\cal D}_\mu, {\cal D}_\nu]$, and we have suppressed
the $SU(4)$ Clebsch-Gordan coefficients corresponding to the products
$\bf{4\times 6 \to {\bar 4}}$, $\bf{4\times {\bar 4} \to 1+15}$ and
$\bf{4\times 4 \to 6}$ in the first three lines of \eno{nfouraction}.

An $\cn=4$ supersymmetric field theory is uniquely determined by
specifying the gauge group, and its field content is a vector
multiplet in the adjoint of the gauge group. Such a field theory is
equivalent to an $\cn=2$ theory with one hypermultiplet in the adjoint
representation, or to an $\cn=1$ theory with three chiral multiplets
$\Phi^i$ in the adjoint representation (in the ${\bf 3}_{2/3}$ of the
$SU(3)\times U(1)_R
\subset SU(4)_R$ which is left unbroken by the choice of a single
$\cn=1$ SUSY generator) and a superpotential of the form $W \propto
\epsilon_{ijk} \tr(\Phi^i \Phi^j \Phi^k)$. The interactions of the
theory include a scalar potential proportional to $\sum_{I,J}
\tr([\phi^I,\phi^J]^2)$, such that the moduli space of the theory is
the space of commuting matrices $\phi^I$ ($I=1,\cdots,6$).

The spectrum of operators in this theory includes all the gauge
invariant quantities that can be formed from the fields described
above. In this section we will focus on local operators which involve
fields taken at the same point in space-time. For the $SU(N)$ theory
described above, properties of the adjoint representation of $SU(N)$
determine that such operators necessarily involve a product of traces
of products of fields (or the sum of such products). It is natural to
divide the operators into single-trace operators and multiple-trace
operators. In the 't Hooft large $N$ limit correlation
functions involving multiple-trace operators are suppressed by powers
of $N$ compared to those of single-trace operators involving the same
fields. We will discuss here in detail only the single-trace
operators; the multiple-trace operators appear in operator product
expansions of products of single-trace operators.

As discussed in section \ref{cft}, it is natural to classify the
operators in a conformal theory into primary operators and their
descendants. In a superconformal theory it is also natural to
distinguish between chiral primary operators, which are in short
representations of the superconformal algebra and are annihilated by
some of the supercharges, and non-chiral primary operators.
Representations of the superconformal algebra are formed by starting
with some state of lowest dimension, which is annihilated by the
operators $S$ and $K_\mu$, and acting on it with the operators $Q$ and
$P_\mu$.  The $\cn=4$ supersymmetry algebra involves 16 real
supercharges.  A generic primary representation of the superconformal
algebra will thus include $2^{16}$ primaries of the conformal algebra,
generated by acting on the lowest state with products of different
supercharges;
acting with additional supercharges always leads to descendants of the
conformal algebra (i.e. derivatives). Since the supercharges have
helicities $\pm 1/2$, the primary fields in such representations will
have a range of helicities between $\lambda-4$ (if the lowest
dimension operator $\psi$ has helicity $\lambda$) and $\lambda+4$
(acting with more than 8 supercharges of the same helicity either
annihilates the state or leads to a conformal descendant). In
non-generic representations of the superconformal algebra a product of
less than 16 different $Q$'s annihilates the lowest dimension
operator, and the range of helicities appearing is smaller. In
particular, in the small representations of the $\cn=4$ superconformal
algebra only up to 4 $Q$'s of the same helicity acting on the lowest
dimension operator give a non-zero result, and the range of helicities
is between $\lambda-2$ and $\lambda+2$. For the $\cn=4$ supersymmetry
algebra (not including the conformal algebra) it is known that medium
representations, whose range of helicities is 6, can also exist (they
arise, for instance, on the moduli space of the $SU(N)$ $\cn=4$ SYM
theory \cite{Bergman:1997yw,Hashimoto:1998zs,Kawano:1998bp,
Bergman:1998gs,Hashimoto:1998nj,Lee:1998nv,Sasakura:1998cx,Tong:1999mg}); 
it is not clear if such medium representations of the
superconformal algebra \cite{Gunaydin:1998jc} 
can appear in physical theories or not
(there are no known examples). More details on the structure of
representations of the $\cn=4$ superconformal algebra may be found in
\cite{Gunaydin:1985fk,Andrianopoli:1998jh,Andrianopoli:1998nc,
Ferrara:1998pr,Gunaydin:1998sw,Andrianopoli:1998ut,Gunaydin:1998jc}
and references therein.

In the $U(1)$ $\cn=4$ SYM theory (which is a free theory), the only
gauge-invariant ``single trace'' operators are the fields of the
vector multiplet itself (which are $\phi^I,\lambda_A,
{\bar \lambda}^A$ and $F_{\mu \nu}
= \del_{[\mu} A_{\nu]}$). These operators form an ultra-short
representation of the $\cn=4$ algebra whose range of helicities is
from $(-1)$ to $1$ (acting with more than two supercharges of the same
helicity on any of these states gives either zero or derivatives,
which are descendants of the conformal algebra). All other local gauge
invariant operators in the theory involve derivatives or products of
these operators. This representation is usually called the doubleton
representation, and it does not appear in the $SU(N)$ SYM theory
(though the representations which do appear can all be formed by
tensor products of the doubleton representation). In the context of
AdS space one can think of this multiplet as living purely on the
boundary of the space \cite{Flato:1978qz,Fronsdal:1980aa,
Flato:1981we,Angelopoulos:1981wg,Nicolai:1984gb,Gunaydin:1985wc,
Gunaydin:1985vz,Flato:1986bf,Ferrara:1997dh,Ferrara:1998bv,Ferrara:1998jm}, as
expected for the $U(1)$ part of the original $U(N)$ gauge group of the
D3-branes (see the discussion in section \ref{correspondence}).

There is no known
simple systematic way to compute the full spectrum of
chiral primary operators of the $\cn=4$ $SU(N)$ SYM theory, so we will
settle for presenting the known chiral primary operators. The lowest
component of a superconformal-primary multiplet is characterized by
the fact that it cannot be written as a supercharge $Q$ acting on any
other operator. Looking at the action of the supersymmetry charges
\eno{nfouraction} suggests that generally operators built from the
fermions and the gauge fields will be descendants (given by $Q$ acting
on some other fields), so one would expect the lowest components of
the chiral primary representations to be built only from the scalar
fields, and this turns out to be correct.

Let us analyze the behavior of operators of the form ${\cal O}^{I_1
I_2 \cdots I_n} \equiv \tr(\phi^{I_1} \phi^{I_2} \cdots \phi^{I_n})$.
First we can ask if this operator can be written as $\{Q,\psi\}$ for
any field $\psi$. In the SUSY algebra
\eno{nfouraction} only commutators of $\phi^I$'s appear on the right
hand side, so we see that if some of the indices are antisymmetric the
field will be a descendant. Thus, only symmetric combinations of the
indices will be lowest components of primary multiplets. Next, we
should ask if the multiplet built on such an operator is a (short)
chiral primary multiplet or not. There are several different ways to
answer this question. One possibility is to use the relation between
the dimension of chiral primary operators and their R-symmetry
representation \cite{Kac:1977hp,Dobrev:1987qz,Dobrev:1985qv,
Seiberg:1997ax,Minwalla:1998ka}, and to check if this relation is
obeyed in the free field theory, where $[{\cal O}^{I_1 I_2 \cdots
I_n}] = n$. In this way we find that the representation is chiral
primary if and only if the indices form a symmetric traceless product
of $n$ $\bf 6$'s (traceless representations are defined as those who
give zero when any two indices are contracted). This is a representation
of weight $(0,n,0)$ of $SU(4)_R$; in this section we will refer to
$SU(4)_R$ representations either by their dimensions in boldface or by
their weights.

Another way to check this is to see if by acting with $Q$'s on these
operators we get the most general possible states or not, namely if
the representation contains ``null vectors'' or not (it turns out that
in all the relevant cases ``null vectors'' appear already at the first
level by acting with a single $Q$, though in principle there could be
representations where ``null vectors'' appear only at higher
levels). Using the SUSY algebra \eno{nfouraction} it is easy to see
that for symmetric traceless representations we get ``null vectors''
while for other representations we do not. For instance, let us
analyze in detail the case $n=2$. The symmetric product of two $\bf
6$'s is given by $\bf{6\times 6 \to 1 + 20'}$. The field in the $\bf
1$ representation is $\tr(\phi^I \phi^I)$, for which $[Q_\alpha^A,
\tr(\phi^I \phi^I)] \sim C^{AJB} \tr(\lambda_{\alpha B} \phi^J)$ where
$C^{AIB}$ is a Clebsch-Gordan coefficient for ${\bf {\bar 4}\times 6
\to 4}$. The right-hand side is in the $\bf 4$ representation, which is
the most general representation that can appear in the product
$\bf{4\times 1}$, so we find no null vectors at this level. On the
other hand, if we look at the symmetric traceless product
$\tr(\phi^{\{I} \phi^{J\}})\equiv \tr(\phi^I \phi^J) - {1\over 6}
\delta^{IJ} \tr(\phi^K \phi^K)$ in the ${\bf 20'}$ representation, we
find that $\{Q_{\alpha}^A, \tr(\phi^{\{I} \phi^{J\}}) \} \sim
\tr(\lambda_{\alpha B} \phi^K)$ with the right-hand side being in the
$\bf 20$ representation (appearing in $\bf{{\bar 4}\times 6 \to 4 +
20}$), while the left-hand side could in principle be in the
$\bf{4\times 20' \to 20+60}$. Since the $\bf 60$ does not appear on
the right-hand side (it is a ``null vector'') we identify that the
representation built on the $\bf 20'$ is a short representation of the
SUSY algebra. By similar manipulations (see \cite{Witten:1998qj,
Ferrara:1998ej,Andrianopoli:1998jh,Gunaydin:1998sw} for more details)
one can verify that chiral primary representations correspond exactly
to symmetric traceless products of $\bf 6$'s.

It is possible to analyze the chiral primary spectrum also
by using $\cn=1$ subalgebras of the $\cn=4$ algebra. If we use an
$\cn=1$ subalgebra of the $\cn=4$ algebra, as described above, the
operators ${\cal O}_n$ include the chiral operators of the form
$\tr(\Phi^{i_1} \Phi^{i_2} \cdots \Phi^{i_n})$ (in a representation of
$SU(3)$ which is a symmetric product of $\bf 3$'s), but for a
particular choice of the $\cn=1$ subalgebra not all the operators
${\cal O}_n$ appear to be chiral (a short multiplet of the $\cn=4$
algebra includes both short and long multiplets of the $\cn=1$
subalgebra).

The last issue we should discuss is what is the range of values of
$n$. The product of more than $N$ commuting\footnote{We can limit the
discussion to commuting matrices since, as discussed above,
commutators always lead to descendants, and we can write any product
of matrices as a product of commuting matrices plus terms with
commutators.} $N\times N$ matrices can
always be written as a sum of products of traces of less than $N$ of
the matrices, so it does not form an independent operator. This means
that for $n > N$ we can express the operator ${\cal O}^{I_1 I_2 \cdots
I_n}$ in terms of other operators, up to operators including
commutators which (as explained above) are descendants of the SUSY
algebra. Thus, we find that the short chiral primary representations
are built on the operators ${\cal O}_n = {\cal O}^{\{I_1 I_2 \cdots
I_n\}}$ with $n=2,3,\cdots,N$, for which the indices are in the
symmetric traceless product of $n$ $\bf 6$'s (in a $U(N)$ theory we would
find the same spectrum with the additional representation
corresponding to $n=1$). The superconformal algebra determines the
dimension of these fields to be $[{\cal O}_n] = n$, which is the same
as their value in the free field theory. We argued above that these
are the only short chiral primary representations in the $SU(N)$ gauge
theory, but we will not attempt to rigorously prove this here.

The full chiral primary representations are obtained by acting on the
fields ${\cal O}_n$ by the generators $Q$ and $P$ of the supersymmetry
algebra. The representation built on ${\cal O}_n$ contains a total of
$256\times {1\over 12}n^2(n^2-1)$ primary states, of which half are
bosonic and half are fermionic. Since these multiplets are built on a
field of helicity zero, they will contain primary fields of helicities
between $(-2)$ and $2$. The highest dimension primary field in the
multiplet is (generically) of the form $Q^4 {\bar Q}^4 {\cal O}_n$,
and its dimension is $n+4$. There is an elegant way to write these
multiplets as traces of products of ``twisted chiral $\cn=4$
superfields'' \cite{Ferrara:1998ej,Andrianopoli:1998jh}; see also
\cite{Ferrara:1998bp} which checks some components of these
superfields against the couplings to supergravity modes predicted on
the basis of the DBI action for D3-branes in anti-de Sitter space
\cite{Das:1998ei}.

It is easy to find the form of all the fields in such a multiplet by
using the algebra \eno{nfouraction}.  For example, let us analyze here
in detail the bosonic primary fields of dimension $n+1$ in the multiplet. To
get a field of dimension $n+1$ we need to act on ${\cal O}_n$ with two
supercharges (recall that $[Q]={1\over 2}$).  If we act with two
supercharges $Q_{\alpha}^A$ of the same chirality, their Lorentz
indices can be either antisymmetrized or symmetrized.
In the first case we get a Lorentz scalar field in the
$(2,n-2,0)$ representation of $SU(4)_R$, which is of the schematic
form 
\eqn{nplusone}{
\epsilon^{\alpha \beta} \{Q_\alpha, [Q_\beta, {\cal O}_n] \} \sim
\epsilon^{\alpha \beta} \tr(\lambda_{\alpha A} \lambda_{\beta B}
\phi^{J_1} \cdots \phi^{J_{n-2}}) + \tr([\phi^{K_1}, \phi^{K_2}]
\phi^{L_1} \cdots \phi^{L_{n-1}}).} 
Using an $\cn=1$ subalgebra some of these operators may be written as
the lowest components of the chiral superfields $\tr(W_{\alpha}^2
\Phi^{j_1} \cdots \Phi^{j_{n-2}})$. In the second case we get an
anti-symmetric 2-form of the Lorentz group, in the $(0,n-1,0)$
representation of $SU(4)_R$, of the form 
\eqn{nplusonet}{
\{Q_{\{\alpha}, [Q_{\beta\}}, {\cal O}_n] \} \sim
\tr((\sigma^{\mu
\nu})_{\alpha \beta} F_{\mu \nu} \phi^{J_1} \cdots \phi^{J_{n-1}}) +
\tr(\lambda_{\alpha A} \lambda_{\beta B} \phi^{K_1} \cdots
\phi^{K_{n-2}}).} 
Both of these fields are complex, with the complex
conjugate fields given by the action of two ${\bar Q}$'s. Acting with
one $Q$ and one ${\bar Q}$ on the state ${\cal O}_n$ gives a (real)
Lorentz-vector field in the $(1,n-2,1)$ representation of $SU(4)_R$,
of the form 
\eqn{nplusonev}{
\{Q_\alpha, [{\bar Q}_{\dot{\alpha}}, {\cal O}_n] \} \sim
\tr(\lambda_{\alpha A} {\bar \lambda}_{\dot{\alpha}}^B
\phi^{J_1} \cdots \phi^{J_{n-2}}) + (\sigma^\mu)_{\alpha \dot{\alpha}}
\tr(({\cal D}_\mu \phi^J) \phi^{K_1} \cdots \phi^{K_{n-1}}).}

At dimension $n+2$ (acting with four supercharges) we find :
\begin{itemize}
\item{} A complex scalar field in the $(0,n-2,0)$ representation,
given by $Q^4 {\cal O}_n$, of the form $\tr(F_{\mu \nu}^2 \phi^{I_1}
\cdots \phi^{I_{n-2}}) + \cdots$.
\item{} A real scalar field in the $(2,n-4,2)$ representation, given
by $Q^2 {\bar Q}^2 {\cal O}_n$, of the form $\epsilon^{\alpha \beta}
\epsilon^{\dot{\alpha} \dot{\beta}} \tr(\lambda_{\alpha A_1}
\lambda_{\beta A_2} {\bar \lambda}_{\dot{\alpha}}^{B_1} {\bar
\lambda}_{\dot{\beta}}^{B_2} \phi^{I_1} \cdots \phi^{I_{n-4}}) +
\cdots$.
\item{} A complex vector field in the $(1,n-4,1)$ representation,
given by $Q^3 {\bar Q} {\cal O}_n$, of the form $\tr(F_{\mu \nu} {\cal
D}^\nu \phi^{J} \phi^{I_1} \cdots \phi^{I_{n-2}}) + \cdots$.
\item{} An complex anti-symmetric 2-form field in the $(2,n-3,0)$
representation, given by $Q^2 {\bar Q}^2 {\cal O}_n$, of the form
$\tr(F_{\mu \nu} [\phi^{J_1},\phi^{J_2}] \phi^{I_1} \cdots
\phi^{I_{n-2}}) + \cdots$.
\item{} A symmetric tensor field in the $(0,n-2,0)$ representation,
given by $Q^2 {\bar Q}^2 {\cal O}_n$, of the form $\tr({\cal D}_{\{\mu}
\phi^J {\cal D}_{\nu\}} \phi^K \phi^{I_1} \cdots \phi^{I_{n-2}}) + \cdots$.
\end{itemize}

The spectrum of primary fields at dimension $n+3$ is similar to that
of dimension $n+1$ (the same fields appear but in smaller $SU(4)_R$
representations), and at dimension $n+4$ there is a single primary
field, which is a real scalar in the $(0,n-4,0)$ representation, given
by $Q^4 {\bar Q}^4 {\cal O}_n$, of the form $\tr(F_{\mu \nu}^4
\phi^{I_1} \cdots \phi^{I_{n-4}}) + \cdots$. Note that fields with
more than four $F_{\mu \nu}$'s or more than eight $\lambda$'s are
always descendants or non-chiral primaries. 

For $n=2,3$ the short multiplets are even shorter since some of the
representations appearing above vanish. In particular, for $n=2$ the
highest-dimension primaries in the chiral primary multiplet have
dimension $n+2=4$. The $n=2$ representation includes the currents of
the superconformal algebra. It includes a vector of dimension 3 in the
$\bf 15$ representation which is the $SU(4)_R$ R-symmetry current, and
a symmetric tensor field of dimension 4 which is the energy-momentum
tensor (the other currents of the superconformal algebra are
descendants of these). The $n=2$ multiplet also includes a complex
scalar field which is an $SU(4)_R$-singlet, whose real part is the
Lagrangian density coupling to ${1\over {4g_{YM}^2}}$ (of the form
$\tr(F_{\mu \nu}^2) + \cdots$) and whose imaginary part is the
Lagrangian density coupling to $\theta$ (of the form $\tr(F\wedge
F)$). For later use we note that the chiral primary multiplets which
contain scalars of dimension $\Delta \leq 4$ are the $n=2$ multiplet
(which has a scalar in the ${\bf 20'}$ of dimension 2, a complex
scalar in the $\bf 10$ of dimension 3, and a complex scalar in the $\bf
1$ of dimension 4), the $n=3$ multiplet (which contains a scalar in
the $\bf 50$ of dimension 3 and a complex scalar in the $\bf 45$ of
dimension 4), and the $n=4$ multiplet which contains a scalar in the
$\bf 105$ of dimension 4.

\subsubsection{The String Theory Spectrum and the Matching}
\label{stringy_spect}

As discussed in section \ref{field_operator}, 
fields on $AdS_5$ are in a one-to-one
correspondence with operators in the dual conformal field
theory. Thus, the spectrum of operators described in section
\ref{fieldspect} should agree with the spectrum of fields of type IIB
string theory on $AdS_5\times S^5$. Fields on AdS naturally lie in the
same multiplets of the conformal group as primary operators; the
second Casimir of these representations is $C_2=\Delta(\Delta-4)$ for
a primary scalar field of dimension $\Delta$ in the field theory, and
$C_2=m^2 R^2$ for a field of mass $m$ on an $AdS_5$ space with a
radius of curvature $R$. Single-trace operators in the field theory
may be identified with single-particle states in $AdS_5$, while
multiple-trace operators correspond to multi-particle states.

Unfortunately, it is not known how to compute the full spectrum of
type IIB string theory on $AdS_5\times S^5$. In fact, the only known
states are the states which arise from the dimensional reduction of
the ten-dimensional type IIB supergravity multiplet. These fields all
have helicities between $(-2)$ and $2$, so it is clear that they all
lie in small multiplets of the superconformal algebra, and we will
describe below how they match with the small multiplets of the field
theory described above. String theory on $AdS_5\times S^5$ is expected
to have many additional states, with masses of the order of the string
scale $1/l_s$ or of the Planck scale $1/l_p$. Such states would
correspond (using the mass/dimension relation described above) to
operators in the field theory with dimensions of order $\Delta \sim
(g_s N)^{1/4}$ or $\Delta \sim N^{1/4}$ for large $N, g_s N$. 
Presumably none of these
states are in small multiplets of the superconformal algebra (at
least, this would be the prediction of the AdS/CFT correspondence).

The spectrum of type IIB supergravity compactified on $AdS_5\times
S^5$ was computed in \cite{Kim:1985ez}. 
The computation involves expanding the ten
dimensional fields in appropriate spherical harmonics on $S^5$,
plugging them into the supergravity equations of motion, linearized
around the $AdS_5\times S^5$ background, and diagonalizing the
equations to give equations of motion for free (massless or massive)
fields\footnote{The fields arising from different spherical harmonics
are related by a ``spectrum generating algebra'', 
see \cite{Berglund:1999}.}. 
For example, the ten dimensional dilaton field $\tau$ may be
expanded as $\tau(x,y) = \sum_{k=0}^{\infty} \tau^k(x) Y^k(y)$ where
$x$ is a coordinate on $AdS_5$, $y$ is a coordinate on $S^5$, and the
$Y^k$ are the scalar spherical harmonics on $S^5$. These spherical
harmonics are in representations corresponding to symmetric traceless
products of $\bf 6$'s of $SU(4)_R$; they may be written as $Y^k(y)
\sim y^{I_1} y^{I_2} \cdots y^{I_k}$ where the $y^I$, for
$I=1,2,\cdots,6$ and with $\sum_{I=1}^6 (y^I)^2 = 1$, are coordinates
on $S^5$. Thus, we find a field $\tau^k(x)$ on $AdS_5$ in each such
$(0,k,0)$ representation of $SU(4)_R$, and the equations of motion
determine the mass of this field to be $m_k^2 = k(k+4)/R^2$. A similar
expansion may be performed for all other fields.

If we organize the results of \cite{Kim:1985ez} into representations
of the superconformal algebra \cite{Gunaydin:1985fk}, 
we find representations of the form
described in the previous section, which are built on a lowest
dimension field which is a scalar in the $(0,n,0)$ representation of
$SU(4)_R$ for $n=2,3,\cdots,\infty$. The lowest dimension scalar field
in each representation
turns out to arise from a linear combination of spherical harmonic
modes of the $S^5$ components of the graviton $h^a_a$ (expanded around
the $AdS_5\times S^5$ vacuum) and the 4-form field $D_{abcd}$, where
$a,b,c,d$ are indices on $S^5$. The scalar fields of dimension $n+1$
correspond to 2-form fields $B_{ab}$ with indices in the $S^5$.  The
symmetric tensor fields arise from the expansion of the
$AdS_5$-components of the graviton. The dilaton fields described above
are the complex scalar fields arising with dimension $n+2$ in the
multiplet (as described in the previous subsection).

In particular, the $n=2$ representation is called the supergraviton
representation, and it includes the field content of $d=5,\cn=8$
gauged supergravity. The field/operator correspondence matches this
representation to the representation including the superconformal
currents in the field theory. It includes a massless graviton field,
which (as expected) corresponds to the energy-momentum tensor in the
field theory, and massless $SU(4)_R$ gauge fields which correspond to
(or couple to) the global $SU(4)_R$ currents in the field theory.

In the naive dimensional reduction of the type IIB supergravity
fields, the $n=1$ doubleton representation, corresponding to a free
$U(1)$ vector multiplet in the dual theory, also appears. However, the
modes of this multiplet are all pure gauge modes in the bulk of
$AdS_5$, and they may be set to zero there. This is one of the reasons
why it seems more natural to view the corresponding gauge theory as an
$SU(N)$ gauge theory and not a $U(N)$ theory. It may be possible (and
perhaps even natural) to add the doubleton representation to the
theory (even though it does not include modes which propagate in the
bulk of $AdS_5$, but instead it is equivalent to a topological theory
in the bulk) to obtain a theory which is dual to the $U(N)$ gauge
theory, but this will not affect most of our discussion in this review
so we will ignore this possibility here.

Comparing the results described above with the results of section
\ref{fieldspect}, we see that we find the same spectrum of chiral
primary operators for $n=2,3,\cdots,N$. The supergravity results
cannot be trusted for masses above the order of the string scale
(which corresponds to $n\sim (g_s N)^{1/4}$) or the Planck scale
(which corresponds to $n\sim N^{1/4}$), so the results agree within
their range of validity. The field theory results suggest that the
exact spectrum of chiral representations in type IIB string theory on
$AdS_5\times S^5$ actually matches the naive supergravity spectrum up
to a mass scale $m^2 \sim N^2 / R^2 \sim N^{3/2} M_p^2$ which is much
higher than the string scale and the Planck scale, and that there are
no chiral fields above this scale. It is not known how to check this
prediction; tree-level string theory is certainly not enough for this
since when $g_s=0$ we must take $N=\infty$ to obtain a finite value of
$g_s N$. Thus, with our current knowledge the matching of chiral
primaries of the $\cn=4$ SYM theory with those of string theory on
$AdS_5\times S^5$ tests the duality only in the large $N$ limit. In
some generalizations of the AdS/CFT correspondence the string coupling
goes to zero at the boundary even for finite $N$, and then classical
string theory should lead to exactly the same spectrum of chiral
operators as the field theory. This happens in particular for 
the near-horizon limit of NS5-branes, 
in which case the exact spectrum was successfully compared in
\cite{Aharony:1998ub}. In other instances of the AdS/CFT
correspondence (such as the ones discussed in
\cite{Witten:1998xy,Klebanov:1998hh,Gubser:1998fp}) 
there exist also additional
chiral primary multiplets with $n$ of order $N$, and these have been
successfully matched with wrapped branes on the string theory side.

The fact that there seem to be no non-chiral fields on $AdS_5$ with a
mass below the string scale suggests that for large $N$ and large $g_s
N$, the dimension of all non-chiral operators in the field theory,
such as $\tr(\phi^I \phi^I)$, grows at least as $(g_s N)^{1/4} \sim
(g_{YM}^2 N)^{1/4}$. The reason for this behavior on the field theory
side is not clear; it is a prediction of the AdS/CFT correspondence.

\subsection{Matching of Correlation Functions and Anomalies}
\label{anomalies}

The classical $\cn=4$ theory has a scale invariance symmetry and an
$SU(4)_R$ R-symmetry, and (unlike many other theories) these
symmetries are exact also in the full quantum theory. However, when
the theory is coupled to external gravitational or $SU(4)_R$ gauge
fields, these symmetries are broken by quantum effects. In field
theory this breaking comes from one-loop diagrams and does not receive
any further corrections; thus it can be computed also in the strong
coupling regime and compared with the results from string theory on
AdS space.

We will begin by discussing the anomaly associated with the $SU(4)_R$
global currents. These currents are chiral since the fermions
$\lambda_{\alpha A}$ are in the $\bf {\bar 4}$ representation while
the fermions of the opposite chirality ${\bar
\lambda}_{\dot{\alpha}}^A$ are in the $\bf 4$ representation. Thus, if
we gauge the $SU(4)_R$ global symmetry, we will find an
Adler-Bell-Jackiw anomaly from the triangle diagram of three $SU(4)_R$
currents, which is proportional to the number of charged fermions.
In the $SU(N)$ gauge theory this number is $N^2-1$. The anomaly can be
expressed either in terms of the 3-point function of the $SU(4)_R$
global currents, 
\eqn{anomcorr}{
\left\langle J_{\mu}^a(x) J_{\nu}^b(y) J_\rho^c(z) \right\rangle_- =
-{{N^2-1}\over {32\pi^6}} id^{abc} {{\tr \left[ 
\gamma_5 \gamma_{\mu} (\not{x} - \not{y})
\gamma_{\nu} (\not{y} - \not{z}) \gamma_{\rho} (\not{z} - \not{x})
\right]}\over {(x-y)^4 (y-z)^4 (z-x)^4}},}
where 
$d^{abc} = 2\tr(T^a \{T^b, T^c\})$ and
we take only the negative parity component of the correlator, 
or in terms of the
non-conservation of the $SU(4)_R$ current when the theory is coupled to
external $SU(4)_R$ gauge fields $F_{\mu \nu}^a$,
\eqn{nonconcorr}{
({\cal D}^\mu J_\mu)^a = {{N^2-1}\over {384 \pi^2}} i d^{abc}
\epsilon^{\mu \nu \rho \sigma} F_{\mu \nu}^b F_{\rho \sigma}^c.}

How can we see this effect in string theory on $AdS_5\times S^5$ ? One
way to see it is, of course, to use the general prescription of
section \ref{correlators} 
to compute the 3-point function \eno{anomcorr}, and indeed
one finds \cite{Freedman:1998tz,Chalmers:1998xr} 
the correct answer to leading order in the large $N$
limit (namely, one recovers the term proportional to $N^2$). It is more
illuminating, however, to consider directly the meaning of the anomaly
\eno{nonconcorr} from the point of view of the AdS theory 
\cite{Witten:1998qj}. In the AdS
theory we have gauge fields $A_\mu^a$ which couple, as explained
above, to the $SU(4)_R$ global currents $J_\mu^a$ of the gauge theory,
but the anomaly means that when we turn on non-zero field strengths
for these fields the theory should no longer be gauge invariant. This
effect is precisely reproduced by a Chern-Simons term which exists in
the low-energy supergravity theory arising from the compactification
of type IIB supergravity on $AdS_5\times S^5$, which is of the form
\eqn{csterm}{{iN^2\over 96\pi^2} \int_{AdS_5} d^5x (d^{abc} \epsilon^{\mu \nu
\lambda \rho \sigma} A_\mu^a \del_\nu A_\lambda^b \del_\rho A_\sigma^c
+ \cdots).} 
This term is gauge invariant up to total derivatives,
which means that if we take a gauge transformation $A_\mu^a \to
A_\mu^a + ({\cal D}_\mu \Lambda)^a$ for which $\Lambda$ does not
vanish on the boundary of $AdS_5$, the action will change by a
boundary term of the form
\eqn{boundary}{
-{{iN^2}\over {384\pi^2}} \int_{\del AdS_5} d^4x \epsilon^{\mu \nu
\rho \sigma} d^{abc} \Lambda^a F_{\mu \nu}^b F_{\rho \sigma}^c.} 
{}From this we can read off the anomaly in $({\cal D}^\mu J_{\mu})$
since, when we have a coupling of the form $\int d^4x A^\mu_a
J_\mu^a$, the change in the action under a gauge transformation is
given by $\int d^4x ({\cal D}^\mu \Lambda)_a J_\mu^a = -\int d^4x
\Lambda_a ({\cal D}^\mu J_\mu^a)$, and we find exact agreement with
\eno{nonconcorr} for large $N$.

The other anomaly in the $\cn=4$ SYM theory is the conformal (or Weyl)
anomaly (see \cite{Deser:1993yx,Duff:1994wm} 
and references therein), indicating the breakdown of
conformal invariance when the theory is coupled to a curved external
metric (there is a similar breakdown of conformal invariance when the
theory is coupled to external $SU(4)_R$ gauge fields, which we will
not discuss here). The conformal anomaly is related to the 2-point and
3-point functions of the energy-momentum tensor 
\cite{Osborn:1994cr,Anselmi:1997mq,Erdmenger:1997yc,Anselmi:1997am}.  
In four dimensions, the general form of the conformal anomaly is
\eqn{confanom}{\vev{g^{\mu \nu} T_{\mu \nu}} = -aE_4 -cI_4,}
where
\eqn{confcoeff}{\eqalign{
E_4 &= {1\over {16\pi^2}} ({\cal R}_{\mu \nu \rho \sigma}^2 - 4 
{\cal R}_{\mu \nu}^2
+ {\cal R}^2), \cr
I_4 &= -{1\over {16\pi^2}} ({\cal R}_{\mu \nu \rho \sigma}^2 - 2 
{\cal R}_{\mu
\nu}^2 + {1\over 3}{\cal R}^2), }}
where ${\cal R}_{\mu \nu \rho \sigma}$ is the curvature tensor, 
${\cal R}_{\mu \nu}
\equiv {\cal R}_{\mu \rho \nu}^{\rho}$ is the Riemann tensor, and 
${\cal R} \equiv
{\cal R}^{\mu}_{\mu}$ is the scalar curvature.
A free field computation in the $SU(N)$ $\cn=4$ SYM theory leads to
$a=c=(N^2-1)/4$. In supersymmetric theories the supersymmetry algebra
relates $g^{\mu \nu}T_{\mu \nu}$ 
to derivatives of the R-symmetry current, so it is
protected from any quantum corrections. Thus, the same result should
be obtained in type IIB string theory on $AdS_5\times S^5$, 
and to leading order in the
large $N$ limit it should be obtained
from type IIB supergravity on $AdS_5\times S^5$. This
was indeed found to be true in
\cite{Henningson:1998gx,Henningson:1998ey,Balasubramanian:1999re,
Mueck:1999nf}\footnote{A generalization with more varying fields may be found
in \cite{Nojiri:1998dh}.}, 
where the conformal anomaly was
shown to arise from subtleties in the regularization of the
(divergent) supergravity action on $AdS$ space. The result of 
\cite{Henningson:1998gx,Henningson:1998ey,Balasubramanian:1999re,Mueck:1999nf}
implies that a computation using 
gravity on $AdS_5$ always gives rise to theories
with $a=c$, so generalizations of the AdS/CFT correspondence which
have (for large $N$) a supergravity approximation
are limited to conformal theories which have $a=c$ in the large
$N$ limit. Of course, if we do not require the string theory to have a
supergravity approximation then there is no such restriction.

For both of the anomalies we described the field theory and string
theory computations agree for the leading terms, which are of order
$N^2$. Thus, they are successful tests of the duality in the large $N$
limit. 
For other instances of the AdS/CFT correspondence there are
corrections to anomalies at order $1/N \sim g_s (\alpha'/R^2)^2$;
such corrections were discussed in \cite{Anselmi:1998zb} and
successfully compared in 
\cite{Aharony:1999rz,Blau:1999vz,Nojiri:1999mh}\footnote{Computing
such corrections tests the conjecture that the correspondence holds
order by order in $1/N$; however, this is weaker than the statement
that the correspondence holds for finite $N$, since the $1/N$
expansion is not expected to converge.}.
It would be interesting to compare other corrections to the large $N$
result.

Computations of other correlation functions 
\cite{Lee:1998bx,D'Hoker:1999tz,Gonzalez-Rey:1999ih}, 
such as 3-point
functions of chiral primary operators and correlation functions which
have only instanton contributions (we will discuss these in section
\ref{baryons}), have suggested that they are
also the same at small $\lambda$ and at large $\lambda$, even though
they are not related to anomalies in any known way. Perhaps there is
some non-renormalization theorem also for these correlation functions,
in which case their agreement would also be a test of the AdS/CFT
correspondence. 
As discussed in \cite{Intriligator:1998ig,
Intriligator:1999ff} (see also \cite{Ferrara:1998zt})
the non-renormalization theorem for 3-point
functions of chiral primary operators would follow from a conjectured
$U(1)_Y$ symmetry of the 3-point functions of $\cn=4$ SCFTs involving
at least two operators which are descendants of chiral 
primaries\footnote{A proof of this, using the analytic harmonic superspace
formalism which is conjectured to be valid in the $\cn=4$ theory,
was recently given in \cite{Eden:1999}.}. This
symmetry is a property of type IIB supergravity on $AdS_5\times S^5$
but not of the full type IIB string theory.

\section{Correlation Functions}
\label{correlators}

A useful statement of the AdS/CFT correspondence is that the partition
function of string theory on $AdS_5 \times S^5$ should coincide with the
partition function of ${\cal N}=4$ super-Yang-Mills theory ``on the
boundary'' of $AdS_5$ \cite{Gubser:1998bc,Witten:1998qj}.  The basic idea
was explained in section~\ref{field_operator}, but before summarizing the
actual calculations of Green's functions, it seems worthwhile to motivate
the methodology from a somewhat different perspective.

Throughout this section, we approximate the string theory partition
function by $e^{-I_{SUGRA}}$, where $I_{SUGRA}$ is the supergravity action
evaluated on $AdS_5 \times S^5$ (or on small deformations of this space).
This approximation amounts to ignoring all the stringy $\alpha'$
corrections that cure the divergences of supergravity, and also all the
loop corrections, which are controlled essentially by the gravitational
coupling $\kappa \sim g_{st} \alpha'^2$.  On the gauge theory side, as
explained in section~\ref{field_operator}, this approximation amounts to
taking both $N$ and $g_{YM}^2 N$ large, and the basic relation becomes
  \eqn{StringGauge}{
   e^{-I_{SUGRA}} \simeq Z_{\rm string} = Z_{\rm gauge} = e^{-W} \ ,
  }
        where $W$ is the generating functional for connected Green's
        functions in the gauge theory.  At finite temperature,
        $W = \beta F$ where $\beta$ is the inverse temperature and
        $F$ is the free energy of the gauge theory.
 When we apply this relation to a Schwarzschild black hole in
$AdS_5$, which is thought to be reflected in the gauge theory by a thermal
state at the Hawking temperature of the black hole, we arrive at the
relation $I_{SUGRA} \simeq \beta F$.  Calculating the free energy of a black
hole from the Euclidean supergravity action has a long tradition in the
supergravity literature \cite{Gibbons:1978ac}, so the main claim that is
being made here is that the dual gauge theory
provides a description of the state of the black hole which is physically
equivalent to the one in string theory. We will discuss the finite
temperature case further in section \ref{FiniteT}, and devote the rest
of this section to the partition function of the field theory on $\IR^4$.

The main technical idea behind the bulk-boundary correspondence is that the
boundary values of string theory fields (in particular, supergravity
fields) act as sources for gauge-invariant operators in the field theory.
From a D-brane perspective, we think of closed string states in the bulk as
sourcing gauge singlet operators on the brane which originate as composite
operators built from open strings.  We will write the bulk fields
generically as $\phi(\vec{x},z)$ (in the coordinate system
(\ref{poinc})), with value $\phi_0(\vec{x})$ for
$z=\epsilon$.  The true boundary of anti-de Sitter space is $z=0$, and
$\epsilon \neq 0$ serves as a cutoff which will eventually be removed.  In
the supergravity approximation, we think of choosing the values $\phi_0$
arbitrarily and then extremizing the action $I_{SUGRA}[\phi]$ in the
region $z > \epsilon$ subject to these boundary conditions.  In short, we
solve the equations of motion in the bulk subject to Dirichlet boundary
conditions on the boundary, and evaluate the action on the solution.  If
there is more than one solution, then we have more than one saddle point
contributing to the string theory partition function, and we must determine
which is most important.  In this section, multiple saddle points will not
be a problem.  So, we can write
  \def\extremum{\mathop{\rm extremum}}
  \eqn{WvsS}{
   W_{\rm gauge}[\phi_0] = -\log \left\langle
    e^{\int d^4 x \, \phi_0(x) {\cal O}(x)} \right\rangle_{CFT} \simeq
    \extremum_{\phi\big|_{z=\epsilon} = \phi_0} I_{SUGRA}[\phi] \ .
  }
 That is, the generator of connected Green's functions in the gauge theory,
in the large $N, g_{YM}^2 N$ limit, is the on-shell supergravity action.

Note that in \WvsS\ we have not attempted to be prescient about inserting
factors of $\epsilon$.  Instead our strategy will be to use \WvsS\ without
modification to compute two-point functions of ${\cal O}$, and then perform
a wave-function renormalization on either ${\cal O}$ or $\phi$ so that the
final answer is independent of the cutoff.  This approach should be
workable even in a space (with boundary) which is not asymptotically
anti-de Sitter, corresponding to a field theory which does not have a
conformal fixed point in the ultraviolet.

A remark is in order regarding the relation of (\ref{WvsS}) to the old
approach of extracting Green's functions from an absorption cross-section
\cite{Gubser:1997se}.  In absorption calculations one is keeping the whole
D3-brane geometry, not just the near-horizon $AdS_5 \times S^5$ throat.
The usual treatment is to split the space into a near region (the throat)
and a far region.  The incoming wave from asymptotically flat infinity can
be regarded as fixing the value of a supergravity field at the outer
boundary of the near region.  As usual, the supergravity description is
valid at large $N$ and large 't~Hooft coupling.  At small 't~Hooft
coupling, there is a different description of the process: a cluster of
D3-branes sits at some location in flat ten-dimensional space, and the
incoming wave impinges upon it.  In the low-energy limit, the value of the
supergravity field which the D3-branes feel is the same as the value in the
curved space description at the boundary of the near horizon region.
Equation~\WvsS\ is just a mathematical expression of the fact that the
throat geometry should respond identically to the perturbed supergravity
fields as the low-energy theory on the D3-branes.

Following \cite{Gubser:1998bc,Witten:1998qj}, a number of papers---notably
\cite{Aref'eva:1998nn, Muck:1998rr, Freedman:1998tz, Liu:1998bu,
Chalmers:1998xr, Muck:1998iz, Solodukhin:1998ec, Lee:1998bx, Liu:1999ty,
D'Hoker:1999tz, Freedman:1998bj, D'Hoker:1998gd, Chalmers:1998wu,
Muck:1998ug, D'Hoker:1998mz, Minces:1999tp, Arutyunov:1999nw}---have 
undertaken the
program of extracting explicit $n$-point correlation functions of gauge
singlet operators by developing both sides of \WvsS\ in a power series in
$\phi_0$.  Because the right hand side is the extremization of a classical
action, the power series has a graphical representation in terms of
tree-level Feynman graphs for fields in the supergravity.  There is one
difference: in ordinary Feynman graphs one assigns the wavefunctions of
asymptotic states to the external legs of the graph, but in the present
case the external leg factors reflect the boundary values $\phi_0$.  They
are special limits of the usual gravity propagators in the bulk, and are
called bulk-to-boundary propagators.  We will encounter their explicit form
in the next two sections.

\subsection{Two-point Functions}
\label{TwoPoint}

For two-point functions, only the part of the action which is
quadratic in the
relevant field perturbation is needed.  For massive scalar fields in
$AdS_5$, this has the generic form
  \eqn{Squad}{
   S = \eta \int d^5 x \, \sqrt{g} 
    \left[ \tf{1}{2} (\partial\phi)^2 + \tf{1}{2} m^2 \phi^2 \right],
  }
 where $\eta$ is some normalization which in principle follows from the
ten-dimensional origin of the action.
The bulk-to-boundary propagator
is a particular 
solution of the equation of motion, $(\square - m^2) \phi = 0$,
which has special asymptotic properties.  We will start by considering the
momentum space propagator, which is useful for computing the two-point
function and also in situations where the bulk geometry loses conformal
invariance; then, we will discuss the position space propagator, which
has proven more convenient for the study of higher point correlators
in the conformal case.  We will always work in Euclidean
space\footnote{The results may be analytically continued to give the
correlation functions of the field theory on Minkowskian $\IR^4$,
which corresponds to the \Poincare\ coordinates of AdS space.}.  A
coordinate system in the bulk of $AdS_5$ such that
  \eqn{AdSCoords}{
   ds^2 = {R^2 \over z^2} \left( d\vec{x}^2 + dz^2 \right) 
  }
 provides manifest Euclidean symmetry on the directions parametrized
by $\vec{x}$.  To avoid divergences associated with the small $z$
region of integration in \Squad, we will employ an explicit cutoff, $z
\geq \epsilon$.

A complete set of solutions for the linearized equation of motion,
$(\square - m^2) \phi = 0$, is given by $\phi = e^{i \vec{p} \cdot
\vec{x}} Z(pz)$, where the function $Z(u)$ satisfies the radial equation
  \eqn{BesselEq}{
   \left[ u^5 \partial_u {1 \over u^3} \partial_u - u^2 - m^2 R^2
    \right] Z(u) = 0 \ .
  }
 There are two independent solutions to \BesselEq, namely $Z(u) = u^2
I_{\Delta-2}(u)$ and $Z(u) = u^2 K_{\Delta-2}(u)$, where $I_\nu$ and
   $K_\nu$ are Bessel functions and
  \eqn{DimMassRel}{
   \Delta = 2 + \sqrt{4 + m^2 R^2} \ .
  }
 The second solution is selected by the requirement of regularity in the
interior: $I_{\Delta-2}(u)$ increases exponentially as $u\to\infty$ and
does not lead to a finite action configuration\footnote{Note that this
solution, when continued to Lorentzian AdS space, generally involves
the non-normalizable mode of the field, with $\lambda_-$ in 
(\ref{hyper}).}.  Imposing the boundary
condition $\phi(\vec{x},z) = \phi_0(\vec{x}) = e^{i \vec{p} \cdot \vec{x}}$
at $z = \epsilon$, we find the bulk-to-boundary propagator 
  \eqn{BoundaryProp}{
   \phi(\vec{x},z) = K_{\vec{p}}(\vec{x},z)
     = {(pz)^2 K_{\Delta-2}(pz) \over 
        (p\epsilon)^2 K_{\Delta-2}(p\epsilon)} e^{i \vec{p} \cdot \vec{x}} \ .
  }
 To compute a two-point function of the operator ${\cal O}$ for which
$\phi_0$ is a source, we write
  \eqn{TwoPtFct}{\eqalign{
   \langle {\cal O}(\vec{p}) {\cal O}(\vec{q}) \rangle
    &= \left. {\partial^2 W\left[ \phi_0 = \lambda_1 e^{i \vec{p} \cdot x} + 
       \lambda_2 e^{i \vec{q} \cdot x} \right] \over
       \partial \lambda_1 \partial \lambda_2} \right|_{\lambda_1 = 
        \lambda_2 = 0}  \cr
    &= \hbox{(leading analytic terms in $(\epsilon p)^2$)}  \cr
    &\qquad {} - \eta \epsilon^{2\Delta-8} (2\Delta-4)
     {\Gamma(3-\Delta) \over \Gamma(\Delta-1)}
     \delta^4(\vec{p}+\vec{q}) \left( {\vec{p} \over 2} 
      \right)^{2\Delta - 4}  \cr
    &\qquad {} + \hbox{(higher order terms in $(\epsilon p)^2$)},  \cr
   \langle {\cal O}(\vec{x}) {\cal O}(\vec{y}) \rangle &=
    \eta \epsilon^{2\Delta-8} {2\Delta-4 \over \Delta} 
     {\Gamma(\Delta+1) \over \pi^2 \Gamma(\Delta-2)}
     {1 \over |\vec{x}-\vec{y}|^{2\Delta}} \ .
  }}
 Several explanatory remarks are in order: 
  \begin{itemize}
   \item
 To establish the second equality in \TwoPtFct\ we have used 
\BoundaryProp , substituted in  \Squad,  performed
the integral  and expanded in
$\epsilon$.  The leading analytic terms give rise to contact terms in
position space, and the higher order terms are unimportant in the
limit where we remove the cutoff.  Only the leading nonanalytic term
is essential.  We have given the expression for generic real values of
$\Delta$.  Expanding around integer $\Delta \geq 2$ one obtains finite
expressions involving $\log \epsilon p$.
   \item{}
 The Fourier transforms used to obtain the last line are singular, but they
can be defined for generic complex $\Delta$ by analytic continuation and
for positive integer $\Delta$ by expanding around a pole and dropping
divergent terms, in the spirit of differential regularization
\cite{Freedman:1992tk}.  The result is a pure power law dependence on the
separation $|\vec{x}-\vec{y}|$, as required by conformal invariance.
   \item{}
 We have assumed a coupling $\int d^4x \, \phi(\vec{x},z=\epsilon) {\cal
O}(\vec{x})$ 
to compute the Green's functions.  The explicit powers of the cutoff in
the final position space answer can be eliminated by absorbing a factor of
$\epsilon^{\Delta-4}$ into the definition of ${\cal O}$.  From here on we
will take that convention, which amounts to inserting a factor of
$\epsilon^{4-\Delta}$ on the right hand side of \BoundaryProp.  
In fact, precise matchings between the normalizations in field theory
and in string theory
for all the chiral primary operators have not
been worked out.  In part this is due to the difficulty of determining the
coupling of bulk fields to field theory operators (or in stringy terms, the
coupling of closed string states to composite open string operators on the
brane).  See \cite{Gubser:1997yh} for an early approach to this problem.
For the dilaton, the graviton, and their superpartners (including gauge
fields in $AdS_5$), the couplings can be worked out explicitly.  In some of
these cases all normalizations have been worked out unambiguously and
checked against field theory predictions (see for example
\cite{Gubser:1998bc,Freedman:1998tz,D'Hoker:1999tz}).
   \item{}
 The mass-dimension relation (\ref{DimMassRel})
 holds even for string states that are not
included in the Kaluza-Klein supergravity reduction: the mass and the
dimension are just different expressions of the second
Casimir of $SO(4,2)$.  For
instance, excited string states, with $m \sim 1/\sqrt{\alpha'}$, are
expected to correspond to operators with dimension $\Delta \sim (g_{YM}^2
N)^{1/4}$.  The remarkable fact is that all the string theory modes with $m
\sim 1/R$ (which is to say, all closed string states which arise from
 massless ten dimensional fields) fall in
short multiplets of the supergroup $SU(2,2|4)$.  All other states have
a much
larger mass.  The operators in short multiplets have algebraically
protected dimensions.  The obvious conclusion is that all operators whose
dimensions are not algebraically protected have large dimension
in the strong 't~Hooft coupling, large $N$ limit to which supergravity
applies.  This is no longer true for theories of reduced supersymmetry: the
supergroup gets smaller, but the Kaluza-Klein states are roughly as
numerous as before, and some of them escape the short multiplets and
live in long multiplets of the smaller supergroups.  They
still have a mass on the order of $1/R$, and typically correspond to dimensions
which are finite (in the large $g_{YM}^2 N$ limit) but irrational.
  \end{itemize}

Correlation functions of non-scalar operators have been widely studied
following \cite{Witten:1998qj}; the literature includes
\cite{Henningson:1998cd, Ghezelbash:1998pf, Arutyunov:1998ve,
Arutyunov:1998xt, l'Yi:1998yt, Volovich:1998tj, l'Yi:1998pi, l'Yi:1998eu,
Koshelev:1998tu, Rashkov:1999ji, Polishchuk:1999nh}.  For ${\cal N}=4$
super-Yang-Mills theory, all correlation functions of fields in chiral
multiplets should follow by application of supersymmetries once those of
the chiral primary fields are known, so in this case it should be enough to
study the scalars.  It is worthwhile to note however that the
mass-dimension formula changes for particles with spin.  In fact the
definition of mass has some convention-dependence.  Conventions seem 
fairly uniform in the literature, and a table of mass-dimension
relations in $AdS_{d+1}$ with unit radius was made in
\cite{Freedman:1999gp} from the various sources cited above (see also
\cite{Ferrara:1998ej}):
  \begin{itemize}
\item scalars:\quad $\Delta_{\pm} = {1 \over 2} (d \pm \sqrt{d^2
+4m^2})$,
\item spinors:\quad  $\Delta = {1 \over 2} (d + 2|m|)$,
\item vectors:\quad
$ \Delta_{\pm} = {1 \over 2} (d \pm \sqrt{(d-2)^2 + 4m^2})$,
\item $p$-forms:\quad
$ \Delta = {1 \over 2} (d \pm \sqrt{(d-2p)^2 + 4m^2})$,
\item first-order $(d/2)$-forms ($d$ even):\quad
$\Delta={1\over 2}(d + 2|m|)$,
\item spin-3/2:\quad
$\Delta = {1 \over 2} (d + 2|m|)$,
\item massless spin-2:\quad
$\Delta = d$.
  \end{itemize}
 In the case of fields with second order lagrangians, we have not attempted
to pick which of $\Delta_\pm$ is the physical dimension.  Usually the
choice $\Delta=\Delta_+$ is clear from the unitarity bound, but in some
cases (notably $m^2 = 15/4$ in $AdS_5$) there is a genuine ambiguity.  In
practice this ambiguity is usually resolved by appealing to some special
algebraic property of the relevant fields, such as transformation under
supersymmetry or a global bosonic symmetry.  
See section~\ref{pfinads} for further discussion.  The scalar case above is 
precisely equation (\ref{branch}) in that section. 

 For brevity we will omit a further discussion of higher spins, and instead
refer the reader to the (extensive) literature.

\subsection{Three-point Functions}
\label{ThreePoint}

Working with bulk-to-boundary propagators in the momentum representation is
convenient for two-point functions, but for higher point functions position
space is preferred because the full conformal invariance is more obvious.
(However, for non-conformal examples of the bulk-boundary correspondence,
the momentum representation seems uniformly more convenient).  The boundary
behavior of position space bulk-to-boundary propagators is specified in a
slightly more subtle way: following \cite{Freedman:1998tz} we require
  \eqn{BBPlimit}{
   K_\Delta(\vec{x},z;\vec{y}) \to z^{4-\Delta} \delta^4(\vec{x}-\vec{y})
    \quad\hbox{as}\quad z \to 0 \ .
  }
 Here $\vec{y}$ is the point on the boundary where we insert the
operator, and $(\vec{x},z)$ is a point
in the bulk.  The unique regular $K_\Delta$ solving the equation of
motion and satisfying \BBPlimit\ is 
  \eqn{BBPvalue}{
   K_\Delta(\vec{x},z;\vec{y}) = {\Gamma(\Delta) \over \pi^2 
    \Gamma(\Delta-2)} \left( {z \over z^2 + (\vec{x}-\vec{y})^2} 
     \right)^{\Delta} \ .
  }
 At a fixed cutoff, $z=\epsilon$, the bulk-to-boundary propagator
$K_\Delta(\vec{x},\epsilon;\vec{y})$ is a
continuous function which approximates $\epsilon^{4-\Delta}
\delta^4(\vec{x}-\vec{y})$ better and
better as $\epsilon\to 0$.  Thus at any finite $\epsilon$, the Fourier
transform of \BBPvalue\ only approximately coincides with \BoundaryProp\
(modified by the factor of $\epsilon^{4-\Delta}$ as explained after
\TwoPtFct).  This apparently innocuous subtlety turned out to be important
for two-point functions, as discovered in \cite{Freedman:1998tz}.  A correct
prescription is to specify boundary conditions at finite $z = \epsilon$,
cut off all bulk integrals at that boundary, and only afterwards take
$\epsilon\to 0$.  That is what we have done in \TwoPtFct.  Calculating
two-point functions directly using the position-space propagators
\BBPlimit, but cutting the bulk integrals off again at $\epsilon$, and
finally taking the same $\epsilon\to 0$ answer, one arrives at a different
answer.  This is not surprising since the $z = \epsilon$ boundary
conditions were not used consistently.  
The authors of \cite{Freedman:1998tz} checked
that using the cutoff consistently (i.e. with the momentum space propagators)
gave two-point functions $\langle {\cal O}(\vec{x}_1) {\cal O}(\vec{x}_2)
\rangle$ a normalization such that Ward identities involving the
three-point function $\langle {\cal O}(\vec{x}_1) {\cal O}(\vec{x}_2)
J_\mu(\vec{x}_3) \rangle$, where $J_\mu$ is a conserved current, were
obeyed.  Two-point functions are uniquely difficult because of the poor
convergence properties of the integrals over $z$.  The integrals involved
in three-point functions are sufficiently benign that one can ignore the
issue of how to impose the cutoff.

\begin{figure}
      \vskip0cm
   \centerline{\psfig{figure=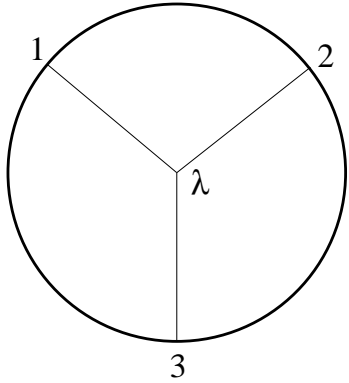,width=1.8in}}
   \vskip0cm
 \caption{The Feynman graph for the three-point function as computed in
supergravity.  The legs correspond to 
factors of $K_{\Delta_i}$, and the cubic vertex to
a factor of $\lambda$.  The position of the vertex is integrated
over $AdS_5$.}\label{figAssg}
  \end{figure}

If one has a Euclidean bulk action for three scalar fields $\phi_1$,
$\phi_2$, and $\phi_3$, of the form
  \eqn{IfAction}{
   S = \int d^5 x \, \sqrt{g} \left[
    \sum_i \tf{1}{2} (\partial\phi_i)^2 + \tf{1}{2} m_i^2 \phi_i^2 + 
    \lambda \phi_1 \phi_2 \phi_3 \right] \ ,
  }
 and if the $\phi_i$ couple to operators in the field theory by interaction
terms $\int d^4 x \, \phi_i {\cal O}_i$, then the calculation of $\langle
{\cal O}_1 {\cal O}_2 {\cal O}_3 \rangle$ reduces, via \WvsS, to the
evaluation of the graph shown in figure~\ref{figAssg}.  
 That is,
  \eqn{ThreeGraph}{\eqalign{
   &\langle {\cal O}_1(\vec{x}_1) {\cal O}_2(\vec{x}_2)
     {\cal O}_3(\vec{x}_3) \rangle = 
     -\lambda \int d^5 x \, \sqrt{g} 
     K_{\Delta_1}(x;\vec{x}_1) K_{\Delta_2}(x;\vec{x}_2)
     K_{\Delta_3}(x;\vec{x}_3)  \cr
   &\qquad\qquad 
    = {\lambda a_1 \over |\vec{x}_1 - \vec{x}_2|^{\Delta_1+\Delta_2-\Delta_3}
     |\vec{x}_1 - \vec{x}_3|^{\Delta_1+\Delta_3-\Delta_2}
     |\vec{x}_2 - \vec{x}_3|^{\Delta_2+\Delta_3-\Delta_1}} \ ,
  }}
 for some constant $a_1$.  The dependence on the $\vec{x}_i$ is dictated by
the conformal invariance, but the only way to compute $a_1$ is by
performing the integral over $x$.  The result \cite{Freedman:1998tz} is
  \eqn{aOneValue}{\eqalign{
   a_1 &= -{\Gamma\left[ {1 \over 2} (\Delta_1 + \Delta_2 -\Delta_3) \right]
    \Gamma\left[ {1 \over 2} (\Delta_1 + \Delta_3 -\Delta_2) \right]
    \Gamma\left[ {1 \over 2} (\Delta_2 + \Delta_3 -\Delta_1) \right] \over
    2\pi^4 \Gamma(\Delta_1 - 2) \Gamma(\Delta_2 - 2) \Gamma(\Delta_3 -
2)} \cdot  \cr
    &\qquad 
     \Gamma\left[ \tf{1}{2} (\Delta_1 + \Delta_2 + \Delta_3) - 2 \right] \ .
  }}
 In principle one could also have couplings of the form $\phi_1
\partial\phi_2 \partial\phi_3$.  This leads only to a modification of the
constant $a_1$.

The main technical difficulty with three-point functions is that one must
figure out the cubic couplings of supergravity fields.  Because of the
difficulties in writing down a covariant action for type IIB supergravity
in ten dimensions (see however
\cite{Dall'Agata:1997ju,Dall'Agata:1998va,Arutyunov:1998hf}), it is most
straightforward to read off these ``cubic couplings'' from quadratic terms
in the equations of motion.  In flat ten-dimensional space these terms can
be read off directly from the original type~IIB supergravity papers
\cite{Schwarz:1983qr,Howe:1984sr}.  For $AdS_5 \times S^5$, one must
instead expand in fluctuations around the background metric and five-form
field strength.  The old literature \cite{Kim:1985ez} only dealt with the
linearized equations of motion; for 3-point functions it is necessary to go
to one higher order of perturbation theory.  This was done for a restricted
set of fields in \cite{Lee:1998bx}.  The fields considered were those dual
to operators of the form $\tr \phi^{(J_1} \phi^{J_2} \ldots \phi^{J_\ell)}$
in field theory, where the parentheses indicate a symmetrized traceless
product.  These operators are the chiral primaries of the gauge theory: all
other single trace operators of protected dimension descend from these by
commuting with supersymmetry generators.  Only the metric and the five-form
are involved in the dual supergravity fields, and we are interested only in
modes which are scalars in $AdS_5$.  The result of \cite{Lee:1998bx} is
that the equations of motion for the scalar modes $\tilde{s}_I$ dual to
  \eqn{lmrsOp}{
   {\cal O}^I = {\cal C}^I_{J_1 \ldots J_\ell} 
     \tr \phi^{(J_1} \ldots \phi^{J_\ell)}
  }
 follow from an action of the form
  \eqn{lmrsAction}{\eqalign{
   S = {4 N^2 \over (2\pi)^5} \int d^5 x \, \sqrt{g} \Bigg\{ &
    \sum_I {A_I (w^I)^2 \over 2} \left[ -(\nabla \tilde{s}_I)^2 - l(l-4) 
      \tilde{s}_I^2 \right]  \cr
    &\qquad {} + 
     \sum_{I_1,I_2,I_3} {{\cal G}_{I_1 I_2 I_3} w^{I_1} w^{I_2} w^{I_3}
       \over 3} \tilde{s}_{I_1} \tilde{s}_{I_2} \tilde{s}_{I_3} \Bigg\} \ .
  }}
 Derivative couplings of the form $\tilde{s} \partial\tilde{s}
\partial\tilde{s}$ are expected {\it a priori} to enter into \lmrsAction,
but an appropriate field redefinition eliminates them.  The notation in
\lmrsOp\ and \lmrsAction\ requires some explanation.  $I$ is an index which
runs over the weight vectors of all possible representations constructed as
symmetric traceless products of the ${\bf 6}$ of $SU(4)_R$.  These are the
representations whose Young diagrams are
$\oalign{\idget\endrow\idget\endyoung}$,
$\oalign{\idget\idget\endrow\idget\idget\endyoung}$,
$\oalign{\idget\idget\idget\endrow\idget\idget\idget\endyoung}$, $\cdots$.
${\cal C}^I_{J_1 \ldots J_\ell}$ is a basis transformation matrix, chosen
so that ${\cal C}^I_{J_1 \ldots J_\ell} {\cal C}^J_{J_1 \ldots J_\ell} =
\delta^{IJ}$.  As commented in the previous section, there is generally a
normalization ambiguity on how supergravity fields couple to operators in
the gauge theory.  We have taken the coupling to be $\int d^4x \,
\tilde{s}_I {\cal O}^I$, and the normalization ambiguity is represented by
the ``leg factors'' $w^I$.  It is the combination $s^I = w^I \tilde{s}^I$
rather than $\tilde{s}^I$ itself which has a definite relation to
supergravity fields.  We refer the reader to \cite{Lee:1998bx} for explicit
expressions for $A_I$ and the symmetric tensor ${\cal G}_{I_1 I_2 I_3}$.
To get rid of factors of $w^I$, we introduce operators ${\cal O}^I =
\tilde{w}^I {\cal O}^I$.  One can choose $\tilde{w}^I$ so that a two-point
function computation along the lines of section~\ref{TwoPoint} leads to
  \eqn{TwoPointO}{
   \langle {\cal O}^{I_1}(\vec{x}) {\cal O}^{I_2}(0) \rangle = 
    {\delta^{I_1 I_2} \over x^{2\Delta_1}} \ .
  }
 With this choice, the three-point function, as calculated using
\ThreeGraph, is 
  \eqn{ThreePointO}{
   \langle {\cal O}^{I_1}(\vec{x_1}) {\cal O}^{I_2}(\vec{x_2})
    {\cal O}^{I_3}(\vec{x_3}) \rangle = 
    {1 \over N} {\sqrt{\Delta_1 \Delta_2 \Delta_3} 
     \langle {\cal C}^{I_1} {\cal C}^{I_2} {\cal C}^{I_3} \rangle \over
     |\vec{x}_1 - \vec{x}_2|^{\Delta_1+\Delta_2-\Delta_3}
     |\vec{x}_1 - \vec{x}_3|^{\Delta_1+\Delta_3-\Delta_2}
     |\vec{x}_2 - \vec{x}_3|^{\Delta_2+\Delta_3-\Delta_1}} \ ,
  }
 where we have defined
  \eqn{CAngleDef}{
   \langle {\cal C}^{I_1} {\cal C}^{I_2} {\cal C}^{I_3} \rangle = 
    {\cal C}^{I_1}_{J_1 \cdots J_i K_1 \cdots K_j}
    {\cal C}^{I_2}_{J_1 \cdots J_i L_1 \cdots L_k}
    {\cal C}^{I_3}_{K_1 \cdots K_j L_1 \cdots L_k} \ .
  }
 Remarkably, \ThreePointO\ is the same result one obtains from free field
theory by Wick contracting all the $\phi^J$ fields in the three operators.
This suggests that there is a non-renormalization theorem for this
correlation function, but such a theorem has not yet been proven (see however
comments at the end of section~\ref{anomalies}). It
is worth emphasizing that the normalization ambiguity in the bulk-boundary
coupling is circumvented essentially by considering invariant ratios of
three-point functions and two-point functions, into which the ``leg
factors'' $w^I$ do not enter.  This is the same strategy as was pursued in
comparing matrix models of quantum gravity to Liouville
theory.

\subsection{Four-point Functions}
\label{FourPointFunctions}

The calculation of four-point functions is difficult because there are
several graphs which contribute, and some of them inevitably involve
bulk-to-bulk propagators of fields with spin.  The computation of
four-point functions of the operators ${\cal O}_\phi$ and ${\cal O}_C$ dual
to the dilaton and the axion was completed in \cite{D'Hoker:1999pj}.  See
also \cite{Muck:1998rr,Liu:1999ty,Freedman:1998bj,D'Hoker:1998gd,
D'Hoker:1999jc,Liu:1998th,D'Hoker:1998mz,Chalmers:1998wu,Chalmers:1999gc,
Gonzalez-Rey:1998tk}
for earlier contributions.  One of the main technical results, further
developed in \cite{D'Hoker:1999ni}, is that diagrams involving an internal
propagator can be reduced by integration over one of the bulk vertices to a
sum of quartic graphs expressible in terms of the functions
  \eqn{DFunction}{\eqalign{
   D_{\Delta_1\Delta_2\Delta_3\Delta_4}(\vec{x}_1,\vec{x}_2,
    \vec{x}_3,\vec{x}_4) &= 
    \int d^5 x \, \sqrt{g} \prod_{i=1}^4 
     \tilde{K}_{\Delta_i}(\vec{x},z;\vec{x}_i),  \cr
   \tilde{K}_\Delta(\vec{x},z;\vec{y}) &= 
    \left( z \over z^2 + (\vec{x} - \vec{y})^2 \right)^\Delta \ .
  }}
 The integration is over the bulk point $(\vec{x},z)$.  There are two
independent conformally invariant combinations of the $\vec{x}_i$:
  \eqn{sAndt}{
   s = {1 \over 2} {\vec{x}_{13}^2 \vec{x}_{24}^2 \over 
    \vec{x}_{12}^2 \vec{x}_{34}^2 + \vec{x}_{14}^2 \vec{x}_{23}^2} \qquad
   t = {\vec{x}_{12}^2 \vec{x}_{34}^2 - \vec{x}_{14}^2 \vec{x}_{23}^2 \over 
        \vec{x}_{12}^2 \vec{x}_{34}^2 + \vec{x}_{14}^2 \vec{x}_{23}^2} \ .
  }
 One can write the connected four-point function as
  \eqn{FourPoint}{\eqalign{
   &\langle {\cal O}_\phi(\vec{x}_1) {\cal O}_C(\vec{x}_2) 
    {\cal O}_\phi(\vec{x}_3) {\cal O}_C(\vec{x}_4) \rangle
    = \left( 6 \over \pi^2 \right)^4 \Bigg[ 
     16 \vec{x}_{24}^2 \left( {1 \over 2s} - 1 \right) D_{4455} + 
     {64 \over 9} {\vec{x}_{24}^2 \over \vec{x}_{13}^2} {1 \over s} D_{3355}  \cr
    &\quad{} + 
     {16 \over 3} {\vec{x}_{24}^2 \over \vec{x}_{13}^2} {1 \over s} D_{2255} -
     14 D_{4444} - {46 \over 9 \vec{x}_{13}^2} D_{3344} - 
     {40 \over 9 \vec{x}_{13}^2} D_{2244} - 
     {8 \over 3 \vec{x}_{13}^6} D_{1144} + 
     64 \vec{x}_{24}^2 D_{4455} \Bigg] \ .
  }}

\begin{figure}
   \vskip0cm
   \centerline{\psfig{figure=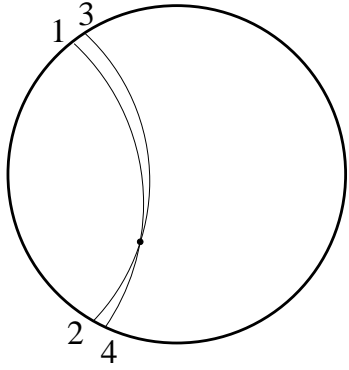,width=1.8in}}
   \vskip0cm
 \caption{A nearly degenerate quartic graph contributing to the four-point
function in the limit $|\vec{x}_{13}|,|\vec{x}_{24}| \ll
|\vec{x}_{12}|$.}\label{degenerate}
  \end{figure}

An interesting limit of \FourPoint\ is to take two pairs of points close
together.  Following \cite{D'Hoker:1999pj}, let us take the pairs
$(\vec{x}_1,\vec{x}_3)$ and $(\vec{x}_2,\vec{x}_4)$ close together while
holding $\vec{x}_1$ and $\vec{x}_2$ a fixed distance apart.  Then the
existence of an OPE expansion implies that
  \eqn{DoubleOPE}{
   \langle {\cal O}_{\Delta_1}(\vec{x}_1) {\cal O}_{\Delta_2}(\vec{x}_2) 
    {\cal O}_{\Delta_3}(\vec{x}_3) {\cal O}_{\Delta_4}(\vec{x}_4) \rangle = 
    \sum_{n,m} {\alpha_n \langle {\cal O}_n(\vec{x}_1) 
      {\cal O}_m(\vec{x}_2) \rangle
     \beta_m \over \vec{x}_{13}^{\Delta_1+\Delta_3-\Delta_m} 
      \vec{x}_{24}^{\Delta_2+\Delta_4-\Delta_n}} ,
  }
 at least as an asymptotic series, and hopefully even with a finite radius
of convergence for $\vec{x}_{13}$ and $\vec{x}_{24}$.  The operators ${\cal
O}_n$ are the ones that appear in the OPE of ${\cal O}_1$ with ${\cal
O}_3$, and the operators ${\cal O}_m$ are the ones that appear in the OPE
of ${\cal O}_2$ with ${\cal O}_4$.  ${\cal O}_\phi$ and ${\cal O}_C$ are
descendants of chiral primaries, and so have protected dimensions.  The
product of descendants of chiral fields is not itself necessarily the descendent
of a chiral field: an appropriately normal ordered product $:{\cal O}_\phi {\cal
O}_\phi:$ is expected to have an unprotected dimension of the form $8 +
O(1/N^2)$.  This is the natural result from the field theory point of view
because there are $O(N^2)$ degrees of freedom contributing to each factor,
and the commutation relations between them are non-trivial only a fraction
$1/N^2$ of the time.  From the supergravity point of view, a composite
operator like $:{\cal O}_\phi {\cal O}_\phi:$ corresponds to a two-particle
bulk state, and the $O(1/N^2) = O(\kappa^2/R^8)$ correction to the mass is
interpreted as the correction to the mass of the two-particle state from
gravitational binding energy.  Roughly one is thinking of graviton exchange
between the legs of figure~\ref{degenerate} that are nearly coincident.

If \DoubleOPE\ is expanded in inverse powers of $N$, then the $O(1/N^2)$
correction to $\Delta_n$ and $\Delta_m$ shows up to leading order as a term
proportional to a logarithm of some combination of the separations
$\vec{x}_{ij}$.  Logarithms also appear in the expansion of \FourPoint\ in
the $|\vec{x}_{13}|, |\vec{x}_{24}| \ll |\vec{x}_{12}|$ limit in which
\DoubleOPE\ applies: the leading log in this limit is ${1 \over
(\vec{x}_{12})^{16}} \log\left( {\vec{x}_{13} \vec{x}_{24} \over
\vec{x}_{12}^2} \right)$.  This is the correct form to be interpreted in
terms of the propagation of a two-particle state dual to an operator whose
dimension is slightly different from $8$.

\section{Isomorphism of Hilbert Spaces} 
\label{isom}

The $AdS$/CFT correspondence is a statement about
the equivalence of two quantum theories: string
theory (or M theory) on \adsp\ $\times$ (compact space)
and CFT$_{p+1}$. The two quantum theories
are equivalent if there is an isomorphism
between their Hilbert spaces, and moreover
if the operator algebras on the Hilbert spaces
are equivalent. In this section, we
discuss the isomorphism of the Hilbert spaces,
following \cite{Horowitz:1998bj,Witten:1998zw,
Maldacena:1998bw,Banks:1998dd}.
Related issues have been discussed 
in \cite{Barbon:1998ix,Barbon:1998cr,Abel:1999rq,
Chamblin:1999pz,Li:1999jy,Peet:1998cr,
Martinec:1998wm,Martinec:1999ja,
Martinec:1999sa}.

States in the Hilbert space of CFT$_{p+1}$
fall into representations of the global conformal
group $SO(2,p+1)$ on $\IR^{p,1}$. At the same time,
the isometry group of \ads\ is also $SO(2,p+1)$, and 
we can use it to classify states in 
the string theory. Thus, it is useful to compare
states in the two theories by organizing them into
representations of $SO(2,p+1)$. 
The conformal group $SO(2,p+1)$ has $\frac{1}{2}(p+2)(p+3)$
generators, $J_{ab} = -J_{ba}$ ($a,b = 0,1, \cdots, p+2$),
obeying the commutation relation
\beq
   [J_{ab}, J_{cd}] = -i ( g_{ac} J_{bd} \pm {\sl permutations})
\label{commutations} 
\eeq
with the metric $g_{ab}={\rm diag} (-1,+1,+1, \cdots, +1, -1)$.
In CFT$_{p+1}$, they are identified with
 the Poincar\'e generators $P_\mu$ and $M_{\mu\nu}$,
the dilatation $D$ and the special conformal generators $K_\mu$
($\mu, \nu = 0, \cdots, p$), by the formulas
\beq
   J_{p+2,p+1} = D, ~ J_{\mu,p+2} = \frac{1}{2}(K_\mu + P_\mu),
~ J_{\mu,p+1} = \frac{1}{2}(K_\mu - P_\mu), ~ J_{\mu\nu} = M_{\mu\nu}.
\eeq

Since the field theory on $\IR^{p,1}$ has no scale,  
the spectrum of the Hamiltonian $P_0$ is
continuous and there is no normalizable
ground state with respect to $P_0$. 
This is also the case for the string theory on 
\adsp. The Killing vector $\partial_t$
corresponding to $P_0$ has the norm
\beq
   || \partial_t || = Ru,
\eeq
and it vanishes as $u \rightarrow 0$. Consequently,
a stationary wave solution of the linearized supergravity
on \ads\ has a continuous frequency spectrum with respect to 
the timelike coordinate $t$. It is not easy to compare the
spectrum of $P_0$ of the two theories.

It is more useful to compare the two Hilbert spaces using the maximum compact
subgroup $SO(2) \times SO(p+1)$ of the conformal group \cite{Horowitz:1998bj}. 
The Minkowski space $\IR^{p,1}$ is conformally embedded in 
the Einstein Universe $\IR \times S^p$, 
and $SO(2) \times SO(p+1)$ is its isometry group. 
In particular, the generator $J_{0,p+2}=\frac{1}{2}(P_0 + K_0)$
of $SO(2)$ is the Hamiltonian 
for the CFT on $\IR \times S^p$. Now we have a scale in the problem, 
which is the radius of $S^p$, and the Hamiltonian $\frac{1}{2}(P_0 + K_0)$
has a mass gap. In string theory on \adsp , 
the generator $\frac{1}{2}(P_0 + K_0)$ corresponds to the global 
time translation 
along the coordinate $\tau$. This is a globally well-defined
coordinate on \ads\ and the Killing vector $\partial_\tau$
is everywhere non-vanishing:
\beq
  || \partial_\tau || = \frac{R}{\cos \theta}. 
\eeq
Therefore, a stationary wave solution with respect to $\tau$
is normalizable and has a discrete frequency spectrum.  
In fact, as we saw in section \ref{kkcatalogue}, the frequency
is quantized in such a way that bosonic fields in the
supergravity multiplet are periodic and their superpartners 
are anti-periodic ($i.e.$ obeying the supersymmetry preserving
Ramond boundary condition) in the $\tau$-direction
with the period $2\pi R$.

\subsection{Hilbert Space of String Theory}

With the techniques that  are currently available, 
we can make reliable  statements 
about the Hilbert space structure of  string theory on \ads\ 
only when the curvature radius $R$ of \ads\ is 
much larger than the string length $l_s$. 
In this section we will study some of the properties of the 
Hilbert space that we can see in the $AdS$ description. We will 
concentrate on the $AdS_5 \times S^5$ case, but it is easy to
 generalize this  to other cases.

We first consider the case that corresponds to the 't Hooft limit
$g_s \to 0 $, $g_s N$ fixed and large, so that we can trust the gravity 
approximation.

\medskip
\noindent
(1) $E \ll  m_s $; {\it Gas of Free Gravitons}

\smallskip

The Hilbert space for low energies is well approximated by the
Fock space of gravitons and their superpartners on $AdS_5\times S^5$.
Since $\tau$ is a globally defined timelike coordinate on \ads ,
we can consider stationary wave solutions in the linearized
supergravity, which are the normalizable states discussed in section
\ref{pfinads}. The frequency $\omega$ of a stationary mode is quantized
in the unit set by the curvature radius $R$ (\ref{energyquanta}),
so  one may effectively view the supergravity particles
in \ads\ as confined in a box of size $R$. 

The operator $H=\frac{1}{2R}(P_0 + K_0)$ 
corresponds\footnote{
The factor $\frac{1}{2R}$ in the relation between $H$ and
$(P_0+K_0)$ is fixed by
the commutation relations (\ref{commutations}).}  to
the Killing vector $\partial_\tau$ on \ads.
Thus, a single particle state of frequency $\omega$
gives an eigenstate of $H$.
Since the supergraviton is a BPS particle, 
its energy eigenvalue $\omega$ 
is exact, free from corrections 
either by first quantized string effects ($\sim l_s/R$) 
or by quantum gravity effects 
($\sim l_P/R$). The energy of multiparticle states
may receive corrections, but they become important only
when the energy $E$ becomes comparable to the gravitational
potential $E^2/(m_P^8 R^7)$, $i.e.$ $E \sim m_P^8 R^7$.
For the energies we are considering  this effect is negligible. 

Therefore, the Hilbert space for $E \ll m_s$ is identified
with the Fock space of free supergravity particles. 
For $E \gg R^{-1}$, the
entropy $S(E)$ ($= \log N(E)$ where $N(E)$ 
is the density of states) behaves as
%
\beq
   S(E) \sim (ER)^{\frac{9}{10}},
 \label{fock}
\eeq
since we effectively have a gas in ten dimensions (we will ignore
multiplicative numerical factors in the entropy in this section).

\medskip
\noindent
(2) $ m_s < E \ll  m_s/g_s^2 $; {\it Gas of Free Strings}

\smallskip

When the energy $E$ becomes comparable to the string scale $m_s$, we
have to take into account excitations on the string worldsheet.
Although we do not know the exact first quantized spectrum 
of string theory on \ads, 
we can estimate the effects of the worldsheet excitations
when $l_s \ll R$.
The mass $m$ of a first quantized string state is a function of 
$l_s$ and $R$.  When $l_s \ll R$, the worldsheet dynamics is 
perturbative and we can expand $m$ in powers of $l_s/R$, with the
leading term given by the string spectrum on flat space 
($R = \infty$). Therefore, for a string state corresponding to the
$n$-th excited level of the string on flat space,
the (mass)$^2$ is given by
\beq
     m^2 = l_s^{-2} \left( n + O(l_s^2/R^2) \right).
\eeq
Unlike the single particle supergravity states discussed in the previous
paragraph, string excitations need  not carry integral
eigenvalues of $H$ (in units of $R^{-1}$). As they
are not BPS particles, they are generically 
unstable in string perturbation theory. 

The free string spectrum in 10 dimensions gives
the Hagedorn density of states 
\beq
  S(E) \simeq  E l_s.
\label{string}
\eeq
Thus, the entropy of supergravity particles (\ref{fock})
becomes comparable to that of excited strings (\ref{string})
when
\beq
 (ER)^{\frac{9}{10}} \sim E l_s, 
\eeq
namely 
\beq
E \sim  m_s^{10}  R^9.
\eeq 
For $ m_s^{10}  R^9 < E $, excited strings dominate
the Hilbert space. 
The free string formula (\ref{string}) 
is reliable until the energy hits another transition
point $E \sim m_s/g_s^2 $. 
We are assuming that $R^9 < l_s^9/g_s^2$, which is true in the 
't Hooft region. 

\medskip
\noindent
(3) $m_s/g_s^2  \ll E \ll m_P^8 R^7$; {\it Small Black Hole}

\smallskip

As we increase the energy, the gas of free strings starts
collapsing to make a black hole. The black hole can be described
by the classical supergravity when the horizon radius $r_+$ 
becomes larger than the string length $l_s$. Furthermore, if the
horizon size $r_+$ is smaller than $R$,
the geometry near the black hole can be approximated by the 10-dimensional
Schwarzschild solution. The energy $E$ and the entropy $S$ of such
a black hole is given by
\begin{eqnarray}
   E &\sim &   m_P^8 r_+^7 
\nonumber \\
  S & \sim & (m_P r_+)^8.
\label{tendformula}
\end{eqnarray}
Therefore, the entropy is estimated to be
\beq
    S(E) \sim  (E l_P )^{\frac{8}{7}}.
\label{smallbh}
\eeq
We can trust this estimate when $l_s \ll r_+ \ll R$,
namely $m_P^8 l_s^7 \ll E \ll m_P^8 R^7$.
Comparing this with the Hagedorn density of states
in the regime (2) given by (\ref{string}), we find that 
the transition to (\ref{smallbh}) takes place at
\beq
 E \sim {m_s \over g_s^2} .
\eeq
For $E \gg m_P^8 l_s^7$, the entropy formula
(\ref{smallbh}) is reliable and the black hole
entropy exceeds that of the gas of free strings. 
Therefore, in this regime, the Hilbert space
is dominated by black hole states. 

\medskip
\noindent
(4) $m_P^8 R^7 < E$; {\it Large Black Hole}

\smallskip
The above analysis assumes that the size of the black hole,
characterized by the horizon radius $r_+$, is small compared 
to the radii $R$ of \ads$_5$ and $S^5$. 
As we increase the energy, the radius $r_+$ grows 
and eventually becomes comparable to $R$. Beyond this point, we 
can no longer use the 10-dimensional Schwarzschild solution to 
estimate the number of states. According to (\ref{tendformula}),
the horizon size becomes comparable to $R$ when the energy of
the black hole reaches the scale $E  \sim m_P^8 R^7$. Beyond 
this energy scale,  we have to use a solution which is asymptotically 
$AdS_5$ \cite{Hawking:1983dh}, 
\beq
   ds^2 = - f(r) d\tau^2
+ \frac{1}{f(r)}dr^2
 + r^2 d\Omega_3^2,
\label{adsbh}
\eeq
where
\beq
  f(r) = 1 + \frac{r^2}{R^2} - \frac{r_+^{2}}{r^{2}}
 \left(1+ \frac{r_+^2}{R^2} \right),
\eeq
and $r=r_+$ is the location of the out-most horizon. 
By studying the asymptotic behavior of the metric,
one finds that the black hole carries the energy
\beq
  E \sim  
  \frac{r_+^2}{{\bf l}_P^3}
\left( 1 + \frac{r_+^2}{R^2} \right).
\label{adsbhenergy}
\eeq
Here ${\bf l}_P$ is the five-dimensional Planck length,
related to the 10-dimensional Planck scale $l_P$
and the compactification scale $R$ as
\beq
  {\bf l}_P^3 = l_P^8 R^{-5}.
\eeq
The entropy of the \ads\ Schwarzschild solution is
given by
\beq
  S \sim \left(\frac{r_+}{{\bf l}_P} \right)^3.
\eeq
For $r_+ \gg R$, (\ref{adsbhenergy}) becomes
$E \sim  r_+^{4}{\bf l}_P^{-3} R^{-2}$, and the 
entropy as a function of energy is
\beq
  S \sim \left(\frac{ER^2}{{\bf l}_P}\right)^{\frac{3}{4}}
= \left(\frac{R}{l_P} \right)^{2} (ER)^{\frac{3}{4}}.
\label{adsbhentropy}
\eeq
As the energy increases, the horizon size expands as
$  R \ll r_+ \rightarrow \infty $,
and the supergravity approximation continues to be reliable.
For $E \rightarrow \infty$, the only stringy and quantum gravity
corrections are due to the finite size $R$ of
the \ads\ radius of curvature and of the compact space, and such
corrections are suppressed by factors of $l_s/R$ and $l_P/R$.
The leading $l_s/R$ corrections to (\ref{adsbhentropy})
were studied in \cite{Gubser:1998nz}, and
found to be of the order of $(l_s/R)^3$.

\medskip
\noindent
$\circ$ Summary

\smallskip
The above analysis gives the following picture about the
structure of the Hilbert space of string theory on \ads\
when $ l_s\ll R $ and $g_s \ll 1$.

\begin{figure}[htb]
\begin{center}
\epsfxsize=3.8in\leavevmode\epsfbox{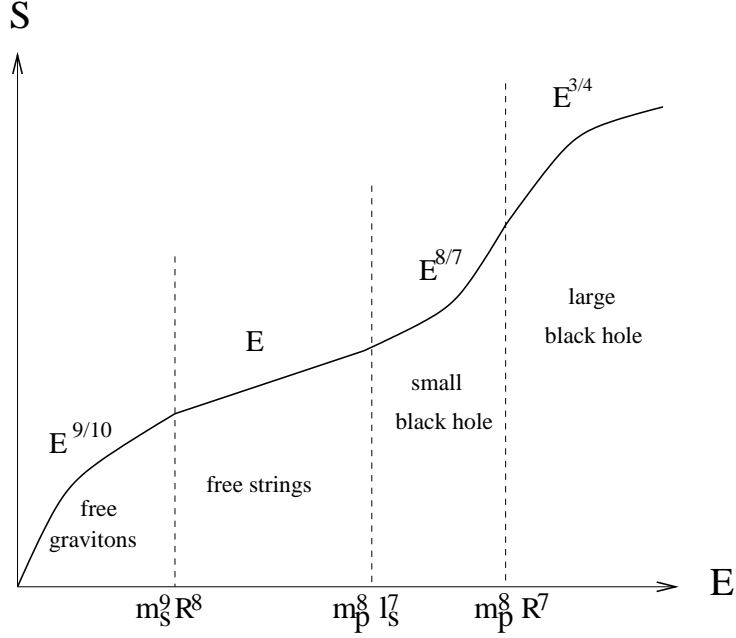}
\end{center}
\caption{The behavior of the entropy $S$ as a function of
the energy $E$ in \ads$_5$ .}
\label{ent}
\end{figure}

\noindent
(1) For energies $E \ll m_s$, the Hilbert space is the
Fock space of supergravity particles and the spectrum is
quantized in the unit of $R^{-1}$. For 
$E \ll m_s^{10} R^9$, the entropy is given by that
of the gas of free supergravity particles in 10 dimensions: 
\beq
S \sim (ER)^{\frac{9}{10}}.
\eeq 

\noindent
(2)  For $m_s^{10} R^9 < E \ll m_P^8 l_s^7 $, stringy
excitations become important, and the entropy grows
linearly in energy: 
\beq
S \sim El_s.
\eeq 

\noindent
(3) 
For $m_P^8 l_s^7 \ll E \ll m_P^8 R^7$,
the black hole 
starts to show up in the Hilbert space. 
For $E \ll m_P^8 R^7$, the size of the black hole
horizon is smaller than $R$, and the entropy is given by
that of the 10-dimensional Schwarzschild solution:
\beq
S \sim  (El_P)^{\frac{8}{7}}.
\eeq

\noindent
(4) For $m_P^8 R^7 < E$, the size of the black hole
horizon becomes larger than $R$. We then have to use
the \adsp\ Schwarzschild solution, and the entropy is
given by:
\beq
S \sim \left(\frac{R}{l_P} \right)^{2}
(ER)^{\frac{3}{4}}.
\label{largebhent}
\eeq

\noindent
The behavior of the entropy is depicted in figure
\ref{ent}. 

In the small black hole regime (3),
the system has a negative specific heat. This
corresponds to the well-known instability
of the flat space at finite temperature \cite{Gross:1982cv}. 
On the other hand, the \ads\ Schwarzschild solution
has a positive specific heat and it is thermodynamically stable.
This means that, if we consider a canonical ensemble, the
free string regime (2) and the small black hole regime
(3) will be missed. When set in contact
with a heat bath of temperature $T \sim m_s$, the
system will continue to absorb heat until its energy
reaches $E \sim m_P^8 R^7$, 
the threshold of the large black hole regime (4). 
In fact the jump from (1) to (4) takes place at much
lower temperature since 
the temperature equivalent of $E \sim m_P^8 R^7$
derived from (\ref{largebhent}) in the regime (4)
is $T \sim R^{-1}$. Therefore, once the temperature is
raised to $T \sim R^{-1}$ a black hole forms.
 The behavior of the canonical
ensemble will be discussed in more detail in section \ref{FiniteT}. 

Finally let us notice that in the case that $g_s \sim 1$ we 
do not have the Hagedorn phase, and we go directly from the gas of
gravitons to the small black hole phase.

\subsection{Hilbert Space of Conformal Field Theory}

Next, let us turn to a discussion of the Hilbert space of
the CFT$_{p+1}$. The generator $J_{0,p+2}=
\frac{1}{2}(P_0+K_0)$ is the Hamiltonian of the CFT on $S^p$
with the unit radius. In the Euclidean CFT, the conformal
group $SO(2,p+1)$ turns into $SO(1,p+2)$ by the Wick rotation, 
and the Hamiltonian $\frac{1}{2}(P_0+K_0)$ and the dilatation 
operator $D$ can be rotated into each other by an internal 
isomorphism of the group. Therefore, if there is a conformal 
field $\phi_h(x)$ of dimension $h$ with respect to the dilatation 
$D$, then there is a corresponding eigenstate $|h \rangle$ of 
$\frac{1}{2}(P_0 + K_0)$
on $S^p$ with the same eigenvalue $h$. In two-dimensional
conformal field theory, this phenomenon is well-known as
the state-operator correspondence, but in fact it holds
for any CFT$_{p+1}$ :
\beq
  \phi_h(x) \rightarrow |h \rangle = \phi_h(x=0) |0 \rangle. 
\eeq

As discussed in section \ref{chiralops}, in maximally 
supersymmetric cases there is a one-to-one 
correspondence between chiral primary operators of CFT$_{p+1}$ and
the supergravity particles on the dual \adsp\ $\times$ (compact
space). This makes it possible to identify a state in the Fock space
of the supergravity particles on \ads\ with a state in the
CFT Hibert space generated by the chiral primary fields. 

To be specific, let us consider the ${\cal N}=4$ $SU(N)$ super Yang-Mills
theory in four dimensions and its dual, type IIB 
string theory on \ads$_5 \times S^5$. 
The string scale $m_s$ and the 10-dimensional
Planck scale $m_P$ are related to the
gauge theory parameters, $g_{YM}$ and $N$, by
\beq
   m_s \simeq (g_{YM}^2 N)^{\frac{1}{4}} R^{-1}, ~~ 
m_P  \simeq N^{\frac{1}{4}} R^{-1} .
\eeq
The four energy regimes of string theory on $AdS_5\times S^5$
 are translated into the gauge theory energy scales 
(measured in the units of the inverse $S^3$ radius) in the 't Hooft limit as
follows:

\medskip
\noindent
(1) $E \ll (g_{YM}^2N)^{\frac{1}{4}}$

The Hilbert space consists of the chiral primary
states, their superconformal descendants and their
products.  Because of the large-$N$ factorization, 
a product of gauge invariant operators 
receives corrections only at subleading orders in the $1/N$
expansion. 
This fits well with the supergravity
description of multi-graviton states, where we
estimated that their energy $E$
becomes comparable to the gravitational potential
when $E \sim m_P^8 R^7$, which in the gauge theory
scale corresponds to $E \sim N^2$. 
The entropy for $1 \ll E \ll (g_{YM}^2 N)^{\frac{1}{4}}$
is then given by
\beq
  S \sim E^{\frac{9}{10}}.
\eeq

\noindent
(2) $(g_{YM}^2N)^{\frac{1}{4}} < E \ll (g_{YM}^2
    N)^{-\frac{7}{2}} N^2$

Each single string state is identified with a single
trace operator in the gauge theory. Supergravity
particles correspond to chiral primary states and
stringy excitations to non-chiral primaries.
Since stringy excitations have an energy $\sim m_s$,
the \ads/CFT correspondence
requires that non-chiral conformal fields have to 
have large anomalous dimensions 
$\Delta \sim m_s R = (g_{YM}^2 N)^{\frac{1}{4}}$.
In the 't Hooft limit ($N \gg (g_{YM}^2 N)^\gamma$
for any $\gamma$), we can consider
the regime $(g_{YM}^2 N)^{\frac{5}{2}} <
E \ll (g_{YM}^2 N)^{-\frac{7}{2}} N^2$ where
the entropy shows the Hagedorn behavior
\beq
  S \sim (g_{YM}^2 N)^{-\frac{1}{4}} E.
\eeq
Apparently, the entropy in this regime is dominated by
the non-chiral fields.

\noindent
(3) $ (g_{YM}^2
    N)^{-\frac{7}{2}} N^2 < E < N^2$

The string theory Hilbert space consists of
states in the small black hole. It would
be interesting to find a gauge theory interpretation 
of the 10-dimensional Schwarzschild black hole.
The entropy in this regime behaves as
\beq
  S \sim N^{-\frac{2}{7}} E^{\frac{8}{7}}.
\eeq

\noindent
(4) $N^2 < E$

The string theory Hilbert space consists of
states in the large black hole. The entropy is given
by 
\beq
 S \sim N^{\frac{1}{2}} E^{\frac{3}{4}}.
\eeq
The $E^{\frac{3}{4}}$ scaling of the entropy is what one expects  
for a conformal field theory in $(3+1)$ dimensions at high energies
(compared to the radius of the sphere). 
It is interesting to note that the $N$ dependence of $S$
is the same as that of $N^2$ free particles in 
$(3+1)$ dimensions, although the precise numerical 
coefficient in $S$ differs from the one that is obtained
from the number of particles in the ${\cal N}=4$
Yang-Mills multiplet by a numerical factor \cite{Gubser:1996de}.

\section{Wilson Loops}
\label{wilsonloops}

In this section we consider Wilson loop operators in the
gauge theory. 
The Wilson loop operator 
\eqn{wilsonlo}{
W({\cal C}) = \tr \left[ P \exp \left(i \oint_{\cal C} A  \right)
\right] } depends on a loop ${\cal
C}$ embedded in four dimensional space, and it involves the
path-ordered integral of the gauge connection along the contour. The
trace is taken over some representation of the gauge group; we will
discuss here only the case of the fundamental representation (see
\cite{Gross:1998gk} for a discussion of other representations).  From
the expectation value of the Wilson loop operator $\langle W({\cal C})
\rangle $ we can calculate the quark-antiquark potential. For this
purpose we consider a rectangular loop with sides of length $T$ and
$L$ in Euclidean space.  Then, viewing $T$ as the time direction, it
is clear that for large $T$ the expectation value will behave as
$e^{-TE}$ where $E$ is the lowest possible energy of the
quark-anti-quark configuration. Thus, we have
\eqn{rectangle}{
\langle W \rangle  \sim e^{ -T V(L)} ~,
}
where $V(L)$ is the quark anti-quark potential. For large $N$ and 
large $g_{YM}^2 N$, the AdS/CFT
correspondence maps the computation of $\langle W \rangle$ in the CFT
into a problem of finding a minimum surface in $AdS$ 
\cite{Maldacena:1998im,Rey:1998ik}. 

\subsection{Wilson Loops and Minimum Surfaces}
 
In QCD, we expect the Wilson loop to be related to the string running
from the quark to the antiquark.
We expect this string to be analogous to the string in our configuration,
which is a superstring
which lives in ten dimensions, and which can stretch between two 
points on the boundary of $AdS$.  In order to motivate this
prescription let us consider the following situation. We start with
the gauge group $U(N+1)$, and we break it to $U(N)\times U(1)$ by
giving an expectation value to one of the scalars. This corresponds,
as discussed in section \ref{correspondence}, to having a D3 brane
sitting at some radial position $U$ in $AdS$, and
at a point on $S^5$.  The off-diagonal states, transforming in the
${\bf N}$ of $U(N)$, get a mass proportional to $U$, $m 
= U / 2 \pi$.  So, from the point of view of the $U(N)$ gauge
theory, we can view these states as massive quarks, which act as a source
for the various $U(N) $ fields. Since they are charged they will act as a
source for the vector fields. In order to get a non-dynamical source
(an ``external quark'' with no fluctuations of its own, which will
correspond precisely to the Wilson loop operator) we need to take $m \to
\infty$, which means $U$ should also go to infinity. Thus, the string
should end on the boundary of AdS space.

These stretched strings will also act as a source for the scalar
fields. The coupling to the scalar fields can be seen qualitatively by
viewing the quarks as strings stretching between the $N$ branes and
the single separated brane. These strings will pull the $N$ branes and
will cause a deformation of the branes, which is described by the
scalar fields.  A more formal argument for this coupling is that these
states are BPS, and the coupling to the scalar (Higgs) fields is
determined by supersymmetry.  Finally, one can see this coupling
explicitly by writing the full $U(N+1)$ Lagrangian, putting in the
Higgs expectation value and calculating the equation of motion for the
massive fields
\cite{Maldacena:1998im}.
The precise definition of the Wilson loop operator corresponding to
the superstring will actually include also the field theory fermions,
which will imply some particular boundary conditions for the
worldsheet fermions at the boundary of $AdS$. However, this will not
affect the leading order computations we describe here.

So, the final conclusion is that the stretched strings
couple to the operator
\eqn{genwil}{
W({\cal C}) = \tr \left[ P \exp\left( \oint ( i A_\mu \dot x^\mu 
+ \theta^I \phi^I \sqrt{
\dot x^2 } ) d\tau  \right) \right], } 
where $x^\mu(\tau)$ is any parametrization of the loop and
$\theta^I$ ($I=1,\cdots,6$) is a unit vector in $\IR^6$ (the point on
$S^5$ where the string is sitting). This is the expression when
the signature of $\IR^4$ is Euclidean. In the Minkowski signature
case, the phase factor associated to the trajectory of the quark
has an extra factor ``$i$'' in front 
of $\theta^I$ \footnote{The difference
in the factor of $i$ between the Euclidean and the Minkowski cases
can be traced to the analytic continuation of 
$\sqrt{\dot x^2}$. A detailed derivation of (\ref{genwil})
can be found in \cite{Drukker:1999zq}.}. 

\begin{figure}[htb]
\begin{center}
\epsfxsize=.4in\leavevmode\epsfbox{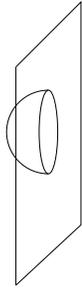}
\end{center}
\caption{
The Wilson loop operator creates a string worldsheet ending 
on the corresponding loop on the boundary of $AdS$.
}
\label{wi2}
\end{figure}

Generalizing the prescription of section \ref{correlators} for computing
correlation functions, the discussion above implies that in order to
compute the expectation value of the operator \genwil\ in $\cn=4$ SYM
we should consider the string theory partition function on
$AdS_5\times S^5$, with the condition that we have a string worldsheet
ending on the loop ${\cal C}$, as in figure
\ref{wi2} \cite{Rey:1998ik,Maldacena:1998im}.
In the supergravity regime, when $g_s N$ is large, the leading
contribution to this partition function will come from the area of the
string worldsheet. This area is measured with the $AdS$ metric, and it
is generally not the same as the area enclosed by the loop ${\cal
C}$ in four dimensions.

The area as defined above is divergent. The divergence arises
from the fact that the string worldsheet is going all the way 
to the boundary of $AdS$. If we evaluate the area up to some 
radial distance $U=r$, we see that for large $r$ it diverges as
$ r  | {\cal C} |$, where $|{\cal C}|$ is the length
of the loop in the field theory \cite{Maldacena:1998im,Rey:1998ik}.
On the other hand, the perturbative computation in the field theory
shows that $\langle W \rangle$, for $W$ given by (\ref{genwil}), is
finite, as it should be since a divergence in the Wilson loop would
have implied a mass renormalization of the BPS particle.  The apparent 
discrepancy between the divergence of the area of the minimum surface 
in $AdS$ and the finiteness of the field theory computation can be 
reconciled by noting 
that the appropriate action for the string worldsheet 
is not the area itself but its Legendre transform with respect to
the string coordinates corresponding to $\theta^I$ 
and the radial coordinate $u$ \cite{Drukker:1999zq}. 
This is because these string coordinates obey
the Neumann boundary conditions rather than the Dirichlet conditions. 
When the loop is smooth, the Legendre
transformation simply subtracts the 
divergent term $r |{\cal C}|$, leaving the resulting action
finite. 


As an example let us consider a circular Wilson loop. Take ${\cal C}$
to be a circle of radius $a$ on the boundary, and let us work in the
\Poincare coordinates (defined in section \ref{adsgeom}).  We could find the
surface that minimizes the area by solving the Euler-Lagrange
equations. However, in this case it is easier to use conformal
invariance. Note that there is a conformal transformation in the field
theory that maps a line to a circle. In the case of the line, the
minimum area surface is clearly a plane that intersects the boundary
and goes all the way to the horizon (which is just a point on the
boundary in the Euclidean case). Using the conformal transformation to
map the line to a circle we obtain the minimal surface we want. It is,
using the coordinates (\ref{poinc}) for $AdS_5$,
\eqn{wilcir}{
\vec x = \sqrt{a^2 -z^2}  (\vec e_1 \cos\theta + \vec e_2 \sin \theta ),} 
where $\vec e_1$, $\vec e_2$ are two orthonormal vectors in four
dimensions (which define the orientation of the circle) and $ 0 \leq z
\leq a $.  We can calculate the area of this surface in $AdS$, and we
get a contribution to the action
\eqn{areacir}{
S \sim { 1 \over 2 \pi \alpha'} {\cal A } = { R^2 \over 2 \pi \alpha'}
\int d\theta \int_\epsilon^a { dz a \over z^2 } = {R^2 \over \alpha'}
( {a \over \epsilon} - 1 ),
}
where we have regularized the area by putting a an IR cutoff at 
$z=\epsilon$ in 
$AdS$, which is equivalent to a UV  cutoff in the field theory
\cite{Susskind:1998dq}.
Subtracting the divergent term we get
\eqn{reswil} {
\langle W \rangle 
\sim e^{ - S} \sim e^{ R^2/\alpha'} = e^{ \sqrt{ 4 \pi g_s N} }.
}
This is independent of $a$ as required by conformal invariance. 

We could similarly consider a ``magnetic'' Wilson loop, which is also
called a 't Hooft loop \cite{'tHooft:1979uj}. This case is related by
electric-magnetic duality to the previous case. Since we identify the
electric-magnetic duality with the $SL(2,\IZ)$ duality of type IIB
string theory, we should consider in this case a D-string worldsheet
instead of a fundamental string worldsheet. We get the same result as
in \reswil\ but with $ g_s \to 1/g_s$.

\begin{figure}[htb]
\begin{center}
\epsfxsize=4.0in\leavevmode\epsfbox{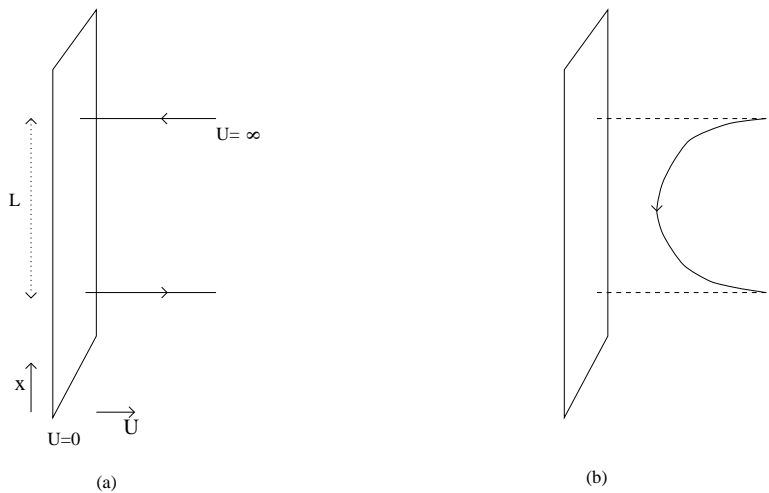}
\end{center}
\caption{ (a) Initial  configuration corresponding to two massive quarks
before we turn on their coupling to the $U(N)$ gauge theory.
(b) Configuration after we consider the 
coupling to the $U(N)$ gauge theory.
This configuration minimizes the action. The quark-antiquark energy
is given by the difference 
of the total length of the strings in (a) and (b).
}
\label{wi1}
\end{figure}

Using (\ref{rectangle}) it is possible to compute the quark-antiquark
potential in the supergravity approximation
\cite{Rey:1998ik,Maldacena:1998im}. In this case we consider a
configuration which is invariant under (Euclidean) time translations.
We take both particles to have the same scalar charge, which means
that the two ends of the string are at the same point in $S^5$ (one
could consider also the more general case
with a string ending at different points on $S^5$ \cite{Maldacena:1998im}). 
We put the quark at $x = -L/2$ and the
anti-quark at $x = L/2$. Here ``quark'' means an infinitely massive
W-boson connecting the $N$ branes with one brane which is (infinitely)
far away.  The classical action for a string worldsheet is
\eqn{string_action}{
S = { 1 \over 2 \pi \alpha'} \int d\tau d\sigma \sqrt{ \det (G_{MN} 
\partial_\alpha X^M \partial_\beta  X^N) },
}
where $G_{MN}$ is the Euclidean $AdS_5\times S^5$ metric.
 Note that
the factors of $\alpha'$ cancel out in (\ref{string_action}), as they should.
Since we are interested in a  static configuration we take
$\tau =t, ~ \sigma = x$, and then the action  becomes
\eqn{act}{
S = { T R^2 \over 2 \pi} \int_{-L/2}^{L/2} dx { \sqrt{ (\partial_x { z})^2+
1 } \over z^2}.
}
We need  to solve the Euler-Lagrange equations for this action.
Since the action does not depend on $x$ explicitly 
 the solution satisfies
\eqn{const}{
{1  \over  z^2 { \sqrt{ (\partial_x { z})^2+
1 } }}  = {\rm constant }.
 }
Defining $z_0$ to be the maximum value of $z(x)$, which by symmetry 
occurs at $x=0$, we find that the solution is\footnote{
All  integrals in this section
can be calculated in terms of elliptic or Beta  functions.}
\eqn{sol}{
 x = { z_0  } \int_{z/z_0}^1 
{ dy y^2 \over \sqrt{1- y^4 } },  
}
where $z_0$ is determined by the condition
\eqn{uzero}{
{ L \over 2 } = z_0
\int_0^1
{ dy y^2 \over  \sqrt{1- y^{4}  } }  =  
z_0 { \sqrt{2} \pi^{3/2} \over  \Gamma( 1/4)^2  }.
}
The qualitative form of the solution is shown in figure \ref{wi1}(b).
Notice that the  string quickly approaches $x =L/2$ 
for small $z$ (close to the boundary),
\eqn{approach}{
{ L\over 2} -x \sim { z^3  }~,~~~~~ z \to 0 ~ .
}
Now we  compute the total energy of the configuration.
We just plug in the solution \sol\ in \act , subtract the infinity
as explained above (which can be interpreted as the energy of two
separated massive quarks, as in figure \ref{wi1}(a)), and we find
\eqn{energy}{
E = V(L) = -  { 4 \pi^2 ( 2 g^2_{YM} N)^{1/2}  \over 
\Gamma({1 \over 4})^4 L}.
}
We see that the energy goes as $1/L$, a fact  which is determined by
 conformal invariance. 
Note that 
the energy is proportional to $(g_{YM}^2 N)^{1/2}$, 
as opposed to $g_{YM}^2 N$ which is the
perturbative result. 
This  indicates some screening of the charges at strong coupling. 
The above calculation makes sense for all distances $L$
when $g_s N$ is large, independently of the value of $g_s$.
Some subleading 
 corrections coming from quantum fluctuations of the worldsheet
were calculated in \cite{Forste:1999qn,Naik:1999bs,Greensite:1999wf}.

In a similar fashion we could compute the potential between two
magnetic monopoles in terms of a D-string worldsheet, 
and the result will be the
same as \energy\ but with $ g_{YM} \to 4\pi/g_{YM}$.  One can also
calculate the interaction between a magnetic monopole and a quark. In
this case the fundamental string (ending on the quark) will attach to
the D-string (ending on the monopole), and they will connect to form a
$(1,1)$ string which will go into the horizon. The resulting potential
is a complicated function of $g_{YM}$
\cite{Minahan:1998xb}, but in the limit that $g_{YM}$ is small (but
still with $g_{YM}^2 N$ large) we get that the monopole-quark
potential is just $ 1/4$ of the quark-quark potential. This can be
understood from the fact that when $g$ is small the D-string is very
rigid and the fundamental string will end almost perpendicularly on
the D-string. Therefore, the solution for the fundamental string will
be half of the solution we had above, leading to a factor of $1/4$ in the
potential. Calculations of Wilson loops in the Higgs phase were done
in \cite{Minahan:1998xq}.

Another interesting case one can study analytically is a surface 
near a cusp on $\IR^4$. In this case, the perturbative computation
in the gauge theory shows a logarithmic divergence with a coefficient
depending on the angle at the cusp. The area of the minimum surface
also contains a logarithmic divergence depending on 
the angle \cite{Drukker:1999zq}.  Other aspects of the gravity 
calculation of Wilson loops were discussed in
 \cite{Kogan:1998ti,Kogan:1999rw,Nojiri:1999gf,Zarembo:1999bu,%
Alvarez:1998dx}.

\subsection{Other Branes Ending on the Boundary}

We could also consider other branes that are ending at the boundary
\cite{Graham:1999pm}.  The simplest example would be a zero-brane
(i.e. a particle) of mass $m$. In Euclidean space a zero-brane
describes a one dimensional trajectory in anti-de-Sitter space which
ends at two points on the boundary.  Therefore, it is associated with
the insertion of two local operators at the two points where the
trajectory ends. In the supergravity approximation the zero-brane
follows a geodesic.  Geodesics in the hyperbolic plane (Euclidean AdS)
are semicircles. If we compute the action we get
\eqn{actizero}{
S = m \int ds = -2m R \int_\epsilon^a { a dz \over z \sqrt{a^2 - z^2 }},
}
where we took the distance between the two points at the boundary
 to be $L = 2a$
and  regulated the result. We find a logarithmic divergence when
$\epsilon \to 0$, proportional to $ \log(\epsilon/a) $. 
If we subtract the logarithmic divergence we get a residual dependence
on $a$. Naively we might have thought that (as in the previous 
subsection) the answer had
to be independent of $a$ due to conformal invariance. 
In fact, the dependence on $a$ is very important, since it leads to 
a result of the form
\eqn{res-zero}{
e^{-S} \sim e^{- 2 m R \log a } \sim { 1 \over a^{2 m R} },
}
which is precisely the result we expect for the two-point function of an
operator of dimension $\Delta = m R$. This is precisely the 
large $mR$ limit of the formula (\ref{dimenmass}), so we reproduce in
the supergravity limit the 2-point function described in section 
\ref{correlators}. 
In general, this sort of logarithmic 
divergence arises when the brane worldvolume is 
odd dimensional \cite{Graham:1999pm},
 and it implies that the expectation
value of the corresponding operator depends on the overall scale. 
In particular one could consider the ``Wilson surfaces'' that 
arise in the six dimensional $\cn=(2,0)$ theory which will be discussed
in section \ref{m5branes}. In that 
case one has to consider a two-brane, with a three
dimensional worldvolume,  ending on a two dimensional
surface on the boundary of $AdS_7$. Again, one gets a logarithmic term,
which is proportional to the rigid string action of the two 
dimensional surface living on the string in the $\cn=(2,0)$ field theory 
\cite{Berenstein:1998ij,Graham:1999pm}.

One can also compute correlation functions involving more than one
Wilson loop. To leading order in $N$ this will be just the product of
the expectation values of each Wilson loop. On general grounds one
expects that the subleading corrections are given by surfaces that end
on more than one loop. One limiting case is when the surfaces look
similar to the zeroth order surfaces but with additional thin tubes
connecting them. These thin tubes are nothing else than massless
particles being exchanged between the two string worldsheets
\cite{Gross:1998gk,Berenstein:1998ij}. We will discuss this further in
section \ref{adsqcd}.


\section{Theories at Finite Temperature}
\label{FiniteT}

As discussed in section \ref{tests}, 
the quantities that can be most successfully compared between gauge
theory and string theory are those with some protection from
supersymmetry and/or conformal invariance --- for instance, dimensions of
chiral primary operators.  Finite temperature breaks both
supersymmetry and conformal invariance, and the insights we gain from
examining the $T>0$ physics will be of a more qualitative nature.
They are no less interesting for that: we shall see in
section~\ref{ConstT} how the entropy of near-extremal D3-branes comes
out identical to the free field theory prediction up to a factor of
a power of $4/3$; 
then in section~\ref{TPhaseT} we explain how a phase transition
studied by Hawking and Page in the context of quantum gravity is
mapped into a confinement-deconfinement transition in the gauge theory,
driven by finite-size effects; and in section~\ref{adsqcd} we will
summarize the attempts to use holographic duals of finite-temperature
field theories to learn about pure gauge theory at zero temperature
but in one lower dimension.

\subsection{Construction}
\label{ConstT}

The gravity solution describing the gauge theory at finite temperature
can be obtained by starting from the general black three-brane solution
(\ref{solution}) and taking the decoupling limit of section 
\ref{correspondence} keeping the energy density above extremality finite.
The resulting metric can be written as
\eqn{NearDThree}{
\eqalign{
   ds^2 &=R^2 \left[ u^2( - h dt^2 + dx_1^2 + dx_2^2 + dx_3^2 ) 
+ { d u^2 \over h  u^2} + d \Omega_5^2 \right]
\cr
 h & = 1 - { u_0^4 \over u^4} ~,~~~~~~~ u_0 = \pi T .
}}
It will often be useful to Wick rotate by setting $t_E = it$, and use
the relation between the finite temperature theory and the Euclidean
theory with a compact time direction.  

The first computation which indicated that finite-temperature $U(N)$
Yang-Mills theory might be a good description of the microstates of $N$
coincident D3-branes was the calculation of the entropy
\cite{Gubser:1996de,sunp}.  On the supergravity side, the entropy of
near-extremal D3-branes is just the usual Bekenstein-Hawking result, $S =
A/{4 G_N}$, and it is expected to be a reliable guide to the entropy of the
gauge theory at large $N$ and large $g_{YM}^2 N$.  There is no problem on
the gauge theory side in working at large $N$, but large $g_{YM}^2 N$ at
finite temperature is difficult indeed.  The analysis of
\cite{Gubser:1996de} was limited to a free field computation in the field
theory, but nevertheless the two results for the entropy agreed up to a
factor of a power of $4/3$.  In the canonical ensemble, where temperature
and volume are the independent variables, one identifies the field theory
volume with the world-volume of the D3-branes, and one sets the field
theory temperature equal to the Hawking temperature in supergravity.  The
result is
  \eqn{CanS}{\eqalign{
   F_{SUGRA} &= -{\pi^2 \over 8} N^2 V T^4,  \cr
   F_{SYM} &= {4 \over 3} F_{SUGRA} \ .
  }}
 The supergravity result is at leading order in $l_s/R$, and it would
acquire corrections suppressed by powers of $T R$ if we had considered the
full D3-brane metric rather than the near-horizon limit, \NearDThree.
These corrections do not have an interpretation in the context of CFT
because they involve $R$ as an intrinsic scale.  Two equivalent methods to
evaluate $F_{SUGRA}$ are a) to use $F = E - TS$ together with standard
expressions for the Bekenstein-Hawking entropy, the Hawking temperature,
and the ADM mass; and b) to consider the gravitational action of the
Euclidean solution, with a periodicity in the Euclidean time direction
(related to the temperature) which eliminates a conical deficit angle at
the horizon.\footnote{The result of \cite{Gubser:1996de}, $S_{SYM} =
(4/3)^{1/4} S_{SUGRA}$, differs superficially from \CanS, but it is only
because the authors worked in the microcanonical ensemble: rather than
identifying the Hawking temperature with the field theory temperature, the
ADM mass above extremality was identified with the field theory energy.}

The $4/3$ factor  is a long-standing puzzle into which we still have
only qualitative insight.  The gauge theory computation was performed
at zero 't~Hooft coupling, whereas the supergravity is supposed to be
valid at strong 't~Hooft coupling, and unlike in the 1+1-dimensional
case where the entropy is essentially fixed by the central charge,
there is no non-renormalization theorem for the coefficient of $T^4$ in the
free energy.  Indeed, it was suggested in \cite{Gubser:1998nz} that
the leading term in the $1/N$ expansion of $F$ has the form
  \eqn{InterpolateS}{
   F = -f(g_{YM}^2 N) {\pi^2 \over 6} N^2 V T^4,
  }
 where $f(g_{YM}^2 N)$ is a function which smoothly interpolates between a
weak coupling limit of $1$ and a strong coupling limit of $3/4$.  It was
pointed out early \cite{Horowitz:1997nw} that the quartic potential
$g_{YM}^2 \tr [\phi^I,\phi^J]^2$ in the ${\cal N}=4$ Yang-Mills action
might be expected to freeze out more and more degrees of freedom as the
coupling was increased, which would suggest that $f(g_{YM}^2 N)$ is
monotone decreasing.  An argument has been given \cite{Itzhaki:1999ge},
based on the non-renormalization of the two-point function of the stress
tensor, that $f(g_{YM}^2 N)$ should remain finite at strong coupling.

The leading corrections to the limiting value of $f(g_{YM}^2 N)$ at
strong and weak coupling were computed in \cite{Gubser:1998nz} and
\cite{Fotopoulos:1999es}, respectively.  The results are 
  \eqn{WeakStrong}{\seqalign{\span\TL & \span\TR \qquad & \span\TT}{
   f(g_{YM}^2 N) &= 1 - {3 \over 2\pi^2} g_{YM}^2 N + \ldots
     & for small $g_{YM}^2 N$,  \cr
   f(g_{YM}^2 N) &= {3\over 4} + {45 \over 32} 
    {\zeta(3) \over (g_{YM}^2 N)^{3/2}} + \ldots
     & for large $g_{YM}^2 N$.
  }}
 The weak coupling result is a straightforward although somewhat tedious
application of the diagrammatic methods of perturbative finite-temperature
field theory.  The constant term is from one loop, and the leading
correction is from two loops.  The strong coupling result follows from
considering the leading $\alpha'$ corrections to the supergravity action.
The relevant one involves a particular contraction of four powers of the
Weyl tensor.  It is important now to work with the Euclidean solution, and
one restricts attention further to the near-horizon limit.  The Weyl
curvature comes from the non-compact part of the metric, which is no longer
$AdS_5$ but rather the AdS-Schwarzschild solution which we will discuss in
more detail in section~\ref{TPhaseT}.  The action including the
   $\alpha'$ corrections no longer has the
Einstein-Hilbert form, and correspondingly the Bekenstein-Hawking
prescription no longer agrees with the free energy computed as $\beta I$
where $I$ is the Euclidean action.  In keeping with the basic prescription
for computing Green's functions, where a free energy in field theory is
equated (in the appropriate limit)
with a supergravity action, the relation $I = \beta F$ is regarded
as the correct one.
(See \cite{Wald:1993nt}.)
It has been conjectured that the interpolating function $f(g_{YM}^2
N)$ is not smooth, but exhibits some phase transition at a finite
value of the 't~Hooft coupling.  We regard this as an unsettled
question.  The arguments in \cite{Li:1999kd,Gao:1998ww} 
seem as yet incomplete.  In
particular, they rely on analyticity properties of the perturbation
expansion which do not seem to be proven for finite temperature field
theories.

\subsection{Thermal Phase Transition}
\label{TPhaseT}

The holographic prescription of \cite{Gubser:1998bc,Witten:1998qj}, applied at
large $N$ and $g_{YM}^2 N$
where loop and stringy corrections are negligible, involves extremizing
the supergravity action subject to particular
asymptotic boundary conditions.  We
can think of this as the saddle point approximation to the path
integral over supergravity fields.  That path integral is ill-defined
because of the non-renormalizable nature of supergravity.  String
amplitudes (when we can calculate them) render on-shell quantities
well-defined.  Despite the conceptual difficulties we can use some
simple intuition about path integrals to illustrate an important point
about the AdS/CFT correspondence: namely, there can be more than one
saddle point in the range of integration, and when there is we should
sum $e^{-I_{SUGRA}}$ over the classical configurations to obtain the
saddle-point approximation to the gauge theory partition function.
Multiple classical configurations are possible because of the general
feature of boundary value problems in differential equations: there
can be multiple solutions to the classical equations satisfying the
same asymptotic boundary conditions.  The solution which globally
minimizes $I_{SUGRA}$ is the one that dominates the path integral.

When there are two or more solutions competing to minimize
$I_{SUGRA}$, there can be a phase transition between them.  An example
of this was studied in \cite{Hawking:1983dh} long before the AdS/CFT
correspondence, and subsequently resurrected, generalized, and
reinterpreted in \cite{Witten:1998qj,Witten:1998zw} as a
confinement-deconfinement transition in the gauge theory.  Since the
qualitative features are independent of the dimension, we will restrict
our attention to $AdS_5$.  It is worth noting however that if the
$AdS_5$ geometry is part of a string compactification, it doesn't
matter what the internal manifold is except insofar as it fixes the
cosmological constant, or equivalently the radius $R$ of anti-de
Sitter space.

There is an embedding of the Schwarzschild black hole solution into
anti-de Sitter space which extremizes the action
  \eqn{EinAct}{
   I = -{1 \over 16 \pi G_5} 
    \int d^5 x \, \sqrt{g} \left( {\cal R} + {12 \over R^2} \right) \ .
  }
 Explicitly, the metric is
  \eqn{AdSSch}{\eqalign{
   ds^2 &= f dt^2 + {1 \over f} dr^2 + r^2 d\Omega_3^2,  \cr
   f &= 1 + {r^2 \over R^2} - {\mu \over r^2} \ .
  }}
 The radial variable $r$ is restricted to $r \geq r_+$, where $r_+$ is the
largest root of $f=0$.  The Euclidean time is periodically identified, 
$t \sim t +
\beta$, in order to eliminate the conical singularity at $r = r_+$.  This
requires
  \eqn{HawkingBeta}{
   \beta = {2\pi R^2 r_+ \over 2 r_+^2 + R^2} \ .
  }
 Topologically, this space is $S^3 \times B^2$, and the boundary is $S^3
\times S^1$ (which is the relevant space for the field theory on $S^3$
with finite temperature).  
We will call this space $X_2$.  Another space with the same
boundary which is also a local extremum of \EinAct\ is given by the metric
in \AdSSch\ with $\mu = 0$ and again with periodic time.  This space, which
we will call $X_1$, is not only metrically distinct from the first (being
locally conformally flat), but also topologically $B^4 \times S^1$ rather
than $S^3 \times B^2$.  Because the $S^1$ factor is not simply connected,
there are two possible spin structures on $X_1$, corresponding to thermal
(anti-periodic) or supersymmetric (periodic) boundary conditions on
fermions.  In contrast, $X_2$ is simply connected and hence admits a unique
spin structure, corresponding to thermal boundary conditions.  For the
purpose of computing the twisted partition function, $\tr (-1)^F e^{-\beta
H}$, in a saddle-point approximation, only $X_1$ contributes.  But, $X_1$
and $X_2$ make separate saddle-point contributions to the usual thermal
partition function, $\tr e^{-\beta H}$, and the more important one is the
one with the smaller Euclidean action.

Actually, both $I(X_1)$ and $I(X_2)$ are infinite, so to compute
$I(X_2)-I(X_1)$ a regulation scheme must be adopted.  The one used in
\cite{Witten:1998zw,Gubser:1998nz} is to cut off both
$X_1$ and $X_2$ at a definite coordinate radius $r=R_0$.  For $X_2$, the
elimination of the conical deficit angle at the horizon fixes the period of
Euclidean time; but for $X_1$, the period is arbitrary.  In order to make
the comparison of $I(X_1)$ and $I(X_2)$ meaningful, we fix the period of
Euclidean time on $X_1$ so that the proper circumference of the $S_1$ at
$r=R_0$ is the same as the proper length on $X_2$ of an orbit of the Killing
vector $\partial/\partial t$, also at $r=R_0$.  In the limit $R_0\to\infty$,
one finds
  \eqn{IDiff}{
   I(X_2)-I(X_1) = {\pi^2 r_+^3 (R^2 - r_+^2) \over
    4 G_5 (2 r_+^2 + R^2)} \ ,
  }
 where again $r_+$ is the largest root of $f=0$.  The fact that \IDiff\ (or
more precisely its $AdS_4$ analog) can change its sign was interpreted in
\cite{Hawking:1983dh} as indicating a phase transition between a black hole
in $AdS$ and a thermal gas of particles in $AdS$ (which is the natural
interpretation of the space $X_1$).  The black hole is the
thermodynamically favored state when the horizon radius $r_+$ exceeds the
radius of curvature $R$ of $AdS$. In the gauge theory we interpret this
transition as a confinement-deconfinement transition.  Since the theory is
conformally invariant, the transition temperature must be proportional to
the inverse radius of the space $S^3$ which the field theory lives on.
Similar transitions, and also local thermodynamic instability due to
negative specific heats, have been studied in the context of spinning
branes and charged black holes in
\cite{Gubser:1998jb,Landsteiner:1999gb,Cai:1998ji,
Cvetic:1999ne,Chamblin:1999tk,Cvetic:1999rb,Caldarelli:1999ar}.  Most of
these works are best understood on the CFT side as explorations of exotic
thermal phenomena in finite-temperature gauge theories.  Connections with
Higgsed states in gauge theory are clearer in
\cite{Kraus:1998hv,Tseytlin:1998cq}.  The relevance to confinement is
explored in \cite{Cvetic:1999rb}.  See also
\cite{Birmingham:1998nr,Louko:1998hc,Hawking:1998kw,Peet:1998cr} for other
interesting contributions to the finite temperature literature.

Deconfinement at high temperature can be characterized by a spontaneous
breaking of the center of the gauge group. In our case the gauge group is
$SU(N)$ and its center is $\IZ_N$.
The order parameter for the breaking of the center is the
expectation value of the Polyakov (temporal) loop $\langle W(C) \rangle$.
The boundary of the spaces $X_1,X_2$ is $S^3 \times S^1$, and 
the path $C$ wraps around the circle.
An element of the center $g \in \IZ_N$ acts on the Polyakov loop
by  $\langle W(C) \rangle \rightarrow g \langle W(C) \rangle$.
The expectation value of the Polyakov loop measures the change of the
free energy of the system $F_q(T)$ induced by the presence
of the external charge $q$, 
$\langle W(C) \rangle \sim exp \left(-F_q(T)/T \right)$.
In a confining phase  $F_q(T)$ is infinite and therefore  
$\langle W(C) \rangle  = 0$.
In the deconfined phase  $F_q(T)$ is finite and therefore  
$\langle W(C) \rangle  \neq 0$.

As discussed in section \ref{wilsonloops}, 
in order to compute $\langle W(C) \rangle$ we have to 
evaluate the partition function of strings with a 
worldsheet $D$ that is bounded
by the loop $C$. 
Consider first the low temperature phase. The relevant space 
is $X_1$ which, as discussed
above,  has
the topology $B^4 \times S^1$.
The contour $C$ wraps the circle and is not homotopic to zero in $X_1$.
Therefore $C$ is not a boundary of any $D$, which immediately
implies that  $\langle W(C) \rangle  = 0$.
This is the expected behavior at low temperatures (compared to the
inverse radius of the $S^3$), where the center
of the gauge group is not broken.

For the high temperature phase the relevant space is $X_2$, which has the
topology $S^3 \times B^2$. The contour $C$ is now a boundary of a
string worldsheet $D=B^2$ (times
a point in $S^3$). This seems to be in agreement with the fact that
in the   high temperature phase $\langle W(C) \rangle  \neq 0$
and the center
of the gauge group is broken. It was pointed out in \cite{Witten:1998zw}
that there is a subtlety with this argument, since the
center should not be broken in finite volume ($S^3$), but only
in the infinite volume limit ($\IR^3$).
Indeed, the solution $X_2$ is not unique
and we can add to it an expectation value for
the integral of the NS-NS 2-form field $B$ on $B^2$, with vanishing
field strength. This is an angular parameter $\psi$
with period $2 \pi$, which contributes $i\psi$ to the string
worldsheet action. The string theory partition function
includes now an integral over all values of
$\psi$, making  $\langle W(C) \rangle  = 0$ on $S^3$.
In contrast, on $\IR^3$ one integrates over the local fluctuations
of $\psi$ but not over its vacuum expectation value. Now
$\langle W(C) \rangle  \neq 0$ and depends on the value of $\psi \in U(1)$,
which may be understood as the dependence on the center $\IZ_N$ in the
large $N$ limit.
Explicit computations of Polyakov loops at finite temperature 
were done in \cite{Rey:1998wp,Brandhuber:1998bs}.

In \cite{Witten:1998zw} the Euclidean black hole solution \AdSSch\ 
was suggested to
be holographically dual to a theory related to
pure QCD in three dimensions.  In the large
volume limit
the solution corresponds to the ${\cal N}=4$ gauge theory on $\IR^3
\times S^1$ with thermal boundary conditions, and when the $S^1$ is made
small (corresponding to high temperature $T$) the theory at distances
larger than $1/T$ effectively reduces to pure Yang-Mills on
$\IR^3$.  Some of the non-trivial successes of this approach to QCD will be
discussed in section \ref{adsqcd}.

\chapter{More on the Correspondence}
\label{ChapCorrAdvanced}

\section{Other AdS$_5$ Backgrounds}
\label{other_backgrounds}

Up to now we have limited our discussion to the $AdS_5\times S^5$
background of type IIB string theory; in section \ref{deformations}
we will describe backgrounds which are
related to it by deformations. However, 
it is clear from the
description of the correspondence in sections \ref{correspondence} and
\ref{correlators} that a similar
correspondence may be defined for any theory of quantum gravity whose
metric includes an $AdS_5$ factor; the generalization of
equation (\ref{genera}) relates such a theory
to a four dimensional conformal field theory. The background does not
necessarily have to be of the form $AdS_5\times X$; it is enough that
it has an $SO(4,2)$ isometry symmetry, and more general possibilities
in which the curvature of $AdS_5$ depends on the position in $X$ are
also possible \cite{vanNieuwenhuizen:1985ri}. It is necessary,
however, for the $AdS$ theory to be a theory of quantum gravity, since
any conformal theory has an energy-momentum tensor operator that is
mapped by the correspondence to the graviton on $AdS_5$\footnote{If we
have a topological field theory on the boundary the bulk theory does
not have to be gravitational, as in \cite{Gopakumar:1998ki}.}. Thus,
we would like to discuss compactifications of string theory or M
theory, which are believed to be consistent theories of quantum
gravity, on backgrounds involving $AdS_5$. For simplicity we will only
discuss here backgrounds which are direct products of the form
$AdS_5\times X$.

Given such a background of string/M theory, it is not apriori clear
what is the conformal field theory to which it corresponds. A special
class of backgrounds are those which arise as near-horizon limits of
branes, like the $AdS_5\times S^5$ background. In this case one can
sometimes analyze the low-energy field theory on the branes by
standard methods before taking the near-horizon limit, and after the
limit this becomes the dual conformal field theory. The most
well-studied case is the case of D3-branes in type IIB string
theory. When the D3-branes are at a generic point in space-time the
near-horizon limit gives the $AdS_5\times S^5$ background discussed
extensively above. However, if the transverse space to the D3-branes
is singular, the near-horizon limit and the corresponding field theory
can be different. The simplest case is the case of a D3-brane on an
orbifold \cite{Kachru:1998ys} or orientifold \cite{Witten:1998xy}
singularity, which can be analyzed by perturbative string theory
methods. These cases will be discussed in sections \ref{orbifolds} and
\ref{orientifolds}. Another interesting case is the conifold
singularity \cite{Klebanov:1998hh} and its generalizations, which will
be discussed in section \ref{conifolds}. In this case a direct
analysis of the field theory is not possible, but various indirect
arguments can be used to determine what it is in many cases.

Not much is known about more general cases of near-horizon limits of
D3-branes, which on the string theory side were analyzed in
\cite{Figueroa-O'Farrill:1998nb,Acharya:1998db,Morrison:1998cs,
Ray:1999qj,Figueroa-O'Farrill:1999va}, and
even less is known about backgrounds which are not describable as
near-horizon limits of branes (several $AdS_5$ backgrounds were discussed
in \cite{Duff:1998us}). An example of the latter is the
$AdS_5\times \CP^3$ background of M theory \cite{Pope:1989xj}, which
involves a 4-form flux on the 4-cycle in $\CP^3$. Using the methods
described in the previous sections we can compute various properties
of such compactifications in the large $N$ limit, such as the mass
spectrum and the central charge of the corresponding field theories
(for the $AdS_5\times \CP^3$ compactification one finds a central
charge proportional to $N^3$, where $N$ is the 4-form flux). However,
it is not known how to construct an alternative description of the
conformal field theory in most of these cases, except for the cases
which are related by deformations to the better-understood orbifold
and conifold compactifications.

Some of the $AdS_5\times X$ backgrounds of string/M theory preserve
some number of supersymmetries, but most of them (such as the
$AdS_5\times \CP^3$ background) do not. In supersymmetric cases,
supersymmetry guarantees the stability of the corresponding solutions.
In the non-supersymmetric cases various instabilities may arise for
finite $N$ (see, for instance, \cite{Berkooz:1999qp,Berkooz:1999ji}) which may
destroy the conformal ($SO(4,2)$) invariance, but the correspondence
is still conjectured to be valid when all quantum corrections are taken
into account (or in the infinite $N$ limit for which the supergravity
approximation is valid). One type of instability
occurs when the spectrum includes a
tachyonic field whose mass is below the Breitenlohner-Freedman
stability bound. Such a field is expected to condense just like a
tachyon in flat space, and generally it is not known what this
condensation leads to. If the classical supergravity spectrum includes
a field which saturates the stability bound, an analysis of the
quantum corrections is necessary to determine whether they raise the
mass squared of the field (leading to a stable solution) or lower it
(leading to an unstable solution). Apriori one would not expect to
have a field which exactly saturates the bound (corresponding to an
operator in the field theory whose dimension is exactly $\Delta=2$) in
a non-supersymmetric theory, but this often happens in orbifold
theories for reasons that will be discussed below. Another possible
instability arises when there is a massless field in the background,
corresponding to a marginal operator in the field theory. Such a field
(the dilaton) exists in all classical type IIB compactifications, and
naively corresponds to an exactly marginal deformation of the theory
even in the non-supersymmetric cases. However, for finite $N$ one
would expect quantum corrections to generate a potential for such a
field (if it is neutral under the gauge symmetries), which could drive
its expectation value away from the range of values where the
supergravity approximation is valid. Again, an analysis of the quantum
corrections is necessary in such a case to determine if the theory has
a stable vacuum (which may or may not be describable in supergravity),
corresponding to a fixed point of the corresponding field theory, or
if the potential leads to a runaway behavior with no stable
vacuum. Another possible source of instabilities is related to the
possibility of forming brane-anti-brane pairs in the vacuum (or,
equivalently, the emission of branes which destabilize the vacuum)
\cite{Brown:1988kg,Dowker:1996sg,Maldacena:1999uz,Seiberg:1999xz}; 
one would expect such an instability to arise, for example,
in cases where we look at the near-horizon limit of $N$ 3-branes which
have a repulsive force between them. For all these reasons, the study
of non-supersymmetric backgrounds usually requires an understanding of
the quantum corrections, which are not yet well-understood either in
M theory or in type IIB compactifications with RR backgrounds. Thus,
we will focus here on supersymmetric backgrounds, for which the
supergravity approximation is generally valid. In the
non-supersymmetric cases the correspondence is still expected to be
valid, and in the extreme large $N$ limit it can also be studied using
supergravity, but getting finite $N$ information usually requires
going beyond the SUGRA approximation. It would be very interesting to
understand better the quantum corrections in order to study
non-supersymmetric theories at finite $N$ using the AdS/CFT
correspondence.

\subsection{Orbifolds of $AdS_5\times S^5$}
\label{orbifolds}

The low-energy field theory corresponding to D3-branes at orbifold
singularities may be derived by string theory methods
\cite{Douglas:1996sw,Douglas:1997de}. 
Generally the gauge group is of the form $\prod_i U(a_i N)$, and there
are various bifundamental (and sometimes also adjoint) matter
fields\footnote{In general one can choose to have the orbifold group
act on the Chan-Paton indices in various ways. We will discuss here
only the case where the group acts as $N$ copies of the regular
representation of the orbifold group $\Gamma$, which is the only case
which leads to conformal theories. Other representations involve also
5-branes wrapped around 2-cycles, so they do not arise in the naive
near-horizon limit of D3-branes. The $AdS_5$ description of this was
given in \cite{Gubser:1998fp}.}. We are interested in the near-horizon
limit of D3-branes sitting at the origin of $\IR^4 \times
\IR^6/\Gamma$ for some finite group $\Gamma$ which is a discrete
subgroup of the $SO(6)
\simeq SU(4)_R$ rotation symmetry \cite{Kachru:1998ys}. 
If $\Gamma \subset SU(3) \subset
SU(4)_R$ the theory on the D3-branes has $\cn=1$ supersymmetry, and if
$\Gamma \subset SU(2) \subset SU(4)_R$ it has $\cn=2$
supersymmetry. The near-horizon limit of such a configuration is of
the form $AdS_5 \times S^5/\Gamma$ (since the orbifold commutes with
taking the near-horizon limit), and corresponds (at least for large
$N$) to a conformal theory with the appropriate amount of
supersymmetry. Note that on neither side of the correspondence is the
orbifolding just a projection on the $\Gamma$-invariant states of the
original theory -- on the string theory side we need to add also
twisted sectors, while on the field theory side the gauge group is
generally much larger (though the field theory 
can be viewed as a projection of the
gauge theory corresponding to ${\rm dim}(\Gamma)\cdot N$ D-branes).

We will start with a general analysis of the orbifold, and then discuss
specific examples with different amounts of supersymmetry\footnote{
We will not discuss here orbifolds that act non-trivially
on the AdS space, as in
\cite{Gao:1999er}.}. The action
of $\Gamma$ on the $S^5$ is the same as its action on the angular
coordinates of $\IR^6$. If the original action of $\Gamma$ had only
the origin as its fixed point, the space $S^5/\Gamma$ is smooth. On
the other hand, if the original action had a space of fixed points,
some fixed points remain, and the space $S^5/\Gamma$ includes orbifold
singularities. In this case the space is not geometrically smooth, and
the supergravity approximation is not valid (though of course in
string theory it is a standard orbifold compactification which is 
generically not
singular). The spectrum of string theory on $AdS_5\times S^5/\Gamma$
includes states from untwisted and twisted sectors of the
orbifold. The untwisted states are just the $\Gamma$-projection of the
original states of $AdS_5\times S^5$, and they include in particular
the $\Gamma$-invariant supergravity states. These states have (in the
classical supergravity limit) the same masses as in the original
$AdS_5\times S^5$ background \cite{Oz:1998of}, 
corresponding to integer dimensions in
the field theory, which is why we often find in orbifolds operators of
dimension 2 or 4 which can destabilize non-supersymmetric
backgrounds. If the orbifold group has fixed points on the $S^5$,
there are also light twisted sector states that are localized near
these fixed points, which need to be added to the supergravity fields
for a proper description of the low-energy dynamics. On the other
hand, if the orbifold has no fixed points, all twisted sector states
are heavy,\footnote{ Note that this happens even when in the original
description there were massless twisted sector states localized at the
origin.} since they involve strings stretching between identified
points on the $S^5$. In this case the twisted sector states decouple
from the low-energy theory in space-time (for large $g_s N$). There is a global
$\Gamma$ symmetry in the corresponding field theory, under which the
untwisted sector states are neutral while the twisted sector states
are charged.

In the 't Hooft limit of $N \to \infty$ with $g_s N$ finite, all the
solutions of the form $AdS_5\times S^5/\Gamma$ have a fixed line
corresponding to the dilaton, indicating that the beta function of the
corresponding field theories vanishes in this limit
\cite{Kachru:1998ys}. In fact, one can prove 
\cite{Lawrence:1998ja,Bershadsky:1998mb,Bershadsky:1998cb} (see also
\cite{Schmaltz:1999bg,Erlich:1998gb}) that in this
limit, which corresponds to keeping only the planar diagrams in the
field theory, all the correlation functions of the untwisted sector
operators in the orbifold theories are the same (up to multiplication by
some power of ${\rm dim}(\Gamma)$) as in the $\cn=4$ SYM theory
corresponding to $AdS_5\times S^5$ \footnote{There is no similar relation for
the twisted sector operators.}. This is the analog of the usual
string theory statement that at tree-level the interactions of
untwisted sector states are exactly inherited from those of the
original theory before the orbifolding. For example, the central
charge of the field theory (appearing in the 2-point function of the
energy-momentum tensor) is (in this limit) just ${\rm dim}(\Gamma)$
times the central charge of the corresponding $\cn=4$ theory. This may
easily be seen also on the string theory side, where the central
charge may be shown \cite{Gubser:1999vd} to be inversely proportional
to the volume of the compact space (and ${\rm Vol}(S^5/\Gamma) = {\rm
Vol}(S^5)/{\rm dim}(\Gamma)$).

The vanishing of the beta function in the 't Hooft limit follows from
this general result (as predicted by the AdS/CFT correspondence). This
applies both to orbifolds which preserve supersymmetry and to those
which do not, and leads to many examples of supersymmetric and
non-supersymmetric theories which have fixed lines in the large $N$
limit. At subleading orders in $1/N$, the correlation functions
differ between the orbifold theory and the $\cn=4$ theory, and in
principle a non-zero beta function may arise. In supersymmetric
orbifolds supersymmetry prevents this\footnote{At least, it prevents a
potential for the dilaton, so there is still some fixed line in the
field theory, though it can be shifted from the $\cn=4$ fixed line
when $1/N$ corrections are taken into account.}, but in
non-supersymmetric theories generically there will no longer be a
fixed line for finite $N$. The dilaton potential is then related to
the appearance of a non-zero beta function in the field theory, and
the minima of this potential are related to the zeros of the field
theory beta function for finite $N$.

As a first example we can analyze the case \cite{Kachru:1998ys} of
D3-branes on an $\IR^4/\IZ_k$ orbifold singularity, which preserves
$\cn=2$ supersymmetry. Before taking the near-horizon limit, the
low-energy field theory (at the free orbifold point in the string
theory moduli space) is a $U(N)^k$ gauge theory with bifundamental
hypermultiplets in the $\bf{(N,{\bar N},1,\cdots,1)+(1,N,{\bar
N},1,\cdots,1)+\cdots+ (1,\cdots,1,N,{\bar N})+({\bar
N},1,\cdots,1,N)}$ representation. The bare gauge couplings $\tau_i$
of all the $U(N)$ theories are equal to the string coupling
$\tau_{IIB}$ at this point in the moduli space. In the near-horizon
(low-energy) limit this field theory becomes the $SU(N)^k$ field
theory with the same matter content, since the off-diagonal $U(1)$
factors are IR-free\footnote{This does not contradict our previous
statements about the beta functions since the $U(1)$ factors are
subleading in the $1/N$ expansion, and the operators corresponding
to the off-diagonal $U(1)$'s come from twisted sectors.} 
(and the diagonal $U(1)$ factor is
decoupled here and in all other examples in this section so we will
ignore it).  This theory is dual to type IIB string theory on
$AdS_5\times S^5/\IZ_k$, where the $\IZ_k$ action leaves fixed an
$S^1$ inside the $S^5$.

This field theory is known (see, for instance, \cite{Witten:1997so})
to be a finite field theory for any value of the $k$ gauge couplings
$\tau_i$, corresponding to a $k$-complex-dimensional surface of
conformal field theories. Thus, we should see $k$ complex parameters
in the string theory background which we can change without destroying
the $AdS_5$ component of this background. One such parameter is
obviously the dilaton, and the other $(k-1)$ may be identified
\cite{Kachru:1998ys} with the values of the NS-NS and R-R 2-form
$B$-fields on the $(k-1)$ 2-cycles which vanish at the $\IZ_k$
orbifold singularity (these are part of the blow-up modes for the
singularities; the other blow-up modes turn on fields which change the
$AdS_5$ background, and correspond to non-marginal deformations of the
field theory).

The low-energy spectrum has contributions both from the untwisted and
from the twisted sectors. The untwisted sector states are just the
$\IZ_k$ projection of the original $AdS_5\times S^5$ states. The
twisted sector states are the same (for large $N$ and at low energies)
as those which appear in flat space at an $\IR^4/\IZ_k$ singularity,
except that here they live on the fixed locus of the $\IZ_k$ action
which is of the form $AdS_5\times S^1$. At the orbifold point the
massless twisted sector states are $(k-1)$ tensor multiplets (these
tensor multiplets include scalars corresponding to the 2-form
$B$-fields described above). Upon dimensional reduction on the $S^1$
these give rise to $(k-1)$ $U(1)$ gauge fields on $AdS_5$, which
correspond to the $U(1)$ global symmetries of the field theory (which
were the off-diagonal gauge $U(1)$'s before taking the near-horizon
limit, and become global symmetries after this limit); see, e.g.
\cite{Hanany:1998it}. The orbifold
point corresponds to having all the $B$-fields maximally turned on
\cite{Aspinwall:1995zi}. 
The spectrum of fields on $AdS_5$ in this background was successfully
compared \cite{Gukov:1998kk} to the spectrum of chiral operators in
the field theory. If we move in the string theory moduli space to a
point where the $B$-fields on some 2-cycles are turned off, the
D3-branes wrapped around these 2-cycles become tensionless, and the
low-energy theory becomes a non-trivial $\cn=(2,0)$ six dimensional
SCFT (see
\cite{Seiberg:1997ax} and references therein). The
low-energy spectrum on $AdS_5$ then includes the dimensional reduction
of this conformal theory on a circle. In particular, when all the
$B$-fields are turned off, we get the $A_{k-1}$ $(2,0)$ theory, which
gives rise to $SU(k)$ gauge fields at low-energies upon
compactification on a circle. Thus, the AdS/CFT correspondence
predicts an enhanced global $SU(k)$ symmetry at a particular point in
the parameter space of the corresponding field theory. Presumably,
this point is in a very strongly coupled regime (the string coupling
$\tau_{IIB} \propto
\sum_i \tau_i$ may be chosen to be weak, but individual $\tau_i$'s can
still be strongly coupled) which cannot be accessed directly in the
field theory. The field theory in this case has a large group of
duality symmetries \cite{Witten:1997so}, which includes (but is not
limited to) the $SL(2,\IZ)$ subgroup which acts on the couplings as
$\tau \to (a \tau + b) / (c \tau + d)$ at the point where they are all
equal. In the type IIB background the $SL(2,\IZ)$ subgroup of this
duality group is manifest, but it is not clear how to see the rest of
this group.

Our second example corresponds to D3-branes at an $\IR^6/\IZ_3$
orbifold point, where, if we write $\IR^6$ as $\IC^3$ with complex
coordinates $z_j$ ($j=1,2,3$), the $\IZ_3$ acts as $z_j \to e^{2\pi
i/3} z_j$. In this case the only fixed point of the $\IZ_3$ action is
the origin, so in the near-horizon limit we get \cite{Kachru:1998ys}
$AdS_5\times S^5/\IZ_3$ where the compact space is smooth. Thus, the
low-energy spectrum in this case (for large $g_s N$) includes only the
$\IZ_3$ projection of the original supergravity spectrum, and all
twisted sector states are heavy in this limit.

The corresponding field theory may be derived by the methods of
\cite{Douglas:1996sw,Douglas:1997de}. 
It is an $SU(N)^3$ gauge theory, with chiral multiplets $U_j$
($j=1,2,3$) in the $\bf{(N,{\bar N},1)}$ representation, $V_j$
($j=1,2,3$) in the $\bf{(1,N,{\bar N})}$ representation, and $W_j$
($j=1,2,3$) in the $\bf{({\bar N},1,N)}$ representation, and a
classical superpotential of the form $W = g\epsilon^{ijk} U_i V_j
W_k$. In the classical theory all three gauge couplings and the
superpotential coupling $g$ are equal (and equal to the string
coupling). In the quantum theory one can prove that in the space of
these four parameters there is a one dimensional line of
superconformal fixed points. The parameter which parameterizes this fixed
line (which passes through weak coupling in the gauge theory) may be
identified with the dilaton in the $AdS_5\times S^5/\IZ_3$
background. Unlike the previous case, here there are no indications of
additional marginal deformations, and no massless twisted sector
states on $AdS_5$ which they could correspond to.

As in the previous case, one can try to compare the spectrum of fields
on $AdS_5$ with the spectrum of chiral operators in the field
theory. In this case, as in all cases with less than $\cn=4$
supersymmetry, not all the supergravity fields on $AdS_5$ are in
chiral multiplets, since the $\cn=4$ chiral multiplets split into
chiral, anti-chiral and non-chiral multiplets when decomposed into
$\cn=2$ (or $\cn=1$) representations\footnote{Note that this means
that unlike the $AdS_5\times S^5$ case, in cases with less SUSY there
are always non-chiral operators which have a finite dimension in the
large $N, g_{YM}^2 N$ limit.} (in general there can also be various sizes of
chiral multiplets). However, one can still compare those of
the fields which are in chiral multiplets (and have the appropriate
relations between their AdS mass / field theory dimension and their
R-charges). The untwisted states may easily be matched since they are
a projection of the original states both in space-time and in the
field theory (if we think of the field theory as a projection of the
$\cn=4$ $SU(3N)$ gauge theory). Looking at the twisted sectors we seem
to encounter a paradox \cite{Morrison:1998cs}. On the string theory
side all the twisted sector states are heavy (they correspond to
strings stretched across the $S^5$, so they would correspond to
operators with $\Delta
\simeq mR \simeq R^2 / l_s^2 \simeq (g_s N)^{1/2}$). On the field
theory side we can identify the twisted sector fields with operators
which are charged under the global $\IZ_3$ symmetry which rotates the
three gauge groups, and naively there exist chiral operators which are
charged under this symmetry and remain of finite dimension in the
large $N,g_{YM}^2 N$ limit. However, a careful analysis shows that all
of these operators are in fact descendants, so their dimensions are
not protected. For example, the operator $\sum_{j=1}^3 e^{2\pi ij/3}
\tr((W_{\alpha}^{(j)})^2)$, where $W_{\alpha}^{(j)}$ is the field
strength multiplet of the $j$'th $SU(N)$ group, seems to be a chiral
superfield charged under the $\IZ_3$ symmetry. However, using linear
combinations of the Konishi anomaly equations
\cite{Konishi:1984hf,Konishi:1985tu} for the three gauge groups, one
can show that this operator (and all other ``twisted sector''
operators) is in fact a descendant, so there is no paradox. The
AdS/CFT correspondence predicts that in the large $N$, $g_{YM}^2 N$
limit the dimension of all these $\IZ_3$-charged operators scales as
$(g_{YM}^2 N)^{1/2}$, which is larger than the scaling $\Delta \sim
(g_{YM}^2 N)^{1/4}$ for the non-chiral operators in the $\cn=4$ SYM
theory in the same limit. It would be interesting to verify this
behavior in the field theory. Baryon-like operators also exist in
these theories \cite{Gukov:1998kn}, which are similar to those
which will be discussed in section \ref{conifolds}.

There are various other supersymmetric orbifold backgrounds which
behave similarly to the examples we have described in detail
here. There are also many non-supersymmetric examples 
\cite{Frampton:1998en,Frampton:1999ti} but, as
described above, their fate for finite $N$ is not clear, and we will
not discuss them in detail here.

\subsection{Orientifolds of $AdS_5\times S^5$}
\label{orientifolds}

The discussion of the near-horizon limits of D3-branes on orientifolds
is mostly similar to the discussion of orbifolds, except for the
absence of twisted sector states (which do not exist for
orientifolds). We will focus here on two examples which illustrate
some of the general properties of these backgrounds. Additional
examples were discussed in \cite{Kakushadze:1998tr,Kakushadze:1998tz,
Kakushadze:1999hb,Kakushadze:1998yq,
Ahn:1998tv,
Park:1998zh,Gukov:1998kt,Gremm:1999jj}.

Our first example is the near-horizon limit of D3-branes on an
orientifold 3-plane. The orientifold breaks the same supersymmetries
as the 3-branes do, so in the near horizon limit we have the full 32
supercharges corresponding to a $d=4, \cn=4$ SCFT. In flat space there
are (see \cite{Giveon:1998sr} and references therein) two types of
orientifold planes which lead to different projections on D-brane
states. One type of orientifold plane leads to a low-energy $SO(2N)$
$\cn=4$ gauge theory for $N$ D-branes on the orientifold, while the
other leads to a $USp(2N)$ $\cn=4$ gauge theory. In the first case we
can also have an additional ``half D3-brane'' stuck on the
orientifold, leading to an $SO(2N+1)$ $\cn=4$ gauge theory. In the
near-horizon limits of branes on the orientifold we should be able to
find string theory backgrounds which are dual to all of these gauge
theories.

The near-horizon limit of these brane configurations is type IIB
string theory on $AdS_5\times S^5/\IZ_2 \equiv AdS_5\times \RP^5$, where
the $\IZ_2$ acts by identifying opposite points on the $S^5$, so there
are no fixed points and the space $\RP^5$ is smooth. The manifestation
of the orientifolding in the near-horizon limit is that when a string
goes around a non-contractible cycle in $\RP^5$ (connecting opposite
points of the $S^5$) its orientation is reversed. In all the cases
discussed above the string theory perturbation expansion had only
closed orientable surfaces, so it was a power series in $g_s^2$ (or in
$1/N^2$ in the 't~Hooft limit); but in this background we can also
have non-orientable closed surfaces which include crosscaps, and the
perturbation expansion includes also odd powers of $g_s$ (or of $1/N$
in the 't~Hooft limit). In fact, it has long been known
\cite{Cicuta:1982fu} that in the 't~Hooft limit the $SO(N)$ and
$USp(N)$ gauge theories give rise to Feynman diagrams that involve
also non-orientable surfaces (as opposed to the $SU(N)$ case which
gives only orientable surfaces), so it is not surprising that such
diagrams arise in the string theory which is dual to these
theories. While in the cases described above the leading correction in
string perturbation theory was of order $g_s^2$ (or $1/N^2$ in the 't
Hooft limit), in the $AdS_5\times \RP^5$ background (and in general in
orientifold backgrounds) the leading correction comes from $\RP^2$
worldsheets and is of order $g_s$ (or $1/N$ in the 't Hooft
limit). Such a correction appears, for instance, in the computation of
the central charge (the 2-point function of the energy-momentum tensor) of
these theories, which is proportional to the dimension of the
corresponding gauge group.

Our discussion so far has not distinguished between the different
configurations corresponding to $SO(2N)$, $SO(2N+1)$ and $USp(2N)$
groups (the only obvious parameter in the orientifold background is
the 5-form flux $N$). In the Feynman diagram expansion it is
well-known \cite{Mkrtchian:1981bb,Cvitanovic:1982bq} that the $SO(2N)$
and $USp(2N)$ theories are related by a transformation taking $N$ to
$(-N)$, which inverts the sign of all diagrams with an odd number of
crosscaps in the 't Hooft limit. Thus, we should look for a similar
effect in string theory on $AdS_5\times \RP^5$. It turns out
\cite{Witten:1998xy} that this is implemented by a ``discrete
torsion'' on $\RP^5$, corresponding to turning on a $B_{NS-NS}$ 2-form
in the non-trivial cohomology class of $H^3(\RP^5, {\tilde \IZ}) =
\IZ_2$. The effect of turning on this ``discrete torsion'' is exactly
to invert the sign of all string diagrams with an odd number of
crosscaps. It is also possible to turn on a similar ``discrete
torsion'' for the RR 2-form $B$-field, so there is a total of four
different possible string theories on $AdS_5\times \RP^5$. It turns
out that the theory with no $B$-fields is equivalent to the $SO(2N)$
$\cn=4$ gauge theory, which is self-dual under the S-duality group
$SL(2,\IZ)$. The theory with only a non-zero $B_{RR}$ field is
equivalent to the $SO(2N+1)$ gauge theory, while the theories with
non-zero $B_{NS-NS}$ fields are equivalent to the $USp(2N)$ gauge
theory \cite{Witten:1998xy}, and this is consistent with the action of
S-duality on these groups and on the 2-form $B$-fields (which are a
doublet of $SL(2,\IZ)$).

An interesting test of this correspondence is the matching of chiral
primary fields. In the supergravity limit the fields on $AdS_5\times
\RP^5$ are just the $\IZ_2$ projection of the fields on $AdS_5\times
S^5$, including the multiplets with $n=2,4,6,\cdots$ (in the notation
of section \ref{tests}). This matches with almost all the chiral
superfields in the corresponding gauge theories, which are described
as traces of products of the fundamental fields as in section
\ref{tests}, but with the trace of a product of an odd number of
fields vanishing in these theories from symmetry arguments. However,
in the $SO(2N)$ gauge theories (and not in any of the others) there is
an additional gauge invariant chiral superfield, called the Pfaffian,
whose lowest component is of the form $\epsilon^{a_1 a_2 \cdots
a_{2N}} \phi^{I_1}_{a_1 a_2} \phi^{I_2}_{a_3 a_4} \cdots
\phi^{I_N}_{a_{2N-1} a_{2N}}$, where $a_i$ are $SO(2N)$ indices and
the $I_j$ are (symmetric traceless) indices in the $\bf 6$ of
$SU(4)_R$. The supersymmetry algebra guarantees that the dimension of
this operator is $\Delta=N$, and it is independent of the other
gauge-invariant chiral superfields. This operator may be identified
with the field on $AdS_5$ corresponding to a D3-brane wrapped around a
3-cycle in $\RP^5$, corresponding to the homology class $H_3(\RP^5,
\IZ) = \IZ_2$. This wrapping is only possible when no $B$-fields are
turned on \cite{Witten:1998xy}, consistent with such an operator
existing for $SO(2N)$ but not for $SO(2N+1)$ or $USp(2N)$. While it is
not known how to compute the mass of this state directly, the
superconformal algebra guarantees that it has the right mass to
correspond to an operator with $\Delta=N$; the naive approximation to
the mass, since the volume of the 3-cycle in $\RP^5$ is $\pi^2 R^3$, 
is $m R \simeq R
\cdot \pi^2 R^3 / (2\pi)^3 g_s l_s^4 = R^4 / 8\pi l_p^4 \simeq N$
(since in the orientifold case $R^4 \simeq 4\pi (2N) l_p^4$ instead of
equation (\ref{dthree})), which leads to the correct dimension for
large $N$. The existence of this operator (which decouples in the
large $N$ limit) is an important test of the finite $N$
correspondence. Anomaly matching in this background was discussed in
\cite{Blau:1999vz}.

Another interesting background is the near-horizon limit of D3-branes
on an orientifold 7-plane, with 4 D7-branes coincident on the
orientifold plane to ensure \cite{Sen:1996vd,Banks:1996nj} 
that the dilaton is constant and
the low-energy theory is conformal (this is the same as D3-branes in
F-theory \cite{Vafa:1996xn} 
at a $D_4$-type singularity). The field theory we get
in the near-horizon limit in this case is
\cite{Aharony:1997en,Douglas:1997js} an $\cn=2$ SQCD
theory with $USp(2N)$ gauge group, a hypermultiplet in the
anti-symmetric representation and four hypermultiplets in the
fundamental representation. In this case the orientifold action has
fixed points on the $S^5$, so the near-horizon limit is
\cite{Fayyazuddin:1998fb,Aharony:1998xz} type IIB string theory on
$AdS_5\times S^5/\IZ_2$ where the $\IZ_2$ action has fixed points on
an $S^3$ inside the $S^5$. Thus, this background includes an
orientifold plane with the topology of $S^3\times AdS_5$, and the
D7-branes stretched along the orientifold plane also remain as part of
the background, so that the low-energy theory includes both the
supergravity modes in the bulk and the $SO(8)$ gauge theory on the
D7-branes (which corresponds to an $SO(8)$ global symmetry in the
corresponding field theories)\footnote{Similar backgrounds were
discussed in \cite{Kehagias:1998gn}.}. The string perturbation expansion
in this case has two sources of corrections of order $g_s$, the
crosscap diagram and the open string disc diagram with strings ending
on the D7-brane, leading to two types of corrections of order $1/N$ in
the 't~Hooft limit. Again, the spectrum of operators in the field
theory may be matched \cite{Aharony:1998xz} with the spectrum of
fields coming from the dimensional reduction of the supergravity
theory in the bulk and of the 7-brane theory wrapped on the $S^3$. The
anomalies may also be matched to the field theory, including $1/N$
corrections to the leading large $N$ result \cite{Aharony:1999rz}
which arise from disc and crosscap diagrams.

By studying other backgrounds of D3-branes with 7-branes (with or
without orientifolds) one can obtain 
non-conformal theories which
exhibit a logarithmic running of the coupling constant
\cite{Aharony:1998xz,deMelloKoch:1999hn}. For instance, by separating the
D7-branes away from the orientifold plane, corresponding to giving a
mass to the hypermultiplets in the fundamental representation, one
finds string theory solutions in which the dilaton varies in a similar
way to the variation of the coupling constant in the field theory, and
this behavior persists also in the near-horizon limit (which is quite
complicated in this case, and becomes singular close to the branes,
corresponding to the low-energy limit of the field theories which is
in this case a free Abelian Coulomb phase). This agreement with the
perturbative expectation, even though we are (necessarily) in a regime
of large $\lambda = g_{YM}^2 N$, is due to special properties of
$\cn=2$ gauge theories, which prevent many quantities from being
renormalized beyond one-loop.

\subsection{Conifold theories}
\label{conifolds}

In the correspondence between string theory on $AdS_5 \times S^5$ and
$d=4$ $\cn=4$ SYM theories, some of the most direct checks, such as
protected operator dimensions and the functional form of two- and
three-point functions, are determined by properties of the supergroup
$SU(2,2|4)$.  Many of the normalizations of two- and three-point
functions which have been computed explicitly are protected by
non-renormalization theorems.  And yet, we are inclined to believe that
the correspondence is a fundamental dynamical principle, valid
independent of group theory and the special non-renormalization
properties of ${\cal N}=4$ supersymmetry.

To test this belief we want to consider theories with reduced
supersymmetry.  Orbifold theories \cite{Kachru:1998ys} provide 
interesting examples;
however, as discussed in the previous sections,
it has been shown \cite{Bershadsky:1998mb,Bershadsky:1998cb} that
at large $N$ these theories are a projection of ${\cal N}=4$
super-Yang-Mills theory; in particular many of
their Green's functions are dictated
by the Green's functions of the ${\cal N}=4$ theory.  
The projection involved is onto
states invariant under the group action that defines the orbifold.
Intuitively,
this similarity with the $\cn=4$ theory arises because
the compact part of the geometry is
still (almost everywhere)
locally $S^5$, just with some global identifications. Therefore, 
to make a more
non-trivial test of models with reduced supersymmetry, we are
more interested in geometries of the form $AdS_5 \times M_5$ where the
compact manifold $M_5$ is not even locally $S^5$.

In fact, such compactifications have a long history in the
supergravity literature: the direct product geometry $AdS_5 \times
M_5$ is known as the Freund-Rubin ansatz \cite{Freund:1980xh}.
The curvature of the anti-de Sitter part of the geometry is supported
by the five-form of type~IIB supergravity.  Because this five-form is
self-dual, $M_5$ must also be an Einstein manifold, but with positive
cosmological constant: rescaling $M_5$ if necessary, we can write
${\cal R}_{\alpha\beta} = 4 g_{\alpha\beta}$.  For simplicity,
we are assuming that only the
five-form and the metric are involved in the solution.

A trivial but useful observation is that five-dimensional Einstein
manifolds with ${\cal R}_{\alpha\beta} = 
4 g_{\alpha\beta}$ are in one-to-one
correspondence with Ricci-flat manifolds $C_6$ whose metric has the
conical form
  \eqn{ConeForm}{
   ds_{C_6}^2 = dr^2 + r^2 ds_{M_5}^2 \ .
  }
 It can be shown that, given any metric of the form \ConeForm, the
ten-dimensional metric
  \eqn{KehagMetric}{
   ds_{10}^2 = \left( 1 + {R^4 \over r^4} \right)^{-1/2} 
     \left( -dt^2 + dx_1^2 + dx_2^2 + dx_3^2 \right) + 
    \left( 1 + {R^4 \over r^4} \right)^{1/2} ds_{C_6}^2
  }
 is a solution of the type IIB supergravity equations, provided one
puts $N$ units of five-form flux through the manifold $M_5$, where
  \eqn{NLRelation}{
   R^4 = {\sqrt{\pi} \over 2} {\kappa N \over \Vol M_5} \ .
  }
 Furthermore, it was shown in \cite{Kehagias:1998gn} that the number of
supersymmetries preserved by the geometry \KehagMetric\ is half the number
that are preserved by its Ricci-flat $R \to 0$ limit.  Preservation of
supersymmetry therefore amounts to the existence of a Killing spinor on
$ds_{C_6}^2$, which would imply that it is a Calabi-Yau metric.  Finally,
the $r \ll R$ limit of \KehagMetric\ is precisely $AdS_5 \times M_5$, and
in that limit the number of preserved supersymmetries doubles.

These facts suggest a useful means of searching for non-trivial
Freund-Rubin geometries: starting with a string vacuum of the form
$\IR^{3,1} \times C_6$, where $C_6$ is Ricci-flat, we locate
a singularity of $C_6$ where the metric locally has the form
\ConeForm, and place a large number of D3-branes at that point.  The
resulting near-horizon Freund-Rubin geometry has the same number of
supersymmetries as the original braneless string geometry.  The
program of searching for and classifying such singularities on
manifolds preserving some supersymmetry was enunciated most completely
in \cite{Morrison:1998cs}.

We will focus our attention on the simplest non-trivial example, which was
worked out in \cite{Klebanov:1998hh}\footnote{Additional aspects and
examples of
conifold theories were discussed in \cite{Uranga:1999vf,Dasgupta:1998su,
Gubser:1999ia,Lopez:1999zf,
vonUnge:1999hc,Erlich:1999rb}.}. 
$C_6$ is taken to be the standard conifold, which
as a complex 3-fold is determined by the equation
  \eqn{ConEq}{
   z_1^2 + z_2^2 + z_3^2 + z_4^2 = 0 \ .
  }
 The Calabi-Yau metric on this manifold has $SU(3)$ holonomy, so one
quarter of supersymmetry is preserved.  We will always count our
supersymmetries in four-dimensional superconformal field theory terms,
so one quarter of maximal supersymmetry (that is, eight real
supercharges) is in our terminology ${\cal N}=1$ supersymmetry
(superconformal symmetry).  The
supergravity literature often refers to this amount of supersymmetry
in five dimensions as ${\cal N}=2$, because in a flat space
supergravity theory with this much supersymmetry, reduction on $S^1$
without breaking any supersymmetry leads to a supergravity theory in
four dimensions with ${\cal N}=2$ supersymmetry.

The Calabi-Yau metric on the manifold \ConEq\ may be derived from 
the K\"ahler
potential $K = \left( \sum_{i=1}^4 |z_i|^2 \right)^{2/3}$, and can be
explicitly written as
  \eqn{ConMet}{
   ds_{C_6}^2 = dr^2 + r^2 ds_{T^{11}}^2,
  }
 where $ds_{T^{11}}^2$ is the Einstein metric on the coset space
  \eqn{TOneOne}{
   T^{11} = {SU(2) \times SU(2) \over U(1)} \ .
  }
In the quotient \TOneOne, the $U(1)$ generator is chosen to be 
the sum $\tf{1}{2}
\sigma_3 + \tf{1}{2} \tau_3$ of generators of 
the two $SU(2)$'s.  The manifolds $T^{pq}$,
where the $U(1)$ generator is chosen to be $\tf{p}{2} \sigma_3 + \tf{q}{2}
\tau_3$, with $p$ and $q$ relatively prime, were studied in
\cite{Romans:1985an}.  The topology of each of these manifolds is $S^2
\times S^3$.  They all admit unique Einstein metrics.  Only $T^{11}$ leads
to a six-manifold $C_6$ which admits Killing spinors.  In fact, besides
$S^5 = SO(6)/SO(5)$, $T^{11}$ is the unique five-dimensional coset space
which preserves supersymmetry.  The Einstein metrics can be obtained via a
rescaling of the Killing metric on $SU(2) \times SU(2)$ by a process
explained in \cite{Romans:1985an}.  The metric on $T^{11}$ satisfying
${\cal R}_{\alpha\beta} = 4 g_{\alpha\beta}$ can be written as
  \eqn{TMet}{
   ds_{T^{11}}^2 = \tf{1}{6} \sum_{i=1}^2 \left( d\theta_i^2 +
    \sin^2 \theta_i d\phi_i^2 \right) + 
    \tf{1}{9} \left( d\psi + \cos\theta_1 d\phi_1 + 
     \cos\theta_2 d\phi_2 \right)^2 \ .
  }
 The volume of this metric is $16\pi^3/27$, whereas the volume of the
unit five-sphere, which also has ${\cal R}_{\alpha\beta} = 4
g_{\alpha\beta}$, is $\pi^3$.

Perhaps the most intuitive way to motivate the conjectured dual gauge
theory \cite{Klebanov:1998hh} is to first consider the $S^5/\IZ_2$
orbifold gauge theory, where the $\IZ_2$ is chosen to flip the signs of
four of the six real coordinates in $\IR^6$, and thus has a fixed $S^1$ on
the unit $S^5$ in this flat space.  This $\IZ_2$ breaks $SO(6)$ down to
$SO(4) \times SO(2)$, which is the same isometry group as for $T^{11}$.  In
fact, it can also be shown that an appropriate blowup of the singularities
along the fixed $S^1$ leads to a manifold of topology $S^2 \times S^3$.
Since $T^{11}$ is a smooth deformation of the blown-up orbifold, one might
suspect that its dual field theory is some deformation of the orbifold's
dual field theory.  The latter field theory is well known
\cite{Kachru:1998ys}, as described in section \ref{orbifolds}.  
It has ${\cal N}=2$ supersymmetry.  The field
content in ${\cal N}=1$ language is
  \eqn{OrbField}{\seqalign{\span\TT \quad & \span\TC \quad & \span\TC}{
   gauge group & SU(N) & SU(N)  \cr
   chirals $A_1$, $A_2$ & \oalign{\idget\endyoung} & 
    \overline{\oalign{\idget\endyoung}}  \cr
   chirals $B_1$, $B_2$ & \overline{\oalign{\idget\endyoung}} &
    \oalign{\idget\endyoung}  \cr
   chiral $\Phi$ & \hbox{adj} & {\bf 1}  \cr
   chiral $\tilde\Phi$ & {\bf 1} & \hbox{adj}.
  }}
 The adjoint chiral fields $\Phi$ and $\tilde{\Phi}$, together with the
${\cal N}=1$ gauge multiplets, fill out ${\cal N}=2$ gauge multiplets.
The chiral multiplets $A_1$, $B_1$ combine to form an ${\cal N}=2$
hypermultiplet, and so do $A_2$, $B_2$.  The superpotential is
dictated by ${\cal N}=2$ supersymmetry:
  \eqn{OrbPot}{
   W = g \tr \Phi (A_1 B_1 + A_2 B_2) + 
       g \tr \tilde\Phi (B_1 A_1 + B_2 A_2) \ ,
  }
where $g$ is the gauge coupling of both $SU(N)$ gauge groups.
 A relevant deformation which preserves the global $SU(2) \times SU(2)
\times U(1)$ symmetry, and also ${\cal N}=1$ supersymmetry, is
  \eqn{DefPot}{
   W \to W + \tf{1}{2} m \left( \tr \Phi^2 - \tr \tilde\Phi^2 \right) \ .
  }
 There is a nontrivial renormalization group flow induced by these mass
terms.  The existence of a non-trivial
infrared fixed point can be demonstrated using
the methods of \cite{Leigh:1995ep}: having integrated out the heavy fields
$\Phi$ and $\tilde\Phi$, the superpotential is quartic in the remaining
fields, which should, therefore, all have dimension $3/4$ at the
infrared fixed point (assuming that we
do not break the symmetry between the two gauge groups).  The anomalous
dimension $\gamma = -1/2$ for the quadratic operators $\tr AB$ is precisely
what is needed to make the exact beta functions vanish.

The IR fixed point of the renormalization group 
described in the previous paragraph is the
candidate for the field theory dual to type IIB string theory on
$AdS_5 \times T^{11}$, or in
weak coupling terms the low-energy field theory of coincident
D3-branes on a conifold singularity.  There are several non-trivial
checks that this is the right theory.  The simplest is to note that
the moduli space of the $N=1$ version of the theory is simply the
conifold.  For $N=1$ the scalar fields $a_i$ and $b_j$ (in the chiral
multiplets $A_i$ and $B_j$) are just
complex-valued.  The moduli space 
can
be parametrized by the
combinations $a_i b_j$, and if we write
  \eqn{RecoverCon}{
   \pmatrix{z_1 + i z_4 & i z_2 + z_3 \cr
            i z_2 - z_3 & z_1 - i z_4} = 
   \pmatrix{a_1 b_1 & a_1 b_2 \cr a_2 b_1 & a_2 b_2} \ ,
  }
 then we recover the conifold equation \ConEq\ by taking the
determinant of both sides.  In the $N>1$ theories, a slight
generalization of this line of argument leads to the conclusion that
the fully Higgsed phase of the theory, where all the D3-branes are
separated from one another, has for its moduli space the $N^{\rm th}$
symmetric power of the conifold.

The most notable prediction of the renormalization group analysis of the
gauge theory is that the operators $\tr A_i B_j$ should have dimension
$3/2$.  This is something we should be able to see from the dual 
description.  As a
warmup, consider first the ${\cal N}=4$ example.  There, as described
in section \ref{tests}, the lowest
dimension operators have the form $\tr \phi^{(I} \phi^{J)}$, 
and their dimension
is two.  Their description in supergravity is a Weyl deformation of the
$S^5$ part of the geometry with $h^a_a \propto Y^2(y)$, 
where
$h^a_a$ is the trace of the metric on $S^5$ and $Y^2(y)$ is a
$d$-wave spherical harmonic on $S^5$.  The four-form potential
$D_{abcd}$ is also involved in the deformation, and
there are two mass eigenstates in $AdS_5$ which are combinations 
of these two 
fields.  A simple way to
compute $Y^2$ is to start with the function $x_i x_j$ on $\IR^6$ and
restrict it to the unit $S^5$.  This suggests quite a general way to find
eigenfunctions of the Laplacian on an Einstein manifold $M_5$: we start by
looking for harmonic functions on the associated conical geometry
\ConeForm.  The Laplacian is
  \eqn{ConeLaplace}{
   \square_{C_6} = {1 \over r^5} \partial_r r^5 \partial_r + 
    {1 \over r^2} \square_{M_5} \ .
  }
 The operator $r^2 \square_{C_6}$ commutes with $r\partial_r$, so we can
restrict our search to functions $f$ on $C_6$ with $\square_{C_6} f = 0$
and $r\partial_r f = \Delta f$ for some constant $\Delta$.  Such
harmonic functions
restricted to $r=1$ have $\square_{M_5} f\Big|_{r=1} = -\Delta(\Delta+4)
f\Big|_{r=1}$.  Following through the analysis of \cite{Kim:1985ez} one
learns that the mass of the lighter of the two scalars in $AdS_5$
corresponding to $h^a_a \propto f\Big|_{r=1}$ is $m^2 R^2 =
\Delta(\Delta-4)$.  So, the dimension of the corresponding operator is
$\Delta$.  In view of \RecoverCon, all we need to do to verify
in the supergravity approximation
the renormalization group prediction $\Delta = 3/2$ for $\tr A_i B_j$ is to
show that $r \partial_r z_i = \tf{3}{2} z_i$.  This follows from scaling
considerations as follows.  The dilation symmetry on the cone is $r \to
\lambda r$.  Under this dilation, $ds_{C_6}^2 \to \lambda^2 ds_{C_6}^2$.
The K\"ahler form should have this same scaling, and that will follow if also
the K\"ahler potential $K \to \lambda^2 K$.  As mentioned above, the
Calabi-Yau metric follows from $K = \left( \sum_{i=1}^4 |z_i|^2
\right)^{2/3}$, which has the desired scaling if $z_i \to \lambda^{3/2}
z_i$.  Thus, indeed $r\partial_r z_i = \tf{3}{2} z_i$.

It is straightforward to generalize the above line of argument to operators
of the form $\tr A_{(i_1} B^{(j_1} \ldots A_{i_\ell)} B^{j_\ell)}$.
Various aspects of the matching of operators in the conformal field theory
to Kaluza-Klein modes in supergravity have been studied in
\cite{Klebanov:1998hh,Gubser:1999vd,Jatkar:1999zk}.  But there is another
interesting type of color singlet operators, which are called dibaryons
because the color indices of each gauge group are combined using an
antisymmetric tensor. The dibaryon operator is
  \eqn{Dibaryon}{
   \epsilon_{\alpha_1 \ldots \alpha_N} 
    \epsilon^{\beta_1 \ldots \beta_N} 
    A^{\alpha_1}{}_{\beta_1} \ldots A^{\alpha_N}{}_{\beta_N} \ ,
  }
 where we have suppressed $SU(2)$ indices.  Let us use the notation
$SU(2)_A$ for the global symmetry
group under which $A_i$ form a doublet, and $SU(2)_B$ for
the group under which $B_j$ form a doublet.  Clearly, \Dibaryon\ is a
singlet under $SU(2)_B$.  This provides the clue to its string theory dual,
which must also be $SU(2)_B$-symmetric: it is a D3-brane wrapped on
$T^{11}$ along an orbit of $SU(2)_B$ \cite{Gubser:1998fp}.  Using the explicit
metric \TMet, it is straightforward to verify that $mR = \tf{3}{4} N$ in
the test brane approximation.  Up to corrections of order $1/N$, the
mass-dimension relation is $\Delta = mR$, so we see that again the 
field theory
prediction for the anomalous dimension of $A$ is born out. The 3-cycle
which the D3-brane is wrapped on may be shown to be the unique
homologically non-trivial 3-cycle of $T^{11}$. There is also
an anti-dibaryon, schematically $B^N$, which is a D3-brane wrapped on an
orbit of $SU(2)_A$.  The two wrappings are opposite in homology, so the
dibaryon and anti-dibaryon can annihilate to produce mesons.  This
interesting process has never been studied in any detail, no doubt because
the dynamics is complicated and non-supersymmetric.  It is possible to
construct dibaryon operators also in a variety of orbifold theories
\cite{Gubser:1998fp,Gukov:1998kn}.

The gauge theory dual to $T^{11}$ descends via renormalization group flow
from the gauge theory dual to $S^5/\IZ_2$, as described after \DefPot.
The conformal anomaly has been studied extensively for such flows (see for
example \cite{Anselmi:1997am}), and the coefficient $a$ in (\ref{confanom}) is
smaller in the IR than in the UV for every known flow that connects UV 
and IR
fixed points.  
        Cardy has conjectured that this must always be the case
        \cite{Cardy:1988cw}.  To describe the field theoretic attempts
        to prove such a c-theorem would take us too far afield, so
        instead we refer the reader to \cite{Forte:1998dx}
        and references therein.
        In section~\ref{cTheorem} we will demonstrate that a limited c-theorem
        follows from elementary properties of gravity if the AdS/CFT
        correspondence is assumed.

\begin{figure}
   \vskip0cm
   \centerline{\psfig{figure=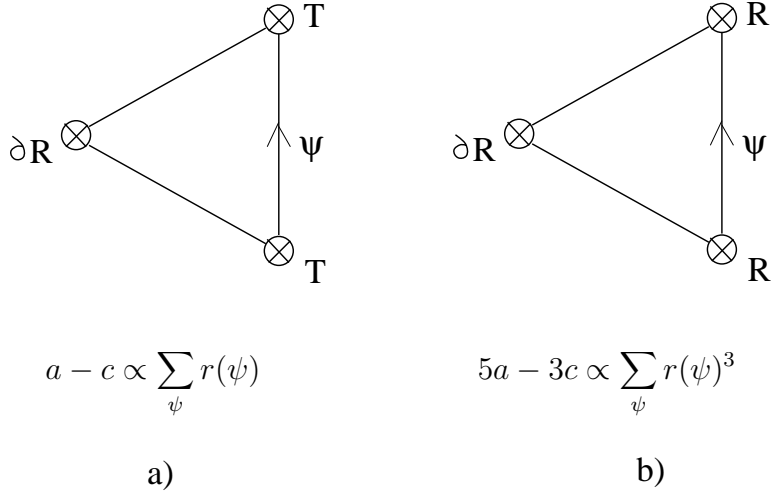,width=4in}}
   \vskip-2cm
   \centerline{$\displaystyle{a-c \propto \sum_\psi r(\psi)
     \qquad\qquad\qquad\quad
    5a-3c \propto \sum_\psi r(\psi)^3}$}
   \vskip1.5cm
 \caption{Triangle diagrams for computing the anomalous contribution to
$\partial_\mu R^\mu$.  The sum is over the chiral fermions $\psi$ which run
around the loop, and $r(\psi)$ is the R-charge of each such
fermion.}\label{figBssg}
  \end{figure}

In the presence of ${\cal N}=1$ superconformal invariance, one can compute
the anomaly coefficients $a$ and $c$ in (\ref{confanom}) if one knows $\langle
\partial_\mu R^\mu \rangle_{g_{\mu\nu}, B_\lambda}$, where $R_\mu$ is the
R-current which participates in the superconformal algebra, and the
expectation value is taken in the presence of an arbitrary metric
$g_{\mu\nu}$ and an external gauge field source $B_\mu$ for the R-current.
The reason $a$ and $c$ can be extracted from this anomalous one-point
function is that $\partial_\mu R^\mu$ and $T^\mu_\mu$ are superpartners in
the ${\cal N}=1$ multiplet of anomalies.  It was shown in
\cite{Anselmi:1997am} via a supergroup argument that
  \eqn{acTR}{\eqalign{
   \langle (\partial_\mu R^\mu) T_{\alpha\beta} T_{\gamma\delta} \rangle
     &= (a-c) \big[ \ \big]_{\alpha\beta\gamma\delta}  \cr
   \langle (\partial_\mu R^\mu) R_\alpha R_\beta \rangle
     &= (5a-3c) \big[ \ \big]_{\alpha\beta} \ ,
  }}
 where now the correlators are computed in flat space.  The omitted
expressions between the square brackets are tensors depending on the
positions or momenta of the operators in the correlator.  Their form is not
of interest to us here because it is the same for any theory: we are
interested instead in the coefficients.  These can be computed
perturbatively via the triangle diagrams in figure~\ref{figBssg}.
 The Adler-Bardeen theorem guarantees that the one loop result is exact,
provided $\partial_\mu R^\mu$ is non-anomalous in the absence of external
sources (that is, it suffers from no internal anomalies).  The constants of
proportionality in the relations shown in figure~\ref{figBssg} can be
tracked down by comparing the complete Feynman diagram amplitude with the
explicit tensor forms which we have omitted from \acTR.  We are mainly
interested in ratios of central charges between IR and UV fixed points, so
we do not need to go through this exercise.

The field theory dual to $S^5/\IZ_2$, expressed in ${\cal N}=1$
language, has the field content described in \OrbField.  The R-current of
the chosen ${\cal N}=1$ superconformal algebra descends from a $U(1)$ in
the $SO(6)$ R-symmetry group of the ${\cal N}=4$ algebra, and it assigns
a $U(1)_R$ charge
$r(\lambda) = 1$ to the $2N^2$ gauginos (fermionic components of the vector
superfield) and $r(\chi) = -1/3$ to the $6N^2$ ``quarks'' (fermionic
components of the
chiral superfields)\footnote{We will ignore here the distinction
between $U(N)$ and $SU(N)$ groups which is subleading in the $1/N$
expansion.}.  We have $\sum_\psi r(\psi) = 0$, which means that the
R-current has no gravitational anomalies \cite{Alvarez-Gaume:1984ig}.

For the field theory dual to $T^{11}$, the R-current described in the
previous paragraph is no longer non-anomalous because we have added a mass
to the adjoint chiral superfields.  
There is, however, a non-anomalous combination $S_\mu$ of
this current, $R_\mu$, with the Konishi currents, $K^i_\mu$, which by
definition assign charge~$1$ to the fermionic
fields in the $i^{\rm th}$ chiral
multiplet and charge~$0$ to the fermionic fields in the vector multiplets:
  \eqn{SDef}{
   S_\mu = R_\mu + \tf{2}{3} \sum_i 
    \left( \gamma_{\rm IR}^i - \gamma^i \right) K^i_\mu \ .
  }
 Here $\gamma^i$ is the anomalous dimension of the $i^{\rm th}$ chiral
superfield.  At the strongly interacting ${\cal N}=1$ infrared fixed point,
$S_\mu$ is the current which participates in the superconformal algebra.
However, to compute correlators $\langle (\partial_\mu S^\mu) \ldots
\rangle$ it is more convenient to go to the ultraviolet, where $\gamma^i =
0$ and the perturbative analysis in terms of fermions running around a loop
can be applied straightforwardly.  Using the fact that $\gamma_{\rm IR}^A =
\gamma_{\rm IR}^B = -1/4$ and $\gamma_{\rm IR}^\Phi = \gamma_{\rm
IR}^{\tilde{\Phi}} = 1/2$, we find that $s_{\rm UV}(\lambda) = 1$ for the
gauginos, $s_{\rm UV}(\chi) = -1/2$ for the quarks which stay light (i.e.,
the bifundamental quarks), and $s_{\rm UV}(\eta) = 0$ for the quarks which
are made heavy (that is, the adjoint quarks).  Note that it is immaterial
whether we include these heavy quarks in the triangle diagram, which is as
it should be since we can integrate them out explicitly.  As before,
$\sum_\psi s_{\rm UV}(\psi) = 0$, so there are no gravitational
anomalies and $a_{IR} = c_{IR}$.
Combining the information in the past two paragraphs, we have a field
theory prediction for the flow from the $S^5/\IZ_2$ theory to the
$T^{11}$ theory:
  \eqn{ChargeRatio}{
   {a_{\rm IR} \over a_{\rm UV}} = {c_{\rm IR} \over c_{\rm UV}} 
     = {5a_{\rm IR} - 3c_{\rm IR} \over 5a_{\rm UV} - 3c_{\rm UV}}
     = {2N^2 + 4N^2 \left( -{1 \over 2} \right)^3 \over
        2N^2 + 6N^2 \left( -{1 \over 3} \right)^3}
     = {27 \over 32} \ .
  }

This analysis was carried out in \cite{Gubser:1999vd}, where it was also
noted that these numbers can be computed in the supergravity
approximation.  To proceed,
let us write the ten-dimensional Einstein metric as
  \eqn{EinForm}{
   ds_{10}^2 = R^2 \widehat{ds}_5^2 + R^2 ds_{M_5}^2,
  }
 where $R$ is given by \NLRelation\ and $\widehat{ds}_5^2$ is the metric of
$AdS_5$ scaled so that $\hat{\cal R}_{\mu\nu} = -4 \hat{g}_{\mu\nu}$.  We will
refer to $\widehat{ds}_5^2$ as the dimensionless $AdS_5$ metric.  Reducing
the action from ten dimensions to five results in
  \eqn{FiveActForm}{
   S = {\pi^3 R^8 \over 2 \kappa^2} \int d^5 x \, \sqrt{\hat{g}} 
    \left( \hat{\cal R} + 12 + \ldots \right)
     = {\pi^2 N^2 \over 8 \Vol M_5} \int d^5 x \, \sqrt{\hat{g}} 
    \left( \hat{\cal R} + 12 + \ldots \right) \ ,
  }
 where $\sqrt{\hat{g}}$ and $\hat{\cal R}$ under the integral sign refer to the
dimensionless $AdS_5$ metric, and in the second equality we have used
\NLRelation.  In \FiveActForm, $\kappa$ is the ten-dimensional
gravitational coupling.  In computing Green's functions using the
prescription of \cite{Gubser:1998bc,Witten:1998qj}, the prefactor ${\pi^2
N^2 \over 8 \Vol M_5}$ multiplies every Green's function.  In particular, it
becomes the normalization factor for the one-point function $\langle
T^\mu_\mu \rangle$ as calculated in \cite{Henningson:1998gx}.  Also, as
pointed out in section \ref{tests}, the supergravity calculation in
\cite{Henningson:1998gx} always leads to $a=c$.  Without further thought we
can write $a = c \propto (\Vol M_5)^{-1}$, and
  \eqn{HenSkenRatio}{
   {a_{\rm IR} \over a_{\rm UV}} = {c_{\rm IR} \over c_{\rm UV}} = 
     \left( {\Vol T^{11} \over \Vol S^5/\IZ_2} \right)^{-1}
    = {27 \over 32} \ ,
  }
 in agreement with \ChargeRatio.  It is essential that the volumes in
\HenSkenRatio\ be computed for manifolds with the same cosmological
constant.  Our convention has been to have ${\cal R}_{\alpha\beta} =
4g_{\alpha\beta}$.

It is possible to do better and pin down the exact normalization of the
central charges.  In fact, literally the first normalization check
performed in the AdS/CFT correspondence was the verification
\cite{Gubser:1998bc} that in the compactification dual to
${\cal N}=4$ $SU(N)$ Yang-Mills theory, the
coefficient $c$ had the value $N^2/4$ (to leading order in large $N$).
Thus, in general
  \eqn{FreundRubinCharge}{
   a = c = {\pi^3 N^2 \over 4 \Vol M_5}
  }
 (again to leading order in large $N$) for the CFT dual to a Freund-Rubin
geometry $AdS_5 \times M_5$ supported by $N$ units of five-form flux
through the $M_5$.  This is in a normalization convention where the CFT
comprised of a single free real scalar field has $c = 1/120$.  See, 
for example,
\cite{Gubser:1999vd} for a table of standard anomaly coefficients per
degree of freedom.  Even more generally, we can consider any
compactification of string theory or M-theory (or any other, as-yet-unknown
theory of quantum gravity) whose non-compact portion is $AdS_5$.  This
would include in particular type~IIB supergravity geometries which involve
the $B_{\mu\nu}^{NS,RR}$ fields, or the complex coupling $\tau$.  Say the
$AdS_5$ geometry has ${\cal R}_{\mu\nu} = 
-\Lambda g_{\mu\nu}$.  If we rescale the
metric by a factor of $4/\Lambda$, we obtain the dimensionless $AdS_5$
metric $\widehat{ds}_5^2$ with $\hat{\cal R}_{\mu\nu} = -4 \hat{g}_{\mu\nu}$.
In defining a conformal field theory through its duality to the $AdS_5$
compactification under consideration, the part of the action relevant to
the computation of central charges is still the Einstein-Hilbert term plus
the cosmological term:
  \eqn{CosmConstC}{
   S = {1 \over 2\kappa_5^2} \int d^5 x \, \sqrt{g} 
    \left( {\cal R} + 3\Lambda + \ldots \right)
     = {4 \over \kappa_5^2 \Lambda^{3/2}} \int d^5 x \, \sqrt{\hat{g}} 
    \left( \hat{\cal R} + 12 + \ldots \right) \ ,
  }
 where $\kappa_5^2 = 8\pi G_5$ is the five-dimensional gravitational
coupling.  Comparing straightforwardly with the special case analyzed in
\FreundRubinCharge, we find that the conformal anomaly coefficients, as
always to leading order in $1/N$, must be given by 
  \eqn{acCos}{
   a = c = {1 \over G_5 \Lambda^{3/2}} \ .
  }

\section{D-Branes in AdS, Baryons and Instantons}
\label{baryons}

A conservative form of the AdS/CFT correspondence would be to say that
classical supergravity captures the large $N$ asymptotics of some
quantities in field theory which are algebraically protected against
dependence on the 't~Hooft coupling.  The stronger form which is usually
advocated, and which we believe is true, is that the field theory is
literally equivalent to the string theory, and the only issue is
understanding the mapping from one to the other.  To put this belief to the
test, it is natural to ask what in field theory corresponds to
non-perturbative objects, such as D-branes, in string theory.  The answer
was found in \cite{Witten:1998xy} for several types of wrapped branes (see
also \cite{Gross:1998gk} for an independent analysis of some cases), and
subsequent papers \cite{Brandhuber:1998xy,Imamura:1998hf,Imamura:1998gk,
Aharony:1998qu,Callan:1998iq,
Witten:1998cd,Morrison:1998cs,Alishahiha:1998ib,Gukov:1998kn,
Craps:1999nc,Callan:1999zf} have extended and elaborated on the story.  See
also \cite{Metsaev:1998hf,Claus:1999fh} for actions for D-branes in anti-de
Sitter space, and \cite{Bilal:1998ck,Pasti:1998tc} for other related
topics.  The connection between D-instantons and gauge theory instantons
has also been extensively studied, and we summarize the results at the end
of this section.

Let us start with wrapped branes which have no spatial extent in
$AdS_5$: they are particles propagating in this space.  The
field theory interpretation must be in terms of some vertex or
operator, as for any other particle in AdS (as described above in the
case of supergravity particles).  If
the compact manifold is $S^5$, then the only topologically stable
possibility is a wrapped 5-brane.  The key observation here is that
charge conservation requires that $N$ strings must run into or out of the
5-brane.  In the case of a D5-brane, these $N$ strings are fundamental
strings (one could also consider $SL(2,\IZ)$ images of this
configuration).  The argument is a slight variant of the ones used in the
discussion of anomalous brane creation
\cite{Hanany:1997ie,Bachas:1997ui,Danielsson:1997wq}.  There are $N$ units
of five-form ($F_5$) flux on the $S^5$, and the coupling ${1 \over
2\pi} a \wedge F_5$ in the D5-brane world-volume translates this flux
into $N$ units of charge under the $U(1)$ gauge field $a$ on the
D5-brane.  Since the D5-brane spatial world-volume is closed, the
total charge must be zero.  A string running out of the D5-brane
counts as $(-1)$ unit of $U(1)$ charge, hence the conclusion.
Reversing the orientation of the D5-brane changes the sign of the
charge induced by $F_5$, and correspondingly the $N$ strings should
run into the brane rather than out.

In the absence of other D-branes, the strings cannot end anywhere in
$AdS_5$, so they must run out to the boundary.  A string ending on the
boundary is interpreted (see section \ref{wilsonloops})
as an electric charge in the fundamental
representation of the $SU(N)$ gauge group: an external (non-dynamical)
quark. 
This interpretation comes from viewing the strings
as running from the D5-brane to a
D3-brane at infinity.  It was shown in
\cite{Bachas:1997ui} that such stretched strings have a unique ground
state which is fermionic, 
and the conclusion is that the D5-brane ``baryon'' is
precisely an antisymmetric combination of $N$ fermionic fundamental string
``quarks.''  The gauge theory interpretation is clear: because the gauge
group is $SU(N)$ rather than $U(N)$, there is a gauge-invariant
baryonic vertex for $N$
external fundamental quarks.  We will return to a discussion of baryonic
objects in section \ref{other_dynamical}.

To obtain other types of wrapped brane objects with no spatial extent in
$AdS_5$, we must turn to compact manifolds with more nontrivial homology
cycles.  Apart from the intrinsic interest of studying such objects and the
gauge theories in which they occur, the idea is to verify the claim that
every object we can exhibit in gauge theory has a stringy counterpart, and
vice versa.


Following \cite{Witten:1998xy} and the discussion in section
\ref{orientifolds}, we now examine wrapped branes in the $AdS_5
\times {\bf RP}^5$ geometry, which is the near-horizon geometry of
D3-branes placed on top of a $\IZ_2$ orientifold three-plane (the
$\IZ_2$ acts as $x_i \to -x_i$ for the six coordinates perpendicular to the
D3-branes).  
%
$H_3({\bf RP}^5,\IZ) = \IZ_2$, and the generator of the homology
group is a projective space ${\bf RP}^3 \subset {\bf RP}^5$.  This seems to
offer the possibility of wrapping a D3-brane on a 3-cycle to get a particle
in $AdS_5$.  However, there is a caveat: as argued in \cite{Witten:1998xy}
the wrapping is permitted only if there is no discrete torsion for the NS
and RR $B$-fields.  In gauge theory terms, that amounts to saying that the
corresponding operator is permitted if and only if 
the gauge group is $SO(N)$ with $N$
even.  Direct calculation leads to a mass $m \simeq N/R$ for the wrapped brane,
so the corresponding gauge theory operator has dimension $N$ (at least to
leading order in large $N$).  A beautiful fact is that a candidate gauge
theory operator exists precisely when the gauge group is $SO(N)$ with $N$
even: it is the ``Pfaffian'' operator, 
  \eqn{PfaffianOp}{
   {1 \over (N/2)!} \epsilon^{a_1 a_2 \ldots a_N} \phi_{a_1 a_2} \ldots
    \phi_{a_{N-1} a_N} \ .
  }
 Here the fields $\phi_{ab}$ are the adjoint scalar bosons which are the
${\cal N}=4$ superpartners of the gauge bosons.  We have suppressed their
global flavor index. A similar wrapped 3-brane was discussed in section
\ref{conifolds}, where the 3-brane was wrapped around the 3-cycle of
$T^{11}$ (which is topologically $S^2\times S^3$).

It is also interesting to consider branes with spatial extent in $AdS_5$.
Strings in $AdS_5$ were discussed in section \ref{wilsonloops}.  A
three-brane in $AdS_5$ (by which we mean any wrapped brane with three
dimensions of spatial extent in $AdS_5$) aligned with one direction
perpendicular to the boundary must correspond to some sort of domain wall
in the field theory.  Some examples are obvious: in $AdS_5 \times S^5$, if
the three-brane is a D3-brane, then crossing the domain wall shifts the
5-form flux and changes the gauge group from $SU(N)$ to $SU(N+1)$ or
$SU(N-1)$.  A less obvious example was considered in \cite{Witten:1998xy}:
crossing a D5-brane or NS5-brane wrapped on some ${\bf RP}^2 \subset {\bf
RP}^5$ changes the discrete torsion of the RR or NS $B$-field, and so one
can switch between $SO(N)$ and $Sp(N/2)$ gauge groups.  D5-branes on
homology 2-cycles of the base of conifolds and orbifolds have also been
studied \cite{Gubser:1998fp,Gubser:1999ia,Gukov:1998kn,Dasgupta:1999wx},
and the conclusion is that they correspond to domain walls across which the
rank of some factor in the product gauge group is incremented.

Another brane wrapping possibility is branes with two dimensions of spatial
extent in $AdS_5$.  These become strings in the gauge theory when they are
oriented with one dimension along the radial direction.  In a particular
model (an $SU(N)^3$ gauge theory whose string theory image is $AdS_5 \times
S^5/\IZ_3$) the authors of \cite{Gukov:1998kn} elucidated their
meaning: they are strings which give rise to a monodromy for the
wave-functions of particles transported around them.  The monodromy belongs
to a discrete symmetry group of the gauge theory.  The familiar example of
such a phenomenon is the Aharonov-Bohm effect, where the electron's
wave-function picks up a $U(1)$ phase when it is transported around a tube
of magnetic flux.  The analysis of \cite{Gukov:1998kn} extends beyond their
specific model, and applies in particular to strings in $SO(N)$ gauge
theories, with $N$ even, obtained from wrapping a D3-brane on a generator
of $H_1({\bf RP}^5,\IZ)$, where the ${\bf RP}^5$ has no discrete
torsion.

Finally, we turn to one of the most familiar examples of a non-perturbative
object in gauge theory: the instanton.  The obvious candidate in string
theory to describe an instanton is the D-instanton, also known as the
D(-1)-brane.  The correspondence in this case has been treated extensively
in the literature
\cite{Banks:1998nr,Chu:1998in,Kogan:1998re,Bianchi:1998nk,Brodie:1998ke,
Balasubramanian:1998de,Dorey:1998xe,Dorey:1999pd}.  The
presentation in \cite{Dorey:1999pd} is particularly comprehensive, and the
reader who is interested in a more thorough review of the subject can find
it there.  Note that the analysis of instantons in large $N$ gauge theories
is problematic since their contribution is (at least naively) highly
suppressed; the $k$ instanton contribution comes with a factor of
$e^{-8\pi^2k/g_{YM}^2} = e^{-8\pi^2 kN/\lambda}$ which goes like $e^{-N}$
in the 't Hooft limit. Therefore, we can only discuss instanton
contributions to quantities that get no other contributions to any order in
the $1/N$ expansion.  Luckily, such quantities exist in the $\cn=4$ SYM
theory, like the one discussed below.

The Einstein metric on $AdS_5 \times S^5$ is unaffected by
the presence of a D-instanton.  The massless fields in five dimensions
which acquire VEV's in the presence of a D-instanton are the axion and the
dilaton: in a coordinate system for the Poincar\'e patch of $AdS_5$ where 
  \eqn{PoincarePatchMetric}{
   ds^2 = {R^2 \over z^2} \left( dx_\mu^2 + dz^2 \right) \ ,
  }
 we have
\cite{Chu:1998in,Kogan:1998re,Bianchi:1998nk,Balasubramanian:1998de}, 
asymptotically as $z \to 0$,
  \eqn{PhiChiVEV}{\eqalign{
   e^\phi &= g_s + {24 \pi \over N^2} {z^4 \tilde{z}^4 \over 
    \left[ \tilde{z}^2 + (x_\mu - \tilde{x}_\mu)^2 \right]^4}
    + \ldots \ ,  \cr  
   \chi &= \chi_\infty \pm (e^{-\phi} - 1/g_s) \ ,
  }}
for a D-instanton whose location in anti-de Sitter space is
$(\tilde{x}_\mu,\tilde{z})$. It can be shown using the general
prescription for computing correlation functions that this
corresponds in the gauge theory to a VEV
  \eqn{VEVF}{
   \langle \tr F^2(x) \rangle = 192 {\tilde{z}^4 \over 
    \left[ \tilde{z}^2 + (x_\mu - \tilde{x}_\mu)^2 \right]^4} \ ,
  }
 which is exactly right for the self-dual background which describes the
instanton in gauge theory.  The action of a D-instanton, $2\pi/g_s$, also
matches the action of the instanton, $8\pi^2/g_{YM}^2$, because of the
relation $g_{YM}^2 = 4\pi g_s$.\fixit{Conventions?  Also in \VEVF?}  The
result \VEVF\ is insensitive to whether the D-instanton is localized on the
$S^5$, since the field under consideration is an $SO(6)$ singlet.  It is a
satisfying verification of the interpretation of the variable $z$ as
inverse energy scale that the position $\tilde{z}$ of the D-instanton
translates into the size of the gauge theory instanton.  In other words, we
understand the $AdS_5$ factor (which appears in the moduli space of an
$SU(2)$ instanton) as merely specifying the position of the D-instanton in
the five-dimensional bulk theory.

In fact, at large $N$, a Yang-Mills instanton is parametrized not only by a
point in $AdS_5$, but also by a point in $S^5$.  The $S^5$ emerges from
keeping track of the fermionic instanton zero modes properly
\cite{Dorey:1999pd}.  The approach is to form a bilinear $\Lambda^{AB}$ in
the zero modes.  $\Lambda^{AB}$ is antisymmetric in the four-valued $SU(4)$
indices $A$ and $B$, and satisfies a hermiticity condition that makes it
transform in the real ${\bf 6}$ of $SO(6)$.  Dual variables $\chi_{AB}$ can
be introduced into the path integral which have the same antisymmetry and
hermiticity properties: the possible values of $\chi_{AB}$ correspond to
points in $\IR^6$.  When the fermions are integrated out, the resulting
determinant acts as a potential for the $\chi_{AB}$ fields, with a minimum
corresponding to an $S^5$ whose radius goes into the determination of
the overall
normalization of correlation functions.

Building on the work of \cite{Banks:1998nr} on $\alpha'$ corrections to the
four-point function of stress-tensors, the authors of \cite{Bianchi:1998nk}
have computed contributions to correlators coming from instanton sectors of
the gauge theory and successfully matched them with D-instanton
calculations in string theory.  It is not entirely clear why the agreement
is so good, since the gauge theory computations rely on small 't~Hooft
coupling (while the string theory computations are for fixed $g_{YM}^2$ in
the large $N$ limit)
and non-renormalization theorems are not known for the relevant
correlators.  The simplest example turns out to be the sixteen-point
function of superconformal currents $\hat{\Lambda}^A_\alpha =
\tr(\sigma^{\mu\nu}{}_\alpha{}^\beta F^-_{\mu\nu} \lambda_\beta{}^A)$,
where $F^-_{\mu\nu}$ is the self-dual part of the field-strength, $A$ is an
index in the fundamental of $SU(4)$, $\alpha$ and $\beta$ are Lorentz
spinor indices, and $\mu$ and $\nu$ are the usual Lorentz vector indices.
One needs sixteen insertions of $\hat\Lambda$ to obtain a non-zero result
from the sixteen Grassmannian integrations over the fermionic zero modes of
an instanton.  The gauge theory result for gauge group $SU(2)$ turns out to
be
  \eqn{SixteenAnswer}{\eqalign{
   \left\langle \prod_{p=1}^{16} g_{YM}^2 \hat{\Lambda}_{\alpha_p}^{A_p}(x_p)
     \right\rangle
   = & {2^{11} 3^{16} \over \pi^{10}} g_{YM}^8 
   e^{-{8\pi^2 \over g_{YM}^2} + i \theta_{YM}} 
   \int {d^4 \tilde{x} \, d\tilde{z} \over \tilde{z}^5} 
   \int d^8 \eta \, d^8 \bar\xi  \cr
   & \prod_{p=1}^{16} \left[ {\tilde{z}^4 \over 
    \left[ \tilde{z}^2 + (x_p - \tilde{x})^2 \right]^4}
    {1 \over \sqrt{\tilde{z}}} \left( \tilde{z} \eta_{\alpha_p}^{A_p} + 
    (x_p - \tilde{x})_\mu \sigma^\mu_{\alpha_p \dot\alpha_p} 
    \bar\xi^{\dot\alpha_p A_p} \right) \right] \ .
  }}
 The superconformal currents $\hat{\Lambda}^A_\alpha$ are dual to
spin~$1/2$ particles in the bulk: dilatinos in ten dimensions which we
denote $\Lambda$.  One of the superpartners of the well-known 
${\cal R}^4$ term in
the superstring action (see for example \cite{Green:1997tv}) is the
sixteen-fermion vertex \cite{Green:1998me}: in string frame,
  \eqn{SixteenLambda}{
   {\cal L} = {e^{-2\phi} \over \alpha'^4} {\cal R} + \ldots + 
    \left( {e^{-\phi/2} \over \alpha'} f_{16}(\tau,\bar\tau) \Lambda^{16} + 
    \hbox{c.c} \right) + \ldots \ ,
  }
 where $f_{16}(\tau,\bar\tau)$ is a modular form with weight $(12,-12)$,
and $\tau$ is the complex coupling of type~IIB theory: 
  \eqn{TauEquation}{
   \tau = \chi + i e^{-\phi} = 
    {\theta_{YM} \over 2\pi} + {4\pi i \over g_{YM}^2} \ .
  }
 There is a well-defined expansion of this modular form in powers of
$e^{2\pi i \tau}$, $e^{-2\pi i \bar\tau}$, and $g_{YM}^2$.  Picking out the
one-instanton contribution and applying the prescription for calculating
Green's functions laid out in section \ref{correlators},
one recovers the form~\SixteenAnswer\ up to an overall factor.  The overall
factor can only be tracked down by redoing the gauge theory calculation
with gauge group $SU(N)$, with proper attention paid to the saddle point
integration over fermionic zero modes, as alluded to in the previous
paragraph.

The computation of Green's functions such as \SixteenAnswer\ has been
extended in \cite{Dorey:1999pd} to the case of multiple instantons.  Here
one starts with a puzzle.  The D-instantons effectively form a bound state
because integrations over their relative positions converge.  Thus the
string theory result has the same form as \SixteenAnswer, with only a
single integration over a point $(\tilde{x},\tilde{z})$ in $AdS_5$.  In view 
of the emergence of an $S^5$ from the fermionic zero modes at large $N$, the 
expectation on the
gauge theory side is that the moduli space for $k$
instantons should be $k$ copies of $AdS_5 \times S^5$.  But through an
analysis of small fluctuations around saddle points of the path integral it
was shown that most of the moduli are lifted quantum mechanically, and what
is left is indeed a single copy of $AdS_5 \times S^5$ as the moduli space,
with a prefactor on the saddle point integration corresponding to the
partition function of the zero-dimensional $SU(k)$ gauge theory which lives
on $k$ coincident D-instantons.  It is assumed that $k \ll N$.  Although
the $k$ instantons ``clump'' in moduli space, their field configurations
involve $k$ commuting $SU(2)$ subgroups of the $SU(N)$ gauge group.  The
correlation functions computed in gauge theory have essentially the same
form as \SixteenAnswer.  In comparing with the string theory analysis, one
picks out the $k$-instanton contribution in the Taylor expansion of the
modular form in \SixteenLambda.  There is perfect agreement at large $N$
for every finite $k$, which presumably means that there is some unknown
non-renormalization theorem protecting these terms.

\section{Deformations of the Conformal Field Theory}
\label{deformations}

In this section we discuss deformations of the conformal field theory, and
what they correspond to in its dual description involving string theory on
AdS space. We will focus on the case of the $\cn=4$ field theory, though
the general ideas hold also for all other examples of the AdS/CFT
correspondence. We start in section \ref{defintro} with a general
discussion of deformations in field theory and in the dual description.
Then in section~\ref{cTheorem} we use the AdS/CFT correspondence to prove a
restricted c-theorem.  In section~\ref{deffield} we discuss the interesting
relevant and marginal deformations of the $\cn=4$ SYM field theory; and in
section \ref{defstring} we review what is known about these deformations
from the point of view of type IIB string theory on $AdS_5\times S^5$. The
results we present will be based on \cite{aks_unpublished,
Girardello:1998pd,Distler:1998gb,Khavaev:1998fb,Karch:1999pv,
Freedman:1999gp}.

 \subsection{Deformations in the AdS/CFT Correspondence}
 \label{defintro}

 Conformal field theories have many applications in their own right,
 but since our main interest (at least in the context of four
 dimensional field theories) is in studying non-conformal field
 theories like QCD, it is interesting to ask how we can learn about
 non-conformal field theories from conformal field theories. One way to
 break conformal invariance, described in section \ref{FiniteT}, is to examine
 the theory at finite temperature. However, it is also possible to
 break conformal invariance while preserving Lorentz invariance, by
 deforming the action by local operators,
 \eqn{deform}{S \to S + h\int d^4x {\cal O}(x),}
 for some Lorentz scalar operator $\cal O$ and some coefficient
 $h$.

 The analysis of such a deformation depends on the scaling dimension
 $\Delta$ of the operator $\cal O$ \footnote{If the operator does not
 have a fixed scaling dimension we can write it as a sum of operators
 which are eigenfunctions of the scaling operator, and treat the
 deformation as a sum of the appropriate deformations.}. If $\Delta <
 4$, the effect of the deformation is strong in the IR and weak in the
 UV, and the deformation is called {\it relevant}. If $\Delta > 4$,
 the deformation is called {\it irrelevant}, and its effect becomes
 stronger as the energy increases. Since we generally describe field
 theories by starting with some UV fixed point and flowing to the IR,
 it does not really make sense to start with a CFT and perform an
 irrelevant deformation, since this would really require a new UV
 description of the theory. Thus, we will not discuss irrelevant
 deformations here. The last case is $\Delta=4$, which is called a
 {\it marginal deformation}, and which does not break conformal
 invariance to leading order in the deformation. Generally, even if
 the dimension of an operator equals 4 in some CFT, this will no
 longer be true after deforming by the operator, and conformal
 invariance will be broken. Such deformations can be either {\it
 marginally relevant} or {\it marginally irrelevant}, depending on the
 dimension of the operator $\cal O$ for finite small values of $h$. In
 special cases the dimension of the operator will remain $\Delta=4$
 for any value of $h$, and conformal invariance will be present for
 any value of $h$. In such a case the deformation is called {\it
 exactly marginal}, and the conformal field theories for all values of
 $h$ are called a {\it fixed line} (generalizing the concept of a
 conformal field theory as a fixed point of the renormalization group
 flow). When a deformation is relevant conformal invariance will be
 broken, and there are various possibilities for the IR behavior of
 the field theory. It can either flow to some new conformal field
 theory, which can be free or interacting, or it can flow to a trivial
 field theory (this happens when the theory confines and there are no
 degrees of freedom below some energy scale $\Lambda$). We will
 encounter examples of all of these possibilities in section
 \ref{deffield}.

 The analysis of deformations in the dual string theory on AdS space
 follows from our description of the matching of the partition
 functions in sections \ref{correspondence} and \ref{correlators}. 
 The field theory with the deformation
 \eno{deform} is described by examining string theory backgrounds in
 which the field $\phi$ on AdS space, which corresponds to the
 operator $\cal O$, behaves near the boundary of AdS space like
 $\phi(x,U) \stackrel{U\to \infty}{\longrightarrow} hU^{\Delta-4}$,
 where $[{\cal O}] = \Delta$ and we use the coordinate system 
 (\ref{metric5}) (with $U$ instead of $u$). In
 principle, we should sum over all backgrounds with this boundary
 condition. Note that, as mentioned in section \ref{correlators}, 
 in Minkowski space
 this involves turning on the non-normalizable solution to the field
 equations for $\phi(x,U)$; turning on the normalizable mode (as done
 for instance in
 \cite{Nojiri:1998yx,
 Kehagias:1999tr,Gubser:1999pk,Girardello:1999hj,Liu:1999fc,
 Kehagias:1999iy,Constable:1999ch}) 
 cannot be understood as a deformation of the field
 theory, but instead corresponds to a different state in the same
 field theory \cite{Balasubramanian:1999sn}\footnote{Some of the
 solutions considered in \cite{Girardello:1999hj} may correspond 
 to actual deformations of the field theory.}. As in the field theory,
 we see a big difference between the cases of $\Delta > 4$ and $\Delta
 < 4$. When $\Delta > 4$, the deformation grows as we approach the
 boundary, so the solution near the boundary will no longer look like
 AdS space; this is analogous to the fact that we need a new UV
 description of the field theory in this case. On the other hand, when
 $\Delta < 4$, the solution goes to zero at the boundary, so
 asymptotically the solution just goes over to the AdS solution, and
 the only changes will be in the interior. For $\Delta=4$ the solution
 naively goes to a constant at the boundary, but one needs to analyze
 the behavior of the string theory solutions beyond the leading order
 in the deformation to see if the exact solution actually grows as we
 approach the boundary (a marginally irrelevant deformation),
 decreases there (a marginally relevant deformation) or goes to a
 constant (an exactly marginal deformation).

 An exactly marginal deformation will correspond to a space of
 solutions of string theory, whose metric will always include an
 $AdS_5$ factor\footnote{The full space does not necessarily have to
 be a direct product $AdS_5\times X$, but could also be a fibration of
 $AdS_5$ over $X$, which also has the $SO(4,2)$ isometry group.}, but
 the other fields can vary as a function of the deformation
 parameters. A relevant (or marginally relevant) deformation will
 change the behavior in the interior, and the metric will no longer be
 that of AdS space. If we start in the regime of large $g_s N$ where
 there is a supergravity approximation to the space, the deformation
 may be describable in supergravity terms, or it may lead to large
 fields and curvatures in the interior which will cause the
 supergravity approximation to break down. The IR behavior of the
 corresponding field theory will be reflected in the behavior of the
 string theory solution for small values of $U$ (away from the
 boundary). If the solution asymptotes to an AdS solution also at
 small $U$, the field theory will flow in the IR to a non-trivial
 fixed point\footnote{Four dimensional field theories are believed
 \cite{Cardy:1988cw} to have a c-theorem analogous to the
 2-dimensional c-theorem \cite{Zamolodchikov:1986gt} which states that
 the central charge of the IR fixed point will be smaller than that of
 the UV fixed point. We will discuss some evidence for this in 
 the AdS context, based
 on the analysis of the low-energy gravity theory, in the next
 subsection.}.
 Note that the
 variables describing this AdS space may be different from the
 variables describing the original (UV) AdS space, for instance the
 form of the $SO(4,2)$ isometries may be different
 \cite{Distler:1998gb}. If the solution is described in terms of a
 space which has a non-zero minimal value of $U$ (similar to the space
 which appears in the AdS-Schwarzschild black hole solution described
 in section \ref{FiniteT}, but in this case with the full $ISO(3,1)$ isometry
 group unbroken) the field theory will confine and be trivial in the
 IR. In other cases the geometrical description of the space could
 break down for small values of $U$; presumably this is what happens
 when the field theory flows to a free theory in the IR.

\subsection{A c-theorem}
\label{cTheorem}

Without a detailed analysis of matter fields involved in non-anti-de Sitter
geometries, there are few generalities one can make about the description
of renormalization group flows in the AdS/CFT correspondence\footnote{See
\cite{Akhmedov:1998vf,Alvarez:1998wr,Gorsky:1998wn,Porrati:1999ew,
Balasubramanian:1999jd}
for general discussions of the renormalization group flow in the context of
the AdS/CFT correspondence.}.  However,
there is one general result in gravity \cite{Freedman:1999gp} (see also
\cite{Girardello:1998pd}) which translates into a c-theorem via the
correspondence.  Let us consider $D$-dimensional metrics of the form
  \eqn{DefMet}{
   ds^2 = e^{2 A(r)} (-dt^2 + d\vec{x}^2) + dr^2 \ .
  }
 Any metric with Poincar\'e invariance in the $t,\vec{x}$ directions can be
brought into this form by an appropriate choice of the radial variable $r$.
Straightforward calculations yield
  \eqn{AInequality}{
      -(D-2) A'' = R^t_t - R^r_r = G^t_t - G^r_r =
     \kappa_D^2 (T^t_t - T^r_r) \geq 0 \ .
  }
 In the second to last step we have used Einstein's equation, and in the
last step we have assumed that the weak energy condition holds in the form
  \eqn{WeakEnergy}{
   T_{\mu\nu} \zeta^\mu \zeta^\nu \geq 0
  }
 for any null vector $\zeta^\mu$.  This form of the weak energy condition
is also known as the null energy condition, and it is obeyed by all fields
which arise in Kaluza-Klein compactifications of supergravity theories to
$D$ dimensions.  Thus, we can take it as a fairly general fact that $A''
\leq 0$ for $D > 2$.  Furthermore, the inequality is saturated precisely
for anti-de Sitter space, where the only contribution to $T_{\mu\nu}$ is
from the cosmological constant.  Thus in particular, any deformation of
$AdS_D$ arising from turning on scalar fields will cause $A$ to be concave
as a function of $r$.  If we are interested in relevant deformations
of the conformal field theory, then we should recover linear behavior in
$A$ near the boundary, which corresponds to the (conformal)
ultraviolet limit in the
field theory.  Without loss of generality, then, we assume $A(r) \sim
r/\ell$ as $r \to \infty$.

The inequality $A'' \leq 0$ implies that the function
  \eqn{CFunction}{
   {\cal C}(r) \equiv {1 \over A'^{D-2}} 
  }
 decreases monotonically as $r$ decreases.  Now, suppose there is a region
where $A$ is nearly linear over a range of $r$ corresponding to many orders
of magnitude of $e^{A(r)}$.  This is the bulk analog of a scaling region in
the boundary field theory.  The asymptotically linear behavior of $A(r)$ as
$r\to\infty$ indicates an ultraviolet scaling region which extends
arbitrarily high in energy.  If $A(r)$ recovers linear behavior as $r \to
-\infty$, there is an infrared scaling region; and there could also be
large though finite scaling regions in between.  Assuming odd bulk
dimension $D$, the perfect $AdS_D$ spacetime which any such scaling region
approximates leads to an anomalous VEV
  \eqn{TVEV}{
   \langle T^\mu_\mu \rangle = {\hbox{universal} \over A'^{D-2}} \ ,
  }
 where the numerator is a combination of curvature invariants which can be
read off from the analysis of \cite{Henningson:1998gx} (see section
\ref{anomalies}).  The point is that
in limits where conformal invariance is recovered, the expression
\CFunction\ coincides with the anomaly coefficients of the boundary field
theory, up to factors of order unity which are universal for all CFT's in a
given dimension.  Thus, ${\cal C}(r)$ is a c-function, and the innocuous
inequality $A'' \leq 0$ amounts to a c-theorem provided that Einstein
gravity is a reliable approximation to the bulk physics.

In geometries such as the interpolating kinks of
\cite{Girardello:1998pd,Distler:1998gb,Freedman:1999gp} (discussed in more
detail in section~\ref{defstring}), the outer anti-de Sitter region is
distinguishable from the inner one in that it has a boundary.  There can
only be one boundary (in Einstein frame) because $A$ gets large and
positive only once.  In fact, the inner anti-de Sitter region has finite
proper volume if the coordinates $t$ and $\vec{x}$ in \DefMet\ are made
periodic.  Supergravity is capable of describing irreversible
renormalization group flows despite the reversibility of the equations,
simply because the basic prescription for associating the partition
functions of string theory and field theory makes use of the unique
boundary.

 \subsection{Deformations of the $\cn=4$ $SU(N)$ SYM Theory}
 \label{deffield}

 The most natural deformations to examine from the field theory point
 of view are mass deformations, that would give a mass to the scalar
 and/or fermion fields in the $\cn=4$ vector multiplet. One is tempted
 to give a mass to all the scalars and fermions in the theory, in order
 to get a theory that will flow to the pure Yang-Mills (YM) theory in
 the IR. Such a deformation would involve operators of the form
 $\tr(\phi^I \phi^I)$ for the scalar masses, and $[\epsilon^{\alpha
 \beta} \tr(\lambda_{\alpha A} \lambda_{\beta B}) + c.c.]$ for the
 fermion masses. In the weak coupling regime of small $\lambda=g_{YM}^2
 N$, such deformations indeed make sense and would lead to a pure
 Yang-Mills theory in the IR. However, the analysis of this region
 requires an understanding of the string theory in the high-curvature
 region which corresponds to small $\lambda$, which is not yet
 available. With our present knowledge of string theory we are limited
 to analyzing the strong coupling regime of large $\lambda$, where
 supergravity is a good approximation to the full string theory. In
 this regime there are two problems with the mass deformation described
 above :
 \begin{itemize}
 \item{} The operator $\tr(\phi^I \phi^I)$ is a non-chiral operator, so
 the analysis of section \ref{chiralops} suggests that for large
 $\lambda$ it acquires a dimension which is at least as large as
 $\lambda^{1/4}$, and in particular for large enough values of
 $\lambda$ it is an irrelevant operator. Thus, we cannot deform the
 theory by this operator for large $\lambda$. In any case this operator
 is not dual to a supergravity field, so analyzing the corresponding
 deformation requires going beyond the supergravity approximation.
 \item{} The pure YM theory is a confining theory which dynamically
 generates a mass scale $\Lambda_{YM}$, which is the characteristic
 mass scale for the particles (glueballs) of the theory. When we deform
 the $\cn=4$ theory by a mass deformation with a mass scale $m$, a
 one-loop analysis suggests that the mass scale $\Lambda_{YM}$ will be
 given by $\Lambda_{YM} \sim m e^{-c/g_{YM}^2(m) N}$, where $c$ is a
 constant which does not depend on $N$ (arising from the one-loop
 analysis) and $g_{YM}^2(m)$ is the coupling constant at the scale
 $m$. Thus, we find that while for small $\lambda$ we have 
 $\Lambda_{YM} \ll
 m$ and there is a separation of scales between the dynamics of the
 massive modes and the dynamics of the YM theory we want to study, for
 large $\lambda$ we have $\Lambda_{YM} \sim m$ and there is no such
 separation of scales (for non-supersymmetric mass deformations the
 one-loop analysis we made is not exact, but an exact analysis is not
 expected to change the qualitative behavior we describe). Thus, we
 cannot really study the pure YM theory, or any other confining theory
 (which does not involve all the fields of the original $\cn=4$ theory)
 as long as we are in the strong coupling regime where supergravity is
 a good approximation.
 \end{itemize}
 We will see below that, while we can find ways to get around the first
 problem and give masses to the scalar fields, there are no known ways
 to solve the second problem and study interesting confining field
 theories using the supergravity approximation. Of course, in the full
 string theory there is no such problem, and the mass deformation
 described above, for small $\lambda$, gives an implicit string theory
 construction of the non-supersymmetric pure YM theory.

 In the rest of this section we will focus on the deformations that
 can arise in the strong coupling regime, and which may be analyzed in
 the supergravity approximation. As described in section
 \ref{chiralops}, the only operators whose dimension remains small for
 large $N$ and large $\lambda$ are the chiral primary operators, so we
 are limited to deformations by these operators. Let us start by
 analyzing the symmetries that are preserved by such
 deformations. Most of the chiral operators are in non-trivial
 $SU(4)_R$ representations, so they break the $SU(4)_R$ group to some
 subgroup which depends on the representation of the operator we are
 deforming by. Generic deformations will also completely break the
 supersymmetry. One analyzes how much supersymmetry a particular
 deformation breaks by checking how many supercharges annihilate
 it. For example, deformations which preserve $\cn=1$ supersymmetry
 are annihilated by the supercharges $Q_\alpha$ and ${\bar
 Q}_{\dot{\alpha}}$ of some $\cn=1$ subalgebra of the $\cn=4$
 algebra. Given the structure of the chiral representations described
 in section \ref{chiralops} it is easy to see if a deformation by such
 an operator preserves any supersymmetry or not. Examples of
 deformations which preserve some supersymmetry are superpotentials of
 the form $W=h\tr(\Phi^{i_1} \Phi^{i_2} \cdots \Phi^{i_n})$, which to
 leading order in $h$ add to the Lagrangian a term of the form
 $[h\epsilon^{\alpha \beta} \tr(\lambda_{\alpha A_1} \lambda_{\beta
 A_2} \phi^{I_1} \cdots \phi^{I_{n-2}}) + c.c.]$. These operators are
 part of the scalar operators described in section \ref{chiralops}
 arising at dimension $n+1$ in the chiral multiplet. In order to
 preserve supersymmetry one must also add to the Lagrangian various
 terms of order $h^2$, so we see that the question of whether a
 deformation breaks supersymmetry or not depends not only on the
 leading order operator we deform by but also on additional operators
 which we may or may not add at higher orders in the deformation
 parameter (note that the form of the chiral operators also changes
 when we deform, so an exact analysis of the deformations beyond the
 leading order in the deformation is highly non-trivial). Another
 example of a supersymmetry-preserving deformation is a superpotential
 of the form $W=h\tr(W_{\alpha}^2 \Phi^{i_1} \cdots \Phi^{i_{n-2}})$,
 which deforms the theory by some of the scalar operators arising at
 dimension $n+2$ in the chiral multiplet (e.g. the dilaton deformation
 for $n=2$, which actually preserves the full $\cn=4$ supersymmetry).

 The list of chiral operators which correspond to marginal or relevant
 deformations was given in section \ref{chiralops}. There is a total
 of 6 such operators, three of which are the lowest components of the
 chiral multiplets with $n=2,3,4$\footnote{In a $U(N)$ theory there is
 an additional scalar operator which is the lowest component of the
 $n=1$ multiplet.}. These operators are traceless
 symmetric products of scalars ${\cal O}_n=\tr(\phi^{\{I_1} \phi^{I_2}
 \cdots \phi^{I_n\}})$, which viewed as deformations of the theory
 correspond to non-positive-definite potentials for the scalar
 fields. Thus, at least if we are thinking of the theory on $\IR^4$
 where the scalars have flat directions before adding the potential,
 these deformations do not make sense since they would cause the
 theory to run away along the flat directions. In particular, the
 deformation in the $\bf 20'$ which naively gives a mass to the
 scalars really creates a negative mass squared for at least some of
 the scalars, so it cannot be treated as a small deformation of the UV
 conformal theory at the origin of moduli space. We will focus here
 only on deformations by the other 3 operators, which all seem to make
 sense in the field theory.

 One marginal operator of dimension 4 is the operator which couples to
 the dilaton, which is a $\bf 1$ of $SU(4)_R$, of the form $[\tr(F_{\mu
 \nu}^2) + i\tr(F \wedge F) + \cdots]$. Deforming by this operator
 corresponds to changing the coupling constant $\tau_{YM}$ of the field
 theory, and is known to be an exactly marginal deformation which does
 not break any of the symmetries of the theory.

 The other two relevant or marginal deformations are the scalars of
 dimension $n+1$ in the $n=2$ and $n=3$ multiplets. Let us start by
 describing the relevant deformation, which is a dimension 3 operator
 in the $\bf 10$ of $SU(4)_R$, of the form 
 \eqn{defthree}{\left[
 \epsilon^{\alpha \beta} \tr(\lambda_{\alpha A} \lambda_{\beta B}) +
 \tr([\phi^I, \phi^J] \phi^K) \right],} 
 where the indices are contracted to be in the $\bf 10$ of $SU(4)_R$
 (which is in the symmetric product of two $\bf {\bar 4}$'s and in the
 self-dual antisymmetric product of three $\bf 6$'s). This operator is
 complex; obviously when we add it to the Lagrangian we need to add it
 together with its complex conjugate. The coefficient
 parametrizing the deformation is a complex number $m^a$ in the $\bf
 10$ of $SU(4)_R$. Deforming by this operator obviously gives a mass
 to some or all of the fermion fields $\lambda$, depending on the
 exact values of $m^a$. For generic values of $m^a$, all the fermions
 will acquire a mass and supersymmetry will be completely broken. The
 scalars will then obtain a mass from loop diagrams in the field
 theory, so that the low-energy theory below a scale of order $m^a$
 will be the pure non-supersymmetric Yang-Mills theory. Unfortunately,
 as described above, for large $\lambda=g_{YM}^2 N$ this is not really
 a good description since this theory will confine at a scale
 $\Lambda_{YM}$ of order $m$. However, for small $\lambda$ this
 deformation does enable us to obtain the pure YM theory as a
 deformation of the $\cn=4$ theory.

 It is interesting to ask what happens if we give a mass only to some
 of the fermions. In this case we may or may not preserve some amount
 of supersymmetry (obviously, preserving $\cn=1$ supersymmetry
 requires leaving at least one adjoint fermion massless). The
 deformations which preserve at least $\cn=1$ supersymmetry correspond
 to superpotentials of the form $W = m_{ij} \tr(\Phi^i
 \Phi^j)$. Choosing an $\cn=1$ subgroup breaks $SU(4)_R$ to
 $SU(3)\times U(1)_R$, and (if we choose the $U(1)$ normalization so
 that the supercharges decompose as ${\bf 4} = {\bf 3}_1 + {\bf
 1}_{-3}$) the ${\bf 10}$ decomposes as ${\bf 10} = {\bf 6}_2 + {\bf
 3}_{-2} + {\bf 1}_{-6}$. The SUSY preserving deformation $m_{ij}$ is
 then in the ${\bf 6}_2$ representation, and it further breaks both
 the $SU(3)$ and the $U(1)$. In a supersymmetric deformation we
 obviously need to also add masses of order $m^2$ to some of the
 scalars; naively this leads to a contradiction because, as described
 above, there are no reasonable scalar masses to add which are in
 chiral operators. However, at order $m^2$ we have to take into
 account also the mixings between operators which occur at order $m$
 in the deformation\footnote{Similar mixings were recently discussed in
 \cite{Intriligator:1999ff}.}; the form of the chiral operators changes after we
 deform, and they mix with other operators (in particular, the form of
 the operator which is an eigenvalue of the scaling operator changes
 when we turn on $m$). In the case of the supersymmetric mass
 deformation, at order $m$ the chiral operator \eno{defthree}
 described above mixes with the non-chiral $\tr(\phi^I \phi^I)$
 operator giving the scalars a mass, so there is no contradiction. The
 simplest way to see this operator mixing in the SUSY-preserving case
 is to note that the $\cn=1$ SUSY transformations in the presence of a
 general superpotential include terms of the form $\{Q_{\alpha},
 \lambda_{\beta i}\} \sim \epsilon_{\alpha \beta} {d{\bar W}\over
 d{{\bar \Phi}^i}}$, which lead to corrections of order $m$ to $[Q^2,
 {\cal O}_2]$ which is the operator that we are deforming by.

 There are two interesting ways to give a mass to only one of the
 fermions. One of them is a particular case of the SUSY-preserving
 deformation described above, of the form $W = m \tr(\Phi^1 \Phi^1)$,
 which is an element of the ${\bf 6}_{2}$ of $SU(3)\times U(1)$, and
 breaks $SU(4)_R \to SU(2)\times U(1)$ while preserving $\cn=1$ SUSY
 (but breaking the conformal invariance). The other possibility is to
 use the deformation in the ${\bf 1}_{-6}$, which breaks SUSY
 completely but preserves an $SU(3)$ subgroup of $SU(4)_R$. To leading
 order in the deformation both possibilities give a mass to one
 fermion, but at order $m^2$ they differ in a way which causes one of
 them to break SUSY while the other further
 breaks $SU(3)\to SU(2)\times
 U(1)$. At weak coupling we can analyze the order $m^2$ terms in
 detail. In the SUSY-preserving deformation at order $m^2$ we turn on a
 scalar mass term of the form $|m|^2 \tr[(\phi^1)^2+(\phi^2)^2]$, which
 may be written in the form 
 \eqn{susymassterm}{{{|m|^2}\over 3}
 \tr[2(\phi^1)^2+2(\phi^2)^2-(\phi^3)^2-(\phi^4)^2-(\phi^5)^2-(\phi^6)^2]
 + {{|m|^2}\over 3}\tr[\phi^I \phi^I],}
 where the first term is one of
 the $\Delta=2$ chiral operators in the $\bf 20'$, and the second term
 is a non-chiral operator which arises from the operator mixing as
 described above (the appearance of the second term allows us to add
 the chiral operator in the first term without destroying the
 positivity of the scalar potential). In the non-SUSY deformation the
 chiral term is not turned on at any order in the deformation (the $\bf
 20'$ representation contains no singlets of $SU(3)$), and all the
 scalars get equal masses from the non-chiral term.

 Which theory do we flow to in the IR after turning on such a
 single-fermion mass term~? In the SUSY-preserving case one can show
 that we actually flow to an $\cn=1$ SCFT (and, in fact, to a fixed
 line of $\cn=1$ SCFTs). Naively, one chiral multiplet gets a mass,
 and we remain with the $\cn=1$ $SU(N)$ SQCD theory with two adjoint
 chiral multiplets, which is expected (based on the amount of matter
 in the theory) to flow to an interacting IR fixed point. In fact, one
 can prove \cite{Karch:1999pv} that there is an exactly marginal
 operator at that fixed point, which (generally) has a non-zero value
 in the IR theory we get after the flow described above. The full
 superpotential with the deformation is of the form $W = h\tr(\Phi^1
 [\Phi^2, \Phi^3]) + m\tr(\Phi^1 \Phi^1)$ (where $h$ is proportional
 to $g_{YM}$), and to describe the low-energy theory we can integrate
 out the massive field $\Phi^1$ to remain with a superpotential $W =
 -{h^2\over 4m} \tr([\Phi^2, \Phi^3]^2)$ for the remaining massless
 fields. Naively this superpotential is irrelevant (its dimension at
 the UV fixed point at weak coupling is 5), but in fact one can show
 (for instance, using the methods of \cite{Leigh:1995ep}) that it is
 exactly marginal in the IR theory, so there is a fixed line of SCFTs
 parametrized by the coefficient $\tilde h$ of the superpotential $W =
 {\tilde h} \tr([\Phi^2,\Phi^3]^2)$. Upon starting from a particular
 value of $g_{YM}$ in the UV and performing the supersymmetric mass
 deformation, we will land in the IR at some particular point on the
 IR fixed line (i.e. some value of $\tilde h$). The unbroken global
 $U(1)$ symmetry of the theory becomes the $U(1)_R$ in the $\cn=1$
 superconformal algebra in the IR.

 It is more difficult to analyze the mass deformation which does not
 preserve SUSY (but preserves $SU(3)$), since we cannot use the
 powerful constraints of supersymmetry. Naively one would expect this
 deformation to lead to masses (from loop diagrams) for all of the
 scalars, but not for the fermions, since the $SU(3)$ symmetry
 prevents them from acquiring a mass. Then, the IR theory seems to be
 $SU(N)$ Yang-Mills coupled to three adjoint fermions, which
 presumably flows to an IR fixed point (this is what happens for
 supersymmetric theories with one-loop beta functions of the same
 order, but it is conceivable also that the theory may confine and
 generate a mass scale). There is no reason for such a fixed point to
 have any exactly marginal deformations (in fact, there are no known
 examples in four dimensions of non-supersymmetric theories with
 exactly marginal deformations), so presumably the flow starting from
 any value of $g_{YM}$ always ends up at the same IR fixed point. We
 assumed that the deformation leads to positive masses squared for the
 scalars; it is also possible that it would give rise to negative
 masses squared for the scalars, in which case the theory on $\IR^4$
 would have no vacuum, as described above.

 If we give a mass to two of the fermions, it is possible to do this
 with a superpotential of the form $W = m\tr(\Phi^1 \Phi^2)$ which in
 fact preserves $\cn=2$ supersymmetry (it gives the $\cn=2$ SQCD
 theory with one massive adjoint hypermultiplet, which was discussed
 in \cite{Donagi:1996cf}). This theory is known to dynamically
 generate a mass scale, at which the $SU(N)$ symmetry is broken (at a
 generic point in the moduli space) to $U(1)^{N-1}$, and the
 low-energy theory is the theory of $(N-1)$ free $U(1)$ vector
 multiplets. The behavior of this theory for large $N$ was discussed
 in \cite{Douglas:1995nw}. At special points in the moduli space there
 are massless charged particles, and at even more special points in
 the moduli space \cite{Argyres:1995jj,Argyres:1996xn,Eguchi:1996vu}
 there are massless electrically and magnetically charged particles
 and the theory is a non-trivial $\cn=2$ SCFT. It is not completely
 clear which point in the moduli space one would flow to upon adding
 the mass deformation to the $\cn=4$ theory. Presumably, without any
 additional fine-tuning one would end up at a generic point in the
 moduli space which corresponds to a free IR theory.

 If we give a mass to two fermions while breaking supersymmetry (as
 above, this depends on the order $m^2$ terms that we add), we
 presumably end up in the IR with Yang-Mills theory coupled to two
 massless adjoint fermions. This theory is expected to confine at some
 scale $\Lambda_{YM}$ (which for large $g_{YM}^2 N$ would be of the
 order of the scale $m$), and lead to a trivial theory in the IR. A
 similar confining behavior presumably occurs if we give a mass to
 three or four of the fermions (for three fermions we can give a mass
 while preserving SUSY, and we presumably flow in the IR to the
 confining $\cn=1$ pure SYM theory).

 The only remaining deformation is the deformation by the $\Delta=4$
 operator in the $\bf 45$ representation, which is in the $n=3$
 multiplet. A general analysis of this deformation is rather difficult,
 so we will focus here on the SUSY preserving case where the
 deformation is a superpotential of the form $W = h_{ijk} \tr(\Phi^i
 \Phi^j \Phi^k)$, with the coefficients $h_{ijk}$ in the ${\bf 10}_0$
 representation in the decomposition ${\bf 45} = {\bf 15}_4+{\bf
 10}_0+{\bf 8}_0+{\bf 6}_{-4}+{\bf {\bar 3}}_{-4}+{\bf 3}_{-8}$. It
 turns out that one can prove (see \cite{Leigh:1995ep} 
 and references therein) that two of these ten deformations
 correspond to exactly marginal operators, that preserve $\cn=1$
 superconformal invariance. This can be done by looking at a general
 $\cn=1$ theory with three adjoint chiral multiplets, a gauge coupling
 $g$, and a superpotential of the form
 \eqn{exmarsup}{W = h_1 \tr(\Phi^1 \Phi^2 \Phi^3 + \Phi^1
 \Phi^3 \Phi^2) + h_2 \tr((\Phi^1)^3+(\Phi^2)^3+(\Phi^3)^3)
 + h_3\epsilon_{ijk}\tr(\Phi^i \Phi^j \Phi^k).}
 This particular superpotential is chosen to preserve a $\IZ_3\times
 \IZ_3$ global symmetry, where one of the $\IZ_3$ factors acts by
 $\Phi^1 \to \Phi^2, \Phi^2 \to \Phi^3, \Phi^3 \to \Phi^1$ and the
 other acts by $\Phi^1 \to \Phi^1, \Phi^2 \to \omega \Phi^2, \Phi^3
 \to \omega^2 \Phi^3$ where $\omega$ is a third root of unity. The
 second $\IZ_3$ symmetry prevents any mixing between the chiral
 operators $\Phi^i$, and the first $\IZ_3$ can then be used to show
 that they all have the same anomalous dimension
 $\gamma(g,h_1,h_2,h_3)$. The beta function may be shown (using
 supersymmetry) to be exactly proportional to this gamma function
 (with a coefficient which is a function of $g$), so that the
 requirement of conformal invariance degenerates into one equation
 ($\gamma=0$) in the four variables $g,h_1,h_2$ and $h_3$, which
 generically has a 3-dimensional space of solutions. This space of
 solutions corresponds to a 3-dimensional space of $\cn=1$ SCFTs. The
 general arguments we used so far do not tell us the form of the
 3-dimensional space, but we can now use our analysis of the $\cn=4$
 theory to learn more about it. First, we know that the $\cn=4$ line
 $g=h_3, h_1=h_2=0$ is a subspace of this 3-dimensional space. We also
 know that at leading order in the deformation away from this
 subspace, $(h_3+g)$, $h_1$ and $h_2$ correspond to marginal operators
 (as described above they couple to chiral operators of dimension 4),
 while $(h_3-g)$ couples to a non-chiral operator (in the $\bf 15$ of
 $SU(4)_R$) whose dimension is corrected away from $g=0$ (and seems to
 be large for large $g_{YM}^2 N$). Thus, we see that to leading order
 in the deformation around the $\cn=4$ fixed line, the exactly
 marginal deformations are given by $h_1$ and $h_2$ (which are two
 particular elements of the ${\bf 10}_0$ representation). It is not
 known if the other deformations in the $\bf 45$ are marginally
 relevant, marginally irrelevant or exactly marginal.

 \subsection{Deformations of String Theory on $AdS_5\times S^5$}
 \label{defstring}

 As described in section \ref{defintro}, to analyze the deformations
 of section \ref{deffield} in the AdS context requires finding
 solutions of string theory with appropriate boundary conditions. For
 the exactly marginal deformation in the $\bf 1$, which corresponds to
 the dilaton, we already know the solutions, which are just the
 $AdS_5\times S^5$ solution with any value of the string coupling
 $\tau_{IIB}$. The other operators discussed above are identified in
 string theory with particular modes of the 2-form field $B_{ab}$ with
 indices in the $S^5$ directions (we view $B$ as a complex 2-form
 field which contains both the NS-NS and R-R 2-form fields). Thus, the
 dimension 3 mass deformation would be related to string theory
 backgrounds in which $B_{ab}(x,U,y) \stackrel{U\to
 \infty}{\longrightarrow} m Y_{ab}^{(1)}(y) / U$ for some spherical
 harmonics $Y_{ab}^{(1)}(y)$ on $S^5$, and the dimension 4
 deformations would be related to backgrounds with $B_{ab}(x,U,y)
 \stackrel{U\to \infty}{\longrightarrow} h Y_{ab}^{(2)}(y)$. It is
 clear from the identification of the superconformal algebra in the
 field theory and in the string theory that these deformations break
 the same supersymmetries in both cases; this can also be checked
 explicitly (say, to leading order in the deformation
 \cite{aks_unpublished,Girardello:1998pd}) by analyzing the SUSY
 variations of the type IIB supergravity fields. The existence of an
 exactly marginal deformation breaking the $\cn=4$ superconformal
 symmetry to $\cn=1$ superconformal symmetry suggests that the theorem
 of \cite{Banks:1988yz}, that forbids flat space compactifications
 with different amounts of supersymmetry from being at a finite
 distance from each other in the string theory moduli space, is not
 valid in AdS compactifications
 \cite{aks_unpublished,Girardello:1998pd}.

 Since we know little about string theory in backgrounds
 with RR fields, our analysis of such solutions is effectively limited
 to the supergravity approximation. This already limits our discussion
 to large $\lambda=g_s N$, and it limits it further to cases where the
 solution does not develop large curvatures in the interior. In the
 supergravity limit one would want to find solutions of type IIB
 supergravity with the boundary conditions described above (with the
 rest of the fields having the same boundary conditions as in the
 $AdS_5\times S^5$ case). Unfortunately, no such solutions are
 known, and they seem to be rather difficult to construct. There are 3
 possible approaches to circumventing this problem of finding exact
 solutions to type IIB supergravity :
\begin{itemize} 
\item{} One can try to construct solutions perturbatively in the
deformation parameter, which should be easier than constructing the
full exact solution. Unfortunately, this approach does not make sense
for the relevant deformations, since already at leading order in the
deformation (corresponding to the linearized equations of motion
around the $AdS_5\times S^5$ solution) we find that the solution
($B_{ab} \sim 1/U$) grows to be very large in the interior, so the
perturbative expansion does not make sense. At best one may hope to
have a perturbative expansion in a parameter like $m/U$ (if $m$ is the
coefficient of a relevant operator of dimension $\Delta=3$),
but this only makes sense near the boundary. On the other hand, for
marginal deformations, and especially for deformations that are
supposed to be exactly marginal, this approach makes sense. Exactly
marginal deformations correspond to solutions which do not depend on
the AdS coordinates at all, so a perturbation expansion in the
parameters of the deformation seems to be well-defined. In practice
such a perturbation expansion is quite complicated, and can only be
done in the first few orders in the deformation. In the case of the
deformation by $h_1,h_2$ which was described in field theory above,
one can verify that it is an exactly marginal deformation to second
order in the deformation, even though additional SUGRA fields need to
be turned on at this order (including components of the metric with
$S^5$ indices). This is in fact true for any deformation in the $\bf
45$. At third order one probably gets non-trivial constraints on which
elements of the $\bf 45$ can be turned on in an exactly marginal
deformation, but the equations of motion of type IIB SUGRA have not
yet been expanded to this order. Verifying that the deformations that
are exactly marginal in the field theory correspond to exactly
marginal deformations also in string theory on $AdS_5\times S^5$ would
be a non-trivial test of the AdS/CFT correspondence.

\item{} There are no known non-trivial solutions of type IIB
supergravity which are asymptotically of the form described above for
the relevant or marginal deformations. However, there are several
known solutions \cite{Romans:1985an,vanNieuwenhuizen:1985ri} of type
IIB supergravity (in addition to the $AdS_5\times S^5$ solution) which
involve $AdS_5$ spaces and have $SO(4,2)$ isometries (these solutions
need not necessarily be direct products $AdS_5\times X$), and one can
try to guess that they would be the end-points of flows arising from
relevant deformations. As long as we are in the supergravity
approximation, only solutions which are topologically equivalent to
$AdS_5\times S^5$ can be related by flows to the $AdS_5\times S^5$
solution, so we will not discuss here other types of $AdS_5$
solutions.

One such solution was found in \cite{Romans:1985an}, which is of the
 form $AdS_5\times X$, where $X$ is an $S^1$ fiber over $CP^2$ (a
 ``stretched five-sphere''), and there is also a 3-form field turned
 on in the compact directions (this is called a Pope-Warner type
 solution \cite{Pope:1985jj}). This solution has an $SU(3)$ isometry
 symmetry (corresponding to an $SU(3)$ global symmetry in the
 corresponding field theory), and it breaks all the
 supersymmetries. Thus, it is natural to try to identify it with the
 deformation by the non-supersymmetric single-fermion mass operator
 described in section \ref{deffield}, which has the same
 symmetries. Unfortunately, as discussed below, this solution seems to
 be unstable.

 An additional solution, found in \cite{vanNieuwenhuizen:1985ri},
 exhibits an $SO(5)$ global symmetry. As discussed below,
 this solution also appears to be unstable.

 \item{} The most successful way (to date) of analyzing the
 appropriate solutions of type IIB supergravity has been to restrict
 attention to the five dimensional $\cn=8$ supergravity
 \cite{Gunaydin:1986cu} sector of the theory, which includes only the
 $n=2$ ``supergraviton'' multiplet from the spectrum described in
 section \ref{chiralops}. Unlike the situation in flat-space
 compactifications, the five dimensional supergravity cannot be viewed
 as a low-energy limit of the ten dimensional supergravity
 compactification in any sense. For instance, the supergraviton
 multiplet contains fields of $m^2=-4/R^2$, while other multiplets (in
 the $n=3,4$ multiplets) which are not included in the truncation to
 the five dimensional supergravity theory involve massless fields on
 $AdS_5$. However, it is conjectured that there does exist a
 consistent truncation of the type IIB supergravity theory on
 $AdS_5\times S^5$ to the five dimensional $\cn=8$ supergravity, in
 the sense that every solution of the latter can be mapped into a
 solution of the full type IIB theory (with the other fields in type
 IIB supergravity being some functions of the five dimensional SUGRA
 fields). A similar truncation is believed to exist (\cite{deWit:1987iy,
 vN:99ct} and references therein)
 for the relation between 11 dimensional supergravity compactified on
 $AdS_4\times S^7$ and the four dimensional $\cn=8$ gauged
 supergravity, and
 for the relation between 11 dimensional supergravity compactified
 on $AdS_7\times S^4$ and the seven dimensional gauged supergravity, 
 and the similarities between the two cases suggest that
 it may exist also in the $AdS_5\times S^5$ case (though this has not
 yet been proven\footnote{Partial evidence for this was given in
 \cite{Cvetic:1999xp}.  See section~\ref{ConsistentTruncation} for
        further discussion.}). In the rest of this section we will assume
 that such a truncation exists and see what we can learn from
 it. Obviously, we can only learn from such a truncation about
 deformations of the theory by fields in the $n=2$ multiplet, so we
 cannot analyze the marginal deformations in the $\bf 45$ in this way.

 The first thing one can try to do with the five dimensional $\cn=8$
 supergravity is to find solutions to the equations of motion with an
 $SO(4,2)$ isometry. These correspond to critical points of the scalar
 potential of $d=5,\cn=8$ supergravity, which is a complicated
 function of the 42 (=${\bf 20'} + {\bf 10}_c + {\bf 1}_c$) scalar
 fields in the $n=2$ multiplet. A full analysis of the critical points
 of this potential has not yet been performed, but there are 4 known
 vacua in addition to the vacuum corresponding to $AdS_5\times S^5$ :

 (i) There is a non-supersymmetric vacuum with an unbroken $SU(3)$
 gauge group. This vacuum is conjectured to correspond to the
 $SU(3)$-invariant vacuum of the full type IIB supergravity theory
 described above, which, as mentioned above, could correspond to a
 mass deformation of the $\cn=4$ field theory. Additional evidence for
 this correspondence was presented in
 \cite{Girardello:1998pd,Distler:1998gb}, which constructed a solution
 of the five dimensional $\cn=8$ supergravity which interpolates
 between the $AdS_5\times S^5$ solution and the $SU(3)$-invariant
 solution, with the leading deformation from the $AdS_5\times S^5$
 solution corresponding exactly to the mass operator in the ${\bf
 1}_{-6}$ in the decomposition ${\bf 10} = {\bf 6}_2 + {\bf 3}_{-2} +
 {\bf 1}_{-6}$, which breaks $SU(4)_R \to SU(3)$. 
 Since this solution is non-supersymmetric, one must
 verify that the classical solution is stable, namely that it does not
 contain tachyons whose mass is below the Breitenlohner-Freedman
 stability bound (in supersymmetric vacua this is guaranteed; using
 equation (\ref{dimenmass}), such tachyons would correspond to operators
 of complex dimension in the field theory which would contradict its
 unitarity). 
 It has recently been shown \cite{KPPrivate} that there are scalars in
 the gauged supergravity multiplet which do violate the
 Breitenlohner-Freedman stability bound in the expansion around the
 $SU(3)$-invariant solution\footnote{
 Except for orbifold constructions, there is no example at the time of
 writing of a non-supersymmetric $AdS_5$ vacuum which is definitely
 known to satisfy the stability bound.  There are however non-orbifold,
 non-supersymmetric $AdS_3$ vacua which are perturbatively stable.}
Thus, this is not a consistent
 vacuum of the supergravity theory. The AdS/CFT correspondence then
 implies that performing this mass deformation at strong coupling
 leads to some instability in the field theory (for instance, it could
 lead to negative masses squared for the scalar fields).

 (ii) There is a non-supersymmetric vacuum with unbroken $SO(5)$ gauge
 symmetry, which is conjectured to be related to the $SO(5)$-invariant
 compactification of type IIB supergravity which we mentioned
 above. The mass spectrum in this vacuum was computed in
 \cite{Distler:1998gb}, where it was found that it has a tachyonic
 particle whose mass is below the stability bound. Thus, even
 classically this is not really a vacuum of the supergravity theory
 (presumably the tachyon would condense and the theory would flow to
 some different vacuum). It was found in
 \cite{Girardello:1998pd,Distler:1998gb} that this ``vacuum'' is
 related to the $AdS_5\times S^5$ vacuum by a deformation involving
 turning on one of the operators in the $\bf{20'}$ representation;
 presumably the instability of the supergravity solution is related to
 the instability of the field theory after performing this
 deformation.

 (iii) There is \cite{Khavaev:1998fb,Karch:1999pv,Freedman:1999gp} a
 vacuum with $SU(2)\times U(1)$ unbroken symmetry and 8 unbroken
 supercharges, corresponding to an $\cn=1$ SCFT in the field
 theory. There is no known corresponding solution of the full type IIB
 theory, but assuming that 5d SUGRA is a consistent truncation, such a
 solution must exist (though it is not guaranteed that all its
 curvature invariants will be small, as required for the consistency
 of the supergravity approximation). It is natural to identify this
 vacuum with the IR fixed point arising from the supersymmetric
 single-chiral-superfield mass deformation described in section
 \ref{deffield}. This is consistent with the form of the 5d SUGRA
 fields that are turned on in this solution, with the global
 symmetries of the solution, and with the fact that on both sides of
 the correspondence we have a fixed line of $\cn=1$ SCFTs (the
 parameter $\tilde{h}$ of the fixed line corresponds to the dilaton on
 the string theory side; supersymmetry prohibits the generation of a
 potential for this field). Recently this identification was supported
 by the construction of the full solution interpolating between the
 $\cn=4$ fixed point and the $\cn=1$ fixed point in the 5d SUGRA theory
 \cite{Freedman:1999gp}. Since we have some supersymmetry left in
 this case, one can also quantitatively test this correspondence by
 matching the global anomalies of the field theory described in
 section \ref{deffield} (the $SU(N)$ $\cn=1$ SQCD theory with two
 adjoint chiral multiplets and a superpotential $W\propto
 \tr([\Phi^2,\Phi^3]^2)$) with those of the corresponding SUGRA
 background, as described in section \ref{anomalies}. The conformal
 anomalies were successfully compared in
 \cite{Karch:1999pv,Freedman:1999gp} in the
 large $N$ limit, giving some evidence for this correspondence (in
 particular, the conformal anomalies of this theory satisfy $a=c$, as
 required for a consistent supergravity approximation). The fact that
 the central charge corresponding to this solution is smaller than
 that of the $AdS_5\times S^5$ solution with the same RR 5-form flux
 (note that the RR flux is quantized and does not change when we
 deform) means that this interpretation is consistent with the
 conjectured four dimensional c-theorem.

 (iv) There is an additional background found in \cite{Khavaev:1998fb}
 with $SU(2)\times U(1)\times U(1)$ unbroken gauge symmetry and no
 supersymmetry. The mass spectrum of this background has not yet been
 computed, so it is not clear if it is stable or not. The SUGRA
 solution involves giving VEVs to fields both in the $\bf 20'$ and in
 the $\bf 10$, but it is not clear exactly what deformation of the
 original $AdS_5\times S^5$ theory (if any) this background
 corresponds to.

In principle, one could also use the truncated five dimensional theory
to analyze other relevant deformations in the $\bf 10$, which are not
expected to give rise to conformal field theories in the
IR. Presumably most of them would lead to high curvatures in the
interior, but perhaps some of them do not and can then be analyzed
purely in supergravity.

\end{itemize}

To summarize, the analysis of deformations in string theory on
$AdS_5\times S^5$ is rather difficult, but the results that are known
so far seem to be consistent with the AdS/CFT correspondence. The only
known results correspond to deformations which lead to conformal
theories in the IR; as discussed in section \ref{deffield}, these are
also the only deformations which we would expect to be able to
usefully study in general in the supergravity approximation. The most
concretely analyzed deformation is the single-chiral-fermion mass
deformation, which seems to lead to another AdS-type background of
type IIB supergravity (though only the truncation of this background
to the five dimensional supergravity is known so far). In
non-supersymmetric cases the analysis of deformations is complicated
(see, for instance, \cite{Berkooz:1999qp}) by the fact that quantum
corrections are presumably important in lifting flat directions, so a
classical supergravity analysis is not really enough and the full
string theory seems to be needed.

\chapter{AdS$_3$}
\label{ChapAdS3}


\def\ra{\rightarrow}
\def\quart{{1\over4}}
\def\p{\partial}
\def\ket#1{|#1\rangle}
\def\bra#1{\langle#1|}
\def\grad{\vec \nabla}
\def\bp{\bar \p}
\def\RN{Reissner-Nordstr\"om}
\def\apm{\alpha'}
\def\at{{\tilde \alpha}}
\def\s42{ 2^{-{1\over 4} } }
\def\shalf{{\scriptscriptstyle{1\over2}}}
\def\sign{{\rm sign}}
\def\csc{closed string channel}
\def\osc{open string channel}
\def\cL{{\cal L}}
\def\goto#1{\mathrel {\mathop {\longrightarrow} \limits_{#1}}}
\def\lr{\goto{r\to\infty}}
\def\exp{{\rm exp}}
\def\lb{\left\langle}
\def\rb{\right\rangle}
\def\ie{i\epsilon}
\def\ra{\rightarrow}
\def\propaga#1{\left({\theta_1(#1) \over \theta_4(#1)} \right) }
\def\propp{\left({\theta'_1(0) \over \theta_4(0)}\right) }
\def\g{\gamma}
\def\gb{\bar{\gamma}}
\def\a{\alpha}
\def\sa{r_0^2 {\rm sinh}^2\alpha }
\def\sg{r_0^2 {\rm sinh}^2\gamma }
\def\ss{r_0^2 {\rm sinh}^2\sigma }
\def\ca{r_0^2 {\rm cosh}^2\alpha }
\def\cg{r_0^2 {\rm cosh}^2\gamma }
\def\cs{r_0^2 {\rm cosh}^2\sigma }
\def\[{\left [}
\def\]{\right ]}
\def\({\left (}
\def\){\right )}
\def\b{\beta}


In this chapter we will study the relation between gravity theories
(string theories) on 
$AdS_3$ and two  dimensional conformal field theories. 
First we are going to describe some generalities which are valid for any
$AdS_3$ quantum gravity theory, and then we will discuss in more detail 
IIB string theory compactified on $AdS_3 \times S^3 \times M^4$ with 
$M^4 = K3$ or $T^4$. 

$AdS_3$ quantum gravity is conjectured to be dual to 
a two dimensional conformal field theory which can be thought of
as living on the boundary of $AdS_3$.
The boundary of $AdS_3$ (in global coordinates)
is a cylinder, so the conformal 
field theory is defined on this cylinder. We choose the cylinder to have 
radius one, which is the usual convention for conformal field theories.
 Of course, all circles are equivalent since
this is a conformal field theory, but we have to rescale energies
accordingly. If the spacetime theory or the conformal field theory contain
 fermions then they have anti-periodic boundary conditions 
 on the circle. The reason is that
the circle is contractible in $AdS_3$, 
 and close to the ``center'' of $AdS_3$ 
a translation by $2 \pi$ on the circle looks like a rotation
by $2 \pi$, and fermions get a minus sign. So, the dual conformal field
theory is in the NS-NS sector. Note that we 
will not sum over sectors as we
do in string theory, since in this case the conformal field theory 
describes string theory on
the given spacetime and all its finite energy excitations,
and we do not  have to 
second-quantize it. 


\section{The Virasoro Algebra}

The  isometry group  of $AdS_3$ is $SL(2,\IR) \times SL(2,\IR)$, or
$SO(2,2)$. 
The conformal group in two dimensions is infinite.  
This seems to be, at first sight, a 
contradiction, since in our previous discussion we identified the
conformal group with the isometry group of $AdS$. 
However, out of the infinite set of generators 
only an $SL(2,\IR)\times SL(2,\IR)$ subgroup leaves the vacuum invariant.
The vacuum corresponds to empty $AdS_3$, and this
 subgroup corresponds to the group of isometries of $AdS_3$.
 The other generators map the vacuum into some
excited states. So, we expect to find that the other generators of the
conformal group map empty $AdS_3$ into $AdS_3$ with (for instance)
a graviton inside.
These other generators are associated to reparametrizations that
leave the asymptotic form of $AdS_3 $ invariant at infinity. 
This problem was analyzed in detail in \cite{Brown:1986nw}
and we will
just sketch the argument here. 
The metric on $AdS_3$ can be
written as
\eqn{metadst}{
ds^2 = R^2 (- \cosh^2  \rho d \tau ^2 + \sinh^2 \rho d\phi^2  + d \rho^2 ).
}
When $\rho$ is large (close to the boundary) this is approximately 
\eqn{metasym}{
ds^2 \sim R^2 \left[
  - e^{2 \rho }  d\tau^+ d \tau^-  + d\rho^2 \right],
}
where $\tau^\pm \equiv \tau \pm \phi$. 
An infinitesimal reparametrization
generated by a 
 general vector field $\xi^\alpha (\tau,\phi,\rho)$ changes the metric
by $ g_{\alpha \beta} \to  g_{\alpha \beta} + \nabla_\alpha \xi_\beta
+ \nabla_\beta \xi_\alpha $. 
If we want to preserve the asymptotic form of the metric \metasym, 
we require that \cite{Brown:1986nw}
\eqn{asymp}{\eqalign{
\xi^+ & = f(\tau^+) +
 {e^{-2\rho} \over 2 } g''(\tau^-) + O(e^{-4 \rho})~, \cr 
\xi^- & = g(\tau^-) +
 {e^{-2\rho} \over 2 } f''(\tau^+) +  O(e^{-4 \rho})~, \cr 
\xi^\rho & = - { f'(\tau^+)  \over 2 } - {  g'(\tau^-) \over 2 }  
+ O(e^{-2\rho})~,
}}
where $f(\tau^+)$ and $g(\tau^-)$ are arbitrary functions.
Expanding the functions $f = \sum L_n e^{n \tau^+ } $, $g = 
\sum {\bar L}_n e^{ n \tau^- } $, we recognize the
Virasoro generators $L_n, \bar L_{n}$. 
For the cases $n =0,\pm1$ one can find some isometries that
reduce to \asymp\ at infinity, are globally defined, and 
leave the metric invariant. These are the $SO(2,2)$ isometries
discussed above. For the other generators it is possible to 
find a globally defined vector field $\xi$, but it does not 
leave the metric invariant. 

It is possible to calculate the classical Poisson brackets 
among these generators, and one finds that 
this classical algebra has a central
charge which is equal to \cite{Brown:1986nw}
\eqn{brown}{
 c = { 3 R \over 2  G_N^{(3)} },
}
where $G_N^{(3)}$ is the three dimensional Newton constant. 
So, this should also be the central charge of the dual conformal
field theory, since \asymp\ implies that
these Virasoro generators are acting on the boundary
as the Virasoro generators of a 1+1 dimensional conformal field 
theory. 

A simple calculation of the central charge term \brown\ 
was given in \cite{Balasubramanian:1999re}.
Under a diffeomorphism of the form \asymp, 
the metric near the boundary changes to 
\eqn{chnmet}{
ds^2 \to  R^2 \left [ - e^{2 \rho} d \tau ^+d\tau^- + d \rho^2  +
 {1 \over 2} (\p_+^3 f) (d\tau^+)^2  +
 {1 \over 2} (\p_-^3 g) (d\tau^-)^2 \right].
}
The metric retains its asymptotic form, but we have kept track of the
subleading correction. This subleading correction changes the 
expectation value of the stress tensor. If we start with 
a zero stress tensor, we get 
\eqn{newexcp}{
\vev{T_{++}} \to { R \over 16 \pi G^{(3)}_N} \p_+^3 f
}
after the transformation.
Under a general conformal transformation, $ \tau^+ \to \tau^+ + f(\tau^+)$,
the stress tensor changes 
as 
\eqn{chstress}{
T_{++} \to T_{++} +  2 \p_+f T_{++} + f \p_+T_{++} + { c \over 24 \pi} 
\p_+^3 f.
}
So, comparing \chstress\ with \newexcp\ we can calculate the central
charge \brown .

It is also possible to show that if we have boundary 
conditions on the metric at infinity that  in the dual 
conformal field theory correspond to considering the theory on a 
curved geometry, then we get the right conformal anomaly 
\cite{Henningson:1998gx} (generalizing the discussion in section 
\ref{anomalies}). 

\section{The BTZ Black Hole}

Three dimensional gravity has no propagating degrees of freedom.
But, if we have a negative cosmological constant, we can have
black hole solutions. 
They are given by \cite{Banados:1992wn,Banados:1993gq}
\eqn{lorbtz}{
ds^2 = - { (r^2 - r^2_+)(r^2 - r^2_-) \over r^2} dt^2 
+{  R^2 r^2 \over  (r^2 - r^2_+)(r^2 - r^2_-)} dr^2 + 
r^2 ( d \phi + {r_+r_- \over r^2} dt)^2,
}
with $\phi \equiv \phi + 2 \pi $. 
We can combine the temperature $T$ and the angular momentum potential
$\Omega$ into 
\eqn{tleftr}{
{ 1 \over T_\pm} \equiv { 1 \over T} \pm { \Omega \over T}, 
}
and their relation to the parameters in \lorbtz\ is 
$ r_\pm = \pi R ( T_+ \pm T_-)$. 
The mass and angular momentum are
\eqn{massang}{
 8 G_N^{(3) } M = R + {(r_+^2 + r_-^2) \over R } ,~~~~~~~~~~J = 
{  r_- r_+   \over 4 G_N^{ (3) }  R   },
}
where we are measuring the mass relative to the 
$AdS_3$ space, which we define to have $M=0$ (the scale of the mass is
set by the radius of the circle in the dual CFT). 
This is not the usual convention, but it is much more
natural in this context since we are measuring energies with 
respect to the NS-NS vacuum.  
 Note that the mass of a black hole is always 
at least
\eqn{massmin}{
M_{min} = {R \over 8 G_N^{(3)}} = { c \over 12 }.
} 
The black hole with this minimum mass (sometimes called the zero mass
black hole) has a singularity at $r= r_+ = r_- =0$. 
All these black holes are locally the same as $AdS_3$ but they differ
by some global identifications \cite{Banados:1992wn,Banados:1993gq},
 i.e. they are quotients of 
$AdS_3$. 
In theories that have supersymmetry it can be checked that the 
zero mass black hole preserves some supersymmetries provided that
we make the fermions periodic as we go around the circle 
\cite{Coussaert:1994jp}, which 
is  something we have the freedom to do once the circle 
is not contractible in the gravity geometry.
These 
supersymmetries commute with the Hamiltonian conjugate to $t$. 
Furthermore, we will see below that if we consider the near horizon
geometry of branes wrapped
on a circle with periodic boundary conditions for the spinors, 
 we naturally obtain the BTZ black hole with 
mass $M_{min}$. This leads us to identify the $M = M_{min}$ 
BTZ black hole with the RR vacuum of the conformal field theory
\cite{Coussaert:1994jp}. 
The energy $M_{min}$ \massmin\  is precisely the energy difference
between the NS-NS vacuum and the RR vacuum. 
Of course, we could still have the $M=M_{min}$ BTZ black hole with
anti-periodic boundary conditions as
an excited state  in the 
NS-NS sector. 

Next, let us calculate the black hole entropy. The Bekenstein-Hawking
entropy formula gives 
\eqn{entgra}{
S = {{\rm Area} \over 4 G_N^{(3)}} =
 { 2 \pi r_+ \over 4 G_N^{(3)} }  =  { \pi^2  c \over 3} 
(T_+ + T_-),
} 
where we used \brown . 
We can also calculate this in the conformal field theory.
All we need is the central charge of the conformal field theory,
which we argued had to be \brown .
 Then, we can use the general formula \cite{Cardy:1986ie} for the 
growth of states in a unitary conformal field theory 
\cite{Strominger:1998eq,Maldacena:1998bw}, which gives
\eqn{entcft}{
S \sim 
{ \pi^2  c \over 3} 
(T_+ + T_-).
}
Thus, we see that the two results agree.
This result if valid for a general conformal field theory 
 as long as we are in the asymptotic high energy
regime (where energies are measured in units of the radius of the
circle), so in particular we need that  $ T \gg 1$. 
When is the result \entgra\ valid? In principle we would say that
it is valid as long as the area of the horizon is much bigger than
the Planck length, $ r_+ \gg G_N^{(3)}$. This gives 
$T \gg 1/c$, which is a much weaker bound on the 
temperature for large $c$.
So, we see that the corresponding
conformal field theory has to be quite special,
since the number of states should grow as determined by the asymptotics
\entcft\ for energies that are much smaller than one would expect 
for a generic  conformal field theory. 

\begin{figure}[htb]
\begin{center}
\epsfxsize=4.5in\leavevmode\epsfbox{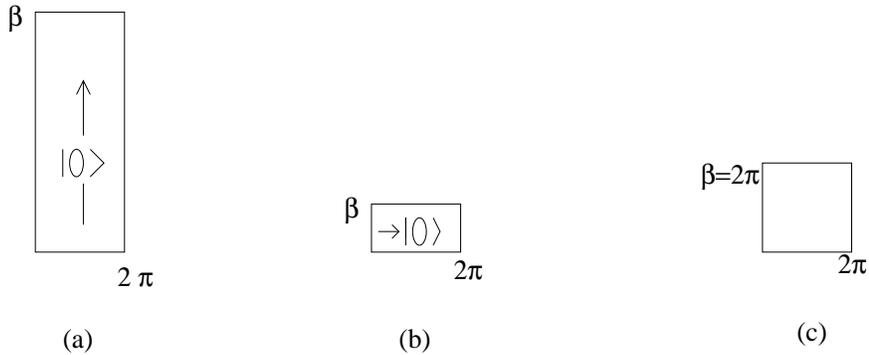}
\end{center}
\caption{
Calculation of the partition function at finite temperature 
through the Euclidean conformal field theory. Since the two directions
are equivalent we can choose the ``time'' direction as we wish.
The partition function is dual under $ \beta \to 4 \pi^2/\beta$.
 (a)~At low temperatures
$ \beta $ is large and only the vacuum propagates in the $\beta $
 direction. (b)~At high temperatures, small $\beta$, only the 
crossed channel vacuum propagates in the $\phi$ direction. 
(c)~When $\beta =2 \pi$ we have a sharp transition according to 
supergravity.
}
\label{rectang}
\end{figure}
 
A related  manifestation of this curious feature of the ``boundary'' 
conformal field theory  is the following. 
We could consider the canonical ensemble by going to Euclidean 
space and  making the Euclidean time coordinate periodic, $ \tau =
\tau + \beta$. We consider the case $\Omega =0$, the general
case is considered in \cite{Maldacena:1998bw}.
 The conformal field theory 
is then defined on a rectangular two-torus, and the free energy will 
be the partition function of the theory on this two-torus. 
Due to the thermal boundary condition in the NS sector, the 
two-torus ends up having   NS-NS boundary conditions on both circles.
In order to calculate the partition function in the dual gravitational 
theory 
 we should find a three-manifold that has the two-torus as its
 boundary (the correspondence tells us to sum over all
 such manifolds).
One possibility is to have the  original $AdS_3$ space but with time
identified, $ \tau = \tau + \beta$. 
The  value of the free energy is then given, to leading order,  
by the ground state energy of $AdS_3$.
This is the expected result 
for large $\beta$, where the torus is very elongated and only the 
vacuum propagates in the $\tau$ channel, see figure \ref{rectang}(a).
For high temperatures, only the vacuum propagates in the crossed 
channel (fig. \ref{rectang}(b)), and this corresponds to the BTZ black hole
in $AdS_3$. Note that the Euclidean BTZ  geometry is the same as $AdS_3$ but 
``on its side'', with $ \tau \leftrightarrow \phi$, so now
the $\tau$ circle is contractible.
 The transition between the two regimes occurs at $ \beta = 2 
\pi $, which corresponds to a square torus (fig. \ref{rectang}(c)).
This is a sharp transition when the gravity approximation is correct, 
i.e. when $ R/G_N^{(3)}\sim c  \gg 1 $.
 This sharp transition will not be present in the partition function of
a generic  conformal field theory, for example it is not present if 
we consider $c$ free bosons. When we discuss in more detail the conformal
field theories that correspond to string theory on $AdS_3$, 
we  will see that they have a  feature that makes it possible 
to explain  this transition. This sharp transition is the
two dimensional version of the large $N$  phase transition
discussed in section \ref{TPhaseT} \cite{Witten:1998zw}
(in this case $c$ plays the role of $N$).

\section{Type IIB String Theory on $AdS_3 \times S^3 \times M^4$}

In this section we study IIB string theory on
 $AdS_3 \times S^3 \times M^4$ \cite{Maldacena:1998bw,deBoer:1998ip}. 
Throughout this section 
 $M^4 = K3$ or $ T^4$. 
In this case we can get some insight on the dual conformal 
field theory  by deriving this duality from D-branes, as we did
for the $AdS_5 \times S^5 $ case. 
We start with type IIB string theory on $M^4$. 
We consider a set of  $Q_1$ D1 branes along a non-compact direction,
and $Q_5$ D5 branes wrapping $M^4$ and sharing the non-compact direction
with the D1 branes. All the branes are coincident in the transverse
non-compact directions. 
The unbroken Lorentz symmetry of this configuration is  $SO(1,1)\times
SO(4)$. $SO(1,1)$ corresponds to boosts along the string, and
$SO(4)$ is the group of rotations in the four non-compact directions
transverse to both branes.
This configuration also  preserves eight supersymmetries, actually
$\cn=(4,4)$
supersymmetry once we decompose them into left and right moving
spinors of $SO(1,1)$\footnote{
If $M^4 = K3$ we need that the sign of the D1 brane charge and the
sign of the D5 brane charge are the same, otherwise we break supersymmetry
(except for the single configuration with charges $(Q_5, Q_1) =
 ( \pm 1 , \mp 1)$).}.
It is possible to find the supergravity solution for this configuration
(see \cite{Youm:1997hw} for a review)
and then  take the near horizon limit as we did in section
\ref{correspondence}
\cite{Maldacena:1997re},
and we get the metric (in string frame)
\eqn{nearhorzn}{
{ d s^2 \over \alpha'} = 
{ U^2 \over g_6 \sqrt{Q_1Q_5} } ( -dt^2 + dx^2_1 ) + 
g_6 \sqrt{Q_1Q_5} { dU^2 \over U^2} + 
g_6 \sqrt{Q_1Q_5} d \Omega^2_3.
}
This is  $AdS_3 \times S^3$ with radius 
$R^2 = R_{AdS}^2 = R_{S^3}^2 = g_6 \sqrt{Q_1 Q_5} l_s^2$, where $g_6$ is
the six dimensional string coupling. The full ten dimensional 
geometry also includes an $M^4$ factor. In this case the volume  of 
the $M^4$  factor in the near-horizon geometry
is proportional to $ Q_1/ Q_5 $, and it is independent
of the volume of the original $M^4$ over which we wrapped the branes.
In the full D1-D5 geometry, which includes the asymptotically flat
region, the volume of $M^4$ varies, and it is equal to the above fixed value
in the near horizon region 
\cite{Andrianopoli:1996ve,Ferrara:1995ih,Ferrara:1996dd,Ferrara:1996um}.

\subsection{The Conformal Field Theory}

The dual conformal field theory is the low energy field theory 
living on the D1-D5 system \cite{Maldacena:1996ky}.
One of the properties of this conformal field theory that we will need
is its central charge, so that we will be able to compare it with 
supergravity. We can calculate this central charge in a way that
is not too
dependent on the precise structure of the conformal field theory. 
The conformal field theory that we are interested in is the IR fixed
point of the field theory living on D1-D5 branes. The field theory
living on D1-D5 branes, before we go to the IR fixed point, is 
some $1+1$ dimensional field theory with $\cn=(4,4)$ supersymmetry. 
This amount of supersymmetry is equivalent to ${\cal N } = 2 $ in
four dimensions, so we can classify the multiplets in a similar fashion.
There is a vector multiplet and a hypermultiplet. In two dimensions
both multiplets have the same propagating degrees of freedom, four
scalars and four fermions, but they have different properties
under the $SU(2)_L \times SU(2)_R$ global R-symmetry. 
Under this group the scalars in the hypermultiplets are in the trivial
representation, while the scalars in the vector multiplet are in the
 ${\bf(2,2)}$. On the fermions these global symmetries act chirally. 
The left moving vector multiplet fermions 
are in the ${\bf (1,2)}$, and
the left moving hypermultiplet fermions are in the ${\bf (2,1)}$.  
The right moving fermions have similar properties with 
$SU(2)_L \leftrightarrow SU(2)_R$. 
The theory can have a Coulomb branch where the scalars in the 
vector multiplets have 
expectation values, and a Higgs branch where the scalars in the
hypermultiplets have expectation
values. 

From the spacetime origin of the supercharges it is clear that the
 $SU(2)_L \times SU(2)_R$ global R-symmetry is the same as the
SO(4) symmetry of spatial rotations in the 4-plane orthogonal to the
D1-D5 system \cite{Breckenridge:1996is,Vafa:1994tf,Vafa:1996bm}.
 The vector multiplets describe motion of the branes
in the transverse directions, this is consistent with their $SO(4)$ 
transformation properties. The vector multiplet ``expectation values''
should be zero if we want the branes to be on top of each other.
We have put quotation marks since expectation values do not  exist in a $1+1$
dimensional field theory. It is possible to show that if $Q_1$ and
$Q_5$ are coprime then,  
by turning on some of the $M^4$ moduli (more precisely some NS B-fields),
one can remove the Coulomb branch altogether, forcing the branes
to be at the same point in the transverse directions 
\cite{Dijkgraaf:1998gf,Seiberg:1999xz}. 

Since the fermions transform chirally under $SU(2)_L$, this theory 
has  a chiral anomaly. The chiral anomaly for 
$SU(2)_L$ is proportional to the number of left moving fermions 
minus the number of right moving fermions that transform under this 
symmetry. The 't Hooft anomaly matching conditions imply that 
this anomaly should be the same at high and low energies
\cite{'tHooft:1980xb}.
At high energies (high compared to the IR fixed point) 
 the anomaly is  $k_a = N_H - N_V$, the difference 
between the number of vector multiplets and hypermultiplets. 
Let us now calculate this, starting with the $T^4$ case. 
On a D1-D5 brane worldvolume there are massless excitations coming from 
(1,1) strings, (5,5) strings and (1,5) (and (5,1)) strings. 
The (1,1) or (5,5) strings come from a vector multiplet of an
 $\cn=(8,8)$ theory,
which gives rise to both a vector multiplet and a hypermultiplet of
 $\cn=(4,4)$ supersymmetry, so they do not contribute to the
anomaly.
 The massless modes of the (1,5) strings 
come only in hypermultiplets, and they contribute to the 
anomaly with $k_a = Q_1 Q_5$. 
For the K3 case the analysis is similar. The D5 branes are 
now wrapped on K3, so the (5,5) strings give rise only to a vector
multiplet. The difference from the $T^4$ case comes from the fact
that in the $T^4$ case the (5,5) hypermultiplet came from Wilson lines on
the torus, and on K3 we do not have one-cycles so we do not have
Wilson lines. On the fivebrane worldvolume there is (when it is
 wrapped on K3) an induced
one-brane charge equal to $Q_1^{ind} = - Q_5 $. The total D1 brane
charge is equal to the sum of the charges carried by 
explicit D1 branes and this negative induced charge,
$Q_1 = Q_1^{ind} + Q_1^{D1}$ \cite{Bershadsky:1996qy}.
Therefore, the number of D1 branes
is really $Q_1^{D1} = Q_1 + Q_5$, and the number of (1,5) strings
is $Q_1^{D1} Q_5 $. So, we conclude that the anomaly is
$k_a = Q_1^{D1} Q_5  - Q_5^2 = Q_1 Q_5 $, which in the end is the same
result as in the $T^4$ case.
Note that in order to calculate this anomaly we only need to know
the massless fields, since all massive fields live in larger
representations which are roughly like a vector multiplet plus a
 hypermultiplet, and
therefore they do not contribute to the anomaly.  

When we are on the Higgs branch all the vectors become massive 
except for the center of mass multiplet, which contains fields
describing the overall motion of all the branes in the four 
transverse directions. This is just a free multiplet, containing
four scalar fields. 
 On the Higgs branch, at the  
 IR fixed point, the $SU(2)_L$ symmetry becomes a current 
algebra with an anomaly $k_{cft}$. 
The total anomaly should be the same, so that $k_a = k_{cft} -1$.
The last term comes from the center of mass $U(1)$ vector multiplet
(which is not included in $k_{cft}$). 
So, we conclude that $k_{cft} = Q_1Q_5 +1 $. 
Since the $U(1)$ vector multiplet is decoupled, we drop it in the rest
of the discussion and we talk only about the conformal field theory
of the hypermultiplets. 
The $\cn=(4,4)$ superconformal
symmetry relates the anomaly in the $SU(2)$ current algebra to the
central charge,  $ c = 6k_{cft} = 6 ( Q_1Q_5 +1) $. 
Using the value for
the $AdS_3$ radius $R = ( g_6^2 Q_1 Q_5)^{1/4} l_s$ and the three dimensional
Newton constant $G_N^{(3)} = g_6^2 l_s^4 / 4 R^3$, we
 can now check that \brown\ is satisfied to leading order for large $k$.
This also ensures, as we saw above, that the black hole entropy comes
out right. 

\begin{figure}[htb]
\begin{center}
\epsfxsize=3in\leavevmode\epsfbox{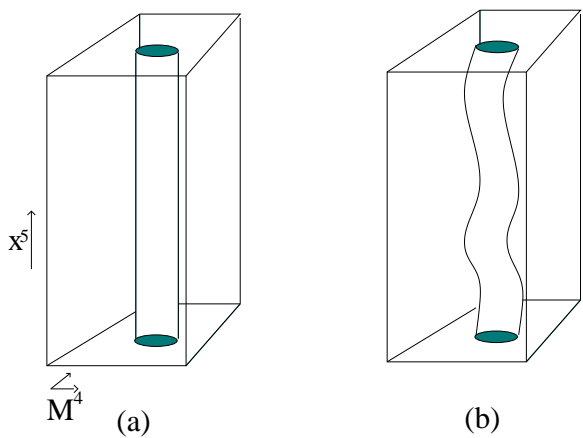}
\end{center}
\caption{
(a) The D1 branes become instantons on the D5 brane gauge theory.
(b)~The instanton moduli can oscillate in time and along $x_5$. 
}
\label{instantons}
\end{figure}

Now we will try to describe this conformal field theory a bit 
more explicitly. 
We start with $ Q_5$ D5 branes, and we view the D1 branes as
instantons of the low-energy SYM theory
on the five-branes \cite{Douglas:1995bn}. 
These instantons live on $M^4$ and
are translationally invariant (actually also $SO(1,1)$ invariant) along
 time and the $x_5$ direction, where $x_5$ is the non compact direction 
along the 
$D5$ branes.  See figure \ref{instantons}(a).
This instanton configuration, with
instanton number $Q_1$, has  moduli, which are the parameters that 
parameterize a continuous family of solutions (classical instanton
configurations). All of these solutions
have the same energy.
Small fluctuations of this configuration (at low energies)
are described by fluctuations of the  instanton moduli. These moduli
can fluctuate in time as well as in the $x_5$ direction. See figure 
\ref{instantons}(b).
So, the low energy dynamics is given by a $1+1$ dimensional sigma model
whose target space is the instanton moduli space. 
Let us be slightly 
more explicit, and choose four coordinates $x^6,...,x^9$
parameterizing $M^4$. The instantons are described in the UV SYM
theory as $SU(Q_5)$ gauge fields
$A_{6,7,8,9}(\xi^a; x^6,...,x^9)$  with field strengths which 
satisfy $F = *_4 F$, where $*_4$ is the epsilon  symbol in $M^4$ and
 $\xi^a $ are the moduli  parameterizing
the family of instantons.
 The dimension of the instanton moduli space
for $Q_1$ instantons in $SU(Q_5)$   is
$4k$, where
\eqn{defofk}{
k \equiv Q_1Q_5 ~~~~{\rm for ~ }T^4, ~~~~~~~~  
k \equiv Q_1Q_5 +1 ~~~~{\rm for ~ }K3.
}
The leading 
behaviour for large $Q$ is the same. In the $T^4$ case we have four
additional moduli coming from the Wilson lines of the $U(1)$ factor
of $U(Q_5)$ \cite{Maldacena:1999bp}. 
It has been argued  in \cite{Vafa:1996bm,Witten:1997yu}
 that the instanton moduli space is 
a deformation of the symmetric product of $k$ copies of $M^4$,
 $Sym(M^4)^k \equiv (M^4)^k/S_k$. The deformation involves blowing up 
the fixed points of the orbifold, as well as modifying the
$B$-fields that live at the orbifold point. We will discuss this 
in more detail later.
The parameter that blows up the singularity can be identified
with one of the supergravity moduli of this solution.
For some particular value of these moduli (which are not to be
confused with the moduli of the instanton configuration)
the CFT will be precisely the
symmetric product, but at that point the gravity approximation will
not be valid, since we will see that the supergravity description
predicts fewer states at low conformal weights than the symmetric
product CFT. When we deform the symmetric product, some of the states 
can get large  corrections and have high energies (i.e. they
correspond to operators 
having high conformal weight). 
Other studies of this D1-D5 system include 
\cite{Costa:1998nq,Hassan:1997ai,Hassan:1997sf}

\subsection{Black Holes Revisited}

We remarked above that the BTZ black hole entropy can be calculated 
just from the  value of the central charge, and therefore the 
gravity result agrees with the conformal field theory result. 
Note that the calculation of the central charge that we did 
above in the CFT is valid for any value of the coupling (i.e. the
moduli), so the field theory 
calculation of the central charge and the entropy
is valid also in the black hole regime (where the gravity
approximation is valid).
This  should be contrasted to the  
 $AdS_5 \times S^5$ case,  where the field theory calculation of the
entropy was only done at weak coupling (in two dimensions the entropy
is determined by the central charge and cannot change as we vary moduli). 
In \cite{Behrndt:1998nt} corrections to the central charge 
in the gravity picture
were analyzed. 

We noticed  above that the gravity description predicted a sharp 
phase transition when the temperature was $ T = 1/(2\pi) $,  and
we remarked that the field theory had to have some special properties
to make this happen.  
We will now explain qualitatively this phase transition.
Our discussion 
will be qualitative because we will work at the orbifold point, and
this is not correct 
if we are in the supergravity regime.  We  will see that
the symmetric product has a feature that makes this sharp 
phase transition possible.

The orbifold theory can be interpreted  
in terms of a gas of strings \cite{Dijkgraaf:1996xw,Dijkgraaf:1997cv}.
These are strings that wind along $x_5$  and move on $M^4$. The total
winding number is $k$. The strings can be singly wound or 
multiply wound. 
In the R-R sector it does not cost any energy to multiply wind the 
strings. 
 If we have NS-NS boundary conditions, which are the appropriate
ones to describe $AdS_3$, 
it will cost some energy to multiply wind the strings. 
The energy cost in the orbifold CFT
is the same as twice the  conformal weight of the 
corresponding twist operator, which is $ h = \bar h = w/4  + O(1/w)$
for a configuration with winding number $w$. 
 If the strings are singly wound and we have a temperature
of order one (or $1/2\pi$), we will not have
many oscillation modes excited on these strings, and the 
entropy will be small. Note that the fact that we have many singly
wound strings
does not help, since we are supposed to symmetrize over all strings,
so most of the strings will be in similar states and they will not
contribute much to the entropy. 
So, the free energy of such a state is basically $F \sim 0  $. 
On the other hand, if we multiply wind all the strings, we raise the 
energy of the system but we also increase the entropy 
\cite{Maldacena:1996ds}, since now the energy gap of the system
will be much lower (the multiply wound strings behave effectively like
a field theory on a circle with a radius which is $w$ times bigger).  
If we multiply 
wind $w$ strings, with $w \gg 1 $, 
 we get an energy $E \sim w/2 + 2 \pi^2 w T^2 $, where
the last term comes from thermal excitations along the string.
The entropy is also  larger, $ S = 4 \pi^2 w T $. 
So, the free energy is $F = E-TS=  w/2 -  2 \pi^2 w T^2 $. Comparing this
to the free energy of the state with all strings singly wound,
we see that the latter  wins when
$T<1/(2\pi)$, and the multiply wound state wins when $T>1/(2 \pi) $.
 This explains the presence of the sharp phase
transition at $T = 1/(2 \pi)$ when we are at  the orbifold point. 

Note that the mass of the black hole at the transition point
is $M = M_{min} + k/2 $, which is (for large $k$) much bigger 
than the minimum mass for a BTZ black hole, like the situation in
other $AdS_{d>3}$. 
We could have black holes which are smaller than this, but they 
cannot be in thermal equilibrium with an external bath.
Of course they could be in equilibrium inside $AdS_3$ if we
do not couple $AdS_3$ to an external bath to keep the temperature
finite. In this case we are considering the microcanonical ensemble,
and there are more black hole solutions that we could be 
considering \cite{Banks:1998dd,Li:1999jy,Martinec:1999ja}. 

\begin{figure}[htb]
\begin{center}
\epsfxsize=4.5in\leavevmode\epsfbox{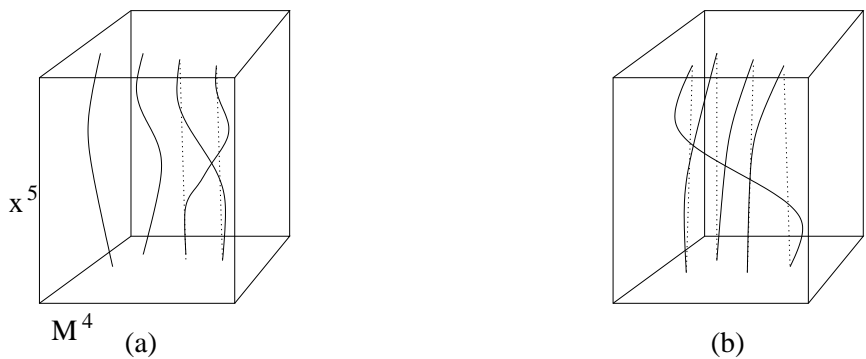}
\end{center}
\caption{Some configurations with winding number four.
(a)~Two singly wound strings and one doubly wound string.
(b)~A maximally multiply wound configuration.
}
\label{wind}
\end{figure}

If we were considering the conformal field theory on a circle with
RR boundary conditions, the corresponding supergravity background 
would be the $M =M_{min}$ BTZ black hole. 
 This follows from the 
fact that we should have preserved supersymmetries that commute 
with the Hamiltonian (in $AdS_3$ the preserved supersymmetries
do not commute with the Hamiltonian generating evolution in
global time). In order to have these supersymmetries we need to
have RR boundary conditions on the circle. Notice that the RR vacuum
is not an excited state on the NS-NS vacuum, it is just in a different
sector of the conformal field theory, even though the $M= M_{min}$ 
BTZ black hole appears in both sectors. 

In the case with RR boundary conditions a black hole forms as soon 
as we raise the temperature (beyond $T \sim 1/k$). 
This seems at first sight
paradoxical, since the temperature could be much smaller than 
one, which would be the natural energy gap for a generic
conformal field theory on a circle. The reason that the 
energy gap is very small for this conformal field theory 
is due to the presence
of ``long'', multiply wound strings. 
In the RR sector all multiply wound strings
have the same energy. But, as we saw before, multiply wound strings
lead to higher entropy states so they are preferred. In fact, one
can estimate the energy gap of the system by saying that 
it will be of the order of the minimum energy excitation that can 
exist on a string multiply wound $k$-times, which is of the order of
$1/k$. This estimate of the energy gap agrees with 
a semiclassical estimate as follows. We can trust the thermodynamic
approximation for black holes as long as the specific heat is 
large enough \cite{Preskill:1991tb}.
 For any system we need a large specific heat, $C_e \equiv
{\partial E \over \partial T}$,  in 
order to trust the thermodynamic approximation.
In this case $E \sim k T^2$, so the condition $C_e \gg1$ boils down to
$E\gg 1/k$ . So, this estimate of the energy gap agrees with the 
conformal field theory estimate. 
Note that in the RR supergravity vacuum (the $M=M_{min}$ black hole)
we could seemingly have arbitrarily low energy excitations as 
waves propagating on this space. The boundary condition on these waves
at the singularity should be such that one gets the above gap, but
in the gravity approximation $k =\infty$ and this gap is not seen. 
Note also that the $M= M_{min}$ black hole does not correspond to
a single state (as opposed to the $AdS_3$ vacuum), but to a large
number of states, of the order of $e^{ 2 \pi \sqrt{2 k} }$ for 
$T^4$ case and  $e^{ 2 \pi \sqrt{4 k} }$ for $K3$\footnote{ 
 An easy way
to calculate this number of BPS states 
is to consider this configuration as a system
of D1-D5 branes on $S^1\times M^4 $ and then do a U-duality 
transformation, transforming this into 
a system of fundamental string momentum and winding.}.

There are other black holes that preserve some supersymmetries,
which are extremal BTZ black holes with $M-M_{min} = J$
\cite{Coussaert:1994jp}. $J$ is the angular momentum in $AdS_3$,
identified with the
momentum along the $S^1$ in the CFT. 
Of course, these black holes will preserve supersymmetry only if the boundary
conditions on $S^1$ are periodic, i.e. only if we are considering
the RR sector of the theory. In the RR sector it becomes more 
natural to measure energies so that the RR vacuum has zero energy.
The extremal black holes correspond to states in the CFT in the 
RR sector with no left moving energy, $\bar L_0=0$, and some right
moving energy, $ L_0 =J>0$. The entropy of these states is
\eqn{entropycft}{
S= 2 \pi \sqrt{ k  J}.
}
This is the entropy as long as $k L_0$ is large, even for
$L_0 = 1$. The reason for this is again the presence of 
multiply wound strings, that ensure that the asymptotic 
formula for the number of states in a conformal field theory
is reached at very low values of $L_0$. In this argument it
is important that we are in the RR sector, and since we 
are counting BPS states we can deform the theory until we 
are at the symmetric product point, and then the argument
we gave in terms of multiply wound strings is  rigorous
\cite{Strominger:1996sh,Maldacena:1999bp}. 

It is possible to consider also black holes which carry angular momentum
on $S^3$. They are characterized by the eigenvalues $J_L, ~J_R$, of 
$J_L^3$ and $J_R^3$ of $SU(2)_L\times SU(2)_R $. 
These rotating black holes can be found by taking the near horizon
limit of rotating black strings  in six dimensions 
\cite{Cvetic:1998xh,Cvetic:1996xz}.
Their metric is locally $AdS_3\times S^3$ but with some
discrete identifications \cite{Cvetic:1998ja}.
Cosmic censorship implies that
their mass has a lower bound 
\eqn{masslb}{
E \equiv M-M_{min} \geq J_L^2/k + J_R^2/k.
}
We can also calculate the entropy for a general configuration 
carrying angular momenta $J_{L,R}$
 on $S^3$, linear momentum $J$ on $S^1$, and energy $E = M-M_{min}$ :
\eqn{entropyrot}{
S = 2 \pi \sqrt{k( E+J)/2 - 
J_L^2} +  2 \pi \sqrt{k( E-J)/2 - 
J_R^2}.
}
We can understand this formula in the following way 
\cite{Breckenridge:1996sn,Breckenridge:1996is}. 
If we bosonize the $U(1)$ currents, $J_{L} \sim {k \over 2} 
\partial \phi $,
and similarly for $J_R$, we can construct the operator
$e^{i J_L \phi}$ with conformal weight $J_L^2/k$. This explains
why the minimum mass is \masslb . This also explains \entropyrot,
since  only a portion of the energy equal to 
 $ L_0-J^2_L/k  = ( E+J)/2-J^2_L/k $ can be distributed freely among
the oscillators\footnote{ Other
black holes were studied in \cite{Kaloper:1998vw}.}.

\subsection{Matching of Chiral-Chiral Primaries }

The CFT we are discussing here, and also its string theory dual, 
have moduli (parameters of the field theory). At some point in the
moduli space
the symmetric product description is valid, and at that point 
the gravity description is strongly coupled and cannot be trusted. 
As we move away from that point we can get to regions in moduli
space where we can trust the gravity description. The energies 
of most states will change when we change the moduli. 
There are, however, states 
that are protected, whose energies are not changed. 
These are chiral primary states \cite{Lerche:1989uy}. 
The superconformal algebra contains  terms of the form\footnote{
Our normalization for $J^3_0$ follows the standard $SU(2)$ practice
and differs by a factor of two from the U(1) current in 
\cite{Lerche:1989uy,Maldacena:1998bw,deBoer:1998ip,deBoer:1998us}.}
\eqn{suprc}{\eqalign{
 \{ Q^{++}_r ,Q^{--}_s \} &= 2 L_{r+s} + 2(r-s) J^3_{r+s} +
{ c\over 3 } \delta_{r+s} (r^2 -{1 \over 4} ), \cr
\{ Q^{+-}_r ,Q^{-+}_s \} &= 2 L_{r+s} +2 (r-s) J^3_{r+s} +
{ c\over 3 } \delta_{r+s} (r^2 -{1 \over 4} ),
 }}
where $Q^{\pm\pm}_r = (Q^{\mp\mp}_{-r})^\dagger $, and $r,s \in
\IZ + {1 \over 2} $. The generators that belong to the global
 supergroup (which leaves the vacuum and $AdS_3$ invariant)
have $r, s = \pm 1/2$.  The first superscript indicates 
the eigenvalues under the global $J_0^3$ generator of $SU(2)$, and
the second superscript corresponds to a global $SU(2)$
 exterior automorphism
of the algebra which is not associated to a symmetry in the theory. 
If we take a state $|h\rangle $
 which has $L_0 = J^3_0$, then we see from \suprc\ 
that $ Q^{+\pm}_{-1/2} |h\rangle $ 
has zero norm, so in a unitary field theory
it should be zero. Thus, these states are annihilated by $Q^{+ \pm}_{-1/2}$.
Moreover, if a state is annihilated by $Q^{+ \pm}_{-1/2}$ then 
$L_0 =  J^3_0 $. These states are called right chiral primaries, and
if  $\bar L_0 = {\bar J}_0^3$ it is a left chiral primary. 
  The possible values of $J_0^3 $ for 
chiral primaries are  bounded by  $J_0^3 \leq c/6 =k $. 
This can be seen by 
 computing  the norm of $ Q_{-3/2}^{+\pm} |h\rangle$.
 Note that $k$ is the
level of the $SU(2)$ current algebra. The values of $J_0^3$ for generic 
states are not bounded. The spins of $SU(2)$ current algebra
primary fields are bounded by $J_0^3 \leq k/2 $, which is {\it not}
 the same as the bound on chiral primaries. 

\begin{figure}[htb]
\begin{center}
\epsfxsize=2in\leavevmode\epsfbox{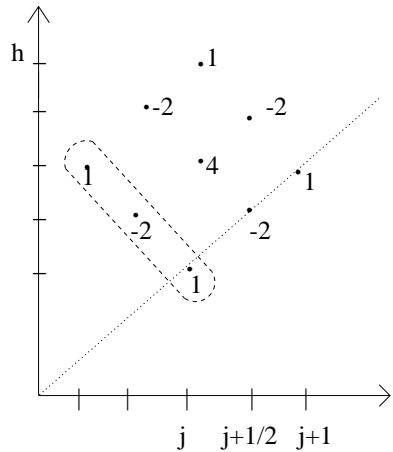}
\end{center}
\caption{ Structure of $SU(1,1|2)$ multiplets. 
We show the spectrum of possible $j$'s and conformal 
weights. We show only the $SL(2,\IR)$ primaries that appear
in each multiplet and their degeneracies. The minus sign
denotes opposite statistics. The full square is a long 
multiplet. The encircled states form a short multiplet. 
Four short multiplets can combine into a long multiplet.
}
\label{multiplet}
\end{figure}

Let us now discuss the structure of the supermultiplets under the
$SU(1,1|2)$ subgroup of the ${\cal N} = 4 $ algebra 
\cite{Gunaydin:1986fe}. This is the subgroup
generated by the supercharges with $r,s = \pm 1/2 $ in \suprc , plus
the global $SU(2)$ generators $J_0^a$ and the $SL(2,\IR)$ subgroup of the
Virasoro algebra. The structure of these multiplets is the following. 
By acting with $Q_{1/2}^{\pm\pm}$ on a state we lower 
its energy, which is the $L_0$ eigenvalue. Energies are all positive
 in a unitary conformal field theory, since  $L_0$ eigenvalues
are related to scaling dimensions of fields which should be
positive. So, we conclude that at some point $Q_{1/2}^{\pm\pm}$ will
annihilate the state. Such a state is also annihilated by $L_1$ \suprc .
 We call such a state a primary, or highest
weight, state. Then, we can generate all other states by acting
with $Q_{-1/2}^{\pm\pm}$. See figure \ref{multiplet}.
This will give in general a set
of $1 +4 + 6 + 4 + 1$ states, where we organized the states according
to their level. On each of these states we can then act with
arbitrary powers of $L_{-1}$. 
However, we could also have a short representation where some of the
$Q_{-1/2}$ operators
annihilate the state. This will happen when $L_0 =\pm J_0^3$,
i.e. only when we have a chiral primary (or an antichiral primary). 
Since by $SU(2)$ symmetry each chiral primary comes with an antichiral
primary, we concentrate on chiral primaries.  
These short multiplets are of the form
\begin{equation}
\begin{array}{rcc}
{\rm states} & J_0^3  &  L_0 \\
|0\rangle & j  & j \\
Q^{-\pm}_{-1/2} |0\rangle & j-1/2 &  j+1/2 \\
Q^{-+}_{-1/2} Q^{--}_{-1/2} |0\rangle & j-1 & j+1. 
\end{array}
\end{equation}
The multiplet includes four states (which are $SL(2,\IR)$ primaries), 
except in the case that $j=1/2$ when
the last state is missing. 
We get a similar structure if we consider the right-moving
part of the supergroup. 

We will first consider states that are left and right moving chiral
primaries, with $L_0 = J^3_0 $ and $\bar L_0 = \bar J_0^3 $.
From now on we drop the indices on $J^3_0, \bar J_0^3$,
and denote the chiral primaries by $(j,\bar j)$. 
By acting with $Q^{-\pm}_{-1/2}$ and $\bar Q^{-\pm}_{-1/2}$
we generate the whole supermultiplet.
We will calculate the spectrum of chiral-chiral primaries both 
in string theory (in the gravity approximation)
and in the conformal field theory at the orbifold point. Since these states
lie in short representations we might expect that
they remain in short representations also
after we deform the theory away from the orbifold point. 
Actually this argument is not enough, since in principle
short multiplets could combine and become long multiplets.
In the K3 case we can give a better argument. We will see that
all chiral primaries  that appear are bosonic in nature, while we
see from figure \ref{multiplet} that we need some bosonic and 
some fermionic chiral primaries to make a long multiplet.
Therefore, all chiral primaries must remain for any value of the
moduli.
  
Let us start with the conformal field theory. Since these states
are protected by supersymmetry we can go to the orbifold point
 $Sym(M^4)^k$. The chiral primaries in this case can be understood 
as follows. In a theory with $\cn=(4,4)$ 
supersymmetry we can do calculations
in the RR sector and then translate them into results about the
NS-NS sector. This process is called ``spectral flow'', and it 
amounts to an automorphism of the ${\cal N} = 4$ algebra. 
Under spectral flow, the chiral primaries of the NS-NS sector (that 
we are interested in) are in one
to one correspondence with the ground states of the RR sector.
It is easier to compute the properties of the RR ground states of
the theory. Orbifold conformal field theories, like $Sym(M^4)^k$, can
be thought of as describing a gas of strings winding on a circle,
the circle where the CFT is defined, with total winding
number $k$ and moving on $M^4$. 
The ground state energies
 of a singly wound string and a multiply wound
string are the same if we are in the RR sector. 
Then, we can calculate a partition function over the RR ground states.
It is more convenient to relax the constraint on the  total winding 
number by introducing a chemical potential for the winding number,
and then we can recover the result with fixed winding number by extracting 
 the appropriate term in the partition function as in 
\cite{Dijkgraaf:1996xw}. 
Since our conformal field theory has fixed $k$ we will be implicitly 
assuming that we are extracting the appropriate term  from the 
partition function. 
The RR ground states for the strings moving on $M^4$ are the same
as the ground states of a quantum mechanical supersymmetric sigma 
model on $M^4$. It was shown by Witten \cite{Witten:1982df}
 that these are in 
one-to-one correspondence with the harmonic forms on $M^4$.
Let us denote by $h_{rs}$ the number of harmonic forms of holomorphic
degree $r$ and antiholomorphic degree $s$. States with
degree $r+s$ odd are fermionic, and states with $r+s$ even are bosonic.
In the case of $K3$ $h_{00} = h_{22} = h_{20}=h_{02}=1$ and 
$h_{11} = 20$. In the case of $T^4$ $h_{00} = h_{22} = h_{20}=h_{02}=1$,
$h_{01}= h_{10} = h_{12} = h_{21} = 2 $, and $h_{11} = 4$. 
A form with degrees $(r,s)$  gives rise to a state with 
angular momenta $ (j,\tilde j) = ( (r-1)/2,(s-1)/2)$.
The partition function in the RR sector  becomes \cite{Dijkgraaf:1996xw}
\eqn{partram}{
\sum_{k\geq 0}  p^kTr_{Sym(M^4)^k} 
[(-1)^{2J + 2 \bar J}  y^{J} {\bar y}^{ \bar J}] = 
{ 1 \over \prod_{n\geq 1}\prod_{r,s} ( 1 -p^n y^{(r-1)/2} 
{\bar y}^{(s-1)/2})^{
(-1)^{r+s} h_{rs} }},
}
where the trace is over the ground states of the RR sector.
Spectral flow boils down to the replacement 
$p\to p y^{1/2} {\bar y}^{1/2}$. Thus, we get the NS-NS partition function,
giving a prediction for the chiral primaries,
\eqn{chiralp}{
\sum_{k}  p^kTr_{Sym(M^4)^k} 
[(-1)^{ 2 J + 2 \bar J }  y^J {\bar y}^{\bar J}] =
{ 1 \over \prod_{n\geq 0}\prod_{r,s} 
( 1 -p^{n+1} y^{(n+r)/2} {\bar y}^{(n+s)/2})^{(-1)^{r+s} h_{rs} }},
}
where here the trace is over the chiral-chiral primaries in the NS-NS
sector.

Now, we should compare this with supergravity. 
In supergravity we start by calculating the spectrum 
of single particle chiral-chiral primaries. 
We then calculate the full spectrum by considering multiparticle states.
Each single particle state  contributes with a factor 
$(1- y^{j} {\bar y}^{\bar j})^{-d(j,\bar j)} $ to the partition function,
were $d(j,\bar j)$ is the total number of single particle states with 
these spins.
The supergravity spectrum was calculated in 
\cite{Maldacena:1998bw,Deger:1998nm,Larsen:1998xm,deBoer:1998ip}.
The number of single particle states is given by 
\eqn{grav}{
\sum_{j,\bar j }   d(j,\bar j ) y^j { \bar y}^{\bar j}  =
 \sum_{n,r,s \geq 0} h_{rs}  y^{ n + r \over 2}
{\bar y}^{n+ s \over 2 }  -1.
}
We  have excluded  the identity, which is not represented by any state
in supergravity.
So, the gravity partition function is given by 
\eqn{gravitypa}{
Tr_{Sugra}[(-1)^{ 2 J + 2 \bar J }  y^J {\bar y}^{\bar J}]_{\rm 
c-c~primaries} =
{ 1 \over \prod_{n\geq 0}\prod'_{r,s} 
( 1 - y^{(n+r)/2} {\bar y}^{(n+s)/2})^{(-1)^{r+s} h_{rs} }},
}
where $\prod'$ means that we are not including the term with $n=r=s=0$. 

Let us discuss some the particles appearing in \grav\ and \gravitypa\
 more explicitly.
Some of them are special because
 they carry only left moving quantum numbers or only right moving
quantum numbers. For example, we have the $(0,1)$ and $(1,0)$ states
that are related to the $SU(2)_L$ and $SU(2)_R $ gauge fields on $AdS_3$.
These $SU(2)$ symmetries come from the $SO(4)$ isometries of the 
3-sphere. 
These gauge fields have a Chern-Simons action 
\cite{Achucarro:1986vz,Achucarro:1989gm} and they 
give rise to $SU(2)$ current algebras on the boundary 
\cite{Witten:1989hf,Elitzur:1989nr}. 
The chiral primary in the current algebra is the operator
$J_{-1}^+$, which has the quantum numbers mentioned above. 
When we apply $Q^{--}_{-1/2} Q^{-+}_{-1/2}$ to this state we 
get the left moving stress tensor. Again, this should correspond to part 
of the physical modes of gravity on $AdS_3$. Pure gravity in three
dimensions is a theory with no local degrees of freedom. 
In fact, it is equivalent to an $SL(2,\IR)\times SL(2,\IR)$ Chern-Simons
theory 
\cite{Achucarro:1986vz,Howe:1996zm,%
Witten:1988hc,Witten:1989sx}.
 This gives rise to some physical degrees of freedom
living at the boundary. It was argued that we get a Liouville
theory at the boundary 
\cite{Coussaert:1995zp,Banados:1998pi,Banados:1998ta,%
Banados:1996tn,Martinec:1998st}, which includes a stress tensor
operator.
In the $T^4$ case we also have some other special particles
which correspond to fermion zero modes $(1/2,0)$ and $(0,1/2)$. 
These fermion zero modes are the supersymmetric partners of the $U(1)$ 
currents associated to isometries of $T^4$.
The six dimensional  theory corresponding to type IIB string theory on
$T^4$ has 16 vector fields transforming
in the spinor  of $SO(5,5)$. From the symmetric product we get only
8 currents ($4_L + 4_R$). The other eight are presumably related to
an extra copy of $T^4$ appearing in the CFT due to the Wilson
lines of the $U(1)$ in $U(Q_5) $ \cite{Maldacena:1999bp,Larsen:1999uk}.  

Besides these purely left-moving or purely right-moving modes,
which are not so easy to see in supergravity,
all other states arise as local bulk  excitations
of  supergravity fields on $AdS_3$ and are clearly present. 
Higher values of $j$ typically correspond to higher Kaluza-Klein 
modes of lower $j$ fields. More precisely, we have 
$n$ $(1/2,1/2) $ states where $n = h_{11} + 1$ 
\cite{Maldacena:1998bw,Deger:1998nm,deBoer:1998ip}. 
By applying $Q$'s, each of these states
gives rise to four $SU(2)$-neutral scalar fields, which
 have conformal weights
 $h = \bar h = 1$. Therefore, they correspond to 
  massless fields in spacetime
by (\ref{dimenmass}).
These are the   $ 4 n$ 
moduli of the supergravity compactification, which are identified with
the moduli of
the conformal field theory. In the conformal field theory 
$4 h_{11}$ of them correspond to deformations of each copy of $M^4$
in the symmetric product, while the extra four are associated 
to a blowup mode, the blowup mode of the $\IZ_2$ singularity that
arises when we exchange two copies of $M^4$. 
Next, we have $n+1$ fields  with 
quantum numbers $(1,1)$, $n $ of these are higher order
Kaluza Klein modes of the $n$ fields we had before, and the new one
corresponds to deformations of the $S^3$. 
Each of these states gives rise to 
$SU(2)$-neutral fields with positive mass, since we have to apply 
$Q$'s twice and we  get  $h =\bar h =2$. 
These are the $n$ fixed scalars of the supergravity background
plus one more field related to changing the size of the $S^3$. 
The fields with  $j,\bar j$'s  above these values are just higher
Kaluza Klein modes of the fields we have already mentioned 
explicitly. See \cite{Maldacena:1998bw,Deger:1998nm,deBoer:1998ip}
for a more systematic treatment and 
derivation of these results. 

Now, we want to compare the supergravity result with the gauge theory
results. 
In \chiralp\ there is an ``exclusion principle'' since the total
power of $p$ has to be $p^k$, thus limiting the total number of 
particles. In supergravity \gravitypa\ we do not have any indication 
of this exclusion principle. Even if we did not know 
about the conformal field theory, from the fact that
there is an ${\cal N } = 4$ superconformal spacetime symmetry
we  get  a bound on the angular momentum of the chiral primaries
$j \leq k $.
However, this bound is less
restrictive than implied by  \chiralp . There are 
multi-particle states with $j< k$ that are excluded from 
\chiralp .  The bounds from \chiralp\  appear for very large angular 
momenta and, therefore, very large energies, where we would not
necessarily trust the gravity approximation.
In fact, the gravity result and the conformal field theory result
match precisely, as long as the conformal weight or spin of 
the chiral primaries  
is $j,{\bar j} \leq k/2 $. 
One can show that the gravity description exactly matches the 
$k \to \infty$ limit of \chiralp\ \cite{deBoer:1998ip}. 
 This limit is extracted from \chiralp\  by noticing that
there is a factor of $(1-p)$ in the denominator, which  is related to the
identity operator. So, we can extract the $k  \to \infty$ limit 
by multiplying \chiralp\ by $(1-p)$ and setting $p\to 1$. 
In principle, we could get precise agreement between the conformal field
theory calculation and the supergravity calculation if we incorporate the
exclusion principle by assigning a ``degree'' to each supergravity
field, as explained in \cite{deBoer:1998us}, and then considering only 
multiparticle states with degree smaller than $k$. 
One can further wonder whether there is something special that
happens at $j=k/2$, when the exclusion principle starts making a
 difference.  Since we are considering states with 
high conformal weight and angular momentum it is natural to wonder 
whether there are any black hole states that could appear. 
There are black holes which carry angular momentum on $S^3$. 
These black holes are characterized by the two angular momenta
$J_L$, $J_R$, of $SU(2)_L\times SU(2)_R$. 
The minimum black hole mass for given angular momenta was given 
in \masslb,
$M_{min}(J_L,J_R) = k/2 + J^2_L/k + J^2_R/k$,
where we used $c  =6k$ and \brown .
We see that these masses are always bigger than the mass 
of the chiral primary states with angular momenta
$(J_L,J_R)$, except when $J_L = J_R = k/2 $. So we see that 
something special is happening at $j=k/2$, since at this
point a black hole appears as a  chiral primary state. 
Connections between this exclusion principle and quantum groups
and non-commutative geometry were studied in 
\cite{Jevicki:1999rr,Chang:1999jm}.

\subsection{Calculation of the Elliptic Genus in Supergravity}

We could now consider states which are left moving chiral 
primaries and anything on the right moving side. 
These states are also in small representations, and 
one might be tempted to compute the spectrum of chiral 
primaries at the orbifold point and then try to match it 
to supergravity. However,
this is not the correct thing to do, and in fact 
the spectrum does not match \cite{Vafa:1998nt}. 
It is not correct because some chiral primary
states could pair up and become very massive 
 non-chiral primaries.
In the case of chiral-anything states, a useful tool to 
count the number of states, which gives a result that is independent
of the deformations of the theory, is the ``elliptic genus'',
which is the partition function 
\eqn{elliptic}{
Z_k = Tr_{RR} [ (-1)^{ 2j + 2 {\bar j} }q^{L_0} {\bar q}^{{\bar L}_0}  
y^{j} ].
}
This receives contributions only from the left  moving ground states,
$\bar L_0 = 0$. These states map into (chiral, anything) under
spectral flow, i.e. states that are chiral primaries on the left moving 
side but are unrestricted on the right moving side.

The number of states contributing to the elliptic genus
goes like $e^{ 2 \pi \sqrt{n k} }$  for  large powers $q^n$.
This raised some doubts that \elliptic\ would agree with supergravity.
The elliptic genus diverges when we take the limit $k\to \infty $.
The origin of this divergence is the contribution of the $(2,0)$ form,
which is a chiral primary on the left but it carries zero 
conformal weight on the right. So, we get a contribution of order $k$ 
from the fact that this state could be occupied $k$ times without 
changing the powers of $q$ or $y$. The function that has a smooth
limit in the $k\to \infty $ limit is then $Z_k^{NS}/k $. 
In the K3 case this function is 
\eqn{ellsu}{
\lim_{k\to \infty} { Z_k^{NS} \over k} =
{  \prod_{m\geq 1} (1-q^{m/2} y^{1/2})^2(1-q^{m/2}y^{-1/2} )^2 
(1-q^{m/2})^{20}
 \over
\prod_{m\geq 1} (1 - q^{m/2} y^{m/2})^{24} (1 -q^{m/2} y^{-m/2} )^{24}  }.
}
We can now compare this expression to the supergravity result. In the
supergravity result we explicitly exclude the contribution of the
$(2,0)$ form, since it is directly related to the factor of $k$ that
we extracted, but we keep the contribution of the $(0,2)$ form and the 
rest of the fields. The supergravity result then agrees precisely with 
\ellsu\ \cite{deBoer:1998us}.
 Both in the supergravity calculation and in the conformal field
theory calculation at the orbifold point there are many fields of the 
form (chiral,anything), but most of them cancel out to give \ellsu.
For example, we can see that the only supergravity single particle
states  that 
contribute for large powers of  $y^{>1/2}$
 are the (chiral, chiral) and (chiral, antichiral) states.
One can further incorporate the exclusion principle in supergravity
by assigning degrees to the various fields, and then one finds
that the elliptic genus agrees up to powers of $q^h$ with $h \leq (k+1)/4$
 \cite{deBoer:1998us}. 
Here again this is the point where a black hole starts contributing
to the elliptic genus. It is an extremal
 rotating black hole with angular 
momentum $J_L = k/2$ and $J_R=0$, which has $L_0 = k/4$ and 
$\bar L_0 = k/2$.

\section{Other $AdS_3$ Compactifications}

We  start by discussing the compactifications discussed in
the last section more broadly, and then we will discuss other
$AdS_3$ compactifications.
In the previous section we started out with type IIB string theory
compactified on 
$M^4$ to six dimensions. 
 The theory has many charges carried by string like objects,
which come from branes wrapping on various cycles of $M^4$.
These  charges transform as vectors 
under the duality group of the theory $SO(5,n)$, where 
$n = 21,5 $ for the $K3$ and $T^4$ cases respectively. 
These $5+n$ strings correspond to the fundamental and D strings, 
the NS and D fivebranes wrapped on $M^4$,
 and to D3 branes wrapped on the 
 $n+1$  two-cycles of $M^4$. 
A general charge configuration is given by a vector $q^I$ transforming
under $SO(5,n)$. The radius of curvature  of the gravity 
solution is proportional to $q^2$, $R^4 \sim q^2$,
 where we use the $SO(5,n)$ metric. 
In the K3 case $q^2 >0$ for  supersymmetric configurations. 
The six dimensional space-time theory
has $5n$ massless scalar fields, which parameterize the coset manifold
$SO(5,n)/SO(5)\times SO(n)$ \cite{Romans:1986er}. 
When we choose a particular charge vector, with $q^2 >0$, we break the duality
group to $SO(4,n)$, and out of the original $5n$ massless scalars
$n$ becomes massive and have values determined by the charges (and the
other scalars) \cite{Andrianopoli:1998qg}. 
The remaining 
$4n$ scalars are massless and represent moduli of the supergravity 
compactification
and, therefore, moduli of the dual conformal
field theory. Note that the conformal field theory involves 
the instanton moduli space, but here the word ``moduli'' refers to the
parameters of the CFT, such as the shape of $T^4$, etc. 

If we start moving in this moduli space we sometimes find that the
gravity solution 
 is best described by doing duality transformations 
\cite{Dijkgraaf:1998gf,Seiberg:1999xz}. One
interesting region in moduli space is when the system is best described
in terms of a 
 system of NS fivebranes and 
fundamental strings. This is 
the S-dual version of the D1-D5 system that we
were considering above.
In this  NS background the
radius of the $S^3 $ and of $AdS_3$ is $R^2 = Q_5 \alpha'$, and it is 
independent of $Q_1$. Actually, $Q_1$ only enters through the 
six dimensional string coupling, which in this case is a fixed scalar
$g_6^2 = Q_5/Q_1 $. The volume of $M^4$ is a free scalar in this case. 
The advantage of this background is that one can 
solve string theory on it to all orders in $\alpha'$,
 since it is a WZW model, actually 
an $SL(2,\IR)\times SU(2)$ WZW model.
String propagation in  $SL(2,\IR)$ WZW models were studied in 
\cite{Balog:1989jb,Petropoulos:1990fc,Mohammedi:1990dp,%
Bars:1991rb,Hwang:1991aq,%
Henningson:1991ua,Henningson:1991jc,Hwang:1992an,Bars:1996mf,%
Bars:1995cn,Satoh:1997xe,Evans:1998wq,Evans:1998qu,%
Giveon:1998ns,deBoer:1998pp,Kutasov:1999xu,Pesando:1999ex,%
Andreev:1999nt,Andreev:1999uh,Ito:1998vd,Hosomichi:1999be}.
Thus,
in this case we can also consider states corresponding to massive
string modes, etc. We can also define the spacetime Virasoro generators
in the full string theory,
and check that they act on  string states as they should 
\cite{Giveon:1998ns,deBoer:1998pp,Kutasov:1999xu}\footnote{
Configurations with NS fluxes that lead to $AdS_{2d+1}$ spaces
where studied in \cite{deBoer:1999ie}. It has also been suggested 
\cite{O'Loughlin:1998qx} that  (2,1) strings can describe $AdS_3$ 
spaces.}.
In the string theory description the Virasoro symmetry appears
directly in the formalism as a spacetime symmetry. One can also 
study D-branes in these $AdS_3$ backgrounds \cite{Stanciu:1999nx}.
Conditions for spacetime supersymmetry for string theory on $SL(2,\IR)$
WZW backgrounds were studied in \cite{Giveon:1999jg,Berenstein:1999gj}. 
In the D1-D5 configuration it is much harder to solve string theory,
since RR backgrounds are involved. Classical actions for 
strings on these backgrounds were written in 
\cite{Pesando:1999wm,Rahmfeld:1998zn,Park:1999un}. 
 However,  a formulation of 
string theory on these backgrounds was proposed in 
\cite{Berkovits:1999im} (see also 
\cite{Bershadsy:1999hk,Berenstein:1999gj,Yu:1998qw}).
For some values of the moduli the CFT is singular. 
What this means is that we will have a continuum of states in the 
cylinder picture. In the picture with NS charges this happens, for
example,
when all RR B-fields on $M^4$
 are zero. This continuum of states
comes from fundamental strings stretching close to the boundary of 
$AdS_3$. These states have finite energy, even though
they are long, due to the interaction 
with the constant three form  field strength, $H =dB_{NS}$, on $AdS_3$
\cite{Maldacena:1999uz,Seiberg:1999xz}.

A simple variation of the previous theme is to quotient (orbifold) 
the three-sphere
by a $\IZ_N \subset SU(2)_L$. 
This preserves $\cn=(4,0)$ supersymmetry.
This quotient changes the central charge of the theory by a factor
of $N$ through \brown\ (since the volume of the $S^3$ is smaller by a
factor of $N$). It is also possible to obtain this 
geometry by considering 
 the near horizon behavior of a D1-D5 + KK monopole 
system, or equivalently a D1-D5 system near an $A_N$ singularity. 
It is possible to analyze the field theory by using the methods
in \cite{Douglas:1996sw}, and using the above anomaly argument one
can calculate the right moving central charge. The left moving 
central charge should be calculated by a more detailed argument. 
When 
we have NS 5 branes and fundamental strings on an $A_N$ singularity,
the worldsheet theory is solvable, and one can calculate the spectrum 
of massive string states, etc. \cite{Kutasov:1998zh}. One can also 
consider also both RR and NS fluxes simultaneously \cite{Duff:1998cr}.
 Other papers analyzing aspects of these
quotients or orbifolds are
 \cite{Sugawara:1999qp,Sugawara:1999qp,Behrndt:1998gr,Yamaguchi:1999gb,%
Balasubramanian:1998ee}.

A related configuration arises if we consider M-theory on $M^6$, 
where $M^6 = T^6, T^2 \times K3$ or $CY_3$, and we wrap  M5 branes
on a four-cycle in $M^6$ with non-vanishing 
 triple self-intersection number.
Then, we get a string in five dimensions, and the near horizon geometry
of the supergravity solution is $AdS_3 \times S^2 \times M^6_f $,
where the subscript on $M^6_f$ indicates that the vector moduli of $M^6$
are fixed scalars. In this case we get again an $\cn=(0,4)$
 theory, and the
$SU(2)_R$ symmetry is associated  to rotations of the sphere. 
It is possible to calculate the central charge by counting 
the number of moduli of the brane configuration. Some of the moduli
correspond to geometric deformations and some of them correspond
to $B$-fields on the fivebrane worldvolume 
\cite{Maldacena:1997de,Minasian:1999qn}. A supergravity analysis of this
compactification was done in \cite{deBoer:1998ip,Fujii:1998tc}.

Another 
 interesting case is string theory compactified on $AdS_3\times S^3 \times
S^3 \times S^1 $, which has a large ${\cal N}=4$ 
symmetry \cite{Boonstra:1998yu,Elitzur:1998mm,deBoer:1999rh}.
 This algebra is sometimes called ${\cal A}_\gamma$.
It includes an $SU(2)_k \times SU(2)_{k'} \times U(1) $ current algebra. 
The relative sizes of the 
levels of the two $SU(2)$ factors are related to the relative sizes
of the  radii of the 
spheres. This case seems to be conceptually simpler than the case
with an $M^4$, since all the spacetime dimensions are 
associated to a symmetry of the system\footnote{In the case of $T^4$
one can show that the $U(1)^4$ symmetries of the torus can 
be viewed as the $ k' \to \infty $ limit of the large $ {\cal N} = 4$ 
algebra \cite{Maldacena:1999bp}.}.
In  \cite{Boonstra:1998yu} 
 a geometry like this was obtained from branes, except that
the $S^1$ was 
replaced by $\IR$, and it is not clear which brane configuration 
gives the geometry with the $S^1$. This makes it more difficult
to guess the dual conformal field theory.
 In \cite{Elitzur:1998mm} a CFT dual was proposed for this system
in the case that $k=k'$. One starts with a theory with a free boson
and four free fermions, which has large ${\cal N} = 4$ symmetry. Let
us call this theory $CFT_3$. Then, we can consider the theory
based on the symmetric product $Sym(CFT_3)^k $. 
The space-time 
 theory has two moduli, which are the radius of the circle and the
value of the RR scalar. These translate into the  radius of the 
compact  
$U(1)$-boson in $CFT_3$ and a  blow up mode of the orbifold.
In \cite{deBoer:1999rh} a dual CFT was proposed for the general case 
($k \not = k'$).

Another interesting example is 
the D1-D5 brane system in Type I string theory
\cite{Johnson:1998ms,Barbon:1998ln,Oz:1999it}.
The $\cN=(0,4)$ theory on the D1 brane worldvolume theory 
encodes in the Yukawa
couplings the ADHM data for the
construction 
of the moduli space of instantons \cite{Douglas:1996gf, Witten:1994sm}.
What distinguishes 
the Type I system from the Type IIB case is the $SO(32)$
gauge group in the open string sector.
When the D5 branes wrap a compact space $M^4$ with $M^4=T^4,K3$, 
the near horizon geometry
of the Type I supergravity solution
is $AdS_3\times S^3\times M^4$ \cite{Oz:1999it}. 
As in the previous examples, one is lead
to conjecture
a duality between Type I string theory on $AdS_3\times S^3\times M^4$ 
and the two-dimensional $(0,4)$ SCFT in the IR limit
of the D1 brane worldvolume theory.
 The supergroup of the Type I compactification
is $SU(1,1|2)\times SL(2,\IR)\times SU(2)$, 
and the Kaluza-Klein spectrum in the supergravity
can be analyzed
as in \cite{deBoer:1998ip}. 
The correspondence to the two-dimensional SCFT has not been much
explored
yet.

The relation between $AdS_3$ compactifications and 
matrix theory 
\cite{Banks:1997vh} was addressed in \cite{Hosomichi:1999uj}.


\section{Pure Gravity}

One might suspect that the simplest theory we could have on 
$AdS_3$ is pure Einstein gravity. 
In higher dimensions this is not possible since pure gravity  is not
renormalizable, so the only known sensible quantum gravity theory is
string theory, but in three dimensions gravity can be rewritten 
as a Chern-Simons theory \cite{Witten:1988hc,Witten:1989sx},
 and this theory is renormalizable.
Gravity in three dimensions has no dynamical degrees of freedom. 
We have seen, nevertheless,  that it has black hole solutions
when we consider gravity with a negative cosmological constant 
\cite{Banados:1992wn}
\lorbtz . So, it should at least describe the dynamics of these
black holes, black hole collisions, etc. 
It has been argued that this Chern-Simons theory 
reduces to a Liouville theory at the boundary 
\cite{Coussaert:1995zp,Banados:1998ta,Banados:1996tn,Oh:1998sv},
 with the right central
charge \brown . Naively, using the Cardy formula, 
this Liouville theory does not seem to 
give the same entropy as the black holes, but the Cardy formula
does not hold in this case (Liouville theory does not satisfy the
assumptions that go into the Cardy formula). 
Hopefully, these problems will be solved
once it is understood how to properly quantize Liouville theory. 
Since we have the right central charge it seems that
we should be able to calculate the BTZ black hole entropy 
\cite{Strominger:1998eq}, but Liouville theory is very peculiar
and the entropy seems smaller \cite{Carlip:1998qw}. 
Other papers studying $AdS$ pure gravity or BTZ black holes in
pure gravity include   
\cite{Banados:1998gg,Martinec:1998wm,Myung:1998fw,Behrndt:1998wg,%
Emparan:1998qp,Lee:1998qe,Navarro-Salas:1998ks,Navarro-Salas:1999sr,%
Nakatsu:1999wt,Chekhov:1999uk,Chandia:1998uf,Banados:1999ir,%
Brotz:1999xx,Park:1999nc,Banados:1998gz,Yoshida:1999sh,Cho:1999xj,%
Mano:1999xs,Banados:1998sm,Birmingham:1998pn,Hyun:1999vg}.

The Chern-Simons approach to gravity has also led to a proposal
for a black hole entropy counting in this pure gravity theory. 
In that approach the black hole entropy is supposed to come
from degrees of freedom in the Chern-Simons theory that become
dynamical when  a horizon is present  \cite{Carlip:1995qv}. 

One interesting question in three dimensional gravity is 
whether we should consider the Chern-Simons theory on a fixed 
topology or whether we should sum over topologies. Naively it
is the second possibility, however it could be that the sum 
over topologies is already included in the Chern-Simons
path integral over a fixed topology. 

In any case, three dimensional pure gravity is part of the 
full string theory compactifications, and it would be interesting
to understand it better. 

The situation is similar if one studies pure $AdS_3$ supergravities 
\cite{Achucarro:1989gm,Banados:1998pi,David:1999nr}.

\section{Greybody Factors}
\label{adsthree_greybody}

In this section we consider  an extremal or  near extremal black string
in six dimensions. We take the direction along the 
string to be  compact, with radius $R_5 \gg l_s $. We need to take
it to be compact since classically an infinite black string is unstable
\cite{Gregory:1994bj,Gregory:1993vy}\footnote{
It might seem that we can avoid the instability of  
\cite{Gregory:1994bj,Gregory:1993vy} by
going very near extremality. Note, however, that for an infinite
 string it is entropically favorable to create  a Schwarzschild black hole
threaded by an extremal string. }.  Here we assume that the 
temperature is small enough 
so that the configuration is classically
stable\footnote{ A general supergravity analysis of the  
various regimes  in  the D1-D5 system was given in 
\cite{Martinec:1999sa}.}. We take a configuration with D1 brane charge
$Q_1$ and D5 brane charge $Q_5$. The general solution with these
charges, and arbitrary energy and momentum along the string, has
the following six dimensional Einstein  metric\footnote{
Throughout this section we use the six dimensional Einstein metric,
related to the six dimensional string metric by $g_E = e^{-\phi_6}
g_{str}$, where $\phi_6$ is the six dimensional dilaton.} 
\cite{Cvetic:1996xz,Horowitz:1996ay} :
\eqn{metricsix}{\eqalign{
ds^2_{E} = &
 \( 1 + { \sa \over r^2}\)^{-1/2} \( 1 + { \sg \over r^2}\)^{-1/2}
\left[ - dt^2 +dx_5^2
\right. \cr
& \quad \quad +\left. {
r^2_0  \over r^2} (\cosh \sigma dt + \sinh\sigma dx_5)^2
 +\( 1 + {\sa \over r^2}\) ds^2_{M^4} \] \cr
 +& \( 1 + { \sa \over r^2}\)^{1/2}\( 1 + { \sg \over r^2}\)^{1/2} 
\left[
\(1-{r_0^2 \over r^2}\)^{-1} dr^2 + r^2 d \Omega_3^2 \right]~.
}}
We consider the case that the internal space $M^4 = T^4$. 
In general we will also have some scalars that are non-constant. 
These  become fixed scalars in the near-horizon $AdS_3$ limit.
In this case there are five fixed scalars, which are
three self-dual NS B-fields, a combination of the RR scalar and the
four-form on $T^4$, and finally the volume of $T^4$. If we take the
first four to zero at infinity they stay zero throughout the 
solution. Then, the physical volume of $T^4$ is
\eqn{voltffix}{
\nu (r) \equiv {{\rm Volume }\over (2\pi)^4 {\alpha'}^2} =
 v 
\(1+   { \sg \over r^2 }\)^{-1}\(1 + {\sa\over r^2 } \),
}
where $v= \nu(\infty)$ 
is the value of the dimensionless volume at infinity.
The solution \metricsix\
 is parameterized by the four  independent quantities
$\alpha,\g,\sigma,r_0$. There are two extra parameters 
which enter through the charge quantization conditions, which are
 the radius of the $x_5$ 
 dimension $R_5$ and the volume $ v$
of $T^4$. The three charges are
\eqn{charges}{
\eqalign{
   Q_1 &= { 1 \over 4\pi^2  \alpha' \sqrt{v}}\int \nu  *H'
   = { \sqrt{v} r_0^2  \over 2  \alpha'} \sinh 2 \a , \cr
 Q_5 &= {1\over 4\pi^2  \sqrt{v} \alpha'} \int H'  =  
{ r_0^2\over 2 \sqrt{v} \alpha'} \sinh 2 \g ,
\cr
  N &= {  R^2  r_0^2 \over 2  \alpha'^2} \sinh 2 \sigma ,
}}
where $*$ is the Hodge dual in the six dimensions $x^0,..,x^5$ and
$H'$ is the RR 3-form field.
The last charge $N$ is related to the  momentum around the $S^1$
by $P_5=  N/R_5$. All three charges are normalized to be integers.

The ADM energy of this solution is
\eqn{mss}{M=  {   R_5  r_0^2  \over 2  {\alpha'}^2}
(\cosh 2 \alpha + \cosh 2 \gamma + \cosh 2 \sigma  )~.
}
The Bekenstein-Hawking entropy is
\eqn{entropyfd}{
S = {A_{10}\over 4 G^{(10)}_N} = {A_6\over 4 G^{(6)}_N} =
{ 2 \pi  R_5  r_0^3 \over  {\alpha'}^2 } 
 \cosh \alpha \cosh \gamma \cosh \sigma,
}
where $A$ is the area of the horizon and
we have used the fact that in the six dimensional Einstein metric
$G_N^{(6),E} =  {\alpha'}^2 \pi^2/2$.
The Hawking temperature is
\eqn{thwk}{T= {1 \over 2 \pi  
r_0 \cosh\alpha \cosh \gamma \cosh \sigma } .}

The near extremal black string corresponds to the case that 
 $R_5$ is large and the total mass is just above the rest 
energy of the branes. By ``rest energy'' of the branes we 
mean the mass given by the BPS bound,
\eqn{energyab}{
E = M -  { Q_5 R_5 \sqrt{v} } -  { Q_1 R_5  \over \sqrt{v}}.
}
Note that this includes the mass due to the excitations carrying 
momentum along the circle.
In the limit that $ \alpha' \to 0 $ with $E , R_5 $ and $N$ fixed we
automatically go into the regime described by the conformal field
theory living on the D1-D5 system which is decoupled.
Instead, we are going to take here $\alpha'$ small but nonzero, so that
we keep some coupling of the CFT to the rest of the degrees of freedom.
The geometry is $AdS_3$ (locally) close to the horizon, but  far away
it is just the flat six dimensional space $\IR^{1,4}\times S^1$. 
In this limit we can approximate the six dimensional geometry by 
\eqn{sixdimmet}{
ds^2_{E} = f^{-1/2}\left[ -dt^2 + dx_5^2 +
 {r_0^2 \over r^2}( \cosh \sigma dt 
+ \sinh \sigma dx_5)^2\right]  + f^{1/2}(dr^2 + r^2 d\Omega_3^2),
}
where 
\eqn{defoff}{
f = \left( 1 + { r_1^2 \over r^2}\right) 
 \left( 1 + { r_5^2 \over r^2}\right)~,~~~~~~~
r_5^2 = \alpha'Q_5 \sqrt{v}~,~~~~
r_1^2 = \alpha' Q_1 /\sqrt{v}.
}

Let us consider a minimally coupled scalar field, $\phi$,  i.e. a 
scalar field that is {\it not} a fixed scalar. 
Let us send a quantum of that field to the black string, and
calculate the absorption cross section for low energies. 
This calculation was already discussed in section~\ref{gbFactorsBH}, but
        for the reader's convenience we resummarize the computations
        here.  The low-energy condition is
\eqn{lowener}{
 \omega \ll 1/r_5 , 1/r_1.
}
We will consider here just an s-wave configuration. 
We also set the momentum in the direction
of the string of the incoming
particle to zero, the general case can be found in 
\cite{Maldacena:1997ix,Gubser:1996xe}.
 Separation of variables,
$\phi = e^{-i \omega t} \chi(r)$, leads to the
radial equation
\eqn{RadialLaplace}{
   \left[ {h \over r^3} \partial_r h r^3 \partial_r + 
     \omega^2 f \right] \chi = 0 ~,~~~~~~~~~ h =  1 -
{ r_0^2 \over r^2}.
}
 Close to the horizon, a convenient radial variable is $z = h = 1 -
r_0^2/r^2$.  The matching procedure can be summarized as follows:
  \eqn{MatchSolnAgain}{\seqalign{\span\TT & \span\TR}{
   far region:  & 
    \eqalign{& \left[ {1 \over r^3} \partial_r r^3 \partial_r + 
      \omega^2 \right] \chi = 0,  
       \cr\noalign{\vskip-0.5\jot}
      & \quad \chi = A {J_1(\omega r) \over r^{3/2}}},
     \cr\noalign{\vskip2\jot}
   near region:  & 
    \eqalign{& \left[ z(1-z) \partial_z^2 + 
      \left( 1 - i {\omega \over 2\pi T_H} \right) (1-z) \partial_z + 
      {\omega^2 \over 16 \pi^2 T_L T_R} \right] 
       z^{i\omega \over 4\pi T_H} \chi = 0,
       \cr\noalign{\vskip-0.5\jot}
      & \quad \chi = z^{-{i \omega \over 4\pi T_H}}
       F\left( -i {\omega \over 4\pi T_L}, -i {\omega \over 4\pi T_R};
        1 - i {\omega \over 2\pi T_H}; z \right),  }
  }}
where $T_L,~T_R$ are defined in terms of the Hawking temperature
$T_H$ and the chemical potential, $\mu$, which is conjugate to momentum
on $S^1$ :
\eqn{tleftright}{
{ 1 \over T_{L,R} } \equiv { 1 \pm \mu \over T_H} ~~,~~~~~~~~
T_{L,R} = { r_0 e^{\pm \sigma} \over 2 \pi  r_1 r_5 }.
}
 After matching the near and far regions together and comparing the
infalling flux at infinity and at the horizon, one arrives at
  \eqn{absor}{
   \sigma_{\rm abs} = \pi^3 r_1^2 r_5^2 \omega
    {e^{\omega \over T_H} - 1 \over 
     \left( e^{\omega \over 2 T_L} - 1 \right) 
     \left( e^{\omega \over 2 T_R} - 1 \right)} \ .
  }
Notice that this has the right form to be interpreted as the 
creation of a pair of particles along the string. 

According to the $AdS_3/CFT_2$ correspondence, we can replace the 
near horizon region by the conformal field theory. The field $\phi$
couples to some operator ${\cal O}$ in the conformal field 
theory \cite{Maldacena:1997ih} :
\eqn{couplingop}{
S_{int} = \int dt dx_5  {\cal O}(t,x_5) \phi(t,x_5,\vec 0).
}
Then, the absorption cross section can be calculated by 
\eqn{abscr}{\eqalign{
\sigma & \sim {1 \over N_i } \sum_i  \sum_{f} \left|
\langle f| \int dt dx_5  {\cal O}(t,x_5)e^{i k_0 t + i k_5 x_5 }
|i \rangle \right|^2 \cr  & \sim 
{1 \over N_i }\sum_i \int e^{i k_0 t + i k_5 x_5 }
 \langle i |  {\cal O}(t,x_5)
 {\cal O}^\dagger(0,0)| i  \rangle \cr
  & \sim 
\int e^{i k_0 t + i k_5 x_5 }
 \langle   {\cal O}(t,x_5)
 {\cal O}^\dagger(0,0)  \rangle_\beta,
}}
where we have summed over final states in the CFT and averaged over
initial states. We will calculate the numerical coefficients later. 
The average over initial states is essentially 
an average over a thermal 
ensemble, since the number of states is very large so the
microcanonical ensemble is the same as a thermal ensemble. 
So, the final result is that we have to compute the two point function
of the corresponding operator over a thermal ensemble. 
This  essentially translates into computing the correlation function
on the Euclidean cylinder, and the result is 
proportional to \absor\ \cite{Das:1996wn,Das:1996jy,Maldacena:1997ih}.
This argument reproduces the functional dependence on $\omega$ of 
\absor . For other fields (non-minimally coupled) the 
functional dependence on $\omega$ is determined just 
in terms of the conformal weight of the associated operator.

Let us emphasize that the matching procedure  \MatchSolnAgain\
is valid only in the low energy regime \lowener .
 In this regime the
typical gravitational size of the configuration, which is of order
$r_5$, is much smaller than the Compton wavelength of the particle. See
figure \ref{nature}. 
In fact, note that in the connecting region 
 $r \sim r_5 $
the 
function $\phi$ does not vary very much.
Let us see this more explicitly. We see from \RadialLaplace\
that we can approximate the equation by something like
$ \omega^2 r_5^2 \phi + \phi'' =0$. From \lowener\ we see that
the variation of 
$\phi$ is very small over this connecting region. 
Furthermore, since absorption will turn out to be small,
we can approximate the value of $\phi $ at the origin by the 
value it has in flat space.
So, we can directly match the values of $\phi$ at the origin 
for a wave propagating in flat space with 
the value of $\phi$ near the boundary of $AdS_3$. 
 
In order to match the numerical coefficient we
need to determine the numerical coefficient in the two-point
function of the operator ${\cal O}$. 
This can be done for minimally coupled scalars using a 
non-renormalization theorem,  as it was done
for the case of absorption of gravitons on a D3 brane. 
The argument is the following.
We first notice that the moduli space of minimally coupled scalars
in supergravity is  $SO(4,5)/SO(4)\times SO(5)$.
This is a homogeneous space
with some metric, so the gravity Lagrangian in spacetime will include
\eqn{lagrangr}{
S = { 1 \over 2 \kappa_6^2 } 
\int d^6x g_{ab}(\phi)\partial \phi^a \partial \phi^b.
}
The fields $\phi^a$ couple to  operators ${\cal O}_a$, and
we are interested in computing
\eqn{metrcft}{
\langle {\cal O}_a(x) {\cal O}_b(0) \rangle = { G_{ab} \over x^4 }.
}
The operators ${\cal O}_a$ are a basis of marginal deformations
of the CFT,
  and  $ G_{ab}$ is the metric
on the moduli space of the CFT. 
Since 
the conformal field theory has $\cn=(4,4)$ supersymmetry, this metric
is highly constrained. In fact, it was shown in  
\cite{Cecotti:1991kz} that it is  the homogeneous metric  on 
$SO(4,5)/SO(4)\times SO(5)$ (up to global identifications). 
Since the CFT moduli space is the same as the supergravity 
moduli space, the two metrics
 could differ only by an overall numerical
factor $G_{ab} = D g_{ab}$, where $D$ is a number. 
In order to compute this number we can go to a point in moduli 
space where the CFT is just the orbifold $Sym(T^4)^k$. 
This point corresponds to having a single D5 brane and $k=Q_5Q_1$ 
D1 branes. We can also choose the string coupling to be arbitrarily 
small. For example, we can choose the scalar $\phi$ to be an off-diagonal 
component of the metric on $T^4$.
The absorption cross section  calculation then reduces to the one  
done in \cite{Das:1996wn}, which we now review. 
We take the metric on the four-torus to be 
$g_{ij} = \delta_{ij} + h_{ij}$, where $h$ is a small perturbation,
and choose $\phi = h_{12}$. 
The bulk action for $\phi$ then reduces to 
\eqn{normac}{ { 1 \over 2 \kappa_6^2} \int d^6x {1\over 2}
 (\partial \phi)^2.
}
The coupling of $h$ to the fields on the D1 branes can be derived
by expanding the Born-Infeld action. The leading term is
  \eqn{donebi}{
   S =  { 1 \over 2 \pi g_s \alpha' } \int dt  dx_5 \,
     \left[ {1 \over 2} (\partial X^i)^2  +   
      h_{12}(\tau,\sigma,\vec{x}=0) \partial X^1 \partial  X^2 
    + {\rm fermions} \right].
}
To extract the cross-section we take $R_5 = \infty$, but the 
volume of the transverse space $V$ finite, 
and we use the usual 2-d S-matrix formulas:
  \eqn{FermiThermal}{\eqalign{
{1  \over \sqrt{2} \kappa_6}   \phi(t,\vec{x}) &=
  \sum_{\vec k} \int {d k_5 \over (2\pi) }  { 1 \over 
    \sqrt{ V  2k^0}} 
\left( a^{12}_k e^{i k \cdot x} + \hbox{h.c.} \right),  \cr
{ 1\over \sqrt{2 \pi g_s \alpha' } } X^i(t,x^5) &= 
\int { dk_5 \over 2 \pi}  {1 \over \sqrt{2   k^0}}
    \left( a_k^i e^{ik \cdot x} + \hbox{h.c.} \right),  \cr
   | \tilde{i} \rangle &= (a_k^{12})^\dagger |0\rangle, \qquad
   | \tilde{f} \rangle = (a_p^1)^\dagger (a_q^2)^\dagger |0\rangle,  \cr
   \langle \tilde{f} | V_{\rm int} | \tilde{i} \rangle &= 
    { \sqrt{2} \kappa_6 p\cdot q \over \sqrt{ V }}, \cr 
\Gamma(k^0) &= {2 Q_1Q_5 \over 2 k^0 2p^02q^0}\int
 {dp^5 \over 2 \pi }  
{ dq^5 \over 2 \pi } 
 \left| \langle \tilde{f} | V_{\rm int} | \tilde{i} \rangle \right|^2
2\pi\delta(p^5+q^5) 
     2\pi \delta(\omega - p^0 - q^0),  \cr
   \sigma_{\rm abs} &= V  \Gamma(\omega) 
     = \pi^3 \alpha'^2 Q_1Q_5  \omega.
 }}

Since we have  put the four transverse dimensions
into a box of volume $V$, 
 the flux of the $h_{ij}$ gravitons on the brane is
${\cal F} = 1/V$. 
 To find the cross-section
we divide the net decay rate   by the flux.
The unusual factors of $ \sqrt{2} \kappa_6$
 and $1/\sqrt{ 2 \pi g_s \alpha'}$ 
come from the coefficients of the kinetic
terms for $h_{12}$ and $X^i$ \normac \donebi . 
  The leading factor of $2$ in the equation for $\Gamma(k^0)$
in \FermiThermal\ is there because there are two distinguishable final
states that can come out of a given $h_{12}$ initial state: an $X^1$ boson
moving left and an $X^2$ boson moving right, or $X^1$ moving right and
$X^2$ moving left. The factor of $Q_1Q_5$ comes from the fact that we
have $Q_1Q_5$ D1 branes.  Note that the delta function constraints
plus the on shell conditions imply that $ p^0 =q^0=p^5=-q^5 =\omega/2$
and $p\cdot q = \omega^2/2$. 

The final answer in \FermiThermal\ agrees with the zero temperature
limit of \absor . As we remarked before, the thermal-looking
factors in \absor\ can be derived just by doing a calculation of
the two point function on the cylinder  \cite{Maldacena:1997ih}. 
Finally, we should remark that this calculation implies that
the metric on the moduli space of the CFT has an overall factor of 
$k=Q_1Q_5$
as compared with the metric that appears in the six dimensional gravity
action \lagrangr . This blends in perfectly with the expectations
from $AdS_3$/CFT${}_2$, since in the $AdS_3$ region, by the time we go 
down to three dimensions, we get factors of the volume of the $S^3$ 
and the radius of $AdS_3$ which produce the correct factor of $k$ 
in the gravity answer for the metric on the moduli space. 

Of course, this  absorption cross section calculation is also 
related to the time reversed process of Hawking emission. Indeed,
the Hawking radiation rates calculated in gravity and in the 
conformal field theory coincide. 

Many other  greybody factors  were calculated and compared with
the field theory predictions 
\cite{Dhar:1996vu,Gubser:1996xe,Gubser:1996zp,Callan:1997tv,%
Klebanov:1997gy,Maldacena:1997ih,Krasnitz:1997gn,Gubser:1997qr,%
Mathur:1997et,Klebanov:1997gt,Birmingham:1997rj,Hosomichi:1997if,%
David:1998ev,Kim:1998yw,Taylor-Robinson:1998tk,%
Cvetic:1997xv,Cvetic:1997uw,Cvetic:1997vp,Cvetic:1998ap,%
Lee:1998vg,Muller-Kirsten:1998mt,Ohta:1998xh,Keski-Vakkuri:1999nw,%
Lee:1998xz}.
In some of these references the ``effective string'' model is mentioned.
This effective string model is essentially the conformal field
theory at the orbifold point $Sym(T^4)^k$. Some of the gravity 
calculations did not agree with the effective string calculation.
Typically that was because either the energies considered were not
low enough, or because one needed to take into account the 
effect of the deformation in the CFT away from the symmetric product
point in the moduli space.

\section{Black Holes in Five Dimensions}
\label{fivedbh}

If we Kaluza-Klein reduce, using \cite{Maharana:1993my,Sen:1994fa},
 the metric \metricsix\
 on the circle along the string,  we get a five 
dimensional charged black hole  solution : 
\eqn{solnfd}{ds_5^2 =  - \lambda^{-2/3} \(1-{r_0^2 \over r^2}\) dt^2 + 
\lambda^{1/3}
\[\(1-{r_0^2 \over r^2}\)^{-1} dr^2 + r^2 d \Omega_3^2 \right]~,}
where
\eqn{deff}{ \lambda 
= \(1+{\sa\over r^2} \)\(1+{\sg\over r^2} \)\(1+{\ss\over r^2} \)~.}
This is just the five-dimensional Schwarzschild metric, with the time
and space components rescaled by different powers of $\lambda$. 
The solution is
manifestly invariant under permutations of the three boost parameters, as
required by U-duality.
The event horizon is clearly at $r=r_0$.
The coordinates we have used present the solution
in a simple and symmetric form, but they do not always cover the entire
spacetime. When all three charges are nonzero, the surface $r=0$ is
a smooth inner horizon. This is analogous to the situation in four
dimensions with four charges 
\cite{Cvetic:1995mx,Cvetic:1995kv}.

The mass, entropy and temperature of this solution are the same
as those calculated above for the black string \mss \entropyfd \thwk .
It is interesting to take the extremal limit $r_0 \to 0$ with
$r_0e^\gamma,~r_0 e^\alpha,~r_0 e^\sigma $ finite and nonzero. 
This is an extremal black hole solution in five dimensions with
a non-singular horizon which has non-zero horizon area. 
The entropy becomes
\eqn{entroextr}{
S = 2 \pi \sqrt{ Q_1Q_5 N},
}
which is independent of all the continuous parameters in the theory, and
depends only on the charges \charges .
We can calculate this entropy as follows \cite{Strominger:1996sh}. 
These black hole states saturate the BPS bound, so they are
BPS states. Thus, we should find an ``index'', which is
a quantity that is invariant under deformations and counts the
number of BPS states. Such an index was computed in 
\cite{Strominger:1996sh} for the case where the internal space was
$M^4=K3$ and in \cite{Maldacena:1999bp} for $M^4=T^4$. 
These indices are also called  helicity supertrace formulas
 \cite{Kiritsis:1997gu}. Once we know that they do not
receive contributions from non-BPS quantities, we can change the
parameters of the theory and go to a point where we can do the
calculation, for example, we can take $R_5$ to be large and then go to 
the point where we have the $Sym(M^4)^k$ description. 

It is interesting that we can also consider near extremal black 
holes, in the approximation that the contribution to the mass
of two of the charges is much bigger than the third and much
bigger than the mass above extremality. This region in parameter
space is sometimes called the ``dilute gas'' regime. 
In the five dimensional context it is natural to take $R_5 \sim l_s$,
and at first sight we would not  expect the CFT description to
be valid. Nevertheless, it is ``experimentally'' observed
that the absorption cross section is still  \absor ,
since the calculation is exactly the same as the one we did above.
This suggests that the CFT description is also valid in this case.
A qualitative explanation of this fact was given in 
\cite{Maldacena:1996ds}, where it was observed that the the strings
could be multiply wound leading to a very low energy gap, much lower
than $1/R_5$, and of the right order of magnitude as expected for
a 5d black hole. 

Almost all that we said in this subsection can be extended to four 
dimensional black holes.


\chapter{Other AdS Spaces and Non-Conformal Theories}
\label{ChapOtherAdS}

\section{Other Branes} 
\label{adsmore}

\subsection{M5 Branes}
\label{m5branes}

There exist six dimensional $\cn=(2,0)$ SCFTs, which have sixteen
supercharges, and are expected to be non-trivial isolated fixed points
of the renormalization group in six dimensions (see
\cite{Seiberg:1997ax} and references therein).  As a consequence, they
have neither dimensionful nor dimensionless parameters.  These
theories have an $Sp(2) \simeq SO(5)$ R-symmetry group.

The $A_{N-1}$ $(2,0)$ theory is realized as the low-energy theory on
the worldvolume of $N$ coincident M5 branes (five branes of M theory).
The $\cn=(2,0)$ supersymmetry algebra includes four real spinors of the
same chirality, in the {\bf 4} of $SO(5)$.  Its only irreducible
massless matter representation consists of a 2-form $B_{\mu\nu}$ with
a self-dual field strength, five real scalars and fermions. It is
called a tensor multiplet. 
For a single 5-brane the five real scalars in the tensor multiplet
define the embedding of the M5 brane in eleven dimensions.  The
R-symmetry group is the rotation group in the five dimensions
transverse to the M5 worldvolume, and it rotates the five scalars.
The low-energy theory on the moduli space
of flat directions includes $r$ tensor multiplets (where for the
$A_{N-1}$ theories $r=N-1$). The moduli space is parametrized by the
scalars in the tensor multiplets.  It has orbifold singularities (for
the $A_{N-1}$ theory it is $\IR^{5(N-1)}/S_N$) and the theory at the
singularities is superconformal.  The self-dual 2-form $B_{\mu\nu}$
couples to self-dual strings. At generic points on the moduli space
these strings are BPS saturated, and at the superconformal point their
tension goes to zero.

The $A_{N-1}$ $(2,0)$ superconformal theory has a Matrix-like DLCQ
description as quantum mechanics on the moduli space of $A_{N-1}$
instantons \cite{Aharony:1997md}.  In this description the chiral
primary operators are identified with the cohomology with compact
support of the resolved moduli space of instantons, which is localized
at the origin
\cite{Aharony:1998lc}.  Their lowest components are scalars 
in the symmetric traceless
representations of the $SO(5)$ R-symmetry group.

The eleven dimensional supergravity metric describing $N$ M5 branes is
given by\footnote{Our conventions are such that the tension of the 
M2 brane is $T_2 = 1 / (2 \pi)^2 l_p^3$. }
\beqar
ds^2 &=& f^{-1/3}(-dt^2 + \sum_{i=1}^5 dx_i^2) + 
f^{2/3}(dr^2 + r^2 d \Omega_4^2) \ , \nonumber\\
f &=& 1 + \frac{\pi N l_p^3}{r^3} \ ,
\label{M5}
\eeqar
and there is a 4-form flux of $N$ units on the $S^4$.

The near horizon geometry of (\ref{M5}) is of the form $AdS_7 \times
S^4$ with the radii of curvature $R_{AdS} = 2R_{S^4} = 2 l_p (\pi
N)^{1/3}$. Note that since $R_{AdS} \neq R_{S^4}$ this background is
not conformally flat, unlike the $AdS_5\times S^5$ background
discussed above. Following similar arguments to those of 
section \ref{correspondence} 
leads to the conjecture that the $A_{N-1}$ $(2,0)$ SCFT is dual to M
theory on $AdS_7 \times S^4$ with $N$ units of 4-form flux on $S^4$
\cite{Maldacena:1997re}.

The eleven dimensional supergravity description is applicable for
large $N$, since then the curvature is small in Planck units.
Corrections to supergravity will go like positive powers of
$l_p/R_{AdS} \sim N^{-1/3}$; the supergravity action itself is of
order $M_p^9 \sim N^3$ (instead of $N^2$ in the $AdS_5\times S^5$
case). The known corrections in M theory are all positive powers of
$l_p^3 \sim 1/N$, suggesting that the $(2,0)$ theories have a $1/N$
expansion at large $N$.  The bosonic symmetry of the supergravity
compactification is $SO(6,2)
\times SO(5)$. The $SO(6,2)$ part is the conformal group of the SCFT,
and the $SO(5)$ part is its R-symmetry.
  
The Kaluza-Klein excitations of supergravity contain particles with
spin less than two, so they fall into small representations of
supersymmetry.  Therefore, their masses are protected from quantum
(M theory) corrections.  As in the other examples of the duality, these
excitations correspond to chiral primary operators of the $A_{N-1}$
$(2,0)$ SCFT, whose scaling dimensions are protected from quantum
corrections.  The spectrum of Kaluza-Klein harmonics of supergravity
on $AdS_7 \times S^4$ was computed in \cite{Nieuwenhuizen:1985tc}.
The lowest components of the SUSY multiplets are scalar fields with
\beq
m^2 R_{AdS}^2 = 4k(k-3),~~~~k=2,3,\cdots \ .
\label{massfor}
\eeq
They fall into the $k$-th order symmetric traceless representation of
$SO(5)$ with unit multiplicity.  The $k=1$ excitation is the singleton
that can be gauged away except on the boundary of $AdS$.  It decouples
from the other operators and can be identified with the free ``center
of mass'' tensor multiplet on the field theory side.

Using the relation between the dimensions of the operators $\Delta$
and the masses $m$ of the Kaluza-Klein excitations $m^2 R_{AdS}^2 =
\Delta(\Delta-6)$, the dimensions of the corresponding operators in
the SCFT are $\Delta=2k,~~k=2,3,\cdots$
\cite{Aharony:1998mt, Minwalla:1998po,Leigh:1998tl,Halyo:1998so}.
These are the dimensions of the chiral primary operators of the
$A_{N-1}$ $(2,0)$ theory as found from the DLCQ
description\footnote{The DLCQ description corresponded to the theory
including the free tensor multiplet, so it included also the $k=1$
operator.}. The expectation values of these operators parametrize the
space of flat directions of the theory, $(\IR^5)^{N-1}/S_N$. The
dimensions of these operators are the same as the naive dimension of
the product of $k$ free tensor multiplets, though there is no good
reason for this to be true (unlike the $d=4$ $\cn=4$ theory, where the
dimension had to be similar to the free field dimension for small
$\lambda$, and then for the chiral operators it could not change as we
vary $\lambda$). For large $N$, the $k=2$ scalar field with $\Delta=4$
is the only relevant deformation of the SCFT and it breaks the
supersymmetry.  All the non-chiral fields appear to have large masses
in the large $N$ limit, implying that the corresponding operators have
large dimensions in the field theory.

The spectrum includes also a family of spin one Kaluza-Klein
excitations that couple to 1-form operators of the SCFT.  The massless
vectors in this family couple to the dimension five R-symmetry
currents of the SCFT.  The massless graviton couples to the
stress-energy tensor of the SCFT.  As in the $d=4$ $\cN=4$ case, the
chiral fields corresponding to the different towers of Kaluza-Klein
harmonics are related to the scalar operators associated with the
Kaluza-Klein tower (\ref{massfor}) by the supersymmetry algebra.  For
each value of (large enough) $k$, the SUSY multiplets include one
field in each tower of Kaluza-Klein states.  Its $SO(5)$
representation is determined by the representation of the scalar
field.  For instance, the R-symmetry currents and the energy-momentum
tensor are in the same supersymmetry multiplet as the scalar
field corresponding to $k=2$ in equation (\ref{massfor}).

As we did for the D3 branes in section \ref{other_backgrounds}, we can place
the M5 branes at singularities and obtain other dual models.  If we
place the M5 branes at the origin of $\IR^6 \times \IR^5/\Gamma$ where
$\Gamma$ is a discrete subgroup of the $SO(5)$ R-symmetry group, we get
$AdS_7 \times S^4/\Gamma$ as the near horizon geometry. With $\Gamma
\subset SU(2) \subset SO(5)$ which is an ADE group we obtain 
theories with $(1,0)$ supersymmetry.  The analysis of these models
parallels that of section \ref{orbifolds}.  In particular, the
matching of the $\Gamma$-invariant supergravity Kaluza-Klein modes and
the field theory operators has been discussed in \cite{Ahn:1998oa}.

Another example is the $D_N$ $(2,0)$ SCFT.  It is realized as the
low-energy theory on the worldvolume of $N$ M5 branes at an
$\IR^5/\IZ_2$ orientifold singularity.  The $\IZ_2$ reflects the five
coordinates transverse to the M5 branes and changes the sign of the
3-form field $C$ of eleven dimensional supergravity.  The near horizon
geometry is the smooth space $AdS_7
\times \RP^4$ \cite{Aharony:1998mt}.  
In the supergravity solution we identify the fields at points on the
sphere with the fields at antipodal points, with a change of the sign
of the $C$ field.  This identification projects out half of the
Kaluza-Klein spectrum and only the even $k$ harmonics remain.  An
additional chiral field arises from a M2 brane wrapped on the
2-cycle in $\RP^4$, which is non-trivial due to the orientifolding;
this is analogous to the Pfaffian of the $SO(2N)$ $d=4$ $\cn=4$ SYM
theories which is identified with a wrapped 3-brane
\cite{Witten:1998xy} (as discussed in section \ref{orientifolds}). The
dimension of this operator is $\Delta=2N$. To leading order in $1/N$
the correlation functions of the other chiral operators are similar to
those of the $A_{N-1}$ SCFT.  The $D_N$ theories also have a DLCQ
Matrix description as quantum mechanics on the moduli space of $D_N$
instantons \cite{Aharony:1997md}. This moduli space is singular.  One
would expect to associate the spectrum of chiral primary operators
with the cohomology with compact support of some resolution of this
space, but such a resolution has not been constructed yet.

A different example is the $(1,0)$ six dimensional SCFT with
$E_8$ global symmetry, which is realized on the worldvolume of M5 branes
placed on top of the nine brane in the Ho\v rava-Witten
\cite{Horava:1996qa} compactification of M theory on $\IR^{10} \times
S^1/\IZ_2$.  The conjectured dual description is in terms of M theory on
$AdS_7 \times S^4/\IZ_2$ \cite{Berkooz:1998as}.  The $\IZ_2$ action has a
fixed locus $AdS_7 \times S^3$ on which a ten dimensional $\cN=1$
$E_8$ vector multiplet propagates.
The chiral operators fall into short representations of the supergroup
$OSp(6,2|2)$. In \cite{Gimon:1999to} $E_8$ neutral and charged
operators of the $(1,0)$ theory were matched with Kaluza-Klein modes
of bulk fields and fields living on the singular locus, respectively.

Correlation functions of chiral primary operators of the large $N$
$(2,0)$ theory can be computed by solving classical differential
equations for the supergravity fields that correspond to the field
theory operators.  Two and three point functions of the chiral primary
operators have been computed in
\cite{Corrado:1999cf}.

The $(2,0)$ SCFT has Wilson surface observables \cite{Ganor:1997nf}, which are
generalizations of the operator given by $W(\Sigma) =
exp (i \int_{\Sigma} B_{\mu\nu} d\sigma^{\mu\nu})$ in the theory of
a free tensor multiplet, where $\Sigma$ is a
two dimensional surface.  A prescription for computing the Wilson
surface in the dual M theory picture has been given in
\cite{Maldacena:1998im}.  It amounts, in the supergravity
approximation, to the computation of the minimal volume of a membrane
bounded at the boundary of $AdS_7$ by $\Sigma$.  The reasoning is
analogous to that discussed in section \ref{wilsonloops}, but here
instead of the strings stretched between D-branes, M2 branes are
stretched between M5 branes.  Such an M2 brane behaves as a string on
the M5 branes worldvolume, with a tension proportional to the distance
between the M5 branes.  By separating one M5 brane from $N$ M5 branes
this string can be used as a probe of the SCFT on the worldvolume of
the $N$ M5 branes, analogous to the external quarks discussed in
section \ref{wilsonloops}.
If we consider two such parallel strings with length $l$ and distance
$L$ and of opposite orientation, the resulting potential per unit length is
\cite{Maldacena:1998im}
\beq
\frac{V}{l} = -c\frac{N}{L^2} \ ,
\eeq
where $c$ is a positive numerical constant.
The dependence on $L$ is as expected from conformal invariance. 
The procedure for Wilson surface computations
has been applied also to the computation of the operator product expansion
of Wilson surfaces, and the extraction of the OPE coefficients of
the chiral primary operators \cite{Corrado:1999cf}.

The six dimensional $A_{N-1}$ theory can be wrapped on various two
dimensional manifolds. At energies lower than the inverse size of the
manifolds, the low-energy effective description is in terms of four
dimensional $SU(N)$ gauge theories.  The two dimensional manifold and
its embedding in eleven dimensions determine the amount of
supersymmetry of the gauge theory.  The simplest case is a wrapping on
$T^2$ which preserves all the supersymmetry.  This results in the $\cN
=4$ $SU(N)$ SCFT, with the complex gauge coupling being the complex
structure $\tau$ of the torus.  In general, when the two dimensional
manifold is a holomorphic curve (Riemann surface), called a
supersymmetric cycle, the four dimensional theory is supersymmetric.
For $\cN=2$ supersymmetric gauge theories the Riemann surface is the
Seiberg-Witten curve and its period matrix gives the low energy
holomorphic gauge couplings $\tau_{ij}$ ($i,j=1,\cdots,N-1$)
\cite{Kachru:1995wm,Kachru:1996fv,Klemm:1996bj,Witten:1997so}.  For $\cN=1$ supersymmetric gauge theories the
Riemann surface has genus zero and it encodes holomorphic properties
of the supersymmetric gauge theory, namely the structure of its moduli
space of vacua \cite{Hori:1998sc}.  For a generic real two dimensional
manifold the four dimensional theory is not supersymmetric.  Some
qualitative properties of the QCD string
\cite{Witten:1997ba} and the $\theta$ vacua follow from the wrapping
procedure.  Of course, in the non-supersymmetric cases the subtle
issue of stability has to be addressed as discussed in section
\ref{other_backgrounds}. In general it is not known how to compute the
near-horizon limit of 5-branes wrapped on a general manifold. At any
rate, it seems that the theory on M5 branes is very relevant to the
study of four dimensional gauge theories.  The M5 branes theory will
be one starting point for an approach to studying pure QCD in section
\ref{adsqcd}.

Other works on M5 branes in the context of the $AdS/$CFT correspondence are 
\cite{Castellani:1998nz,Russo:1998ze,Kallosh:1998qs,Grojean:1998zt,Ahn:1999qe,
Claus:1998mw,Claus:1999yw,
Awata:1998qy,Gutowski:1999iu,Forste:1999yj,
Fayyazuddin:1999zu, Bastianelli:1999bm}.

\subsection{M2 Branes}
\label{m2branes}

$\cN=8$ supersymmetric gauge theories in three dimensions can be
obtained by a dimensional reduction of the four dimensional $\cN=4$
gauge theory.  The automorphism group of the $\cN=8$ supersymmetry
algebra is $SO(8)$.  The fermionic generators of the $\cN=8$
supersymmetry algebra transform in the real two dimensional
representation of the $SO(2,1)$ Lorentz group, and in the ${\bf 8_s}$
representation of the $SO(8)$ automorphism algebra.  The massless
matter representation of the algebra consists of eight bosons in the
${\bf 8_v}$ and eight fermions in the ${\bf 8_c}$ of $SO(8)$.  Viewed
as a dimensional reduction of the vector multiplet of the four
dimensional $\cN=4$ theory which has six real scalars, one extra
scalar is the component of the gauge field in the reduced dimension
and the second extra scalar is the dual to the vector in three
dimensions.

An $\cN=8$ supersymmetric Yang-Mills Lagrangian does not posses the
full $SO(8)$ symmetry.  It is only invariant under an $SO(7)$
subgroup. At long distances it is expected to flow to a superconformal
theory that exhibits the $SO(8)$ R-symmetry (see \cite{Seiberg:1997ax}
and references therein). The flow will be discussed in the next
section.  This IR conformal theory is realized as the low-energy theory
on the worldvolume of $N$ overlapping M2 branes. For a single M2
brane, the eight real scalars define its embedding in eleven
dimensions.  The R-symmetry group is the rotation group in the eight
transverse dimensions to the M2 worldvolume, which rotates the eight
scalars.

The eleven dimensional supergravity metric describing $N$ M2 branes
is given by
\beqar
ds^2 &=& f^{-2/3}(-dt^2 + dx_1^2 + dx_2^2 ) + f^{1/3}(dr^2 + r^2 d
\Omega_7^2) \ , \nonumber\\ f &=& 1 + \frac{32 \pi^2 N l_p^6}{r^6} \ ,
\label{M2}
\eeqar
and there are $N$ units of flux of the dual to the 4-form field on $S^7$. 

The near horizon geometry of (\ref{M2}) is of the form $AdS_4 \times
S^7$ with the radii of curvature $2R_{AdS} = R_{S^4} = l_p (32 \pi^2
N)^{1/6}$.  One conjectures that the three dimensional $\cN=8$ SCFT
on the worldvolume of $N$ M2 branes is dual to M theory on $AdS_4
\times S^7$ with $N$ units of flux of the dual to the 4-form field on
$S^7$
\cite{Maldacena:1997re}.

The supergravity description is applicable for large $N$.  Corrections
to supergravity will be proportional to positive powers of
$l_p/R_{AdS} \sim N^{-1/6}$; the known corrections are all
proportional to powers of $l_p^3 \sim N^{-1/2}$. The supergravity
action itself is in this case proportional to $M_p^9 \sim N^{3/2}$, so
this will be the leading behavior of all correlation functions in the
large $N$ limit.  The bosonic symmetry of the supergravity
compactification is $SO(3,2) \times SO(8)$.  As is standard by now,
the $SO(3,2)$ part is identified with the conformal group of the three
dimensional SCFT, and the $SO(8)$ part is its R-symmetry.  The
fermionic symmetries may also be identified.  We can relate the chiral
fields of the SCFT with the Kaluza-Klein excitations of supergravity
whose spectrum was analyzed in \cite{Biran:1984tf,Castellani:1984tb}.

The lowest component of the supersymmetry multiplets is a
family of scalar excitations with
\beq
m^2 R_{AdS}^2 = \frac{1}{4}k(k-6),~~~~k=2,3,\cdots \ .
\label{massfor2}
\eeq
They fall into the $k$-th order symmetric traceless representation of
$SO(8)$ with unit multiplicity.  The dimensions of the corresponding
operators in the $\cn=8$ SCFT are $\Delta=k/2,~~k=2,3,\cdots$ 
\cite{Aharony:1998mt, Minwalla:1998po, Halyo:1998so}.  Their
expectation values parametrize the space of flat directions of the
theory, $(\IR^8)^{N-1}/S_N$.  When viewed as the IR limit of the three
dimensional $\cN=8$ Yang-Mills theory, some of these operators can be
identified as $\tr(\phi^{I_1}...\phi^{I_k})$ where $\phi^I$ are the seven
scalars of the vector multiplet.  As noted above, the eighth scalar
arises upon dualizing the vector field, which we can perform
explicitly only in the abelian case. The other chiral fields are all
obtained by the action of the supersymmetry generators on the fields
of (\ref{massfor2}).


Unlike the $(2,0)$ SCFTs, the $d=3$ $\cn=8$ theories do not have a
simple DLCQ description (see \cite{Ganor:1998jx}), and the spectrum of
their chiral operators is not known.  The above spectrum is the
prediction of the conjectured duality, for large $N$.
 
We can place the M2 branes at singularities and obtain other dual
models, as in section \ref{other_backgrounds}.  If we place the M2 branes at
the origin of $\IR^3 \times
\IR^8/\Gamma$ with $\Gamma$ a discrete subgroup of the $SO(8)$
R-symmetry group, we get $AdS_4 \times S^7/\Gamma$ as the near horizon
geometry.  One class of models is when $\Gamma \subset SU(2) \times
SU(2)$ is a cyclic group.  It is generated by multiplying the complex
coordinates $z_{1,2,3,4}$ of $\IC^4 \simeq 
\IR^8$ by $diag(e^{2 \pi i/k},e^{-2 \pi
i/k},e^{2 \pi i a/k},e^{-2 \pi i a/k})$ for relatively prime integers
$a,k$.  When $a=1, k=2$ the near horizon geometry is $AdS_4 \times
\RP^7$ with a dual $\cN=8$ theory, which is the IR limit of the 
$SO(2N)$ gauge theory \cite{Aharony:1998mt}. As in section
\ref{orientifolds}, one can add a discrete theta angle to get
additional theories \cite{Sethi:1998zk,Berkooz:1999sn}. When $a=\pm 1,
k > 2$ one gets $\cN=6$ supersymmetry, while for $a\neq \pm 1$ the
supersymmetry is reduced to $\cN=4$.  Other models are obtained by non
cyclic $\Gamma$.  As for the D3 branes \cite{Oz:1998of} and the M5
branes \cite{Ahn:1998oa}, the $\Gamma$-invariant supergravity
Kaluza-Klein modes and the field theory operators of some of these
models have been analyzed in
\cite{Entin:1998so}.

Another class of models is obtained by putting the M2 branes at hypersurface singularities
defined by the complex equation
\beq
x^2 + y^2 + z^2 + v^3 + w^{6k-1} =0  \ ,
\eeq
where $k$ is an integer.  The near horizon geometry is of the form
$AdS_4 \times H$, where $H$ is topologically equivalent to $S^7$ but in
general not diffeomorphic to it.  Some of these examples,
$k=1,\cdots,28$, correspond to the known exotic seven-spheres.  The
expected supersymmetry is at least $\cN=2$ and may be $\cN=3$, depending
on whether the R-symmetry group corresponding to the isometry
group of the metric on the exotic seven spheres is $SO(2)$ or $SO(3)$.
An example with $\cN=1$ supersymmetry is when $H$ is the squashed
seven sphere which is the homogeneous space $(Sp(2) \times Sp(1))/(Sp(1)
\times Sp(1))$.  In this case the R-symmetry group is trivial ($SO(1)$).

A general classification of possible near horizon geometries of
the form $AdS_4 \times H$ and related SCFTs in three dimensions is
given in \cite{Morrison:1998cs, Acharya:1998db}.  Most of these SCFTs
have not been explored yet.

Other works on M2 branes in the context of the $AdS/$CFT correspondence are 
\cite{Ferrara:1998vf,Gomis:1998xj,Ahn:1998sv,deWit:1998yu,Claus:1999fh,
Oh:1999qi,Ahn:1998vm,Duff:1999gh,
Furuuchi:1999tn,Fabbri:1999mk,Berkooz:1999ji,
Dall'Agata:1999hh}.

\subsection{Dp Branes}
\label{dpbranes}

Next, we discuss the near-horizon limits of other Dp~branes. They give
spaces which are different from AdS, corresponding to the fact that
the low-energy field theories on the Dp~branes are not conformal.

The Dp branes of the type II string are charged under the
Ramond-Ramond $p+1$-form potential.  Their tension is given by $T_p
\simeq 1/g_sl_s^{p+1}$ and is equal to their Ramond-Ramond charge.
They are BPS saturated objects preserving half of the 32 supercharges
of Type II string theories.  The low energy worldvolume theory of $N$
flat coinciding Dp branes is thus invariant under sixteen
supercharges. It is the maximally supersymmetric $p+1$ dimensional
Yang-Mills theory with $U(N)$ gauge group.  Its symmetry group is
$ISO(1,p) \times SO(9-p)$, where the first factor is the $p+1$
dimensional \Poincare group and the second factor is the R-symmetry
group.  The theory can be obtained as a dimensional reduction of
$\cN=1$ SYM in ten dimensions to $p+1$ dimensions.  Its bosonic fields
are the gauge fields and $9-p$ scalars in the adjoint
representation of the gauge group.  The scalars parametrize the
embedding of the Dp branes in the $9-p$ transverse dimensions.  The
$SO(9-p)$ R-symmetry group is the rotation group in these dimensions,
and the scalars transform in its vector representation.  In the
following we will discuss the decoupling limit of the brane
worldvolume theory from the bulk and the regions of validity of
different descriptions.

The Yang-Mills gauge coupling in the Dp~brane theory is given 
by
\beq
g_{YM}^2 =2  (2 \pi)^{p-2} g_s l_s^{p-3} \ .
\label{dp_coupling}
\eeq
The decoupling from the bulk (field theory) limit is the limit $l_s
\rightarrow 0$ where we keep the Yang-Mills coupling constant and the
energies fixed.  For $p \leq 3$ this
implies that the theory decouples from the bulk and that the higher
$g_s$ and $\alpha'$ corrections to the Dp brane action are suppressed.
For $p > 3$, as seen from (\ref{dp_coupling}), the string coupling goes
to infinity and we need to use a dual description to analyze this
issue.

Let $u \equiv r/\alpha'$ be a fixed expectation value of a scalar. 
At an energy scale $u$, the dimensionless effective coupling constant
of the Yang-Mills theory is 
\beq
g_{eff}^2 \sim g_{YM}^2Nu^{p-3} \ .
\label{effective}
\eeq
The perturbative Yang-Mills description is applicable when $g_{eff}^2 \ll 1$.

The ten dimensional supergravity background describing $N$ Dp branes
is given by the string frame metric
\beqar
ds^2 &=& f^{-1/2}(-dt^2 + \sum_{i=1}^p dx_i^2) + 
f^{1/2}\sum_{i=p+1}^9 dx_i^2 \ , \nonumber\\
f &=& 1 + \frac{c_p g_{YM}^2 N }{l_s^4 u^{7-p}} \ ,
\label{Dp}
\eeqar
with a constant $c_p=2^{6-2p}\pi^{(9-3p)/2}\Gamma((7-p)/2)$.
The background has
a Ramond-Ramond $p+1$-form potential $A_{0...p} = (1-f^{-1})/2$, and
a dilaton
\beq
e^{-2(\phi-\phi_{\infty})} = f^{(p-3)/2} \ .
\eeq

After a variable redefinition 
\beq 
 z = {2 \sqrt{c_p g^2_{YM} N}  \over (5-p)  u^{ 5-p \over 2 } },
\label{ztou}
\eeq
the field theory limit of the metric (\ref{Dp}) for $p < 5$ takes the form 
\cite{Itzhaki:1998sa,Boonstra:1999mp}
\beq
ds^2 = \alpha' \left( 2 \over 5-p\right)^{7-p \over 5-p}
\left(c_p g^2_{YM} N \right)^{1\over 5-p} z^{3-p \over 5-p} 
\left\{ {-dt^2 + d{\vec x}^2 + dz^2 \over z^2 } + { (5-p)^2 \over 4 } 
d \Omega_{8-p}^2 \right\} ~,
%
\label{Dpnear}
\eeq
with the dilaton
\beq
e^{\phi} \sim \frac{g_{eff}^{\frac{7-p}{2}}}{N} \ .
\label{effectivephi}
\eeq
The curvature associated with the metric (\ref{Dpnear})
is
\beq
{\cal R} \sim \frac{1}{l_s^2 g_{eff}} \ .
\label{curvature}
\eeq
In the form of the metric (\ref{Dpnear}) it is easy to see that
the UV/IR correspondence, as described in section \ref{holography},
leads to the relationship $\lambda \sim z$ between wavelengths
in the dual field theories and distances in the gravity solution. 
Through (\ref{ztou}) we can then relate energies in the field theory
to distances in the $u$ variable. 

In the limit of infinite $u$ the effective string coupling
(\ref{effectivephi}) vanishes for $p < 3$. This corresponds to the UV
freedom of the Yang-Mills theory.  For $p > 3$ the coupling increases
and we have to use a dual description.  This corresponds to the fact
that the Yang-Mills theory is non renormalizable and new degrees of
freedom are required at short distances to define the theory.  The
isometry group of the metric (\ref{Dpnear}) is $ISO(1,p) \times
SO(9-p)$. The first factor corresponds to the
\Poincare symmetry group of the Yang-Mills theory
and the second factor corresponds to its R-symmetry group.

For each Dp brane we can plot a phase diagram as a function
of the two dimensionless parameters $g_{eff}$ and $N$ \cite{Itzhaki:1998sa}.
Different regions in the phase diagram have a good description
in terms of different variables.  
As an example consider the D2 branes in Type IIA string theory.   
The dimensionless effective gauge coupling 
(\ref{effective})
is now $g_{eff}^2 \sim g_{YM}^2N/u$.
The perturbative Yang-Mills description is valid for $g_{eff}\ll 1$.
When  $g_{eff} \sim 1$ we have a transition from the perturbative Yang-Mills
description to the Type IIA supergravity description.
The Type IIA supergravity description is valid when both the curvature 
is string units (\ref{curvature})
and the effective string coupling (\ref{effectivephi})
are small. This implies that $N$ must be large.

When $g_{eff} > N^{2/5}$ the effective string coupling becomes large.
In this region we grow the eleventh dimension $x_{11}$ and the good
description is in terms of an eleven dimensional theory.  We can
uplift the D2 brane solution (\ref{Dpnear}) and (\ref{effectivephi})
to an eleven dimensional background that reduces to the ten
dimensional background upon Kaluza-Klein reduction on $x_{11}$.  This
can be done using the relation between the ten dimensional Type IIA
string metric $ds_{10}^2$ and the eleven dimensional metric
$ds_{11}^2$,
\beq
ds_{11}^2 = e^{4\phi/3} (dx_{11}^2 + A^{\mu}dx_{\mu})^2 +
e^{-2\phi/3}ds_{10}^2 \ .
\label{11d}
\eeq
$\phi$ and $A_{\mu}$ are the Type IIA dilaton and RR gauge field.
The 4-form field strength is independent of $x_{11}$.

The curvature of the eleven dimensional metric 
in eleven dimensional Planck units $l_p$
is given by 
\beq
{\cal R} \sim \frac{e^{2\phi/3}}{l_p^2 g_{eff}} \sim
\frac{g_{eff}^{2/3}}{l_p^2 N^{2/3}} \ .
\label{R11}
\eeq
When the curvature (\ref{R11})
is small we can use the eleven dimensional
supergravity description.    

\begin{figure}[htb]
\begin{center}
\epsfxsize=6in\leavevmode\epsfbox{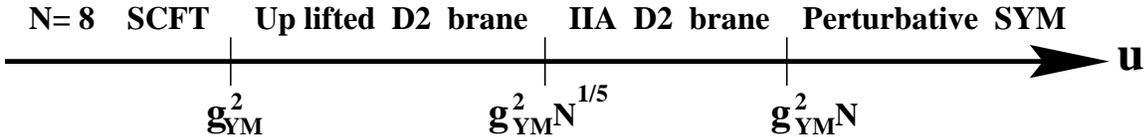}
\end{center}
\caption{The different descriptions of the D2 brane theory as
a function of the energy scale $u$. 
We see the flow from the high energy $\cN=8$
super Yang-Mills theory to
the low energy $\cN=8$ SCFT. 
}
\label{regions}
\end{figure}

The metric (\ref{11d}) corresponds to the M2 branes solution smeared
over the transverse direction $x_{11}$.  The near-horizon limit of the
supergravity solution describing M2 branes localized in the compact
dimension $x_{11}$ has the form (\ref{M2}), but with a harmonic
function $f$ of the form
\beq
f = \sum_{n=-\infty}^{\infty} \frac{32 \pi^2 l_p^6 N}{\left(r^2 + (x_{11}-x_{11}^0 + 
2 \pi n R_{11})^2 \right)^3} \ ,
\label{f}
\eeq
where $r$ is the radial distance in the seven non-compact transverse
directions and $x_{11} \sim x_{11} + 2 \pi R_{11}$.  $x_{11}^0$
corresponds to the expectation value of the scalar dual to the vector
in the three dimensional gauge theory.  The expression for the
harmonic function (\ref{f}) can be Poisson resummed at distances much
larger than $R_{11}= g_{YM}^2 l_s^2$, leading to
\beq
f = \frac{6 \pi^2 N g_{YM}^2}{l_s^4 u^5} + O( e^{-u/g_{YM}^2}) \ .
\label{local}
\eeq 
The difference between the localized M2 branes solution and the
smeared one is the exponential corrections in (\ref{local}). They can
be neglected at distances  $u \gg g_{YM}^2$,  or in terms of the
dimensionless parameters when $g_{eff} \ll N^{1/2}$. 
According to (\ref{ztou}) this corresponds to distance scales in 
the field theory of order $\sqrt{N}/g^2_{YM}$. In this region we
can still use the up lifted D2 brane solution since it is the same
as the one coming from (\ref{f}) up to exponentially small corrections.
 When $g_{eff} \gg N^{1/2}$, which
corresponds to very low energies $u \ll g_{YM}^2$, the sum in
(\ref{f}) is dominated by the $n=0$ contribution.  This background is
of the form (\ref{M2}) (with $f=32\pi^2 Nl_p^6/r^6$), 
namely the near-horizon limit of M2 branes in
eleven non-compact dimensions.  This is the superconformal theory
which we discussed in the previous section.  In figure \ref{regions}
we plot the transition between the different descriptions as a
function of the energy scale $u$.  We see the flow from the high
energy $\cN=8$ super Yang-Mills theory realized on the worldvolume of
D2 branes to the low energy $\cN=8$ SCFT realized on the worldvolume
on M2 branes.

A similar analysis can be done for the other Dp branes of the Type II
string theories.  In the D0 branes case one starts at high energies
with a perturbative super quantum mechanics description. At
intermediate energies the good description is in terms of the Type IIA
D0 brane solution. At low energies the theory is expected to describe
matrix black holes \cite{Banks:1998hz}. 
 In the D1 branes case one starts in
the UV with a perturbative super Yang-Mills theory in two dimensions.
In the intermediate region the good description is in terms of the
Type IIB D1 brane solution. The IR limit is described by the $Sym^N
(\IR^8)$ orbifold SCFT.  The D3 branes correspond to the $\cn=4$ SCFT
discussed extensively above.

In the D4 branes case, the UV definition of the theory is obtained by
starting with the six dimensional $(2,0)$ SCFT discussed in section
\ref{m5branes}, and compactifying it on a circle. At high energies,
higher than the inverse size of the circle, we have a good description
in terms of the $(2,0)$ SCFT (or the $AdS_7\times S^4$ background of M
theory). The intermediate description is via the background of the
Type IIA D4 brane. Finally at low energies we have a description in
terms of perturbative super Yang-Mills theory in five dimensions.  In
the D5 branes case we have a good description in the IR region in
terms of super Yang-Mills theory. At intermediate energies the system
is described by the near-horizon background of the Type IIB D5 brane,
and in the UV in terms of the solution of the Type IIB NS5~branes.
We will discuss the NS5~brane theories in the next section.

Consider now the system of $N$ D6 branes of Type IIA string theory.
As before, we can attempt at a decoupling of the seven dimensional
theory on the D6 branes worldvolume from the bulk by taking the string
scale to zero and keeping the energies and the seven dimensional
Yang-Mills coupling fixed.  The effective Yang-Mills coupling
(\ref{effective}) is small at low energies $u \ll
(g_{YM}^2N)^{-1/3}$ and super Yang-Mills is a good
description in this regime.  The curvature in string units
(\ref{curvature}) is small when $u \gg (g_{YM}^2N)^{-1/3}$
while the effective string coupling (\ref{effectivephi}) is small when
$u \ll N/g_{YM}^{2/3}$. In between these limits we can use the
Type IIA supergravity solution.

When $u \sim N/g_{YM}^{2/3}$ the effective string coupling is
large and we should use the description of D6~branes in terms of
eleven dimensional supergravity compactified on a circle with $N$
Kaluza-Klein monopoles. Equivalently, the description is in terms of
eleven dimensional supergravity on an ALE space with an $A_{N-1}$
singularity.  When $u \gg N/g_{YM}^{2/3}$ the curvature of the
eleven dimensional space vanishes and, unlike the lower dimensional
branes, there does not not exist a seven dimensional field theory that
describes the UV.  In fact, the D6 brane worldvolume theory does not
decouple from the bulk.

A simple way to see that the D6 brane worldvolume theory does not
decouple from the bulk is to note that now in the decoupling limit we
keep $g_{YM}^2 \sim g_s l_s^3$ fixed.  When we lift the D6 branes
solution to M theory, this means that the eleven dimensional Planck
length $l_p^3 = g_s l_s^3$ remains fixed, and therefore gravity does not
decouple.  Another way to see that gravity does not decouple is to
consider the system of D6 branes at finite temperature in the
decoupling limit. For large energy densities above extremality, $E/V
\gg N/l_p^7$, we need the eleven dimensional description.  This is
given by an uncharged Schwarzschild black hole at the ALE
singularity. The associated Hawking temperature is $T_H \sim
1/\sqrt{Nl_p^9E/V}$ and there is  Hawking radiation to the asymptotic
region of the bulk eleven dimensional supergravity.  Generally, the
worldvolume theories of D$p$~branes with $p >5$ do not decouple from
the bulk.

The supergravity computation of the Wilson loop, discussed in section
\ref{wilsonloops}, can be carried out for the Dp brane theories.  For
instance for the $N$ D2 branes theory one gets for the quark antiquark
potential, using the type IIA SUGRA D2~brane solution
\cite{Maldacena:1998im},
\beq
V = - c \frac{(g_{YM}^2N)^{1/3}}{L^{2/3}} \ ,
\label{VV}
\eeq
where $c$ is a positive numerical constant.  In view of the discussion
above, this result should be trusted only for loops with sizes
$  1/g_{YM}^2N \ll L \ll \sqrt{N}/g^2_{YM}$. For smaller 
loops the computation fails because we go into the perturbative regime, 
where the potential becomes logarithmic. For larger loops we 
get into the $AdS_4 \times S^7$ region.

Other works on Dp branes in the context of the $AdS/$CFT correspondence are 
\cite{Itzhaki:1998uz,Boonstra:1999mp,Itzhaki:1998ka,Pelc:1999ms,
Hashimoto:1999xx,Youm:1999zs,Lu:1999uc,Lu:1999uv,Barbon:1999zp}.

\subsection{NS5 Branes}
\label{ns5branes}

The NS5 branes of Type II string theories couple magnetically to the
NS-NS $B_{\mu\nu}$ field, and they are magnetically dual to the
fundamental string.  Their tension is given by $T_{NS} \simeq
1/g_s^2l_s^6$.  Like the Dp branes, they are BPS objects that preserve
half of the supersymmetry of Type II theories.  A fundamental string
propagating in the background of $N$ parallel NS5 branes is described
far from the branes by a conformal field theory with non trivial
metric, $B$ field and dilaton, constructed in
\cite{Callan:1991ss}. The string coupling grows as the string
approaches the NS5~branes.  At low energies the six dimensional theory
on the worldvolume of $N$ Type IIB NS5 branes is a $U(N)$ $\cn=(1,1)$
super Yang-Mills theory, which is free in the IR. However, it is an
interacting theory at intermediate energies.  At low energies the
theory on the worldvolume of $N$ Type IIA NS5 branes is the $A_{N-1}$
$(2,0)$ SCFT discussed above.

The six dimensional theories on the worldvolume of NS5 branes of Type
II string theories were argued \cite{Seiberg:1997md} to decouple from
the bulk in the limit
\beq
g_s\rightarrow 0,~~~~~l_s=fixed \ .
\label{decoupling}
\eeq
This is because the effective coupling on the NS5 branes (e.g. the
low-energy Yang-Mills coupling in the type IIB case) is $1/l_s$, while
the coupling to the bulk modes goes like $g_s$.  However, the
computation of \cite{Maldacena:1997sd} showed that in this limit there
is still Hawking radiation to the tube region of the NS5~brane
solution, suggesting a non decoupling of the worldvolume theory from
the bulk.  In the spirit of the other correspondences discussed
previously, one can reconcile the two by conjecturing
\cite{Aharony:1998ub} that string or M theory in the NS5 brane
background in the limit (\ref{decoupling}), which includes the tube
region, is dual to the decoupled NS5 brane worldvolume theory
(``little string theory'').  In particular, the fields in the tube
which are excited in the Hawking radiation correspond to objects in
the decoupled NS5 brane theory.  In the following we will mainly
discuss the Type IIA NS5 brane theory\footnote{Type IIB NS5~branes at 
orbifold singularities are discussed in    \cite{Diaconescu:1998pj}.}.

The Type IIA NS5 brane may be considered as the M5 brane localized on
the eleven dimensional circle.  Therefore its metric is that of an
M5 brane at a point on a transverse circle.  In such a configuration
the near horizon metric of $N$ NS5 branes can be written as
\cite{Itzhaki:1998sa,Aharony:1998ub}
\beqar
ds^2 &=& l_p^2 \left(f^{-1/3}(-dt^2 + \sum_{i=1}^5 dx_i^2) + 
f^{2/3}(dx_{11}^2  + du^2 + u^2 d\Omega_3^2) \right) \ , \nonumber\\
f &=& \sum_{n=-\infty}^{\infty}
\frac{\pi N}{(u^2+(x_{11}-2\pi n/l_s^2)^2)^{3/2}} \ .
\label{NS5}
\eeqar
The $x_{11}$ coordinate is periodic and has been rescaled by $l_p^3$
($x_{11} \equiv x_{11} + 2\pi/l_s^2$).  The background also has a 4-form
flux of $N$ units on $S^1 \times S^3$.

At distances larger than $l_s\sqrt{N}$ the NS5 brane theory is
described by the $A_{N-1}$ $(2,0)$ SCFT.  Indeed, in the extreme low
energy limit $l_s \rightarrow 0$ the sum in (\ref{NS5}) is dominated
by the $n=0$ term and the background is of the form $AdS_7 \times
S^4$.  This reduces to the conjectured duality between M theory on
$AdS_7 \times S^4$ and the $(2,0)$ SCFT, discussed previously.
However, the NS5 brane theory is not a local quantum field theory at
all energy scales since at short distances it is not described by a UV
fixed point.  To see this one can take $l_s$ to infinity (or $u$ to
infinity) in (\ref{NS5}) and get a Type IIA background with a linear
dilaton.  It has the topology of $\IR^{1,5} \times \IR \times S^3$
with $g_s^2(\phi) = e^{-2\phi/l_s\sqrt{N}}$, where $\phi$ is the $\IR$
coordinate.
This is in accord with the fact that the NS5 brane theory exhibits a
T-duality property upon compactification on tori (note that in this
background a finite radius in field theory units corresponds to a
finite radius in string theory units on the string theory side of the
correspondence, unlike the previous cases we discussed).

The NS5 brane theories have an A-D-E classification. This can be seen
by viewing them as Type II string theory
on K3 with A-D-E singularities in the
decoupling limit (\ref{decoupling}).  The NS5 brane theories have an
$SO(4)$ R-symmetry which we identify with the $SO(4)$ isometry of $S^3$.
The IIA NS5~brane theories have a moduli space of vacua of the form
$(\IR^4 \times S^1)^r/\cW$ where $r$ is the rank of the A-D-E gauge
group and $\cW$ is the corresponding Weyl group.  It is parametrized
by the $\cW$-invariant products of the $5r$ scalars in the $r$ tensor
multiplets.  They fall into short representations of the supersymmetry
algebra.  
We can match these chiral operators with the string excitations in the
linear dilaton geometry describing the large $u$ region of
(\ref{NS5}). The string excitations, in short representations of the
supersymmetry algebra, in the linear dilaton geometry were analyzed in
\cite{Aharony:1998ub}. Indeed, they match the spectrum of the chiral
operators in short representations of the NS5 brane theories. Actually,
due to the fact that the string coupling goes to zero at the boundary
of the linear dilaton solution, one can compute here the precise
spectrum of chiral fields in the string theory, and find an agreement
with the field theory even for finite $N$ (stronger than the large $N$
agreement that we described in section \ref{tests}). 

As in the dualities with local quantum field theories, also here one
can compute correlation functions by solving differential equations on
the NS5 branes background (\ref{NS5}). Since in this case the boundary
is infinitely far away, it is more natural to compute correlation
functions in momentum space, which correspond to the S-matrix in the
background (\ref{NS5}). The computation of two point functions 
of a scalar field was sketched in \cite{Aharony:1998ub} and described
more rigorously in \cite{Minwalla:1999xi}.
The NS5~brane theories are non-local, and this
causes some differences in the matching between M theory and the
non-gravitational NS5~brane theory in this case. One difference
from the previous cases we discussed is that in the linear dilaton
backgrounds if we put a cutoff at some value of the radial coordinate
(generalizing the discussion of \cite{Susskind:1998dq} which we
reviewed in section \ref{holography}), the volume enclosed by the
cutoff is not proportional to the area of the boundary (which it is in
AdS space). Thus, if holography is valid in these backgrounds (in the
sense of having a number of degrees of freedom proportional to the
boundary area) it is more remarkable than holography in AdS space.

\section{QCD} 
\label{adsqcd}

The proposed extension of the duality conjecture between field
theories and superstring theories to field theories at finite 
temperature, as described in section \ref{FiniteT}, opens up the
exciting possibility of studying the physically relevant non
supersymmetric gauge theories.  Of particular interest are non
supersymmetric gauge theories that exhibit asymptotic freedom and
confinement.  In this section, we will discuss an approach to studying
pure (without matter fields) QCD$_p$ in $p$ dimensions using a dual
superstring description.  We will be discussing mainly the cases
$p=3,4$.

The approach proposed by Witten \cite{Witten:1998zw} was to start with
a maximally supersymmetric gauge theory on the $p+1$ dimensional
worldvolume of $N$ $Dp$ branes. One then compactifies the
supersymmetric theory on a circle of radius $R_0$ and imposes
anti-periodic boundary conditions for the fermions around the
circle. Since the fermions do not have zero frequency modes around the
circle they acquire a mass $m_f \sim 1/R_0$.  The scalars then acquire
a mass from loop diagrams, and at energies much below $1/R_0$ they
decouple from the system. The expected effective theory at large
distances compared to the radius of the circle is pure QCD in $p$
dimensions.  Note that a similar approach was discussed in the
treatment of  gauge theories at finite
temperature $T$ in section \ref{FiniteT}, where the radius of the circle is
proportional to $1/T$.  The high temperature limit of the
supersymmetric gauge theory in $p+1$ dimensions is thus described by a
non supersymmetric gauge theory in $p$ dimensions.

The main obstacle to the analysis is clear from the discussions of the
duality between string theory and quantum field theories in the
previous sections.  The string approach to weakly coupled gauge
theories is not yet developed.  Most of the available tools are
applicable in the supergravity limit that describes the gauge theory
with a large number of colors and large 't Hooft parameter. In this
regime we cannot really learn directly about QCD, since the typical
scale of candidate 
QCD states (glueballs) is of the same order of magnitude (for
QCD$_4$, or a larger scale for QCD$_3$) as the scale $1/R_0$ of the
mass of the ``extra'' scalars and fermions. A related issue is that
at
short distances asymptotically free gauge theories are weakly coupled
and the dual supergravity description is not valid.  Therefore, we
will be limited to a discussion in the strong coupling region of the
gauge theories and in particular we will not be able to exhibit
asymptotic freedom.

One may hope that a full solution of the classical ($g_s=0$) string
theory will provide a description of large $N$ gauge theories for all
couplings (in the 't Hooft limit).  To study the gauge theories with a
finite number of colors requires the quantum string theory.  However,
there is also a possibility that the gauge description is valid for
weak coupling and the string theory description is valid for strong
coupling with no smooth crossover between the two descriptions.  In
such a scenario there is a phase transition at $\lambda = \lambda_c$ 
\cite{Li:1999kd,Gao:1998ww}.
This will prevent us from using the string description to study QCD,
and will prevent classical string theory from being the master field
for large $N$ QCD.

In the last part of this section we will briefly discuss another
approach, based on a suggestion by Polyakov \cite{Polyakov:1998ju}, to
study non supersymmetric gauge theories via a non supersymmetric
string description. In this approach one can exhibit asymptotic
freedom qualitatively already in the gravity description.  In the IR
there are gravity solutions that exhibit confinement at large
distances as well as strongly coupled fixed points.

\subsection{QCD$_3$}

The starting point for studying QCD$_3$ is the $\cN=4$ superconformal
$SU(N)$ gauge theory in four dimensions which is realized as the low
energy effective theory of $N$ coinciding parallel D3 branes.  As
outlined above, the three-dimensional non-supersymmetric theory is
constructed by compactifying this theory on $\IR^3 \times {\bf S}^1$
with anti-periodic boundary conditions for the fermions around the
circle. The boundary conditions break supersymmetry explicitly and as
the radius $R_0$ of the circle becomes small, the fermions decouple
from the system since there are no zero frequency modes.  The scalar
fields in the four dimensional theory will acquire masses at one-loop,
since supersymmetry is broken, and these masses become infinite as
$R_0 \rightarrow 0$.  Therefore in the infrared we are left with only
the gauge field degrees of freedom and the theory should be
effectively the same as pure QCD$_3$.

We will now carry out the same procedure in the dual superstring
(supergravity) picture.  As has been extensively discussed in the
previous sections, the ${\cal N}=4$ theory on $\IR^4$ is
conjectured to be dual to type IIB superstring theory on AdS$_5 \times
{\bf S}^5$ with the metric (\ref{nearhor}) or (\ref{metricu}).
 
Recall that the dimensionless gauge coupling constant $g_4$ of the
$\cN=4$ theory is related to the string coupling constant $g_s$ as
$g_4^2 \simeq g_s$.  In the 't Hooft limit, $N \rightarrow \infty$
with $g_4^2N \simeq g_s N$ fixed, the string coupling constant
vanishes, $g_s \rightarrow 0$.  Therefore, we could study the $\cN=4$
theory using the tree level string theory in the AdS space
(\ref{metricu}). If also $g_s N \gg 1$, the curvature of the AdS space is
small and the string theory is approximated by classical supergravity.

Upon compactification on ${\bf S}^1$ with
supersymmetry breaking boundary conditions, (\ref{metricu})
is replaced by the Euclidean black hole geometry \cite{Horowitz:1998pq, 
Witten:1998zw} \footnote{The stability issue of this background 
is discussed in \cite{Horowitz:1999ha}.} 

\beq
ds^2 = \alpha' \sqrt{4 \pi g_s N} 
\left( u^2 (h(u)
 d\tau^2 + \sum_{i=1}^3 dx_i^2)+ h(u)^{-1}
 \frac{du^2}{u^2} + d\Omega_5^2 \right) \ ,
\label{3dmetric}
\eeq
where
 $\tau$ parametrizes the compactifying circle (with radius 
$R_0$ in the field theory) and
\beq
h(u)=1-{u_0^4\ov u^4} \ .
\eeq

The $x_{1,2,3}$ directions correspond to the $\IR^3$
coordinates of QCD$_3$. The horizon of this geometry is
located at $u=u_0$ with
\beq
u_0 = {1 \over 2 R_0} \ .
\label{horizonlocation}
\eeq
The supergravity approximation is applicable
for $N \rightarrow \infty$ and $g_s N \gg 1$, so that all the curvature
invariants are small.
The metric (\ref{3dmetric}) describes the Euclidean theory, the 
Lorentzian theory is obtained by changing $\sum_{i=1}^3 dx_i^2 \to
 -dt^2 + dx_1^2 + dx_2^2 $. Notice that this is not the same
as the Wick rotation that leads to the near extremal 
black hole solution
\NearDThree .

{}From the point of view of QCD$_3$, the radius $R_0$ of the
compactifying circle provides  the ultraviolet cutoff
scale. To obtain large $N$
QCD$_3$ itself (with infinite cutoff), 
one has to take $g_4^2 N \rightarrow 0$  
as  $R_0 \rightarrow 0$ so that the three dimensional effective
coupling $g_3^2 N = g_4^2N/( 2\pi R_0) $ remains
at the intrinsic energy scale of QCD$_3$. 
$g_3^2$ is the classical dimensionful coupling of  QCD$_3$. The effective 
dimensionless gauge coupling of QCD$_3$ at the 
distance scale $R_0$ is therefore $g_sN$.

The proposal is that Type IIB string theory on the AdS black hole background
(\ref{3dmetric}) provides a dual description to QCD$_3$ (with the UV
cutoff described above).
The limit in which the classical supergravity description is valid, 
$g_sN \gg 1$,
is the limit where the typical mass scale of QCD$_3$, $g_3^2 N$, is much
larger than the cutoff scale $1/R_0$.
It is the opposite of the limit that is required in order 
to see the ultraviolet freedom
of the theory.
Therefore, with the currently available techniques, we 
can only study large $N$ QCD$_3$ with a fixed ultraviolet
cutoff $R_0^{-1}$ in the strong coupling regime.
It should be emphasized that by strong coupling we mean here that the
coupling is large compared to the cutoff scale, so we really have many
more degrees of freedom than just those of QCD$_3$.
QCD$_3$ is the theory which we would get 
in the limit of vanishing bare coupling, which
is the opposite limit to the one we are taking.

This is analogous to, but not the same as, 
the lattice strong coupling expansion with a fixed 
cutoff given by the
 lattice spacing $a$ (which is analogous to $R_0$ here).
There, QCD$_3$ is
 obtained in the limit $g_3^2 a \rightarrow 0$ while strong 
coupling lattice QCD$_3$ is the theory at large $g_3^2 a$.
An important  difference in the approach that we take, 
compared to the lattice description,
is that 
we have full Lorentz 
invariance in the three gauge theory coordinates.  
The regularization of the gauge theory
in the dual string theory description
is provided by a one higher dimensional theory, the theory on D3 branes.

In the limit 
$R_0 \rightarrow 0$ the geometry (\ref{3dmetric}) is singular.
As discussed 
above, in this limit the supergravity description is not valid
and we have to use the string theory description.

\subsubsection{Confinement}

As we noted before,
the gauge coupling of QCD$_3$ $g_3^2$ has dimensions of mass, and 
it provides a scale
already for the classical theory.
The effective dimensionless expansion parameter at a 
length scale $l$, $g_3^2(l)\equiv
 l g_3^2$, goes to zero as $l \rightarrow 0$.
Therefore, like QCD$_4$, the theory is free at short distances.
Similarly, at a large length scale $l$ the effective coupling 
becomes strong. Therefore,
the interesting IR physics is non-perturbative.

In three dimensions the Coulomb potential is already confining. This is a 
logarithmic confinement $V(r) \sim ln(r)$. 
Lattice simulations provide evidence that in QCD$_3$
at large distances there is confinement
with a linear potential $V(r) \sim \sigma r$.

\begin{figure}[htb]
\begin{center}
\epsfxsize=1.5in\leavevmode\epsfbox{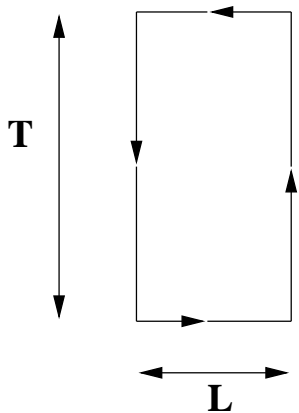}
\end{center}
\caption{A confining quark-antiquark linear potential $V(L) \sim \sigma L$ 
can be extracted from the Wilson loop obeying an area law
$\langle W(C) \rangle \sim  exp(-\sigma TL)$.
}
\label{arealaw}
\end{figure}

To see confinement in the dual description
we will consider the spatial Wilson loop.
In a confining theory the vacuum expectation value
of the Wilson loop operator exhibits an area law behavior \cite{Wilson:1974co}
\beq
\langle W(C) \rangle \simeq exp(-\sigma A(C)) \ ,
\label{area}
\eeq
where $A(C)$ is the area enclosed by the loop $C$. The constant 
$\sigma$ is called the string tension.
The area law (\ref{area}) is equivalent to the quark-antiquark confining
linear potential $V(L) \sim \sigma L$.
This can be simply seen by considering a rectangular loop $C$ with sides of 
length $T$ and $L$ in Euclidean
space as in figure \ref{arealaw}. 
For large $T$ we have, when $V(L) \sim \sigma L$ and interpreting $T$ as
the time direction,
\beq
\langle W(C) \rangle \sim  exp(-TV(L)) \sim exp(-\sigma A(C))  \ .
\eeq 

The prescription to evaluate the vacuum expectation value
of the Wilson loop operator in the dual string description
has been introduced in section \ref{wilsonloops}.
It amounts to computing 
\beq
\langle W(C) \rangle = \int exp(-\mu(D)) \ ,
\label{sumarea}
\eeq
where $\mu(D)$ is the regularized area of the worldsheet of a string
$D$ bounded at infinity by $C$. 

We will work in the supergravity approximation in which (\ref{sumarea})
is approximated 
by
\beq
\langle W(C) \rangle = exp(-\mu(D)) \ ,
\label{minarea}
\eeq
where $\mu(D)$ is the minimal area of a string worldsheet $D$ 
bounded at infinity by $C$.

\begin{figure}[htb]
\begin{center}
\epsfxsize=3in\leavevmode\epsfbox{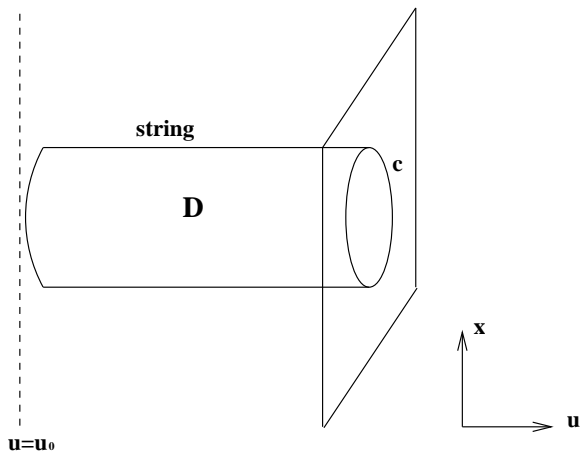}
\end{center}
\caption{The worldsheet of the string $D$ is bounded at infinite $u$ by the 
loop $C$. The string tends to minimize its length by going to the 
region with smallest
metric component $g_{ii}$, which in this case 
is near the horizon $u=u_0$. The energy between the quark and the antiquark
is proportional to the distance $L$ between them and to the string tension
which is $\sigma = \frac{1}{2\pi}g_{ii}(u_0)$.
}
\label{string_fig}
\end{figure}

This prescription has been applied in section \ref{wilsonloops} 
to the calculation of the
Wilson loop in the $\cN=4$ theory which is not a confining theory. 
Indeed, it has been
found there that it exhibits a Coulomb like behavior.
The basic reason was that when we scaled up the loop $C$ by 
$x^i \rightarrow \alpha x^i$ with a positive number $\alpha$, 
we could use conformal invariance to 
scale up $D$ without changing its (regularized) area. 
Therefore $D$ was not proportional
to $A(C)$.
When scaling up the loop the surface D bends in the interior of the AdS space. 
In the case when such a bending is limited by the range of the radial 
coordinate
one gets an area law. This is the case in the models at hand, in which
the coordinate $u$ in (\ref{3dmetric}) 
is bounded from below by $u_0$ as in figure \ref{string_fig}.

The evaluation of the classical action of the string worldsheet 
bounded by the loop $C$ at infinite $u$ is straightforward, 
as done in section \ref{wilsonloops}
\cite{Rey:1998wp,Brandhuber:1998wl}.
The string minimizes its length by going to the region with the
smallest possible metric component $g_{ii}$ (where $i$ labels the
$\IR^3$ directions), from which it gets the contribution to the string
tension.  The smallest value of $g_{ii}$ in the metric
(\ref{3dmetric}) is at the horizon.  Thus, we find that the Wilson
loop exhibits an area law (\ref{area}), where the string tension is
given by the $g_{ii}$ component of the metric (\ref{3dmetric})
evaluated at the horizon $u=u_0$ times a numerical factor
$\frac{1}{2\pi}$ :
\beq
\sigma = \frac{1}{2 \pi} \sqrt{4 \pi g_s N} u_0^2 = 
\frac{(g_sN)^{1/2}}{4\sqrt{\pi} R_0^2} \ .
\label{tension}
\eeq

\begin{figure}[htb]
\begin{center}
\epsfxsize=1.5in\leavevmode\epsfbox{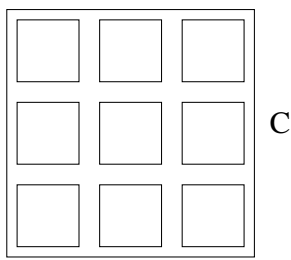}
\end{center}

\caption{The leading contribution in the lattice 
strong coupling expansion to the string tension
is the minimal tiling by plaquettes of the Wilson loop $C$.
}
\label{strong}
\end{figure}

The way supergravity exhibits confinement has an analog in the lattice
strong coupling expansion, as first demonstrated by Wilson
\cite{Wilson:1974co}.  The leading contribution in the lattice strong
coupling expansion to the string tension is the minimal tiling by
plaquettes of the Wilson loop $C$ as we show in figure \ref{strong}.
This is analogous to the minimal area of the string worldsheet $D$
ending on the loop $C$ in figure \ref{string_fig}. One important
difference is that in the supergravity description the space is
curved.  Of course, a computation analogous to the Wilson loop
computations we described in section \ref{wilsonloops}
which would be done in flat
space would also
exhibit confinement, since the minimal area of the string worldsheet
$D$ ending on the loop $C$ is simply the area enclosed by the loop
itself.

\begin{figure}[htb]
\begin{center}
\epsfxsize=2in\leavevmode\epsfbox{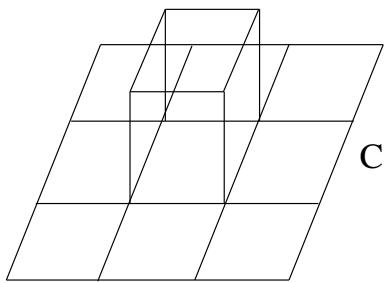}
\end{center}
\caption{Subleading contribution in the lattice strong coupling 
expansion to the string tension,
which is 
a non minimal tiling of the Wilson loop. This is the lattice analog
of the fluctuations of the string worldsheet. 
}
\label{strong2}
\end{figure}

The quark-antiquark linear potential $V = \sigma L$ can have
corrections arising from the fluctuations of the thin tube (string)
connecting the quark and antiquark.  L\"uscher studied a leading
correction to the quark-antiquark potential at large separation $L$.
Within a class of bosonic effective theories in flat space that
describe the vibrations of the thin flux tubes he found a universal
term, $-c/L$, called a L\"uscher term \cite{Luscher:1981sb}~:
\beq
V = \sigma L - c/L \ .
\eeq
For a flux tube in $d$ space-time dimensions $c= (d-2)/24 \pi$.
Lattice QCD calculations of the heavy quark potential have not
provided yet a definite confirmation of this subleading term.  This
term can also not been seen order by order in the lattice strong
coupling expansion.  Subleading terms is this expansion are of the non
minimal tiling type, as in figure \ref{strong2}, and correct only the
string tension but not the linear behavior of the potential.

The computation of the vacuum expectation value of the Wilson loop
(\ref{minarea}) based on the minimal area of the string worldsheet $D$
does not exhibit the L\"uscher term \cite{Greensite:1998ro}.  This is
not surprising. Even if the the L\"uscher term exists in QCD$_3$, it
should originate from the fluctuations of the string worldsheet
(\ref{sumarea}) that have not been taken into account in
(\ref{minarea}).  Some analysis of these fluctuations has been done
in \cite{Greensite:1999wf}, 
but the full computation has not been carried out yet.

Other works on confinement as seen by a dual supergravity description are
\cite{Dorn:1998tu,Kinar:1998xx,Danielsson:1999wt,
Naik:1999bs}.

\subsubsection{Mass Spectrum}

If the dual supergravity description is in the same universality class
as QCD$_3$ it should exhibit a mass gap.  In the following we will
demonstrate this property. We will also compute the spectrum of lowest
glueball masses in the dual supergravity description.  They will
resemble qualitatively the strong coupling lattice picture.  We will
also discuss a possible comparison to lattice results in the continuum
limit.

The mass spectrum in pure QCD can be obtained by computing the
correlation functions of gauge invariant local operators (glueball
operators) or Wilson loops, and looking for the particle poles.  As we
discussed extensively before, correlation functions of local operators
are related (in some limit)
to tree level amplitudes in the dual supergravity
description.  We will consider the two point functions of glueball
operators $\cO$ (for instance, we could take $\cO = \tr(F^2)$).  
For large $|x-y|$ it has an expansion of the form
\beq
\langle \cO(x) \cO(y) \rangle \simeq \sum c_i exp(-M_i |x-y|) \ ,
\label{corr}
\eeq
where $M_i$ are called the glueball masses.

We will classify the spectrum of
glueballs by $J^{PC}$ where $J$ is the glueball spin, $P$ its parity
and $C$ its charge conjugation eigenvalue. 
The action of $C$ on the gluon fields is \cite{Mandula:1983cl}
\beq
C: A_{\mu}^a T^a_{ij} \rightarrow -A_{\mu}^a T^a_{ij} \ ,
\eeq
where the $T^a$'s are the hermitian generators of the gauge group.
In string theory, charge conjugation corresponds to the worldsheet parity
transformation changing the orientation of the open strings attached to the
D-branes.

Consider first the lowest mass glueball state.
It carries $0^{++}$ quantum numbers.
One has to identify a corresponding glueball
operator, namely a local gauge invariant operator with these quantum numbers.
The lowest  dimension operator with these properties is $\tr(F^2)$,
and we have to compute its two point function. 
To do that we need to identify first
the corresponding supergravity field that couples to it as a source
at infinite $u$.
This is the Type IIB dilaton field $\Phi$. 

The correspondence  between the gauge theory and the dual string 
theory picture asserts that in the SUGRA limit the 
computation of the correlation function amounts
to solving the field equation for $\Phi$ in the AdS black hole background
(\ref{3dmetric}),
\beq
\partial_{\mu}(\sqrt{g}g^{\mu\nu}\partial_{\nu} \Phi) = 0 \ .
\label{Laplace}
\eeq

In order to find the lowest mass modes we consider solutions of $\Phi$
which are independent of the angular coordinate $\tau$ and take the
form $\Phi = f(u)e^{ikx}$.  Plugging this in (\ref{Laplace}) we obtain
the differential equation
\beq
\partial_u [u(u^4-u_0^4)\partial_u f(u)] + M^2 uf(u) = 0,~~~~~ M^2 = -k^2 \ .
\label{difeq}
\eeq
The eigenvalues $M^2$ of this equation are the glueball masses squared.

At large $u$ equation (\ref{difeq}) has two independent solutions, 
whose asymptotic behavior is $f
\sim constant$ and $f\sim 1/u^4$. We consider normalizable solutions
and choose the second one.  Regularity requires the vanishing of the
derivative of $f(u)$ at the horizon.  The eigenvalues $M^2$ can be
determined numerically \cite{Csaki:1998gm,Koch:1998eo, Zyskin:1998an}, 
or approximately via WKB techniques \cite{Csaki:1998gm, Minahan:1999tm}.

One finds that:\\
(i) There are no solutions with eigenvalues $M^2 \leq 0$.\\
(ii) There is a discrete set of eigenvalues $M^2 > 0$.

This exhibits the mass gap property of the supergravity picture.  In
fact, even without an explicit solution of the eigenvalues $M^2$ of
equation (\ref{difeq}), the properties (i) and (ii) can be deduced
from the structure of the equation and the requirement for
normalizable and regular solutions \cite{Witten:1998zw}.

The $0^{++}$ mass spectrum in the WKB approximation closely agrees with the
more accurate numerical solution.
It takes the form
\beq
M^2_{0^{++}} \approx \frac{1.44 n(n+1)}{R_0^2},~~~~ n=1,2,3,\cdots \ .
\label{spectrum}
\eeq

The mass spectrum (\ref{spectrum}), that corresponds to a massless
mode of the string in ten dimensions, is proportional to the cutoff
$1/R_0$ and not to $\sigma^{1/2}$, which is bigger by a power of $g_s
N$ (\ref{tension}).  This is qualitatively similar to what happens in
strong coupling lattice QCD with lattice spacing $a$. As we will
discuss in the next section, in the strong coupling lattice QCD
description the lowest masses of glueballs are proportional to $1/a$.
Note that in a stringy description of QCD we would expect the
glueballs to correspond to
string excitations, which  are expected to have masses of order
$\sigma^{1/2}$. Therefore in the supergravity limit, $g_sN \gg 1$, the
glueballs that correspond to the string excitations are much heavier
than the ``supergravity glueballs'' which we analyzed.

The natural scale for the glueball masses of continuum QCD$_3$ is  
$g_3^2N$. Therefore to get to the continuum QCD$_3$ region we have to
require $g_3^2N \ll 1/R_0$ which implies $g_s N \ll 1$.
As discussed above, our computation is performed in
the opposite limit $g_s N \gg 1$.
In particular, we do not have control 
over possible mixing between glueball states
and the other scalars and fermionic degrees of freedom which are 
at the same mass scale $1/R_0$ in the field theory.

We can attempt a numerical comparison of the supergravity computations
with the continuum limit of lattice QCD, obtained by taking the bare
coupling to zero. Since these are computations at two different limits
of the coupling value (of the original $\cn=4$ theory) there is
apriori no reason for any agreement.  Curiously, it turns out that
ratios of the glueball excited state masses with $n > 1$ in
(\ref{spectrum}) and the lowest mass $n=1$ state are in reasonably
good agreement with the lattice computations (within the systematic
and statistical error bars)
\cite{Csaki:1998gm,Teper:1998sg}.

As a second example consider the spectrum of $0^{--}$ glueball masses.
It can be computed via the field equations of the NS-NS
2-form field. 
The details of the computation can be found in \cite{Csaki:1998gm} and,
as in the $0^{++}$ case, the
ratios of the glueball masses 
are found to be in good agreement
with the lattice computations. 

In closing the numerical comparison we note another curious agreement
between the supergravity computation and the weak coupling lattice
computations. This is for the
ratio of the lowest mass
$0^{++}$ and $0^{--}$ glueball
states,
\beqar
&\left(\frac{M_{0^{--}}}{M_{0^{++}}}\right)_{{\rm supergravity}}&= 1.50, 
\nonumber\\
&\left(\frac{M_{0^{--}}}{M_{0^{++}}}\right)_{{\rm lattice~~~~~}}& =
 1.45\pm 0.08 \ .
\eeqar

As stressed above, the regime where we would have liked to compute the
mass spectrum is in the limit of small $g_s N$ (or large ultraviolet
cutoff $1/R_0$).  In this limit the background is singular and we have
to use the string theory description,  which we lack.  We can compute
the subleading correction in the strong coupling expansion to the
masses. This requires the inclusion of the $\alpha'^3$ corrections to
the supergravity action.  The typical form of the masses is
\beq
M^2 = \frac{c_0 + c_1 \alpha'^3/R^6}{R_0^2}  \ ,
\eeq
with $c_0$ as in (\ref{spectrum}).
The background metric is modified by the inclusion of the $\alpha'^3 \cR^4$
string correction to the supergravity action. The modified
metric has been derived in
\cite{Gubser:1998nz,Pawelczyk:1998pb}.
Based on this metric
the corrections to the masses $c_1$ have been computed in \cite{Csaki:1998gm}.
While these corrections significantly change the glueball masses, 
the corrections to the
mass ratios turn out to be relatively small.

Lattice computations may exhibit lattice artifacts due to the finite
lattice spacing. Removing them amounts to taking a sufficiently small
lattice spacing such that effectively the right physics of the
continuum is captured. Getting close to the continuum means, in
particular, that deviations from Lorentz invariance are minimized.

Analogous ``artifacts'' are seen in the dual supergravity description.
They correspond to Kaluza-Klein modes that are of the same mass scale
as the glueball mass scale.  There are Kaluza-Klein modes from the
circle coordinate $\tau$ in (\ref{3dmetric}) that provides the cutoff
to the three dimensional theory.  They have a typical mass scale of
order $1/R_0$.  There are also $SO(6)$ non-singlet Kaluza-Klein modes
from the five-sphere in (\ref{3dmetric}). In the field theory they
correspond to operators involving the $SO(6)$ non-singlet scalar and
fermion fields of the high-energy theory. They have a mass scale of
order $1/R_0$ too.

The inclusion of the subleading $\alpha'^3$ correction does not make
the Kaluza-Klein modes sufficiently heavy to decouple from the
spectrum \cite{Ooguri:1998ga,Csaki:1998gm}.  This means that the dual
supergravity description is also capturing physics of the higher
dimensions, or of the massive scalar and fermion fields from the point
of view of QCD$_3$.  One hopes that upon inclusion of all the
$\alpha'$ corrections, and taking the appropriate limit of small
$g_sN$ (or large cutoff $1/R_0$), these Kaluza-Klein modes will
decouple from the system and leave only the gauge theory degrees of
freedom.  Currently, we do not have control over the $\alpha'$
corrections, which requires an understanding of a two dimensional sigma
model with a RR background.  In section \ref{diffsuba} 
we will use an analogy with
lattice field theory to improve on our supergravity description and
remove some of the Kaluza-Klein modes.

\subsection{QCD$_4$}

One starting point for obtaining QCD$_4$ is the $(2,0)$ superconformal
theory in six dimensions realized on $N$ parallel coinciding
M$5$-branes, which was discussed in section \ref{adsmore}.  The
compactification of this theory on a circle of radius $R_1$ gives a
five-dimensional theory whose low-energy effective theory is the
maximally supersymmetric $SU(N)$ gauge theory, with a gauge coupling
constant $g_5^2 = 2 \pi R_1$.  To obtain QCD$_4$, one compactifies
this theory further on another ${\bf S}^1$ of radius $R_0$. The
dimensionless gauge coupling constant $g_4$ in four dimensions is
given by $g_4^2 = g_5^2/(2 \pi R_0) =  R_1/R_0$. As in the previous case,
to break supersymmetry one imposes the anti-periodic boundary
condition on the fermions around the second ${\bf S}^1$. And, as in
the previous case, to really get QCD$_4$ we need to require that the
typical mass scale of QCD states, $\Lambda_{QCD}$, will be much smaller
than the other mass scales in our construction ($1/R_1$ and $1/R_0$),
and this will require going beyond the supergravity
approximation. However, one can hope that the theory obtained from the
supergravity limit will be in the same universality class as QCD$_4$,
and we will give some evidence for this.

As discussed in section \ref{adsmore}, the large $N$ limit of the
six-dimensional theory is $M$ theory on AdS$_7 \times {\bf S}^4$. Upon
compactification on the two circles and imposing anti-periodic
boundary conditions for the fermions on the second ${\bf S}^1$, we
get $M$ theory on a black hole background \cite{Witten:1998zw}.
Taking the large $N$ limit while keeping the 't Hooft parameter $2 \pi
\lambda = g_4^2 N$ finite requires $R_1 \ll R_0$. We can now use the
duality between M theory on a circle and Type IIA string theory, and
the M$5$ brane wrapping on the ${\bf S}^1$ of radius $R_1$ becomes a
D$4$ brane. The large $N$ limit of QCD$_4$ then becomes Type IIA
string theory on the black hole geometry given by the metric
\beq
ds^2 = \frac{2 \pi \lambda}{3u_0}u
\left(4u^2 \sum_{i=1}^4 dx_i^2
+ \frac{4}{9u_0^2}u^2(1- \frac{u_0^6}{u^6})  d \tau^2 +
4\frac{du^2}{u^2(1- \frac{u_0^6}{u^6})} + 
d\Omega_4^2 \right) \ ,
\label{D4}
\eeq
with a non constant dilaton background
\beq
e^{ 2\phi} = \frac{8 \pi \lambda^3 u^3}{27 u_0^3  N^2} \ .
\label{dilaton_qcd4}
\eeq
The coordinates $x_i, i=1,..,4$, parametrize the 
$\IR^4$ gauge theory space-time, the coordinate
$u_0 \leq u \leq \infty$, 
and $\tau$ is an angular coordinate with period $2 \pi$.   
The location of the
horizon is  at $u=u_0$, 
which is related to the radius $R_0$
of the compactifying circle as
\beq
u_0 = {1 \over 3 R_0} \ .
\eeq

Equivalently, we could have started with the five dimensional 
theory on the worldvolume 
of $N$ D4 branes and heated it up to a finite 
temperature $T=1/2 \pi R_0$. 
Indeed, the geometry (\ref{D4}) with the dilaton
background (\ref{dilaton_qcd4}) is the near horizon geometry
of the non-extremal D4 brane background. But again,  when we 
Wick rotate (\ref{D4}) back to Lorentzian signature we take one of 
the coordinates $x_i$ as time. Notice that the string coupling
(\ref{dilaton_qcd4}) goes as $1/N$.

\subsubsection{Confinement}

QCD$_4$ at large distances is expected to confine with a linear
potential $V(r) \sim \sigma r$ between non-singlet states.  Therefore,
the vacuum expectation value of the Wilson loop operator is expected
to exhibit an area law behavior.  In order to see this in the dual
description we follow the same procedure as in QCD$_3$.
 
The string tension $\sigma$ is given by the coefficient of the term
$\sum_{i=1}^4 dx_i^2$ in the metric (\ref{D4}), evaluated at the
horizon $u=u_0$, times a $\frac{1}{2\pi}$ numerical factor :
\beq
\sigma = \frac{4}{3} \lambda u_0^2 = \frac{ 4\lambda}{27 R_0^2} \ .
\label{tension4}
\eeq

In QCD$_4$ it is believed that confinement is a consequence of the
condensation of magnetic monopoles via a dual Meissner effect.  Such a
mechanism has been shown to occur in supersymmetric gauge theories in
four dimensions
\cite{Seiberg:1994mc}.
This has also been demonstrated to some extent on the lattice via the
implementation of the 't Hooft Abelian Projection \cite{Hooft:1981to}.
We will now see that this appears to be the mechanism also in the dual
string theory description \cite{Gross:1998gk}.

\begin{figure}[htb]
\begin{center}
\epsfxsize=3in\leavevmode\epsfbox{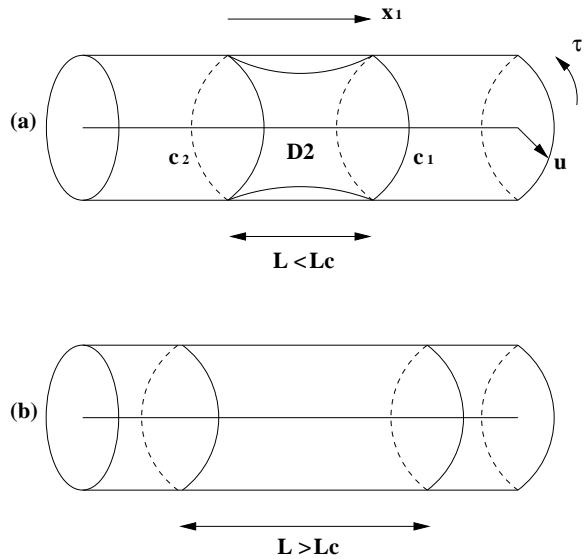}
\end{center}
\caption{The magnetic monopole is a string in five dimensions and
the four dimensional monopole is obtained by wrapping
the string on the circle. 
The potential between a monopole (wrapped on $c_1$) and an anti-monopole
(wrapped in the opposite orientation on $c_2$),
separated by a distance $L$ in the $x_1$ direction,
amounts to computing the action of a D2-brane which 
mediates the force between them
as in figure (a). 
For $L > L_c$ 
there is no minimal volume D2 brane configuration that connects
the monopole and the anti-monopole and the energetically favorable
configuration is as in figure (b), and then
the magnetic charge is screened.
}
\label{Monopole}
\end{figure}

Consider the five dimensional theory on the world volume of the D4
branes.  A magnetic monopole is realized as a D2 brane ending on the
D4 brane \cite{Douglas:1996du}. It is a string in five dimensions.
Upon compactification on a circle, the four dimensional monopole is
obtained by wrapping the string on the circle. We can now compute the
potential between a monopole and anti-monopole.  This amounts to
computing the action of a D2-brane interpolating between the monopole
and the anti-monopole, which mediates the force between them as in
figure \ref{Monopole}(a).  This is the electric-magnetic dual of the
computation of the quark-anti-quark potential described above.

If the pair is separated by a distance $L$ in the $x_1$ direction,
and stretches along the $x_2$ direction (which we can interpret as the
Euclidean time), the D2 brane coordinates are
${\tau,x_1, x_2}$. 
The action per unit length in the $x_2$ direction is given by 
\beq
V= \frac{1}{(2 \pi)^2 {\alpha'}^{3/2}}\int_0^L d\tau dx_1 e^{-\phi} 
\sqrt{det G} \ ,
\label{V}
\eeq
where $G$ is the induced metric on the D2 brane worldvolume.
We have to find a configuration of the D2-brane that minimizes (\ref{V}).
For $L > L_c$ where (up to a numerical constant) $L_c \sim R_0$,
there is no minimal volume D2 brane configuration that connects
the monopole and the anti-monopole and the energetically favorable
configuration is as in figure \ref{Monopole}(b).
Therefore there is no force between the monopole and the anti-monopole, which 
means that the magnetic charge is screened.
At length scales $L \gg R_0$ 
we expect pure QCD$_4$ as the effective description.
We see that in this region confinement is accompanied by 
monopole condensation, as we expect.

\subsubsection{$\theta$ Vacua}

In addition to the gauge coupling, four dimensional gauge theories
have an additional parameter $\theta$ which is the coefficient of the
$\tr(F\wedge F)$ term in the Lagrangian.  The $\theta$ angle
dependence of asymptotically free gauge theories captures non trivial
dynamical information about the theory.  Unlike in spontaneously broken
gauge theories, it cannot be analyzed by an instanton expansion.  What
is required is an appropriate effective description of the theories at
long wavelengths.  Such an effective description is provided by the
lattice.  However, since the Lorentz invariance is lost by the
discretization of space time, it is very difficult to study questions
such as the behavior of the system under $\theta \rightarrow \theta +
2\pi$.  Also, the construction of instantons which are the relevant
objects in the analysis of the $\theta$ dependence is a rather non
trivial task and involves delicate cooling techniques.

Another effective description may be provided by the description of
the four dimensional gauge theories by the M5 brane wrapping a non
supersymmetric cycle.  Indeed, in this formalism, one sees that the
vacuum energy exhibits the correct $\theta$ angle behavior in softly
broken supersymmetric gauge theories \cite{Oz:1998ba}.

In this subsection we use the dual string theory description to
analyze the $\theta$ angle dependence in large $N$ $SU(N)$ gauge
theory \cite{Witten:1998td}.  Since the amplitude for an instanton is
weighted by a factor $exp(-8 \pi^2 N/ \lambda)$ where $\lambda$ is the
't Hooft parameter (which we keep fixed), it naively seems that the
instanton effects vanish as $N\rightarrow \infty$.  However, unlike
the $\cN=4$ gauge theory for instance, here one expects this not to be
the case due to IR divergences in the theory.

Let us first review what we expect the behavior of the $\theta$
dependence to be from the field theory viewpoint.
The Yang-Mills action is
\beq 
I_{YM} = \int d^4x \tr( {N \over 4 \lambda} F^2 + {\theta
\over 16 \pi^2} F \tilde{F} ) \ .
\label{YM}
\eeq
At large $N$ we expect the energy of the vacuum 
to behave like
$E(\theta) = N^2 C(\theta/N)$.
The $N^2$ factor is due to the fact that this is the order of the number
of degrees of freedom (this also follows from the standard scaling of
the leading diagrams in the 't Hooft limit).
The dependence on $\theta/N$ follows from (\ref{YM}) 
as is implied by the large $N$ limit.
$\theta$ is chosen to be periodic with period $2 \pi$. Since the physics
should not change under 
$\theta \rightarrow \theta + 2 \pi$ we require
that 
$E(\theta + 2 \pi ) = E(\theta)$.

These conditions cannot be satisfied by a smooth function
of $\theta/N$. They can be satisfied by a multibranched function with 
the interpretation 
that there are $N$ inequivalent
vacua, and all of them are stable in the large $N$ limit.
The vacuum energy is then given by a minimization of 
the energy of the $k^{th}$ vacuum $E_k$
with respect 
to $k$
\beq
E(\theta) =  \min_k E_k(\theta) =  N^2 \min_k C((\theta+ 2 \pi k)/N) \ ,
\eeq
for some function $C(\theta)$ which is quadratic in $\theta$ for small
values of $\theta$.

\begin{figure}[htb]
\begin{center}
\epsfxsize=4in\leavevmode\epsfbox{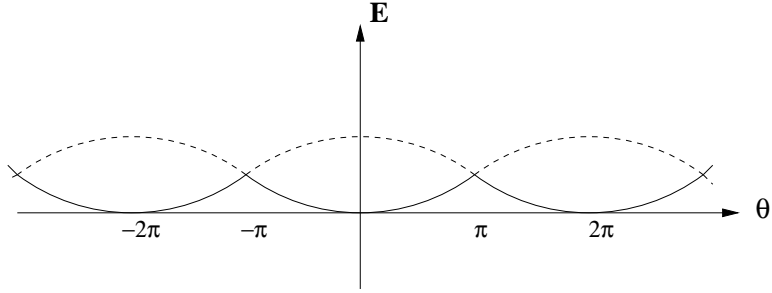}
\end{center}
\caption{The energy of the vacuum is expected to be a multibranched function.
}
\label{theta}
\end{figure}

The function $E(\theta)$ is periodic in $\theta$ and jumps at some
values of $\theta$ between different branches. The CP transformation
acts by $\theta \rightarrow -\theta$ and is a symmetry only for
$\theta=0,\pi$. Therefore, $C(\theta) = C(-\theta)$.  One expects an
absolute minimum at $\theta=0$ and a non vanishing of the second
derivative of $E(\theta)$ with respect to $\theta$, which corresponds
to the topological susceptibility $\chi_t$ of the system as we will
discuss later.  Taking all these facts into account one conjectures in
the leading order in $1/N$ that \cite{Witten:1980ln}
\beq
E(\theta) =  \chi_t \min_k (\theta+ 2 \pi k)^2 + O(1/N) \ ,
\label{vacuumenergy}
\eeq 
where $\chi_t$ is positive and independent of $N$.  At $\theta=\pi$
the function exhibits the jump between the vacua at $k=0$ and $k=-1$
and the spontaneous breaking of CP invariance.

In order to analyze the $\theta$ dependence in the dual string theory
description with the background (\ref{D4}) we have to identify the
$\theta$ parameter.  This is done by recalling that the effective
Lagrangian of $N$ D4 branes in Type IIA string theory has the coupling
\beq
\frac{1}{16 \pi^2}\int d^5x \varepsilon^{\rho\alpha\beta\gamma\delta}
\cA_{\rho} \tr(F_{\alpha\beta} F_{\gamma\delta}) \ ,
\label{FFcoupling}
\eeq
where $\cA$ is the Type IIA RR 1-form and $F$ is the $U(N)$ gauge
field strength on the five dimensional brane worldvolume.  Upon
compactification of the D4 brane theory on a circle we see that the
four dimensional $\theta$ parameter is related to the integral of the
RR 1-form on the circle.  Since it is a ten dimensional field it is 
a parameter from the worldvolume  point of view.

In the dual description we define the parameters at infinite $u$.  The
$\theta$ parameter is defined as the integral of the RR 1-form
component on the circle at infinite $u$
\beq
\theta = \int d \tau \cA_{\tau} = 2 \pi \cA_{\tau}^{\infty} \ ,
\eeq
which is defined modulo $2\pi k, k \in \IZ$.

The action for the RR 1-form takes the form
\beq
I = \frac{1}{2 \kappa_{10}^2} \int d^{10}x \sqrt{g} \frac{1}{4}
g^{\alpha\alpha'}g^{\beta\beta'}(\partial_{\alpha}
\cA_{\beta} - \partial_{\beta}\cA_{\alpha})
(\partial_{\alpha'}
\cA_{\beta'} - \partial_{\beta'}\cA_{\alpha'}) \ ,
\label{actionRRone}
\eeq
and the equation of motion for $\cA$ is 
\beq
\partial_{\alpha}[\sqrt{g}g^{\beta\gamma}g^{\alpha\delta}(\partial_{\gamma}
A_{\delta} - \partial_{\delta}A_{\gamma})] = 0 \ .
\label{mode}
\eeq

The required solution $A_{\tau}(u)$ to (\ref{mode}), regular at $u=u_0$ and 
with vanishing field strength
at infinite $u$ (in order to have finite energy), takes the form 
\beq
A_{\tau}(u) = A_{\tau}^{\infty}(1- \frac{u_0^6}{u^6}) \ .
\label{solution_a}
\eeq

Evaluating the Type IIA action for the RR 1-form
(\ref{actionRRone}) 
with the solution (\ref{solution_a}) and recalling the $2 \pi 
\IZ$ ambiguity
we get the vacuum energy (\ref{vacuumenergy})
where $\chi_t$ is independent of $N$ \cite{Witten:1998td}.

In order to check that the vacua labeled by $k$ are all stable in the
limit $N\rightarrow \infty$ we need a way to estimate their lifetime.
The domain wall separating two adjacent vacua is constructed by
wrapping a D6 brane of Type IIA string theory on the $S^4$ part of the
metric \cite{Witten:1998td}.  Since the energy density of the brane at
weak coupling is of order $1/g_s$ where $g_s$ is the Type IIA
string coupling, as $N \rightarrow \infty$ (with fixed $g_s N$)
it is of order $N$. If we
assume a mechanism for the decay of a $k$-th vacuum via a D6 brane
bubble, its decay rate is of the order of $e^{-N}$.  Thus, there is an
infinite number of stable vacua in the infinite $N$ limit.
 
One can repeat the discussion of confinement in the previous subsection
for $\theta \neq 0$.
When $\theta = 2 \pi p/q$ with co-prime integers $p,q$ the confinement is
associated with a condensation of $(-p,q)$ dyons and realizes the
mechanism of oblique confinement.

\subsubsection{Mass Spectrum}

The analysis of the mass spectrum of QCD$_4$ as seen by  the dual
description in the supergravity limit is similar to the one we carried
out for QCD$_3$.
It is illuminating to consider an analogous picture of 
strong coupling lattice QCD \cite{Gross:1998gk}.

\begin{figure}[htb]
\begin{center}
\epsfxsize=1.6in\leavevmode\epsfbox{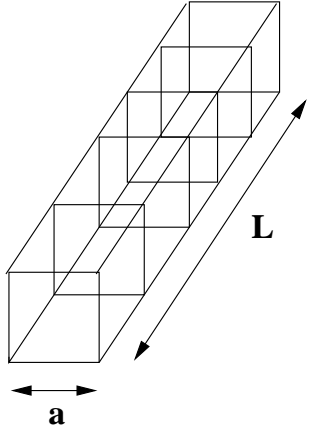}
\end{center}
\caption{The leading contribution in strong coupling lattice
QCD to the correlator of
two Wilson loops, separated by distance $L$,
is from a tube with the size of one plaquette
that connects the loops. 
This leads to the lowest mass glueballs having a mass of the order of
$1/a$, where $a$ is the lattice spacing.
}
\label{glueball}
\end{figure}

In strong coupling lattice QCD
the masses of the lightest glueballs 
are of order $1/a$ where $a$ is the lattice spacing.
The reasoning is that in strong coupling lattice
QCD the leading contribution to the correlator of
two Wilson loops separated by distance $L$
is from a tube with the size of one plaquette, as in figure \ref{glueball},
that connects the loops. 
With the Wilson lattice action the $0^{++}$ glueball mass is given by 
\cite{Creutz:1987xx}
\beq
M_{0^{++}}  =  -4 \log(g_{4}^2 N) a^{-1} \ .
\label{lattice}
\eeq

To make the connection with continuum QCD$_4$ we would like to sum the
lattice strong coupling expansion $M_{0^{++}} = F(g_{4}^2 N) a^{-1}$,
and take the limit $a \to 0$ and $g_4 \to 0$ with
\beq
g_{4}^2 N \simeq 
\frac{1}{b log(1/a\Lambda_{QCD})}~~~as~~~~a \rightarrow 0 \ ,
\label{limit}
\eeq
where $g_4$ is the four dimensional coupling and $b$ is the first coefficient
of the $\beta$-function.
We hope that in the limit (\ref{limit}) we will get a finite glueball mass
measured in $\Lambda_{QCD}$ units.

\begin{figure}[htb]
\begin{center}
\epsfxsize=5in\leavevmode\epsfbox{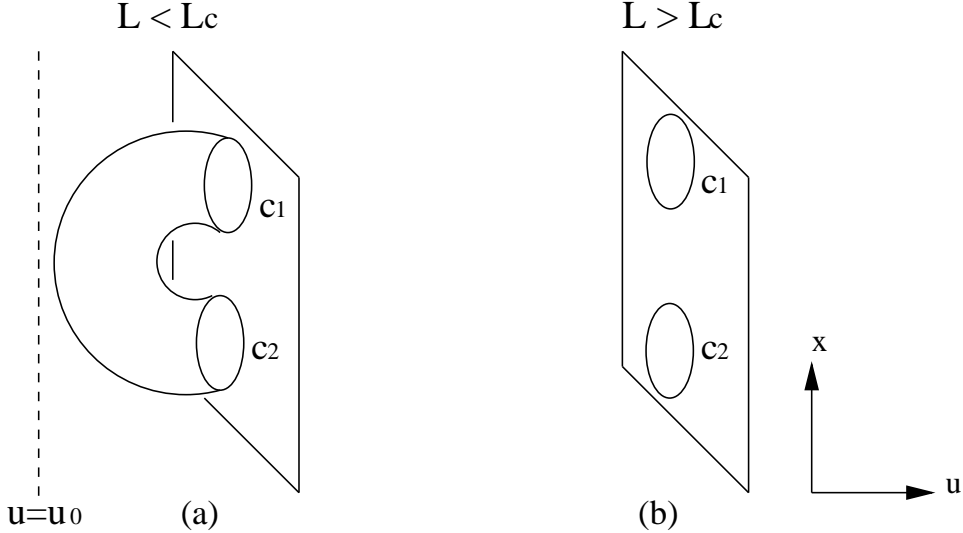}
\end{center}
\caption{The Wilson loop correlation function
in figure (a) is computed by minimization of the string worldsheet
that interpolates between them.  When the distance between the loops
$L$ is larger than $L_c$ there is no stable string worldsheet
configuration connecting the two loops as in figure (b).  }
\label{wilson}
\end{figure}

In the dual string theory description the analog of $a$ is $R_0$.
The strong coupling expansion is analogous to
the $\alpha'$ expansion of string theory.
Supergravity is the leading contribution in this expansion. 
The lowest glueball masses $M_g$ correspond to the zero modes
of the string, and their mass is proportional to $1/R_0$.
Another way to see that this limit 
resembles the strong coupling lattice QCD picture is 
to consider the Wilson loop correlation function
$\langle W(C_1) W(C_2) \rangle$ as in figure \ref{wilson}(a).

\begin{figure}[htb]
\begin{center}
\epsfxsize=2.8in\leavevmode\epsfbox{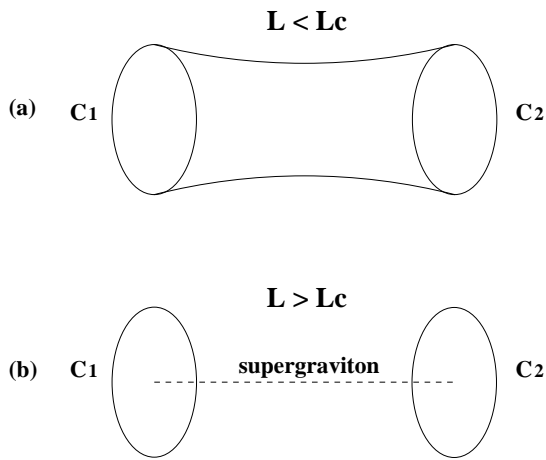}
\end{center}
\caption{The string worldsheet that connects the loops in figure (a)
collapses and 
the two disks are now connected by a tube of a string scale as in figure (b).
The correlation
function is mediated by a supergraviton exchange between the disks
and the  supergravitons are identified with the glueball states.
}
\label{graviton}
\end{figure}

For $L > L_c$, where $L$ is the distance
between the loops and $L_c$ is determined by the size of the
loops, there is no stable 
string worldsheet configuration
connecting the two loops, as in figure \ref{wilson}(b).
The string worldsheet that connects the loops as in figure \ref{graviton}(a)
collapses and 
the two disks are now connected by a tube of string scale size as in 
figure \ref{graviton}(b),
resembling the strong coupling lattice QCD picture.
The correlation
function is then mediated by a supergraviton exchange between the disks.
Thus, the  supergravitons are identified with the glueball states and the 
lowest glueball masses turn out to be proportional to $1/R_0$ 
\cite{Gross:1998gk}.

As in strong coupling lattice QCD,
to make the connection with the actual QCD$_4$ theory we need 
to sum the strong coupling expansion $M_g= F(g_{4}^2 N) / R_0$
and take the limit of $R_0 \to 0$ and $g_4 \to 0$ with
\beq
g_{4}^2 N \rightarrow \frac{1}{b log(1/R_0 \Lambda_{QCD})}~~~
as~~~~R_0\rightarrow 0 \ .
\label{limitsg}
\eeq
Again, 
we hope that in the limit (\ref{limitsg}) we will get a finite glueball mass
proportional to $\Lambda_{QCD}$.

In the limit (\ref{limitsg}) the background (\ref{D4}) is singular.
Thus, to work at large $N$ in this limit we need the full
tree level string theory description and not just the SUGRA limit.
The supergravity description will provide us with information analogous
to that of strong coupling lattice QCD with a finite cutoff.
However, since as discussed before
the regularization here is done via a higher dimensional theory,
we will have the advantage of a full Lorentz invariant description
in four dimensions.
What we should be worried about is whether we capture the physics of the
higher dimensions as well (which from the point of view of QCD$_4$ 
correspond to additional charged fields).

In order to compute the mass gap 
we consider the scalar glueball $0^{++}$.
The  $0^{++}$ glueball mass  spectrum is obtained by solving
the supergravity equation for any mode $f$ that couples
to $0^{++}$ glueball operators; we expect (and this is verified by the
calculation) that the lightest glueball will come from a mode that couples
to the operator
$\tr(F^2)$. There are several steps to be taken 
in order to identify this mode and its supergravity equation.
First, we consider small
fluctuations of the supergravity 
fields on the background (\ref{D4}), (\ref{dilaton_qcd4}).
The subtlety that arises is the need to disentangle the 
mixing between the dilaton field and
the volume factor which has been done in \cite{Hashimoto:1998ao}. 
One then plugs the appropriate ``diagonal''
combinations of these fields into the supergravity equations of motion.
The field/operator identification can then be 
done by considering the Born-Infeld action of
the D4 brane in the gravitational background. 

To compute the lowest mass modes
we consider solutions of the form $f = f(u) e^{ikx}$ 
which satisfy the equation
\beq
\frac{1}{u^3}\partial_{u}[u(u^6-u_0^6)\partial_{u}f(u)] + M^2 f(u) = 0 \ .
\eeq
The eigenvalues $M^2$ are the glueball masses.  The required solutions
are normalizable and regular at the horizon.  The eigenvalues $M^2$
can be determined numerically \cite{Hashimoto:1998ao} or approximately
via WKB techniques \cite{Minahan:1999tm}.

As in QCD$_3$ one finds that:\\
(i) There are no solutions with eigenvalues $M^2 \leq 0$.\\
(ii) There is a discrete set of eigenvalues $M^2 > 0$.

This exhibits the mass gap property of the supergravity picture.

The $0^{++}$ mass spectrum in the WKB approximation closely agrees with the
more accurate numerical solution.
It takes the form
\beq
M^2 \approx \frac{0.74 n(n+2)}{R_0^2},~~~~ n = 1,2,3,\cdots \ .
\label{spectrum4}
\eeq
As in QCD$_3$,  the ratios of the glueball excited state masses with $n > 1$
in (\ref{spectrum4}) and the lowest mass $n=1$ state are in good agreement
with the available lattice computations \cite{Hashimoto:1998ao, Csaki:1998gm}.

As another example consider the $0^{-+}$ glueballs.
The lowest dimension operator with these quantum numbers is $\tr(F \tilde{F})$.
As we discussed previously, 
on the D4 brane worldvolume it couples to the RR 1-form
$\cA_{\tau}$ (\ref{FFcoupling}).
Its equation of motion is given by (\ref{mode}).
We look for solutions of the form $\cA_{\tau} = f_{\tau}(u)e^{ikx}$.
Plugging this into (\ref{mode}) we get
\beq
\frac{1}{u^5}(u^6-u_0^6) \partial_u[u^7\partial_u f_{\tau}(u)] 
+ u^4 M^2 f_{\tau}(u)=0 \ .
\eeq
As for the $0^{++}$ glueball states, the 
ratios of the $0^{-+}$ glueball masses 
are found to be in good agreement
with the lattice computations 
\cite{Hashimoto:1998ao}.

Finally, we note that the
ratio of the lowest masses
$0^{++}$ and $0^{-+}$ glueball
states \cite{Hashimoto:1998ao}
\beqar
\left(\frac{M_{0^{-+}}}{M_{0^{++}}}\right)_{{\rm supergravity}}&= &1.20, 
\nonumber\\
\left(\frac{M_{0^{-+}}}{M_{0^{++}}}\right)_{{\rm lattice~~~~~~}}& =&
1.36 \pm 0.32 \ ,
\eeqar
agrees with the lattice results too.
Similar types of agreements 
in mass spectrum computations were claimed 
in strong coupling lattice QCD \cite{Munster:1983ps}.
However, note that (as discussed above for QCD$_3$)
other ratios, such as the ratio of the glueball masses
to the square root of the string tension, are very different in the SUGRA
limit from the results in QCD.

The computation of the mass gap in the dual supergravity picture is in
the opposite limit to QCD.  As in the supergravity description of
QCD$_3$, also here the Kaluza-Klein modes do not decouple.  In this
approach, in order to perform the computation in the QCD regime we
need to use string theory.  The surprising agreement of certain mass
ratios with the lattice results may be a coincidence. Optimistically,
it may have an underlying dynamical reason.

\subsubsection{Confinement-Deconfinement Transition}

We will now put the above four dimensional QCD-like theory 
at a finite temperature $T$ (which should not be confused with $\frac{1}
{2 \pi R_0}$).
We will  see that there is a deconfinement transition. 
In order to consider the theory at finite temperature we go to Euclidean
space and we compactify the time direction $t_E$ on a circle of radius 
$\beta$ with antiperiodic fermion 
boundary conditions. Since we already had
one circle (labeled by $\tau$ in (\ref{D4})), we now have two circles
with antiperiodic boundary conditions. So, we can have several possible
gravity solutions. One is the original extremal D4 brane, another
is the solution  (\ref{D4}) and a third one is the same solution
 (\ref{D4}) but with $\tau$ and $t_E$ interchanged. 
These last two solutions are possible only when the fermions have
antiperiodic boundary conditions on the corresponding circles. 
One of the last two solutions always has lower free energy than 
the first, so we concentrate on these last two.
 
It turns out that the initial solution (\ref{D4}) has the lowest free
energy for low temperatures, when $\beta = 1/T > 2 \pi R_0$,
 while the one with 
 $\tau \leftrightarrow t_E$  has the lowest free energy for 
$\beta = 1/T < 2 \pi R_0$ (high temperatures). 
The entropy of these two solutions is very different, and therefore
there is a first order phase transition, in complete analogy with
the discussion in section \ref{FiniteT}.
We do not know of a proof that there are no other solutions, but 
these two solutions have different topological properties, so
there cannot be a smoothly interpolating solution. In any case,
for very low and very high temperatures they are expected to be
the dominant configurations (see \cite{Horowitz:1999ha})\footnote{
There are other singular solutions  
 \cite{Russo:1998ze},
 but the general philosophy
is that we do not  allow singular solutions unless we can give a 
physical interpretation for the singularity.}.
 The entropy of the 
the high  temperature  phase is of order $N^2$, while the entropy of the 
low temperature  phase is essentially zero since the number of states in
the gravity picture 
is independent of the Newton constant. 

If we compute the potential between a quark and an antiquark then
in the low temperature phase it grows linearly, so that we have 
confinement, while in the high temperature phase the strings
coming from the external quarks can end on the horizon, so that
the potential vanishes beyond a certain separation. Thus, this
is  a confinement-deconfinement transition. It might seem a bit
surprising at first sight that essentially the same solution can 
be interpreted as a confined and a deconfined phase at the same time. The 
point it that quark worldlines are timelike, therefore they select
one of the two circles, and the physical properties depend crucially 
on whether this circle
is contractible or not in the full ten-dimensional geometry.

\subsubsection{Other Dynamical Aspects}
\label{other_dynamical}

In this subsection we comment 
on various aspects of QCD$_4$  as seen by the
string description. 
We first show  how the baryons appear in the dual string theory (M
theory) picture.
We will then compute other properties of the QCD vacuum, the topological
susceptibility and the gluon condensate, as seen in the dual description.

\medskip

{\it\bf Baryons}

\begin{figure}[htb]
\begin{center}
\epsfxsize=2.8in\leavevmode\epsfbox{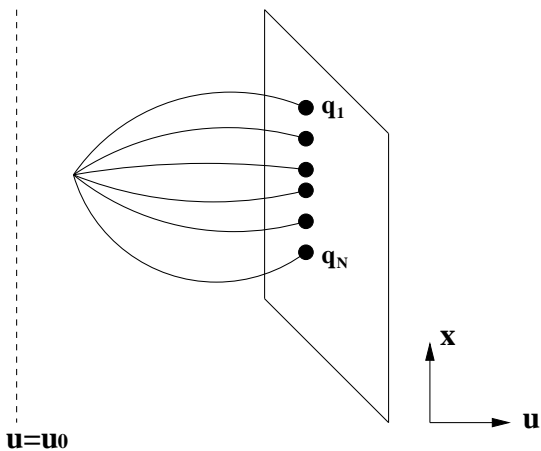}
\end{center}
\caption{The baryon is an $SU(N)$ singlet bound state of 
$N$ quarks. It is constructed as $N$ open strings that join together
at a point in the bulk AdS black hole geometry.
}
\label{baryon}
\end{figure}

\begin{figure}[htb]
\begin{center}
\epsfxsize=2in\leavevmode\epsfbox{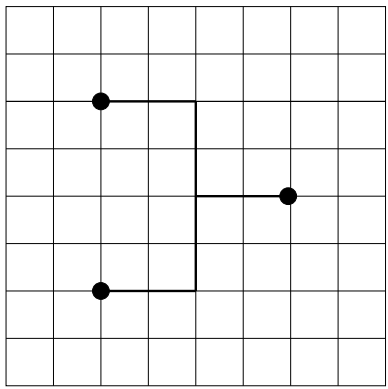}
\end{center}
\caption{A baryon state in strong coupling lattice QCD.
The quarks located at lattice sites 
are connected by flux links to a vertex. A similar picture is obtained
by projecting the baryon vertex in figure \ref{baryon} on $x$ space.}
\label{baryonlat}
\end{figure}

The baryon is an $SU(N)$ singlet bound state of 
$N$ quarks. Since we do not have quarks in our theory, we need to put
in external quarks as described in section \ref{wilsonloops}, 
and then there is a
baryon operator coupling $N$ external quarks.
As in the conformal case, also here it can be constructed as $N$ 
open strings that end on a D4 brane that is wrapped on 
$S^4$ 
\cite{Gross:1998gk,Witten:1998xy}, 
as in figure \ref{baryon}.  
If we view this geometry as arising from M-theory, then the strings
are M2 branes wrapping the circle with periodic fermion 
boundary conditions
 and the D4 brane is an M5 fivebrane also wrapping 
this circle. Then, $N$ M2 branes can end on this M5 brane as in 
\cite{Witten:1998xy}.
%
%
There is a very similar picture of a baryon in strong coupling
lattice QCD as is depicted in 
figure \ref{baryonlat}, where quarks are connected by flux links to a vertex.

Several aspects of baryon physics can be seen from the string picture of
figure \ref{baryon} 
\cite{Witten:1998xy, Gross:1998gk}.
The baryon energy is proportional to the string tension (\ref{tension4})
and (in the limit of large distances between the quarks) to the sum of
the distances between the $N$ quark locations and the location of the
baryon vertex in the four dimensional $x$-space
\cite{Gross:1998gk,Brandhuber:1998xy,Imamura:1998hf}. 
(There is some subtlety
in evaluating the baryon energy, and it was clarified
in \cite{Imamura:1998gk} in the case of ${\cal N}=4$ gauge theory. 
See also \cite{Callan:1998iq,Callan:1999zf}.)  
We may consider the baryon vertex 
as a fixed (non-dynamical) point in the Born Oppenheimer
approximation. In such an approximation, the $N$ quarks move
independently in the potential due to the string stretched between
them and the vertex.  The baryon mass spectrum can be computed by
solving the one body problem of the quark in this potential.
Corrections to this spectrum can be computed by taking into account
the potential between the quarks and the dynamics of the vertex.  A
similar analysis has been carried out in the flux tube model
\cite{Isgur:1985fs} based on the Hamiltonian strong coupling lattice
formulation \cite{Kogut:1975xx}.

In a confining theory we do not expect to see a baryonic configuration
made from $k < N$ quarks. This follows for the above description. If
we want to separate a quark we will be left with a string running to 
infinity, which has infinite energy.
%

\medskip

{\it\bf Topological Susceptibility}

The topological susceptibility $\chi_t$ measures the fluctuations of the
topological charge of the QCD vacuum. 
It is defined by
\begin{equation}
\chi_t=\frac{1}{(16\pi^2)^2} \int d^4x \langle \tr(F\tilde{F}(x))
  \tr(F\tilde{F}(0))\rangle \ .
  \label{topol}
\end{equation}
  
At large $N$ the Witten-Veneziano formula \cite{Witten:1979ca,
Veneziano:1979uv} relates the mass $m_{\eta'}$ in $SU(N)$ Yang-Mills
gauge theory with $N_f$ quarks to the topological susceptibility of
$SU(N)$ Yang-Mills theory without quarks:
\beq m_{\eta'}^2 = \frac{4N_f}{f_{\pi}^2} \chi_t \ .
\label{mass}
\eeq
Equation (\ref{mass})
is applicable at large $N$ where $f_{\pi}^2 \sim N$. In this limit
$m_{\eta'}$ goes to zero and we have the $\eta'-\pi$ degeneracy.

Nevertheless, plugging the phenomenological values $N_f=3, N=3,
m_{\eta'}\sim 1~ GeV,f_{\pi} \sim 0.1~ GeV$ in (\ref{mass}) leads to a
prediction $\chi_t \sim (180~ MeV)^4$, which is in surprising
agreement with the lattice simulation for a finite number of colors
\cite{Teper:1997pf}.

Evaluating the 2-point function from the type IIA SUGRA 
action for the RR 1-form
(\ref{actionRRone}) with the solution (\ref{solution_a}), we get the topological
susceptibility
\beq
\chi_t = \frac{2 \lambda^3}{729 \pi^3 R_0^4} \ .
\label{chit}
\eeq

The supergravity result (\ref{chit})
 depends on two parameters, $\lambda$ and
$R_0$. This is the leading  asymptotic
behavior in $1/\lambda$ of the full string theory expression 
$\chi_t \sim (F(\lambda) / R_0)^4$.
We would have liked to compute $F(\lambda)$, take the limit (\ref{limitsg})
and compare to the lattice QCD result.
However, this goes beyond the currently available calculational tools. 

It may be instructive, though, to consider the following comparison.
Let us assume that there is a cross-over between the supergravity
description and the continuum QCD description. We can estimate the
cross-over point.  In perturbative QCD we find $F(\lambda) \sim e^{-12 \pi/
11 \lambda}$, therefore the cross-over point (to the $F \sim
\lambda^{3/4}$ behavior of (\ref{chit})) can be estimated to be at
$\lambda \sim 12\pi/11$.  Also, since the mass scale in the
QCD regime is $\Lambda_{QCD}$, at the cross-over point $T = 1/2 \pi R_0 \sim
\Lambda_{QCD}\sim 200~ MeV$.  Of course, we should bear in mind that
at the cross-over point both the supergravity and perturbative QCD are
not applicable descriptions.  If we compare the topological
susceptibility (\ref{chit}) at the correspondence point with the
lattice result we get
\beq
\left({\chi_t^{{\rm SUGRA}}  \over 
\chi_t^{{\rm Lattice}}}\right)^{1/4} =  1.7 \ .
\eeq
It may be an encouraging sign that the number we get is of order one,
though its level of agreement is not as good as the mass ratios of the
glueball spectrum.

\medskip

{\it\bf Gluon Condensation}

The gluon condensate $\langle {1 \over 4 g_{4}^2}\tr(F^2(0)) \rangle$
is related by the trace anomaly to the energy density $T_{\mu\mu}$ of
the QCD vacuum.  In the supergravity picture the one point function of
an operator corresponds to the first variation of the supergravity
action. This quantity is expected to vanish by the equations of
motion. However, the first variation is only required to vanish up to
a total derivative term. Since asymptotically anti-de Sitter space has
a time-like boundary at infinity, there is a possible boundary
contribution. Indeed, unlike the $\cN=4$ case, the one point function
of the $\tr(F^2)$ operator in the dual string theory description of
QCD does not vanish.

It can be computed either directly or by using the relation between
the thermal partition function and the free energy $Z(T)=\exp(-{\cal
F}/T)$. This relates the free energy associated with the string theory
(supergravity) background to the expectation value of the operator
$\tr(F^2) $.  One gets \cite{Hashimoto:1998ao}
\beq
\langle {1\over 4g_{\rm 4}^2 }\tr(F^2_{\mu\nu}(0)) \rangle =
{1\over 8\pi } {N^2\over\lambda} \sigma^2  \ .
\label{ggl2}
\eeq

The relation (\ref{ggl2}) between the gluon condensate and the string
tension is rather general and applies for other regular backgrounds
that are possible candidates for a dual description
\cite{Csaki:1999ln}. 

If we attempt again a numerical comparison with the lattice computation 
\cite{Campostrini:1989gc,DElia1997fs} we find at the
cross-over point
\beq
\left({{({\rm Gluon\ condensate})}^{{\rm SUGRA}}  
\over {({\rm Gluon\ condensate})}^{{\rm Lattice}}}
\right)^{1/4}  =  0.9 \ .
\eeq
We should note that in field theory the gluon condensate is divergent, 
and there are subtleties (which are not
completely settled) as to the relation between the lattice regularized result
and the actual property of the QCD vacuum.

Finally, for completeness of the numerical status,
we note that if we compare the string tension (\ref{tension4}) 
at the cross-over point and the lattice result 
we get 
\beq
\left({{({\rm QCD\ string\ tension})}^{{\rm SUGRA}}  
\over {({\rm QCD\ string\ tension})}^{{\rm Lattice}}}\right)^{1/2}  = 2 \ .
\eeq
\subsection{Other Directions}

In this subsection we briefly review other possible ways of describing
non supersymmetric asymptotically free gauge theories via a dual
string description. Additional possibilities are described in section
\ref{deformations}.

\subsubsection{Different Background Metrics}
\label{diffsuba}

The string models dual to QCD$_p$ that we studied exhibit the required
qualitative properties, such as confinement, a mass gap and the
$\theta$ dependence of the vacuum energy, already in the supergravity
approximation.  We noted that besides the glueball mass spectrum there
exists a spectrum of Kaluza-Klein modes at the same mass scale.  This
indicates that the  physics of the higher
dimensions is not decoupling from the four dimensional 
physics\footnote{From the field
theory point of view it indicates that $SU(4)$-charged fields and KK
modes of five dimensional fields contribute
in addition to the four dimensional gluons.}.
The 
Kaluza-Klein states did not decouple upon the inclusion of the
$\alpha'^3$ correction, but one hopes that they do decouple in the
full string theory framework.  In the following we discuss an approach
to removing some of them already at the supergravity level. It should
be stressed, however, that this does not solve the issue of a possible
mixing between the glueball states and states that
correspond to the scalar and fermion fields,
which for large
$\lambda$ are at the same mass scale in the field theory.

Again, the analogy with lattice gauge theory is useful.  It is well
known in the lattice framework that the action one starts with has a
significant effect on the speed at which one gets to the continuum
limit.  One can add to the lattice action deformations which are
irrelevant in the continuum limit and arrive at an appropriate
effective description of the continuum theory while having a larger
lattice spacing.  Such actions are called improved actions.

A similar strategy in the dual supergravity description amounts to a
modification of the background metric. The requirement is that the
modification will better capture the effective description of the
gauge theory while still having a finite cutoff (corresponding to
finite $\lambda$ in our case).  On the lattice a criterion for
improvement is Lorentz invariance.  Here, since the cutoff is provided
by a higher dimensional theory we have the full Lorentz invariance in
any case. The improvement will be measured by the removal of the
Kaluza-Klein modes.  Note that we are attempting at an improvement in
the strong coupling regime.  Such ideas have only now begun to be
explored on the lattice \cite{Dalley:1998gc}. Till now, the effort of
lattice computations was directed at the computation of the strong
coupling expansion series.

Models that generalize the above background by the realization of the
gauge theories on non-extremal rotating branes have been studied in
\cite{Russo:1998nc, Csaki:1999ln,Cvetic:1999rb}.  The deformation of the background
is parametrized by the angular momentum parameter.  Kaluza-Klein modes
associated with the circle have the form $\Phi = f(u)
e^{ikx}e^{in\tau}, n > 0$.  It has been shown that as one varies the
angular momentum one decouples these Kaluza-Klein modes, while
maintaining the stability of the glueball mass spectrum.  This
deformation is not sufficient to decouple also the Kaluza-Klein modes
associated with the sphere part of the metrics (\ref{3dmetric}) and
(\ref{D4}), so we are still quite far from QCD.

The number of non-singular backgrounds is limited by the no hair
theorem. One may consider more angular momenta, for instance.
However, this does not seem to be sufficient to decouple all the
Kaluza-Klein states \cite{Russo:1999rd,Russo:1999sm}.  It is possible
that we will need to appeal to non regular backgrounds in order to
fully decouple the higher dimensional physics.  Some non
supersymmetric singular backgrounds of Type II supergravity that
exhibit confinement were constructed and discussed in
\cite{Kehagias:1999tr,Gubser:1999pk,Girardello:1999hj,Constable:1999ch}.

\subsubsection{Type 0 String Theory}

The Type 0 string theories have worldsheet supersymmetry but no
space-time supersymmetry as a consequence of a non-chiral GSO
projection \cite{Dixon:1986ba,Seiberg:1986ss}.  Consider two types of
such string theories, Type 0A and Type 0B.  They do not have
space-time fermions in their spectra. Nevertheless, they have a
modular invariant partition function.  The bosonic fields of these
theories are like those of the supersymmetric Type IIA and Type IIB
string theories, with a doubled set of Ramond-Ramond fields.  Type 0
string theories can be formally viewed as the high temperature limit of the
Type II string theories.  They contain a tachyon field $\cT$.

Type 0 theories have D-branes. As in the Type II case, we can consider
the gauge theories on the worldvolume of $N$ such branes.  These
theories do not contain an open string tachyon.  Moreover,  the usual
condensation of the tachyon could be avoided in the near horizon region
as we explain below.

One particular example studied in \cite{Klebanov:1998db} is the theory
on $N$ flat D3 branes in Type 0B theory.  Since there is a doubled set
of RR 4-form fields in Type 0B string theory, the D3 branes can carry
two charges, electric and magnetic.  The worldvolume theory theory of
$N$ flat electric D3 branes is a $U(N)$ gauge theory with six scalars
in the adjoint representation of the gauge group.  There are no
fermionic fields.  The classical action is derived by a dimensional
reduction of the pure $SU(N)$ gauge theory action in ten dimensions.
The six scalars are the components of the gauge fields in the reduced
dimensions. The
 classical theory has an $SO(6)$ global symmetry that rotates the
six scalars.  This allows several possible parameters (from the point
of view of renormalizable field theory)~: a gauge coupling $g_{YM}$, a
mass parameter for the scalars $m$ and various scalar quartic
potential couplings $g_i$, one of which appears in the classical 
  Lagrangian.  In the classical worldvolume 
 action, the mass
parameter is zero and the $g_i$ are fixed in terms of $g_{YM}$, it
is just the dimensional reduction of the ten dimensional 
bosonic Yang-Mills theory.
Quantum mechanically, the parameters are corrected differently and can
take independent values.  The theory has a phase diagram depending on
these parameters.  Generically we expect to see in the diagram
Coulomb-like (Higgs) phases, confinement phases and maybe non trivial
RG fixed points arising from particular tunings of the parameters.

As in the case of D branes in Type II theories, one
can conjecture here that the low-energy theory on the electric D3 branes 
has a dual non supersymmetric
string description. At first sight this should   involve a
solution of $AdS_5\times S^5$ type. 
The closed string tachyon might be allowed in $AdS$ if the curvature
is of the order of the string scale, since in that case the 
tachyon would obey the  Breitenlohner-Freedman bound (\ref{posbound}).
The fact that the curvatures are of the order of the string scale
renders the gravity analysis invalid. In principle we should
solve the worldsheet string theory. Since we  do not know how to 
do that at present
we can just do a gravity analysis and hope that the full string
theory analysis will give similar results. 
It was observed  in \cite{Klebanov:1998db} that the tachyon 
potential includes the terms
\beq
 {1 \over 2} m^2 e^{-2 \Phi} {\cT}^2 + 
|\cF|^2 \left( 1 + \cT  + { {\cT}^2 \over 2}  \right) \ ,
\label{Tcouple}
\eeq
where $\cF$ is the electric  RR five form field strength (the magnetic
one  
couples in a similar way but with $\cT \to - \cT$). 
The fact that the RR fields contribute positively to the mass
allows  curvatures which, numerically, 
are a bit less than the string scale. Furthermore, 
it has been noticed  in 
\cite{Klebanov:1998yy} that the first string correction to this
background seems to vanish. These conditions on the curvature 
translate into the condition  $g_sN < O(1)$ which is precisely what
we expect to get in QCD.
%

An interesting feature is that,
due to the potential (\ref{Tcouple}) the tachyon would have a nonzero
expectation value and that 
 induces a variation of the dilaton field $\Phi$
in the radial coordinate via the equation 
\cite{Minahan:1999tm,Klebanov:1998yy}
\beq
\nabla^2 \Phi = \frac{1}{8}m^2 e^{\Phi/2}\cT^2,~~~~~~m^2 = 
-\frac{2}{\alpha'} \ .
\eeq
Since the radial coordinate is associated with the energy scale of the
gauge theory, this variation may be interpreted as the flow of the
coupling.  In the UV (large radial coordinate) the tachyon is constant
and one finds a metric of the form $AdS_5 \times S^5$. This indicates
a UV fixed point. The coupling vanishes at the UV fixed point, and
this makes the curvature of the gravity solution infinite in the UV, but
that is precisely what is expected since the field theory is UV free.
The running of the coupling is logarithmic, though it  goes
like $ 1/(\log E)^2$. However,  the quark-antiquark potential goes as
 $1/\log E$ due to the square root in \energy .  


In the IR (small radial coordinate) the tachyon vanishes and one finds
again a solution of the form $AdS_5 \times S^5$.  In the IR the
coupling is infinite.  Therefore this solution seems to exhibit a
strong coupling IR fixed point.  However, since the dilaton is large,
classical string theory is not sufficient to study the fixed point
theory.  The gravity solution at all energy scales $u$ has not been
constructed yet.

Generically one expects
the gauge theory to have different phases parametrized by the possible  
couplings.
The IR fixed point should occur as a particular tuning of the 
couplings.
Indeed, other solutions at small radial coordinate were constructed
in \cite{Minahan:1998af} that  exhibit confinement and a mass gap.
Moreover they were argued to be more generic 
than the IR fixed point solution.

It was pointed out in \cite{Nekrasov:1999mn} that the theories on
 the D3 branes
of Type 0B string theory are particular examples of the orbifold models 
of $\cN=4$ theory that we studied in section \ref{orbifolds}.
The R-symmetry of $\cN=4$ theory is $SU(4)$, the spin cover of $SO(6)$.
It has a center $\IZ_4$ and one can orbifold with respect to it or 
its subgroups $\Gamma$.
The theory on $N$ flat electric D3 branes arises when the action
 of $\Gamma$
on the Chan-Paton (color) indices is in a trivial representation.
This orbifold is not in the class of ``regular representations'' which
we discussed in section \ref{orbifolds}; in particular, in this case the
beta function does not vanish in the planar diagram limit. If we study
instead the theory arising on $N$ self-dual D3-branes of type 0 (which may
be viewed as bound states of electric and magnetic D3-branes) we find
a theory which is in the class of 
``regular representation orbifolds'' \cite{Klebanov:1999ch},
and behaves similarly to type II D3-branes in the large $N$ limit. We will
not discuss this theory here.
    
As with the D branes in Type II string theory, we can 
construct a large number of non supersymmetric
models in Type 0 theories by placing the D branes at singularities.
One example is the theory of
D3 branes of Type 0B string theory at a conifold singularity.
As discussed in section \ref{conifolds},
when placing $N$ D3 branes of Type IIB string theory at a conifold
the resulting low-energy worldvolume theory is $\cN=1$ supersymmetric
$SU(N) \times SU(N)$ gauge theory  
with chiral superfields $A_k,k=1,2$ transforming in the $(N,\bar{N})$
representation and $B_l, l=1,2$ transforming in the $(\bar{N},N)$ 
representation, and with some superpotential.

On the worldvolume of $N$ electric D3 branes of Type 0B string theory
at a conifold there is a truncation of the fermions and one gets an
$SU(N) \times SU(N)$ gauge theory with complex scalar fields
$A_k,k=1,2$ transforming in the $(N,\bar{N})$ representation and $B_l,
l=1,2$ transforming in the $(\bar{N},N)$ representation.  This theory
(at least if we set to zero the coefficient of the scalar potential
which existed in the supersymmetric case) is asymptotically free.  The
gravity description of this model has been analyzed in
\cite{Mohsen:1999ba}.  In the UV one finds a solution of the form
$AdS_5 \times T^{1,1}$ which indicates a UV fixed point.  The
effective string coupling vanishes in accord with the UV freedom of
the gauge theory.  In the IR one finds again a solution of the form
$AdS_5 \times T^{1,1}$ with infinite coupling that points to a strong
coupling IR fixed point.  Of course, one expects the gauge theory to
have different phases parametrized by the possible couplings.  Indeed,
there are other more generic solutions that exhibit confinement and
a mass gap \cite{Mohsen:1999ba}.

Other works on dual descriptions of gauge theories via the Type 0 D branes
are \cite{Ferretti:1998xu,Zarembo:1999hn,Kogan:1999gi,%
Tseytlin:1999ii,Armoni:1999fb,%
Ferretti:1999gj,Costa:1999qx}.
\newpage

\chapter{Summary and Discussion}
\label{ChapSummary}

\label{summary}

We conclude by summarizing some of the successes and remaining 
open problems  of the AdS/CFT correspondence.

From the field theory point of view we have learned and understood
better many properties
of the large $N$ limit. Since 't Hooft's work \cite{'tHooft:1974jz}
 we knew 
that the large $N$ limit of gauge theories should be described
by strings, if the parameter $g_{YM}^2 N$ is kept fixed. 
Through the correspondence we have learned that not only does this
picture really work (beyond perturbation theory where it was first
derived), but that the Yang-Mills strings (made from gluons) are the same
as the fundamental strings. 
 Moreover, these strings move in higher
dimensions, as was argued in \cite{Polyakov:1997tj}. 
These extra dimensions arise dynamically in the gauge theory. 
For some field theories the curvatures in the higher dimensional space
could be small. The prototypical example is ${\cal N} = 4$ 
super-Yang-Mills with large $N, g_{YM}^2 N$. 
From this example we can obtain others by taking 
quotients, placing branes at various singularities, etc. (section 
\ref{other_backgrounds}). 
In all cases for which we can find a low-curvature gravity description we 
can do numerous calculations in the large $N$ limit. 
We can calculate the spectrum of operators
and states (sections \ref{tests}, \ref{isom}).
We can calculate correlation functions of operators and of 
Wilson loops (sections \ref{correlators}, \ref{wilsonloops}). 
We can calculate thermal properties, like the equation of state 
(section \ref{FiniteT}), and so on.

If the field theory is conformal the gravity solution will include
an $AdS$ factor.   
It is possible, in principle, to deform the theory by any
relevant operators.
In some cases fairly explicit solutions have been found for 
flows between different conformal field theories 
(section \ref{deformations}). A ``$c$-theorem'' for field theories in
more than two dimensions was
proven within the gravity approximation. It would be very interesting
to generalize this beyond this approximation. 
It would also be interesting to understand better 
exactly what it is the class of field
theories which have a gravity approximation. One constraint on such
four dimensional conformal
field theories, described in section \ref{anomalies},
is that they must have $a=c$.

It is possible to give a field theory 
interpretation to various branes that one 
can have in the AdS description (section \ref{baryons}).
 Some correspond to baryons in the 
field theory, others to various defects like domain walls, etc. 
In the $AdS_5$ case D-instantons in the string theory correspond 
to gauge theory instantons in the field theory. 

In general, the large $N$ limit of a gauge theory should have 
a string theory
description. Whether it also has a gravity description depends on how
large the curvatures in this string theory
are. If the curvatures are small, we can 
have an approximate classical gravity description. 
Otherwise, we should consider all string 
modes on the same footing. This involves solving the worldsheet
theories for strings in 
Ramond-Ramond backgrounds. This is a problem that only now is beginning
to be elucidated \cite{Berkovits:1999im,Metsaev:1998it,%
Pesando:1998fv,Kallosh:1998nx,Kallosh:1998ji,Polyakov:1998pm,%
Dolan:1999pi,Rajaraman:1999rc,Berenstein:1999jq}.
For non-supersymmetric QCD, or other theories which are weakly coupled
(as QCD is at high energies), we expect to 
have curvatures at least
of the order of the string scale, so that a proper understanding of
strings on highly curved spaces seems crucial. 

It is also possible to deform the $\cn=4$ field theory,
breaking supersymmetry and conformal invariance, by giving  a mass
to the fermions or by compactifying the theory on a circle with
supersymmetry breaking boundary conditions.
Then, we have a theory that should describe pure
 Yang-Mills theory at low energies (sections \ref{deformations},
\ref{FiniteT}, \ref{adsqcd}). 
In the case of field theories compactified on a circle with
supersymmetry breaking boundary conditions and  
 large $g_{YM}^2 N$ at the compactification scale,
one can show that the theory is confining, has a 
mass gap, has $\theta$-vacua with the right qualitative properties and
has a confinement-deconfinement transition at finite temperature.
However, in the regime where the analysis
can be done (small curvature) this theory includes many
additional degrees of freedom beyond those in the standard bosonic
Yang-Mills theory.
In order to do quantitative calculations in bosonic Yang-Mills
one would have to do calculations when 
the curvatures are large, which goes beyond the gravity approximation
and requires understanding the propagation of strings in Ramond-Ramond
backgrounds. Unfortunately, this is proving to be very difficult, and
so far we have not obtained new results in QCD from the correspondence.
As discussed in section \ref{adsqcd}, the gravity approximation
resembles the strong coupling lattice QCD description 
\cite{Wilson:1974co},
where the $\alpha'$ expansion of string theory corresponds to the strong
coupling expansion.
The gravity description has an advantage over the  strong  
coupling lattice QCD description
by being fully Lorentz invariant. This allows, for instance, 
the analysis of topological  
properties 
of the vacuum which is a difficult task in the lattice description.
The AdS/CFT correspondence does provide direct
evidence that QCD is describable as some sort of string theory (to the
extent that we can use the name string theory for strings propagating
on spaces whose radius of curvature is of the order of  the string scale
or smaller).

One of the surprising things we learned about field theory is that 
there are various new large $N$ limits which had not been considered
before. For instance, 
we can take $N\to \infty$ keeping $g_{YM}$ fixed, and the AdS/CFT
correspondence implies that many properties of the field theory (like
correlation functions of chiral primary operators) have a
reasonable limiting behavior in this limit, though there is no good
field theory argument for this. Similarly, we find that there exist
large $N$ limits for theories which are not gauge theories, like the 
$d=3,\cn=8$ and $d=6,\cn=(2,0)$ superconformal field theories, and for
various theories with less supersymmetry. The existence of these
limits cannot be
derived directly in field theory.

The correspondence has also been used to learn about the properties of
field theories which were previously only poorly understood. For
instance, it has been used \cite{Seiberg:1999xz} to understand
properties of two dimensional field theories with singular target
spaces, and to learn properties of ``little string theories'', like
the fact that they have a Hagedorn behavior at high energies. The
correspondence has also been used to construct many new conformal
field theories, both in the large $N$ limit and at finite $N$.

Another interesting case is topological Chern-Simons 
theory in three dimensions, which is related to a topological
string theory in six dimensions \cite{Gopakumar:1998ki}. In this
case one can solve exactly both sides of the correspondence and
see explicitly that it works.

The correspondence is also useful for studying non-conformal 
gauge theories,
as we discussed in section \ref{dpbranes}. A particularly interesting
case is the maximally supersymmetric quantum mechanical $SU(N)$ gauge
theory, which is related
to Matrix theory \cite{Banks:1997vh,Balasubramanian:1997kd,
Hyun:1998bi,Itzhaki:1998sa,McCarthy:1998uw,
deAlwis:1999ki,Silva:1998nk,Chepelev:1999pm,
Yoneya:1999zi,Townsend:1998qp,
Polchinski:1999br}.

From the quantum gravity point of view we have now an explicit
holographic description for gravity in many backgrounds involving 
an asymptotically AdS space. 
The field theory effectively sums over all geometries which are 
asymptotic to $AdS$. This defines the theory non-perturbatively. 
This also implies that gravity in these spaces is unitary, giving the
first explicit non-perturbative
construction of a unitary theory of quantum gravity,\footnote{In
the context of Matrix theory \cite{Banks:1997vh} we need to take a
large $N$ limit which is not well understood in order 
to describe a theory of gravity in a space with no
closed light-like curves.} albeit in a curved space background. 
Black holes are some  mixed states in the field theory Hilbert space.
Explicit microscopic 
calculations of black hole entropy and greybody factors
can be done in the $AdS_3$ case (chapter \ref{ChapAdS3}). 

Basic properties of quantum gravity, such as approximate causality and
locality at low energies, 
are far from clear in this description \cite{Horowitz:1998pq,Banks:1998dd,
Das:1999fx,Horowitz:1999gf,Kabat:1999yq,Bak:1999iq,Lowe:1999pk}, and it would
be interesting to understand them better. We are also still far from
having a precise mapping between general configurations in the
gravitational theory and in the field theory (see \cite{Berkooz:1998wv,
Balasubramanian:1999ri}
for some attempts to go in this direction).

In principle one can extract the physics of quantum gravity 
in flat space by taking
the large radius limit of physics in $AdS$ space. 
Since we have not discussed this yet in the review, let us expand on
this here, following
\cite{Polchinski:1999ry,Susskind:1998vk,Giddings:1999qu,Polchinski:1999yd}
(see also \cite{Balasubramanian:1999ri,Aref'eva:1999mi,Lee:1999ua,
Li:1999vi}). 
We would like to be able to describe processes in flat space which
occur, for instance, at some fixed string coupling, with the energies and
the size of the interaction region kept fixed in string (or Planck)
units. Computations on AdS space are necessarily done with some finite
radius of curvature; however, we can
view this radius of curvature as a 
regulator, and take it to infinity at the end of any calculation, 
in such a way that the local physics remains the same.
Let us discuss what this means for the $AdS_5\times S^5$ case (the
discussion is similar for other cases). We need to keep the string
coupling fixed, and take $N \to \infty$ since the radius of curvature
in Planck units is proportional to $N^{1/4}$. Note that this is
different from the 't Hooft limit, and involves taking $\lambda \to
\infty$. In order to describe a scattering process in space-time which
has finite energies in this limit, it turns out that the energies in
the field theory must scale as $N^{1/4}$ (measured in units of
the scale of the
$S^3$ which the field theory is compactified on; we need to work in
global AdS coordinates to describe flat-space scattering). In this
limit the field theory is very strongly coupled and the energies are
also very high, and there are no known ways to do any computations on
the field theory side.
It would be interesting to compute anything explicitly in this limit. 
For example, it would be interesting to compute the entropy of a small
Schwarzschild black hole, much smaller than the radius of $AdS$, to
see flat-space Hawking radiation, and so on. If we start with
$AdS_5\times S^5$ this limit gives us the physics in flat ten
dimensional space, and similarly starting with $AdS_4\times S^7$ or
$AdS_7\times S^4$ we can get the physics in flat eleven dimensional
space. It would be interesting to understand how the correspondence
can be used to learn about theories with lower dimension, where some
of the dimensions are compactified. A limit of string theory on
$AdS_3\times S^3\times M^4$ may be used to give string theory on
$\IR^{5,1}\times M^4$, but it is not clear how to get four dimensional
physics out of the correspondence. 

One could, in principle, get four dimensional flat space by
starting from   $AdS_2\times
S^2$ compactifications.  However, the correspondence in the case of
$AdS_2$ spaces is not well understood. 
$AdS_2$ spaces 
arise as the near horizon geometry of extremal
charged Reissner-Nordstrom black holes. Even though fields propagating in 
$AdS_2$ behave similarly to  the higher dimensional cases 
\cite{Strominger:1999yg},
the problem is that any finite energy excitation seems to 
destroy the $AdS_2$ boundary conditions \cite{Maldacena:1999uz}. 
This is related to the fact that black holes (as opposed to black
$p$-branes, $p>0$) have an energy gap 
(see section \ref{fivedbh}), so that in the extreme low energy limit
we seem to have no excitations. 
One possibility is that the correspondence works only for the ground
states. Even then, there are instantons that can lead to a fragmentation
of the spacetime into several pieces 
\cite{Brill:1992rw}. 
Some conformal quantum mechanics systems that are, or 
could be,  related to 
$AdS_2$ were studied in
\cite{Kumar:1999fx,Kallosh:1999mi,Gibbons:1998fa,Townsend:1998qp}. 
Aspects of Hawking radiation in $AdS_2$ were studied in 
\cite{Spradlin:1999bn}.

In all the known cases of the correspondence the gravity solution 
has a timelike boundary\footnote{This is not precisely true in the
linear dilaton backgrounds described in section \ref{ns5branes}
\cite{Aharony:1998ub}.}. It would be interesting to understand how the
correspondence works when the boundary is light-like, as in Minkowski
space. It seems that holography must work quite differently in these
cases (see \cite{Aharony:1999tt,Hashimoto:1999yc} for discussions of
some of the issues involved).
In the cases we understand, the asymptotic space close to the boundary 
has a well defined notion of time, which is the one that is associated
to the gauge theory.  It would be interesting to understand how 
holography works in other spacetimes, where we do not have this
notion of time. Interesting examples are spatially closed universes, 
expanding universes, de-Sitter spacetimes, etc. See
 \cite{Horowitz:1998xk,Hull:1998vg,Hull:1998ym,Hull:1998fh}
 for some
attempts in this direction.
The precise meaning of holography in the cosmological context is still
not clear \cite{Fischler:1998st,Bak:1998vj,%
Dawid:1998ip,Rama:1998pk,Easther:1999gk,%
Bak:1999hd,Kaloper:1999tt}.

To summarize, the past 18 months have seen much progress in our
understanding of string/M theory compactifications on AdS and related
spaces, and in our understanding of large $N$ field theories. However,
the correspondence is still far from realizing the hopes that it
initially raised, and much work still remains to be done. The
correspondence gives us implicit ways to describe QCD and related
interesting field theories in a dual ``stringy'' description, but so
far we are unable to do any explicit computations in the field
theories that we are really interested in. The main hope for progress
in this direction lies in a better understanding of string theory in
RR backgrounds. The correspondence also gives us an explicit example
of a unitary and holographic theory of quantum gravity. We hope this
example can be used to better understand quantum gravity in flat
space, where the issues of unitarity (the ``information problem'') and
holography are still quite obscure. Even better, one could hope that
the correspondence would hint at a way to formulate string/M theory
independently of the background. These questions will apparently have
to wait until the next millennium.

\section*{Acknowledgements}

We would like to thank T. Banks, M. Berkooz, A. Brandhuber,
M. Douglas, D. Freedman, 
S. Giddings, D. Gross, G. Horowitz, K. Jansen, S. Kachru,
D. Kutasov, E. Martinec, G. Moore, A. Rajaraman, N. Seiberg,
E. Silverstein, L. Susskind, and E. Witten for many useful discussions.
The research of O.A. was supported in part by DOE grant
DE-FG02-96ER40559. O.A.,
J.M. and Y.O. would like to thank the Institute for Advanced
Studies at the Hebrew University of Jerusalem for hospitality during
part of this work.
J.M. wants to thank the hospitality of the 
 Institute for Advanced Study at Princeton, where   he was a 
 Raymond and  Beverly Sackler Fellow. 
 The research of J.M. 
was supported in part by DOE grant DE-FGO2-91ER40654,
 NSF grant PHY-9513835, the Sloan Foundation and the 
 David and Lucile Packard Foundations. 
The research of S.S.G.\ was supported by the
Harvard Society of Fellows, and also in part by the NSF under grant
number PHY-98-02709, and by DOE grant DE-FGO2-91ER40654.  S.S.G.\ also
thanks the Institute for Theoretical Physics at Santa Barbara for
hospitality. The research of H.O.\ was supported in part by 
National Science Foundation under Contract PHY-95-14797 and in part by
the Director, Office of Science, Office
of High Energy and Nuclear Physics, Division of High Energy Physics, 
of the U.S. Department of Energy under Contract DE-AC03-76SF00098.

\newpage
\renewcommand{\baselinestretch}{0.87}
\footnotesize

\bibliography{review,more}
\bibliographystyle{ssg}

\end{document}